\PassOptionsToPackage{dvipsnames}{xcolor}
\documentclass[11pt, a4paper, BCOR=2cm, DIV=9, british]{scrbook}
\usepackage{scrhack}
\KOMAoptions{numbers=noenddot, cleardoublepage=plain, captions=tableheading} %
\setcapindent{0em} %
\usepackage{setspace} %
\AfterTOCHead{\singlespacing} %
\usepackage[singlespacing=true]{scrlayer-scrpage} %
\usepackage[a-3u]{pdfx}
\usepackage[T1]{fontenc}\usepackage{lmodern}
\usepackage{mathtools}
\usepackage{amssymb}
\usepackage{dsfont}
\usepackage{mathrsfs}
\usepackage{babel}
\recalctypearea %
\newlength{\CharsLX}%
\newlength{\LinesXXX}%
\areaset[current]{\CharsLX}{\LinesXXX}
\usepackage[style=numeric-comp,sorting=none,sortcites,alldates=comp,giveninits=true]{biblatex}%
\usepackage{graphicx}

\usepackage{csquotes}
\usepackage{xpatch}

\usepackage[textfont=normalsize]{subcaption} %
\usepackage{booktabs}

\usepackage[uncertainty-mode=separate, per-mode=symbol]{siunitx}
\DeclareSIUnit{\nuclearmagneton}{\mu_{\text{N}}}
\DeclareSIUnit\barn{b}

\usepackage{tensor}
\usepackage{braket}

\usepackage[nameinlink,noabbrev]{cleveref}
\DeclareRobustCommand{\abbreviatedcleverreferences}{%
		\crefname{section}{sec.}{secs.}%
		\crefname{equation}{eq.}{eqs.}%
		\crefname{figure}{fig.}{figs.}%
	}%
	\DeclareRobustCommand{\scref}[1]{{\abbreviatedcleverreferences\cref{#1}}}%

\let\Re\relax%
\DeclareMathOperator{\Re}{Re}%
\let\Im\relax%
\DeclareMathOperator{\Im}{Im}%

\newcommand{\campo}[1]{\ensuremath{\mathbb{#1}}}%
\renewcommand{\d}{\mathop{}\!\mathrm{d}}%
\renewcommand{\v}[1]{\vec{#1}}%
\newcommand{\op}{\widehat}
\newcommand{\coniugato}[1]{{#1}^*}
\newcommand{\m}[1]{\left|#1\right|}
\newcommand{\media}[1]{\operatorname{<} \! {#1} \! \operatorname{>}} %
\renewcommand{\(}{\left(}\renewcommand{\)}{\right)}
\renewcommand{\[}{\left[}\renewcommand{\]}{\right]}

\newcommand{\transfAngMom}{j} %
\newcommand{\intrSpin}{J}
\DeclareMathOperator{\Fourier}{\mathscr{F}}

\numberwithin{equation}{section}
\begin{document}%
\frontmatter%
\begin{titlepage}\textsc{\begin{center}%
\begin{minipage}{0.45\textwidth}%
	\centering%
	\includegraphics[keepaspectratio=true,height=5em]{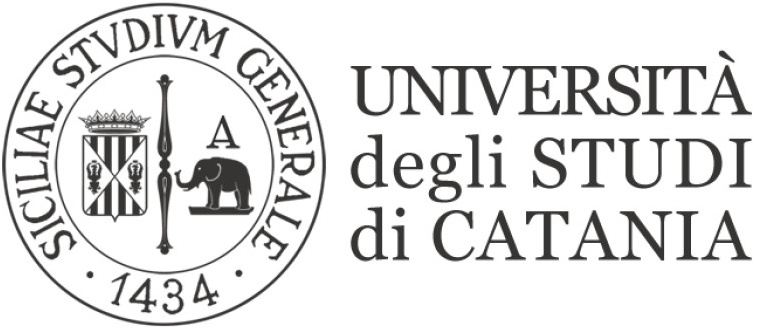}\\%
	\vspace{0.2truecm}%
	Dottorato di Ricerca in Fisica%
\end{minipage}%
\begin{minipage}{0.10\textwidth}\ \end{minipage}%
\begin{minipage}{0.45\textwidth}%
	\centering%
	\includegraphics[keepaspectratio=true,height=5em]{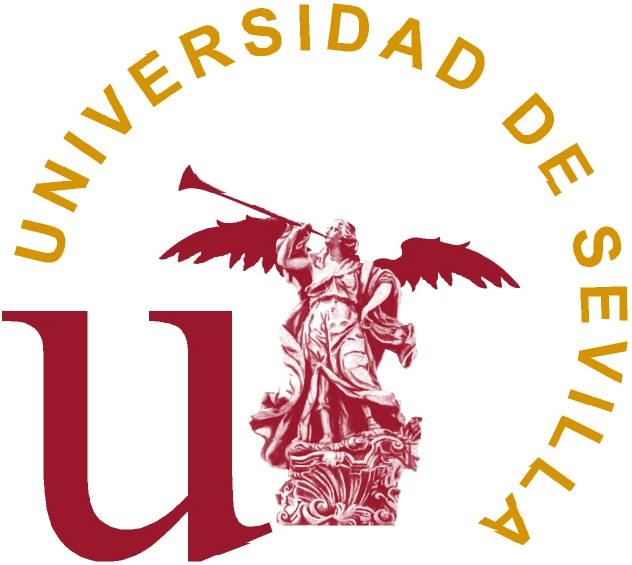}\\%
	\vspace{0.2truecm}%
	Doctorado en\\Ciencias y Tecnologías Físicas%
\end{minipage}%
\\\vspace{0.5truecm}%
\hbox to \textwidth{\hrulefill}%
\vfill%
{\large
Salvatore Simone Perrotta%
\vfill%
Reaction dynamics in clustered nuclear systems%
}%
\vfill%
\centerline{\hbox to 3.5truecm{\hrulefill}}%
\medskip%
PhD thesis\\%
\centerline{\hbox to 3.5truecm{\hrulefill}}%
\vfill%
{\large
\begin{minipage}{0.65\textwidth}\ \end{minipage}%
\begin{minipage}{0.35\textwidth}\begin{tabbing}%
		S\=upervisors:\\%
		\>Maria Colonna\\%
		\>José Antonio Lay\\%
		\>Vincenzo Greco
\end{tabbing}\end{minipage}%
}
\vfill%
\hbox to \textwidth{\hrulefill}%
XXXIV ciclo -- 2018 - 2022%
\end{center}}\end{titlepage}
\clearpage{}%
\vspace*{\fill}
© 2022. Licensed under a \href{http://creativecommons.org/licenses/by-nc-nd/4.0}{Creative Commons Attribution-NonCommercial-NoDerivatives 4.0 International License}.
\clearpage{}%
	\tableofcontents%
	\listoffigures%
	\begingroup\let\cleardoublepage\relax\let\clearpage\relax%
	\listoftables%
	\endgroup%
\clearpage{}%
\mainmatter\pagenumbering{arabic}%
\chapter*{Introduction}\addcontentsline{toc}{chapter}{Introduction}\markboth{Introduction}{}%
	
	Since the dawn of the field, %
	with the first experiments %
	related to \nuclide{\alpha}-particle decay, and up until the most recent advances in e.g.~ab initio methods,
	a wide fraction of nuclear physics studies
	revolve around
	the interplay between the structure adopted to depict nuclei and/or nucleons and the models describing their interactions, both bound to the role of approximated representations due to the complex and non-perturbative nature of the underlying fundamental interactions. %
	This also caused %
	the developments in the two directions to be %
	correlated to a certain extent. %
	Very heavy %
	nuclear systems %
	more easily lend themselves to a description in terms of spatial density and other properties that may be commonly encountered in macroscopic physics (or at least their quantum counterpart), as they show a behaviour comparable to that of a fluid, including the possibility of undergoing phase transitions, and often %
	respond well to semi-classical treatments.
	In the ``intermediate'' mass regime, shell-model descriptions of the nuclear structure proved to be a great success, as the mutual interaction between a %
	set of nucleons, thanks to the effects of anti-symmetrisation, can in fact be reasonably accounted for as the result of a self-generated mean-field, which provides an accurate description of the system when appropriately corrected by the residual correlations. %
	Nuclear reactions are then routinely employed to deduce relevant properties regarding such structure, and vice-versa. %

	In very light nuclei, which are the subject of study of the present work, %
	the available degrees of freedom are sufficiently small in number that more microscopic techniques can be attempted, %
	both for the structure of isolated systems (as the No-Core Shell Model \cite{Navratil2000} or Quantum Monte Carlo \cite{Forest1996} methods) and even for nuclear reactions (as the Resonating Group Method \cite{Arai2002}). These approaches bear considerable predictive power, being based on the elementary interaction at the nucleon %
	level with little reliance on phenomenological constraints.
	At the same time, %
	such lack of flexibility complicates the physical interpretation of %
	disagreements with expected or measured properties, and makes it harder to identify which ingredients or features of the model are responsible of a given result. %
	More macroscopic %
	models, in contrast,
	find their roots in general %
	features observed in the regime of interest, %
	and need to be supplied with several phenomenological ingredients, mainly in the form of %
	model Hamiltonians %
	and general structure properties, %
	adapted to the specific situation at hand.
	This also %
	implies that the impact of each ingredient on the model predictions can be explored with relative ease, %
	opening for the possibility of a qualitative (and not merely numerical) understanding of the observations. %
	In this sense, macroscopic approaches can complement and complete ab-initio studies, %
	and their application thus %
	maintains significant scientific interest.
	
	Specifically, in regard to %
	the description of scattering and reaction phenomena between light particles, the same sort of techniques often adopted for somewhat heavier systems can be applied with profit: %
	optical models, distorted-wave Born approximation calculations and coupled-channels schemes (an in-depth treatment of such concepts may be found for instance in \cite{Satchler1983Direct}) are routinely employed in this regime, %
	for instance in the description of reactions involving %
	exotic nuclei (see for instance \cite{Casal2017}).
	As for the representation of the structure of %
	light nuclear systems, an excellent framework is provided by cluster models. In view of the general expectation that the characteristics of the nuclear interaction and anti-symmetrisation can favour the formation of correlated clusters of particles within a nucleus (see for instance \cite[sec.~3.2]{Wildermuth1977}), a system might be described as a bound state of a very small number of sub-systems (rarely more than three, in practice), %
	each with properties resembling those of the corresponding nuclide in isolation.
	Most or all the %
	non-essential degrees of freedom regarding the relative motion of particles forming each cluster %
	are then neglected, simplifying the problem to a tractable level.
	Such description is especially fruitful at relatively small excitation energies, where resonances are sparse and there is often sufficient experimental information to pinpoint the main characteristics of each excited state and the clustered structure %
	which best describes it.
	
	This work is in particular concerned with the study of nuclear reactions between light charged ions at incident energies around and below the reactants Coulomb barrier, %
	with a %
	focus on the energy range of astrophysical interest for Big Bang and quiescent stellar processes. %
	Cross-sections of reactions pertaining to this regime are %
	of great interest in the modelling of associated astrophysical scenarios, %
	in particular for nucleosynthesis and stellar evolution, where they appear as key physical ingredients. For instance, reactions destroying \nuclide[6]{Li} %
	are interesting with regard to the lithium-depletion problem, and in the study of pre-main-sequence stars \cite{Lamia2013}.
	Other examples include several reactions involving lithium, beryllium and boron isotopes, which are all nuclei known to exhibit a pronounced cluster structure.
	Some of these processes can also be important %
	in the field of controlled nuclear energy production. %
	The investigation of this class of %
	collision processes is a challenging pursuit from both the theoretical and experimental point of view. %
	Regarding the former, despite ongoing %
	efforts current microscopic calculations are often not able to provide a fully %
	satisfactory account of available experimental data %
	regarding cross-sections (see e.g.~\cite{Solovyev2018} and references therein for the case of the $\nuclide[6]{Li} + \nuclide{p} \to \nuclide[3]{He} + \nuclide{\alpha}$ reaction, and \cite{Gnech2019} and references therein for the $\nuclide[6]{Li} + \nuclide{p} \to \nuclide[7]{Be} + \nuclide{\gamma}$), and in some cases even %
	ground-state structure properties (see \cite{Tilley2002} for a review, although not so recent,
	in particular regarding \nuclide[6]{Li}). %
	In this respect, %
	the possible influence of clustering effects %
	on the low-energy reaction dynamics has received some attention (see e.g.~\cite{Vasilevsky2009} for a Faddeev %
	$\nuclide[6]{Li} + \nuclide{p} \to \nuclide[3]{He} + \nuclide{\alpha}$ calculation including %
	dynamic excitations in the inter-cluster motion) %
	and still represents an interesting field of study.  %
	From the experimental point of view, the reactants electrostatic repulsion causes the cross-section to depend exponentially on energy, being reduced to %
	values of the order of the nanobarn or less within the energy range of interest. %
	Direct measurements of these reactions are consequently very difficult and time-consuming, and normally %
	require heavy shielding from all background sources (see e.g.~\cite{Broggini2010} for a review). A valuable contribution in the field is given by indirect measurement methods, in which the theoretical understanding on nuclear reaction processes is employed to devise techniques yielding %
	the desired information through experiments which are easier to perform than the corresponding direct measurement (see e.g.~\cite{Tribble14} for a review).
	
	Another complication affecting the processes of interest in this work %
	is that, when approaching the stellar %
	energy range, %
	the structure of matter above the nuclear level (atoms, plasmas, solids,\,\dots) %
	starts playing a relevant role. For instance, it is well known %
	that the cross-section of a reaction between charged particles directly measured in a fixed target experiment at low energies shows an enhancement due to %
	the screening of the nuclear charge operated by atomic electrons. In order to correctly %
	describe %
	the process of astrophysical interest, %
	a sufficiently accurate prediction %
	of the screening effects acting %
	both in laboratory and in the relevant astrophysical site %
	is thus required. %
	Extensive theoretical work exist in this direction (see \cite{Salpeter1954,Bracci1990}, just to mention two relevant examples).
	From the experimental perspective, a direct measurement of screening effects in plasma environments with conditions comparable with the astrophysical %
	ones is challenging.
	Concerning %
	standard laboratory measurements employing atomic targets, %
	the situation is rather %
	involved, as
	the cross-section enhancement determined experimentally for several reactions is larger than the upper limit predicted by atomic theory (see \cref{tabPotScreeningAltreReazioni} or \cite{Spitaleri2016} for some examples). This anomalous observation is known as the %
	atomic electron screening problem.
	It was proposed in \cite{Spitaleri2016} that these deviations %
	might be connected to clustering phenomena. %
	The main goal of this work %
	is to investigate %
	the %
	sensitivity of nuclear reaction dynamics, and more specifically
	the cross-section predictions, %
	to the description of
	the structure of each reactant, and in particular to clustering phenomena.
	Emphasis is given to results concerning the astrophysical %
	energy range, %
	given the interest of its applications, %
	and the reactions %
	affected by the electron screening problem, with the aim of advancing its understanding. %
	The practical applications are focused on the \nuclide[6]{Li} nucleus, and precisely the $\nuclide[6]{Li} + \nuclide{p} \to \nuclide[3]{He} + \nuclide{\alpha}$ transfer reaction and the barrier penetrability of the $\nuclide[6]{Li} + \nuclide{p}$ system. %

	This thesis is structured as follows.
	After this introduction, %
	\cref{chaPhenomenology} treats some phenomenological aspects of the class of nuclear reactions of interest in this work, both taking place in vacuum (``bare-nuclei'') or immersed in an external medium. %
	The tools developed in this \namecref{chaPhenomenology}, in spite of (and thanks to) %
	their simplicity, allow to
	understand the main features of the non-resonant sub-Coulomb reaction cross-section, %
	providing analytical expressions %
	that can be compared to the results of more advanced computations to discuss their qualitative meaning and implications. %
	The formalism presented in \cref{secPhenomenologyScreeningEffects} can be employed to %
	factor out the expected screening effects from experimental low-energy data, %
	so that the latter %
	can be sensibly compared with bare-nucleus cross-sections, as those computed in \cref{secCalcoliDiTransfer}. %
	The same \namecref{secPhenomenologyScreeningEffects} also includes a quantitative analysis on %
	the electron screening anomalies %
	observed in the $\nuclide[6]{Li} + \nuclide{p} \to \nuclide[3]{He} + \nuclide{\alpha}$ reaction.
	Furthermore, some of the concepts reviewed in this \namecref{chaPhenomenology} will find direct application in the study presented in \cref{secLiDeformation}. %
	
	\Cref{chaLinkingCLusterModelToObservables} is a dissertation on some characteristics %
	of the static structure of an isolated nucleus, with a focus on clustered systems.
	In particular, \cref{secOverlapFunctionSpectroscopicFactors} presents the formalism of overlap functions, which are important for treating direct transfer reactions and encode the required information to represent %
	reactants %
	within a cluster model.
	\Cref{secLegameProprietaCompositoEFunzioneDonda} instead is dedicated to the explicit computation of %
	the root-mean-square radius and the electric quadrupole moment of a nucleus described by a cluster-model wave-function. To the author's knowledge, the expressions regarding systems of three or more clusters are not available in literature. 

	\Cref{secReactionTheory} reviews the theory of direct transfer reactions that is later employed in the practical calculations shown in \cref{secCalcoliDiTransfer}. %
	In particular, %
	the discussion is focused on the formalism of first- and second-order distorted-wave Born approximation (DWBA) and of coupled-channels %
	approaches.

	\Cref{secCalcoliDiTransfer} is concerned with %
	the study of the $\nuclide[6]{Li} + \nuclide{p} \to \nuclide[3]{He} + \nuclide{\alpha}$ reaction, treated as the direct transfer of either a structureless deuteron or a generic $\nuclide{p}+\nuclide{n}$ system, through %
	the whole sub-Coulomb energy range and up to the resonance corresponding to the second $\frac{5}{2}^-$ state of \nuclide[7]{Be} (namely, at centre-of-mass incident energies between few \si{\keV} to about \SI{1.5}{\MeV}). The excitation function of the process %
	is explicitly evaluated within a fully quantum framework and without adjusting the calculation parameters on transfer experimental data. %
	Different descriptions of the reactants structure are considered in the evaluation of the cross-section. First- and second-order DWBA %
	calculations are employed to study the role %
	of the system ground state in the initial and final partition, %
	and in particular the model adopted for the transferred particle internal structure, %
	and the corresponding strength of clustered configurations. %
	Virtual excitations %
	in the relative motion between core and transferred systems, which can account for a dynamical deformation of the reactants during the reaction, are investigated in a preliminary calculation using a coupled-channels %
	scheme.
	The physical ingredients required to perform the calculations are also discussed here. %
	
	\Cref{secLiDeformation} presents an investigation on the %
	quadrupole deformation of the ground state of \nuclide[6]{Li}, %
	described within a di-cluster model. Such deformation is required to explain the measured electric quadrupole moment of \nuclide[6]{Li}, %
	and %
	induces tensor components in the interaction of \nuclide[6]{Li} with other particles, breaking the angular symmetry of the Coulomb barrier. The semi-classical approach presented in \cite{Spitaleri2016} is here improved and expanded, and employed to evaluate and compare the impact %
	of %
	the classical and quantum-mechanical cluster model %
	on the Coulomb-barrier penetrability within a scattering process. %
	The deformed state of \nuclide[6]{Li}
	constructed here is also used as input in the deuteron-transfer calculations discussed in \cref{secCalcoliDiTransfer}.
	Finally, the text closes with some concluding remarks and perspectives. %

\chapter{Phenomenology of nuclear reactions induced by light charged particles below the Coulomb barrier}\markboth{\Cref{chaPhenomenology}. Phenomenology of sub-Coulomb nuclear reactions}{\Cref{chaPhenomenology}. Phenomenology of sub-Coulomb nuclear reactions}\label{chaPhenomenology}%

	In the scattering process of electrically charged nuclei at energies well below their Coulomb barrier, the energy trend of the cross-section is dominated by the initial stage of quantum tunnelling through the repulsive electrostatic potential, necessary for the reactants to feel the short-ranged strong interaction. %
	An appropriate description %
	of such stage is thus important both to acquire %
	a qualitative understanding %
	of nuclear reactions in this regime, and to allow a sensible discussion of finer effects connected to the other properties of the system under study.

	Since the incident energy is low, open nuclear reaction channels almost exclusively involve exothermic processes, thus the role of the Coulomb repulsion in the final state tends to be less critical.
	With the relevant exception of resonant processes, %
	it is consequently possible to provide a phenomenological description %
	of a given nuclear reaction focusing just on the system initial state, and disregarding both the reactants internal structure and the details of the nuclear interaction. This is done in \cref{secPhenomenologyBarenucleusnonresonantcrosssection}.
	
	Moreover, due to
	the prominent position held %
	by the Coulomb barrier at low incident energies, %
	and the exponential energy trend it generates on the cross-sections,
	any alteration of the electrostatic interaction, even relatively small, may induce sizeable effects on the reaction dynamics. Similar outcomes could be caused also by %
	long-range mechanisms of different nature altering %
	the reactants effective collision energy.
	This sort of alterations can be sometimes observed when reactants %
	collide while immersed in an active medium, %
	which perturbs the scattering process.
	The issue is discussed in \cref{secPhenomenologyScreeningEffects}, focusing in particular on the phenomenon of charge screening operated by atomic electrons, which enhances nuclear cross-sections measured in laboratory at astrophysical energies.

\section{Bare-nucleus non-resonant cross-section}\label{secPhenomenologyBarenucleusnonresonantcrosssection} %

	This section discusses the phenomenological modelling of the non-resonant component of the cross-section for a nuclear reaction between two isolated charged particles (hence the name ``bare nucleus''). %
	The process is described in terms of the probability (transmission coefficient) for quantum barrier penetration, which is defined %
	in \cref{secCoulombbarrierpenetrabilityastrophysicalFactor}
	starting from general principles.
	The case of a barrier generated by the combination of Coulomb repulsion and a sharp-edge nuclear potential is discussed in some detail, and is employed as starting point to derive 
	some formulas often found in literature, %
	also discussing the approximations involved in their use.

	When considering more general shapes for the nuclear potential, the barrier penetration problem is often treated in Wentzel-Kramers-Brillouin-Jeffreys approximation, as it yields a solution which is well suited for numerical computations, and will be employed in \cref{secLiDeformation}.
	The approximation %
	is discussed in \cref{secPenetrabilityWKBApproximation}, where it is also employed to deduce some qualitative features of the reaction cross-section.
	The developed formalism is in conclusion applied in \cref{secPenetrabilityFittingExperimentalData} to the phenomenological description of the $\nuclide[6]{Li}+\nuclide{p}\to\nuclide[3]{He}+\nuclide{\alpha}$ bare-nucleus cross section. Its application will also be very important in \cref{secScreeningExperimentalData}, where it will be combined to the model adopted for %
	screening effects to attempt a complete description of experimental data.

\subsection{Transmission through a potential barrier} %
\label{secCoulombbarrierpenetrabilityastrophysicalFactor}

	Consider a system of two nuclei with charge numbers $Z_1$ and $Z_2$, spin%
	\footnote{This is the total spin of the given state of the nucleus, including any internal orbital motion, and not just the coupling of all nucleons intrinsic spins.}
	modulus quantum number $s_1$ and $s_2$, and reduced mass%
	\footnote{If $m_1$ and $m_2$ are the reactants proper masses, their reduced mass is $m_1 m_2 /(m_1+m_2)$.}
	$m$,
	colliding in vacuum at a non-relativistic centre-of-mass collision energy $E$, which is the kinetic energy at infinite distance in vacuum. %
	$E$ is assumed to be significantly smaller than the maximum of the Coulomb barrier between the pair of reactants. %
	It is of interest to evaluate the non-polarised angle-integrated cross-section for barrier penetration regardless of the particular exit channel for the reaction, here labelled $\sigma(E)$. %
	This is connected to the probability for the reactants to overcome their electrostatic repulsion and come into ``close contact'' to interact strongly, at some distance $R_n$, referred to as the ``effective nuclear radius'', whose definition is discussed later in this \namecref{secCoulombbarrierpenetrabilityastrophysicalFactor}. %
	The details of the interaction at distances smaller than $R_n$ are normally neglected. %
	The process is %
	described considering only the initial partition and state of the reactants, %
	assigning them %
	a real phenomenological projectile-target %
	potential, $V(\v r)$ (thus disregarding the reactants internal structure and degrees of freedom), evaluating the associated transmission coefficient, $T$, %
	between $R_n$ and infinite distances,
	and writing the barrier-penetration cross-section as in %
	e.g.~\mbox{\cite[eq.~(4.107)]{Satchler1990Introduction}}~%
		\footnote{The factor $g_\alpha$ in~\cite[eq.~(4.107)]{Satchler1990Introduction} is here 1 because only the penetrability in the entrance channel (regardless of what state is reached afterwards) %
			is of interest. %
		See also the text commenting~\cite[eq.~(3.84)]{Satchler1990Introduction}.}, %
	\begin{equation}
		\sigma(E)
		= \frac{\pi}{k^2} \, T(E)
	\end{equation}
	where $k$ is the entrance channel wave-number, such that $E = \hbar^2 k^2 / 2m$.

	The problem can be approached imposing boundary conditions in two ways (see also \cite{Hagino2012} for some discussion). ``Scattering'' boundary conditions, employed for instance in~\cite[sec.~3.6.1]{Satchler1990Introduction}, %
	involve a fixed incoming wave plus scattered outgoing waves in the exterior region (as in standard scattering problems), and only ingoing waves for $r < R_n$, which is treated as a strongly absorptive region. The sought $T(E)$ is then the transmission coefficient of the incoming beam into the absorptive region: namely, the ratio of the ingoing %
	probability current density %
	flowing through %
	the surface of the sphere with radius $R_n$, %
	to the incoming flux from infinite distances. %
	``Decay'' boundary conditions, on the contrary, prescribe that the $r < R_n$ region generates outgoing flux (and will also host some scattered ingoing flux), %
	while at sufficiently high distances, where the nuclear potential is negligible, there are only outgoing waves (and no incoming beam). The transmission coefficient of interest in this case is the ratio between the outgoing flux reaching infinity and the outgoing flux generated from $r = R_n$.
	In both cases, for $r \geq R_n$ the system is described as a stationary state of the chosen Hamiltonian.
	Let $\psi(\v r)$ be the system wave-function in position representation%
	\footnote{The system %
		may include additional degrees of freedom (as the reactant spin projections), %
		here omitted for brevity. To discuss spatial probability, expressions such as $\m{\psi(\v r)}^2$ are interpreted as including a scalar product over these additional degrees of freedom.}.
	In some cases it will be relevant to consider only the ingoing or outgoing components of $\psi(\v r)$. %
	The probability current density $\v{\mathcal J}(\v r)$ associated to $\psi(\v r)$ %
	is defined as in~\cite[eq.~(2.4.16)]{SakuraiModern1994},
	to satisfy the continuity equation, which, if the Hamiltonian is hermitian %
	(in particular includes %
	only real potentials) is given by~\cite[eq.~(2.4.15)]{SakuraiModern1994}:
	\begin{equation}\label{eqEquazioneDiContinuitaPotenzialeReale}
		\partial_t \m{\psi(\v r)}^2 = - \v\nabla \cdot \v{\mathcal J}(\v r)
	\end{equation}
	where the time derivative of the probability density, $\partial_t \m{\psi(\v r)}^2$, is computed from the Schrödinger equation.
	Remind that in the present case, for $r < R_n$ the system additionally generates and/or absorbs probability; this may be implemented formally allowing the system to access negative radii, or adding an imaginary component $W(\v r)$ to the potential in the interior region, which would generate a term $+ \frac{2}{\hbar} \m{\psi(\v r)}^2 W(\v r)$ in the right-hand-side of \cref{eqEquazioneDiContinuitaPotenzialeReale}. %

	As mentioned above, the quantity of interest here (see e.g.~\cite[sec.~$\text{B}_\text{VIII}$]{Cohen1977}) is the outgoing or ingoing flux going across a sphere of some radius $r$: this coincides with the ``amount of probability'' transmitted from or into the sphere volume.
	From \cref{eqEquazioneDiContinuitaPotenzialeReale},
	\begin{equation}\label{eqEquazioneDiContinuitaPotenzialeRealeIntegrataSuSfera}
		\partial_t \int_{\frac{4}{3} \pi r^3} \m{\psi(\v r)}^2 \d^3 r = - r^2 \int_{4 \pi} \mathcal J_r(\v r) \d\Omega
	\end{equation}
	where $r^2 \d\Omega$ is the surface element ($\Omega$ can be practically expressed, for instance, as a pair of angles in spherical coordinates), and $\mathcal J_r$ is the component of $\v{\mathcal J}$ along the radial \emph{outgoing} direction,
	which (expressing the gradient in spherical coordinates, see~\cite[sec.~VIII, eq.~(B-20)]{Cohen1977}) is%
	\footnote{$\Im z$ denotes the imaginary part of $z$.}
	\begin{equation}
		\mathcal J_r(\v r) = \frac{\hbar}{m} \Im\(\coniugato{\psi}(\v r) \, \partial_r \psi(\v r) \)
	\end{equation}

	For the systems of interest, $V(\v r)$ is always the sum of the nuclei electrostatic repulsion %
	and of a %
	``nuclear potential'' which is short-ranged. Hence, in the limit of high distances ($r \to +\infty$), the potential always goes asymptotically as $1/r$. %
	The %
	potential %
	is expected to not diverge negatively anywhere, i.e.\ $\inf_{\v r} V(\v r) > -\infty$. %
	A positive divergence can be accepted only toward $r \to 0$: this is found in some descriptions of the Coulomb or nuclear core repulsion%
	\footnote{When decomposing in partial waves (see below), such a divergence always appears in the effective radial potential for any orbital angular momentum greater than 0, %
		thus a diverging $V(\v r)$ does not complicate the model significantly.}.
	Regarding the definition of $R_n$, the nuclear radius %
	should arguably depend only on $V(\v r)$ itself (and not, for instance, on the collision energy). If $V$ is not isotropic, an angle-dependent (or angular-momentum dependent) radius may in principle be considered.
	For simplicity, in the following qualitative discussion let $V_0(r)$ be the central part of $V(\v r)$, and restrict to only $V_0$ to define $R_n$.
	If %
	$V_0$ has a very sharp %
	profile, $R_n$ may be defined as the radius where the nuclear part of the potential becomes appreciably different from zero. %
	In general, the effective nuclear radius could be simply set independently of the specific form of $V$ through some convention.
	A common choice in this direction (see e.g.~\cite[eq.~(4-105)]{Clayton1983}), in view of the general systematics regarding nuclear sizes, is
	\begin{equation}\label{eqScalingRaggioNucleareLiquidDrop}
	R_n = r_0 \cdot (A_1^{1/3} + A_2^{1/3})
	\end{equation}
	where $A_i$ is reactant $i$ mass number, and
	the reduced radius $r_0$ is taken as a phenomenological constant, %
	usually of the order of \SI{1}{\femto\metre}.
	The approach favoured in this work %
	is instead to set $R_n$ to %
	the outermost local maximum of %
	$V(r)$: %
	at radii smaller than this threshold, the nuclear potential must be attractive, and growing fast enough to counter the trend given by the Coulomb repulsion. %
	The choice is also advantageous in case a WKB solution is sought (see \cref{secPenetrabilityWKBApproximation}), because it helps meeting the validity conditions of the approximation. %

\subsubsection{Partial-wave expansion} %
	
	If $V(\v r)$ is central, %
	or more generally includes only terms which conserve both the projectile-target orbital angular momentum modulus and the total angular momentum (as spin-orbit and spin-spin couplings),
	$\sigma(E)$ may be decomposed in partial waves
	(see for instance~%
	\cite[sec.~4.3]{Satchler1983Direct}, %
	\cite[eq.~(3.2.29)]{ThompsonNunes2009}) 
	with definite projectile-target relative orbital angular momentum modulus quantum number $l$,
	total reactants spin modulus quantum number $S$, total angular momentum modulus $J$ and total projection $M$ (a different coupling order for the spins may be more appropriate, depending on the potential), yielding
	\begin{equation}\label{eqDecomposizioneSezioneDurtoPenetrabilitaPotenzialeCentrale}
	\sigma(E) %
	= \frac{\pi}{k^2} \sum_{l,S,J} \frac{2 J + 1}{(2 s_1 + 1)(s_2 + 1)} T_{l,S,J}(E)
	\end{equation}
	where $T_{l,S,J}(E)$ is the radial transmission coefficient for the given partial wave, independent of $M$, %
	which can be found in terms of the reduced radial wave-function, $u_{l,S,J}(r)$, for the partial wave of interest.
	The general complete solution of the system, $\Ket{\Psi}$, can be expanded in %
	partial waves (see \cref{secAppendiceSphericalHarmonics}),
	\begin{equation}\label{eqScritturaSoluzioneCompletaEspansaInOndeParziali}
	\Ket{\Psi}
	= \sum_{l,S,J,M} c_{l,S,J,M} \frac{u_{l,S,J}(r)}{r} \Ket{l,S,(J,M)}
	\end{equation}
	where %
	the coefficients $c$ are determined by %
	the boundary conditions,
	and the angular-momentum state $\Ket{l,S,(J,M)}$ can be expanded using \cref{eqDecomposizioneClebschGordan}, if useful.
	Each reduced radial wave-function $u(r)$ is %
	an eigenstate
	of the effective radial projectile-target Hamiltonian of interest, $\mathcal H_{l,S,J}$, with the desired boundary conditions and eigenvalue $E$.
	Taking into account \cref{eqApplicazioneLaplacianoScritturaRadialeRidotta}, it is %
	\begin{equation}\label{eqEffectiveRadialHamiltonianDefinition}
	\mathcal H_{l,S,J} = - \frac{\d^2}{\d r^2} + \mathcal V_{l,S,J}(r)
	\quad , \quad
	\mathcal V_{l,S,J}(r) = V_{l,S,J}(r) + \frac{\hbar^2}{2m} \frac{l (l+1)}{r^2}
	\end{equation}
	where $V_{l,S,J}(r)$ is the relevant component of the ``actual'' potential of the system, and $\mathcal V$ the corresponding effective radial potential. %
	Note that
	\begin{equation}
	\sum_{S = \m{s_1-s_2}}^{s_1+s_2} \sum_{J=\m{l-S}}^{l+S} (2 J + 1) = (2 s_1 + 1)(s_2 + 1) (2l+1)
	\end{equation}
	which confirms that, if $V_{l,S,J}(r)$ actually does not depend on $S$ and $J$, the spins can be ignored in \cref{eqDecomposizioneSezioneDurtoPenetrabilitaPotenzialeCentrale}.
	Finally, perform the integral on the right-hand-side of \cref{eqEquazioneDiContinuitaPotenzialeRealeIntegrataSuSfera} using the complete solution $\Psi$ written in \cref{eqScritturaSoluzioneCompletaEspansaInOndeParziali} as wave-function: employing \cref{eqOrtonormalitaArmonicheSferiche,eqRelazioniOrtonormalitaCoefficientiClebschGordan}, and noting that some terms cancel out as they are purely real, it is %
	\begin{equation}
	- r^2 \int_{4 \pi} \mathcal J_r(\v r) \d\Omega
	= - \frac{\hbar}{m} \sum_{l,S,J,M} \m{c_{l,S,J,M}}^2 \Im\(\coniugato{u}_{l,S,J}(r) \, \partial_r u_{l,S,J}(r) \)
	\end{equation}
	As a result, it can %
	be convenient to define a ``reduced'' radial current density for each partial wave as
	\begin{equation}\label{eqDefinizioneReducedRadialCurrent}
	\mathcal J[u_{l,S,J}](r)
	= \frac{\hbar}{m} \Im\(\coniugato{u}_{l,S,J}(r) \, \partial_r u_{l,S,J}(r) \)
	\end{equation}
	absorbing the $r^2$ factor in \cref{eqEquazioneDiContinuitaPotenzialeRealeIntegrataSuSfera} through the use of reduced functions, so that the contribution from each partial wave appears as a purely one-dimensional problem. More explicitly, \cref{eqEquazioneDiContinuitaPotenzialeRealeIntegrataSuSfera} is rewritten as
	\begin{equation}\label{eqEquazioneDiContinuitaPotenzialeRealeIntegrataSuSferaRadialeRidotta}
		\partial_t \int_{\frac{4}{3} \pi r^3} \m{\psi(\v r)}^2 \d^3 r = - \sum_{l,S,J} \m{c_{l,S,J}}^2 \mathcal J[u_{l,S,J}](r)
	\end{equation}
	where $\m{c_{l,S,J}}^2 = \sum_M \m{c_{l,S,J,M}}^2$.

	\paragraph{}
	Since the nuclear potential is short-ranged, at high distances the effective radial potential %
	is central (does not depend on $S$ and $J$) and includes only the Coulomb and centrifugal repulsion:
	\begin{equation}\label{eqDefinizionePotenzialeEfficaceARaggiAlti}
		\mathcal V_{l}(r \gg R_n) = \hbar c \alpha_e \frac{Z_1 Z_2}{r} + \frac{\hbar^2}{2m} \frac{l (l+1)}{r^2}
	\end{equation}
	where $\alpha_e$ is the fine-structure constant and $c$ the speed of light.
	In the same region, the reduced radial solution %
	is a combination of the spherical Coulomb wave-functions,
	$H^+_l(k r)$ and $H^-_l(k r)$, %
	defined in~\cite[sec.~33]{NISTdlmf}%
	\footnote{These functions depend also on the Sommerfeld parameter $\eta$, defined in \cref{eqDefinizioneParametroSommerfeld}, but the associated index is omitted for brevity.}. %
	Note that these functions %
	are %
	dimensionless, while the square modulus of a physical reduced radial wave-function bears the dimensions of a one-dimensional spatial density: such dimension is %
	included in the normalisation factor. %
	In practice, results are independent of the chosen normalisation. %
	The reduced radial probability current density, defined in \cref{eqDefinizioneReducedRadialCurrent},
	associated to the wave-function $A H_l^-(kr) + B H_l^+(kr)$, with $A$ and $B$ arbitrary constants, is found to be, using~\cite[eq.~33.2.13, 33.4.4]{NISTdlmf},
	\begin{equation}\label{eqDensitaCorrenteProbabilitaFunzioneCoulomb}
		\mathcal J[A H_l^- + B H_l^+](r) = \( \m{B^2} - \m{A^2} \) \frac{\hbar k}{m}
		\end{equation}

	Impose %
	decay boundary conditions (as defined in \cref{secCoulombbarrierpenetrabilityastrophysicalFactor}). %
	In the region where \cref{eqDefinizionePotenzialeEfficaceARaggiAlti} holds,
	the solution %
	is proportional just to the spherical outgoing Coulomb wave-function:
	\begin{equation}\label{eqSoluzioneAsintoticaPotenzialeCoulombianoBoundaryConditionsDecadimento}
		u_{l,S,J}(r \gg 1) \to B \, H^+_l(k r) %
	\end{equation}
	with arbitrary normalisation $B$. %
	Asymptotically, in the limit of $r \to +\infty$ (and fixed energy), $H_l^+$ becomes %
	an outgoing plane-wave (as the potential vanishes). %

\subsubsection{Transmission for sharp-edge nuclear potentials}\label{eqPenetrabilitaPotenzialeSharpEdgeEsatta} %
	
	The outgoing radial probability current at $r = R_n$ depends on the details of the nuclear potential $V_{lSJ}(r)$.
	Consider a ``sharp-edge'' nuclear potential, namely the limit where $V_{lSJ}(r)$ is exactly the point-like Coulomb potential for all $r > R_n$. %
	This implies that, in the exterior region, the electrostatic repulsion %
	can be treated as the Coulomb potential between two point sources, and the nuclear interaction is zero.
	Then, the problem is reduced to properly matching the exterior solution %
	with the interior one.
	For the choice of $R_n$ to be sensible, it is expected that %
	at the same time the nuclear potential is significantly different from zero for $r < R_n$.
	In the spirit of the picture where the system source is formally an outgoing wave coming from negative radii (see text commenting \cref{eqEquazioneDiContinuitaPotenzialeReale}),
	a simple possibility to model the internal region, at least near $r = R_n$, is to consider a combination of an outgoing plane wave with wave-number $\tilde k$, %
	generated by the source, plus an ingoing plane wave produced from the scattering at the interface in $r = R_n$: %
	\begin{equation}\label{eqEspressioneFunzioneDonaZonaInternaModellinoPenetrabilitaCondDecay}
	u(r) = D e^{i \tilde k r} + A e^{- i \tilde k r} %
	\end{equation}
	where $D$ and $A$ are constants to be determined by matching with the external wave-function.
	$\tilde k$ in general depends on $k$ and on the features of the potential (including the quantum numbers $l,S,J$). For brevity, $\tilde k$ and $u$ are written with no index.
	The reduced radial current at $R_n$ connected to the outgoing wave, $\mathcal J[D e^{i \tilde k r}]$, %
	is $\m{D}^2 \hbar \tilde k / m$.
	It is stressed that scattered waves (either in the internal or external solution, depending on the boundary conditions) appear whenever the transmission is not complete ($T_{lSJ}<1$ if and only if $A \neq 0$).

	The wave-function in \cref{eqEspressioneFunzioneDonaZonaInternaModellinoPenetrabilitaCondDecay} %
	can be seen as the solution for a real and negative constant total potential, $\mathcal V_{lSJ}(r<R_n) = \mathcal V_<$, %
	with
	\begin{equation}\label{eqDefinizioneImpulsoBucaInterna}
	\tilde k = \frac{1}{\hbar} \sqrt{2 m (E - \mathcal V_<)}
	\end{equation}
	If instead a real constant $V_n$ is assumed just for the nuclear component of the potential %
	(``spherical-well potential''), the internal solution is a combination of spherical Coulomb functions, with $\tilde k = \sqrt{2 m (E - V)}/\hbar$, which however reduces again to \cref{eqEspressioneFunzioneDonaZonaInternaModellinoPenetrabilitaCondDecay} in the limit of $\tilde k R_n \to +\infty$, which at the low energies of interest corresponds to $V \to -\infty$.
	It is also possible to model the nuclear potential adding a positive imaginary part, %
	in which case the wave-function in the interior region, near the interface, acquires an expression analogous to that shown in~\cite[eq.~(3.73)]{Satchler1990Introduction}. This approach is in principle more consistent, but the final qualitative result shown in the following %
	would be the same, provided that $\m{\tilde k}$ is appropriately redefined.
	
	Imposing continuity of the wave-function and of its first-derivative %
	between \cref{eqSoluzioneAsintoticaPotenzialeCoulombianoBoundaryConditionsDecadimento,eqEspressioneFunzioneDonaZonaInternaModellinoPenetrabilitaCondDecay} at the interface $r = R_n$, %
	one finds
	\begin{equation}
	\frac{B}{D} = \frac{ 2 e^{i \tilde k R_n} }{ H^+_l(k R_n) - i \partial_r H^+_l(k R_n) / \tilde k }
	\end{equation}
	The transmission coefficient is the ratio between the outgoing radial current densities associated to the wave-functions in \cref{eqSoluzioneAsintoticaPotenzialeCoulombianoBoundaryConditionsDecadimento,eqEspressioneFunzioneDonaZonaInternaModellinoPenetrabilitaCondDecay}: %
	\begin{equation}\label{eqRisultatoTrasmissioneFunzionidiCoulombTildekEsplicito}
		T_{l,S,J}(E) = \frac{\mathcal J[B H_l^+]}{\mathcal J[D e^{i \tilde k r}]} = \frac{k}{\tilde k / 4} \frac{1}{\m{H^+_l(k R_n) - i \partial_r H^+_l(k R_n) / \tilde k }^2} %
	\end{equation}
	In the limit of $k \to \infty$, it is $\tilde k \to k$ and $H_l^+(k r) \to \exp(i k r)$, thus as expected $T_{l,S,J}(E \to \infty) = 1$.

	In the more interesting limit of strong nuclear potential, %
	more precisely %
	of $\tilde k \to +\infty$, %
	the term involving the derivative of the Coulomb function may be
	neglected. %
	Furthermore,
	within the rather simple model employed here, it is often reasonable to choose $\mathcal V_<$ to be independent of the reactants spins, %
	so that $\tilde k$ depends only on $k$ and $l$, and the transmission coefficient only on $l$, $E$ and $R_n$.
	Finally, if $k \ll \tilde k$, the momentum in the internal region can be expected to not depend strongly on $k$ (see the previous expressions given for $\tilde k$),
	and $4/\tilde k$ may thus be approximated as $C_l R_n$, where $C_l$ is a dimensionless constant. When using the formula to fit experimental data, $C_l$ can be absorbed into another constant appearing in the expression (see \cref{secPenetrabilityFittingExperimentalData}), reducing the number of free parameters.
	In literature (see e.g.~\cite[eq.~(7.4.17)]{ThompsonNunes2009}), the result is %
	normally presented as:
	\begin{equation}\label{eqPenetrabilityFactorRMatrix}
	T_{l} = \frac{ k R_n }{ \m{H^+_l(k R_n)}^2 } %
	\end{equation}
	Note that this $T_l$ is just the penetrability factor appearing in R-Matrix theory (see e.g.~\cite[eq.~(10.2.5)]{ThompsonNunes2009}).
	It is emphasised again that, as far as a transmission coefficient is of interest, the approximation in \cref{eqPenetrabilityFactorRMatrix} is accurate only for small $k$
	(for instance, it %
	admits $T_l > 1$ for $k$ large enough).
	If an approximated expression was required for high values of $k$, %
	setting $\tilde k = C_l k$ %
	would be more appropriate. In fact, sometimes \cref{eqPenetrabilityFactorRMatrix} is quoted without the $k R_n$ coefficient (see e.g.~\cite[eq.~(4-124)]{Clayton1983}).
	However, it also pointed out that a model describing only barrier penetrability, as the present one, is useful only at energies %
	below the Coulomb barrier height, which can be defined as the Coulomb potential at $r = R_n$. %
	This in turn is, typically, much smaller than the depth of the total potential $\mathcal V$ in the interior region near $r = R_n$, with the exception of partial waves with $l$ so high that the penetrability is negligible anyway at the energies of interest. As a result, \cref{eqPenetrabilityFactorRMatrix} can be expected to be accurate in the relevant regime. %

	Finally, consider the limit of vanishing nuclear radius. %
	Using \cite[eq.~5.4.3, 5.5.1, 33.2.5, 33.5.1, 33.5.2]{NISTdlmf}, %
	\begin{equation}\label{eqCoefficienteTrasmissioneEsattoPerRaggioNucleareZero}
	\lim_{R_n \to 0} T_{l,S,J}(E) = \left\lbrace \begin{aligned}
		& 0 &&\text{if} \ l>0 \\
		& %
			\frac{8 \pi k \eta}{\tilde k} \frac{1}{\m{ 1 + \frac{2 \pi k \eta}{\tilde k} \frac{1}{e^{2 \pi \eta} - 1} }^2} \frac{1}{e^{2 \pi \eta} - 1}
			&&\text{if} \ l=0
	\end{aligned} \right.
	\end{equation}
	where $\eta$ is the Sommerfeld parameter, defined as in \cite[eq.~33.22.4]{NISTdlmf},
	\begin{equation}\label{eqDefinizioneParametroSommerfeld}
		\eta = \alpha_e Z_1 Z_2 \sqrt{\frac{m c^2}{2 E}}
		= \alpha_e Z_1 Z_2 \frac{m c}{\hbar k}
	\end{equation}
	In particular, note that $k \eta$ depends only on the reactants mass and charge.
	If $k$ is very small, so that $k \ll \tilde k$, with $\tilde k$ approximately independent of $k$, and $\exp(2 \pi \eta) \gg 1$, %
	the result can be approximated as $T_{l=0,SJ}(E) \approx C_{SJ} \, e^{-2 \pi \eta}$, where $C_{SJ}$ is an appropriate constant. Assuming a nuclear potential independent of the reactants spin, the corresponding cross-section for barrier penetration, $\sigma(E)$ in \cref{eqDecomposizioneSezioneDurtoPenetrabilitaPotenzialeCentrale}, is $\frac{\pi}{k^2} C e^{-2 \pi \eta}$. This result motivates the definition of the \emph{astrophysical $S$-factor}
	\footnote{In some older references, see e.g.~\cite{Shinozuka1979,Kwon1989}, the same name is employed for a slightly different quantity, which depends on the effective nuclear radius (see also \cite[eq.~(4-157)]{Clayton1983}). %
	The definition given here is the most common, especially in recent literature, see e.g.~\cite[eq.~(4-36)]{Clayton1983} or \cite[eq.~(3.71)]{Iliadis2007}, and involves only model-independent parameters known with good accuracy.}
	for any given nuclear reaction between charged particles from an initial state $i$ to a final state $f$, %
	$S_{i\to f}(E)$, in terms of the corresponding non-polarised angle-integrated cross-section $\sigma_{i\to f}(E)$:
	\begin{equation}\label{eqDefinizioneFattoreAstrofisico}
		\sigma_{i\to f}(E) = \frac{1}{E} e^{- 2 \pi \eta(E)} S_{i\to f}(E)
	\end{equation}
	At energies above the Coulomb barrier, the the astrophysical factor %
	and the integrated cross-section %
	essentially differ only by the $1/E$ coefficient.
	Below the barrier, $S(E)$ %
	is a much more slowly varying function of energy than $\sigma(E)$, %
	because the main contribution connected to the barrier penetrability has been factored out.
	For this reason, in this work all graphics of angle-integrated (non-elastic) cross-sections are shown %
	as astrophysical factor plots.

\subsection{Barrier penetrability in WKB approximation}\label{secPenetrabilityWKBApproximation}

	A convenient expression for the transmission coefficient %
	is obtained employing the Wentzel-Kramers-Brillouin-Jeffreys (``WKB'') approximation %
	(see %
		e.g.~\cite[sec.~VI.II]{Messiah2014quantum}, \cite[sec.~2.4]{SakuraiModern1994} or \cite[sec.~4-5]{Clayton1983} %
		and references therein, %
		or e.g.~\cite[sec.~2.4.3]{Iliadis2007} for the analogous barrier-slicing approach). %
	For any given partial-wave, let
	\begin{equation}\label{eqDefinizioneImpulsoEfficacePerWKB}
	k(r) = \sqrt{2 m [E - \mathcal V_{l,S,J}(r)]} / \hbar
	\end{equation}
	where, as before, %
	$E$ is the system collision energy, $m$ the reduced mass, and $\mathcal V_{lSJ}(r)$ the effective radial potential, %
	defined in \cref{eqEffectiveRadialHamiltonianDefinition}. $k(r)$ can be either positive or purely imaginary. %
	In a region where the WKB approximation is valid,
	the solution of a one-dimensional problem (as the reduced radial wave-function for a given partial wave) %
	can be written as a combination of two functions, $u^{+}_{\text{WKB}}$ and $u^{-}_{\text{WKB}}$, %
	as in \cite[eq.~(2.4.35), (2.4.38)]{SakuraiModern1994}. %
	Here, these are defined to be dimensionless, and the trivial time-dependence (the state is stationary) was dropped: %
	\begin{equation}\label{eqEspressioneSoluzioneWKBZonaSingola}
	u^{\pm}_{\text{WKB}}(r)
	= \sqrt{\m{\frac{k_\infty}{k(r)}}} \exp\( \pm i \int_{r_0}^r k(x) \d x \)
	\end{equation}
	where %
	$k_\infty = \lim_{r\to+\infty} k(r)$ %
	(in \cref{secCoulombbarrierpenetrabilityastrophysicalFactor} %
	this was simply denoted by $k$), %
	and
	$r_0$ is an arbitrary point within the region under study.
	The reduced radial current, defined in \cref{eqDefinizioneReducedRadialCurrent}, associated to a generic combination $A u^{-}_{\text{WKB}}(r) + B u^{+}_{\text{WKB}}(r)$, with arbitrary constants $A$ and $B$, is
	(the ``WKB'' subscript is dropped for brevity)
	\begin{equation}
	\mathcal J[A u^{-} + B u^{+}](r) = \left\lbrace \begin{aligned}
		& \frac{\hbar}{m} k_\infty \( \m{B}^2 - \m{A}^2 \) &&\text{if} \ k(r) \in \campo{R} \\
		& \frac{\hbar}{m} k_\infty 2 \Im\( A \coniugato{B} \) &&\text{if} \ i k(r) \in \campo{R}
	\end{aligned} \right. \end{equation}
	Following \cite[eq.~(VI.47)]{Messiah2014quantum}, the WKB approximation may be considered to be valid when
	\begin{equation}\label{eqCondizioneValiditaWKB}
	\m{\partial_r k(r)} \ll \m{k^2(r)}
	\end{equation}
	For instance, \cref{eqEspressioneSoluzioneWKBZonaSingola} is exact in a region where $\mathcal V_{lSJ}$ is constant. %
	On the contrary, the approximation is not accurate around classical turning points, i.e.~points where $E = \mathcal V_{lSJ}(r)$. %
	In that case, it is still possible to write WKB solutions in the regions where the approximation is valid, and then match them through specific connection formulas,
	discussed for instance in \cite[sec.~VI.9]{Messiah2014quantum} and \cite[sec.~2.4]{SakuraiModern1994} and references therein.
	Such situation is in fact expected
	for the problem at hand at sufficiently low energies. %
	In particular, suppose that, for $r > R_n$, there is only one classical turning point, at position $R_c$, %
	and that the WKB approximation is valid around $R_n$: for this to hold, it is necessary (but not sufficient) that $R_n \ll R_c$ and %
	$E \ll \mathcal V_{lSJ}(R_n)$. %
	At $r \to +\infty$, if the potential is given by \cref{eqDefinizionePotenzialeEfficaceARaggiAlti} the WKB is certainly valid as well.
	The appropriate connection formulas are then %
	given by \cite[eq.~(VI.49), (VI.50)]{Messiah2014quantum} (see also \cite[sec.~VI.10]{Messiah2014quantum}).
	Using decay boundary conditions, and conveniently setting the normalisation factor, in the region $r \gg R_c$ the WKB wave-function can be written as $B e^{- i \pi/4} u^{+}_{\text{WKB}}(r)$, setting $r_0 = R_c$ in \cref{eqEspressioneSoluzioneWKBZonaSingola}. The matching solution for $r \ll R_c$ is
	\begin{equation}
	u(r \ll R_c) = B \[ \frac{1}{2} u^{-}_{\text{WKB}}(r) - i u^{+}_{\text{WKB}}(r) \]
	\end{equation}
	The real part of this function, compared to the imaginary one, %
	is expected to be very small in module for $r \ll R_c$, but it is in general necessary to properly match the wave-function at $r = R_n$. %
	If the system wave-function for $r < R_n$ is \cref{eqEspressioneFunzioneDonaZonaInternaModellinoPenetrabilitaCondDecay}, with steps analogous to those followed to obtain \cref{eqRisultatoTrasmissioneFunzionidiCoulombTildekEsplicito}, the transmission coefficient, $T_{l,S,J}(E)$, is
	\begin{equation}
	\frac{ 4 \exp\( - 2 \int_{R_n}^{R_c} \m{k(x)} \d x \) \m{k(R_n)} / \tilde k }{ \m{ \[ \frac{1}{2} + i \frac{ \partial_r \m{k(R_n)} }{ 4 \m{k(R_n)} \tilde k } - i \frac{\m{k(R_n)}}{2 \tilde k} \] e^{- 2 \int_{R_n}^{R_c} \m{k(x)} \d x} + \[ \frac{ \partial_r \m{k(R_n)} }{ 2 \m{k(R_n)} \tilde k } + \frac{\m{k(R_n)}}{\tilde k} - i \] }^2 }
	\end{equation}
	If the WKB approximation is valid at $r = R_n$, from the conditions given above it can be expected that the term involving the exponential in the denominator can be neglected. Furthermore, unless $\tilde k \ll k$, the terms involving $\partial_r \m{k(r)}$ %
	can be neglected by virtue of \cref{eqCondizioneValiditaWKB}. The expression then simplifies to
	\begin{multline}\label{eqDefinizionePenetrabilitaWKB}
	T_{l,S,J}(E)
	= \frac{ 4 \m{k(R_n)} \tilde k }{ \m{k(R_n)}^2 + \tilde k^2} \exp\( - 2 \int_{R_n}^{R_c} \m{k(x)} \d x \) = \\
	= 4 \frac{ \sqrt{ (\mathcal V_l(R_n) - E) (E - \mathcal V_<) } }{ \mathcal V_l(R_n) - \mathcal V_< } \exp\( - 2 \int_{R_n}^{R_c} \m{k(x)} \d x \)
	\end{multline}
	as in \cite[sec.~VI.10]{Messiah2014quantum}.
	In the last equality, $\tilde k$ was defined through \cref{eqDefinizioneImpulsoBucaInterna}.
	For small collision energies, the factor before the exponential may be approximated as independent of $E$, %
	thus writing $T_{lSJ} = C_{lSJ} \exp(\dots)$.

	Finally, consider the case where $E \gg \max_{r > R_n} \mathcal V_{lSJ}(r)$ (the process is taking place ``above the barrier''),
	and \cref{eqCondizioneValiditaWKB} holds for all $r > R_n$. %
	Then, %
	the exterior WKB solution $B u^{+}_{\text{WKB}}(r)$ can be applied in the same region, and the reasoning yielding \cref{eqRisultatoTrasmissioneFunzionidiCoulombTildekEsplicito} can be repeated identically changing $H_l^+$ with $u^{+}_{\text{WKB}}$, regardless of the precise form of the potential. The transmission coefficient is consequently
	\begin{equation}
		T_{l,S,J}(E)
		= \frac{k(R_n)}{\tilde k / 4} \frac{1}{\m{ 1 + k(R_n)/\tilde k + i \frac{ \partial_r k(R_n) }{ 2 k(R_n) \tilde k } }^2}
	\end{equation}
	Again, unless $\tilde k \ll k$, the term involving $\partial_r \m{k(r)}$ can be neglected using \cref{eqCondizioneValiditaWKB},
	so that $T_{l,S,J} = \frac{1}{4} k(R_n) \tilde k / \[k(R_n) + \tilde k\]^2$, which coincides with \cref{eqDefinizionePenetrabilitaWKB} setting $R_c = R_n$, and also matches \cref{eqRisultatoTrasmissioneFunzionidiCoulombTildekEsplicito} in the limit of high collision energy, where $k(R_n) \approx k_\infty$ and $H^+_l \approx \exp(i k_\infty r)$.

\subsubsection{Transmission for central sharp-edge nuclear potentials} %

	\Cref{eqDefinizionePenetrabilitaWKB} is especially useful to numerically estimate the penetrability connected to a generic potential. However, it is also instructive to
	apply it to %
	the case of
	a central sharp-edge nuclear potential of the same kind treated exactly in \cref{eqPenetrabilitaPotenzialeSharpEdgeEsatta}: %
	for each partial wave, the effective radial potential is \cref{eqDefinizionePotenzialeEfficaceARaggiAlti} for $r > R_n$, and a constant $\mathcal V_<$ %
	for $r < R_n$ (remind that the region at $r \ll R_n$ is irrelevant). %
	The depth of the inner well is allowed to depend on $l$ to account for the contribution due to the centrifugal barrier. %
	This configuration admits analytical results, which can be manipulated formally with greater ease than \cref{eqRisultatoTrasmissioneFunzionidiCoulombTildekEsplicito}, and are thus better suited to analyse some qualitative features.
	There is a single
	classical turning point $R_c(l)$, which is the point where $\mathcal V_{l}(R_c) = E$.
	Let $R_c^0 = R_c(l=0) = \hbar c \alpha_e Z_1 Z_2 / E$. Then
	\begin{equation}\label{eqEspressioneRaggioCoulombianoRGenerico}
		R_c(l) = \frac{1}{2} R_c^0 + \sqrt{\frac{1}{4} (R_c^0)^2 + \frac{l (l+1)}{k^2} }
	\end{equation}
	where $k = \sqrt{2 m E} / \hbar$ as before.
	As required to apply \cref{eqDefinizionePenetrabilitaWKB}, restrict to the case where $R_c(l=0) > R_n$, so that the classically forbidden region overlaps with the domain of interest. %
	The integral in \cref{eqDefinizionePenetrabilitaWKB} can be simplified %
	applying the variable change $r = R_c^0 x$. %
	Let $\gamma$ and $\beta$ be the new lower and upper bounds of the integration domain.
	It is $\gamma = R_n/R_c^0$, which is also the ratio between the collision energy and the Coulomb potential at $r = R_n$ (see \cref{eqDefinizionePotenzialeEfficaceARaggiAlti}), %
	hence $0 \leq \gamma \leq 1$ (by hypothesis). %
	Note that $\gamma \eta = k R_n / 2$, where $\eta$ is the Sommerfeld parameter defined in \cref{eqDefinizioneParametroSommerfeld}. %
	Similarly, $\beta = R_c / R_c^0 \geq 1$.
	Furthermore, %
	let $B %
	= l (l+1) / (4 \eta^2)$, which also equals $\beta^2 - \beta$.
	The argument of the exponential in \cref{eqDefinizionePenetrabilitaWKB}, labelled $- \mathcal G_l$ (as in \cite[eq.~(7.4.22)]{ThompsonNunes2009}) for brevity, %
	can be rewritten as
	\begin{equation}%
	-\mathcal G_l = -4 \eta \int_{\gamma}^{\beta} \sqrt{ \frac{1}{x} + \frac{B}{x^2} - 1 } \d x
	\end{equation}
	It can be verified explicitly that a primitive of the given integrand %
	is %
	\begin{multline}\label{eqIntegraleIndefinitoLogCoefficienteTrasmissioneWKBBucaSferica}
	- \int \sqrt{ \frac{1}{x} + \frac{B}{x^2} - 1 } \d x =
	- \sqrt{B + x - x^2} +\\+ \sqrt{B} \ln\(\frac{ 2 B + x + 2 \sqrt{B} \sqrt{B+x-x^2} }{ x \, \sqrt{1+4B} }\) + \frac{1}{2} \arctan\( \frac{\frac{1}{2}-x}{ \sqrt{B + x - x^2} } \)
	\end{multline}
	which is always real within the integration domain. %
	Therefore,
	\begin{multline}\label{eqEspressioneGeneraleFattoreGamowGeneralizzatoWKB}
		- \mathcal G_l = - \pi \eta - 4 \eta \[ \sqrt{B} \ln\(\frac{ 2 B + \gamma + 2 \sqrt{B} \sqrt{B+\gamma-\gamma^2} }{ \gamma \, \sqrt{1+4 B} }\) +\right.\\\left.+ \frac{1}{2} \arctan\( \frac{1-2\gamma}{ 2 \sqrt{B + \gamma - \gamma^2} } \) - \sqrt{B + \gamma - \gamma^2} \]
	\end{multline}
	Finally, it is convenient to define
	\begin{equation}
		b = \gamma/B = \frac{4 \gamma \eta^2}{l(l+1)} = \frac{2 \eta k R_n}{l(l+1)} = \frac{2 \alpha_e Z_1 Z_2 m c^2 R_n}{\hbar c \, l(l+1)}
	\end{equation}
	which is the ratio between the Coulomb and centrifugal potential at $r = R_n$ (see \cref{eqDefinizionePotenzialeEfficaceARaggiAlti}). $b$ is thus dimensionless and does not depend on the collision energy.

	Only for $1 - 2 \gamma > 0$ (i.e.~$R_n < R_c^0 / 2$), which is satisfied at sufficiently low energies, the inverse tangent can be rewritten using $\arctan(1/x) = \pi/2 - \arctan(x)$. Restrict to this case in the following.
	Furthermore, for simplicity, in \cref{eqDefinizionePenetrabilitaWKB} approximate the scaling factor before the exponential to be independent of $E$ (see text commenting the equation). The corresponding astrophysical factor $S(E)$, defined as in \cref{eqDefinizioneFattoreAstrofisico} with respect to the cross-section in \cref{eqDecomposizioneSezioneDurtoPenetrabilitaPotenzialeCentrale}, for WKB barrier penetration with a sharp-edge central nuclear potential, is then
	\begin{multline}\label{eqFattoreAstrofisicoWKBPerEnergiePiccole}
		S(E) %
		= \sum_l C_l e^{2 \pi \eta -\mathcal G_l}
		= \sum_l C_l \(\frac{ b \, \sqrt{1+4 \gamma/b} }{ 2 + b + 2 \sqrt{1+b-b\gamma} }\)^{ 2 \sqrt{l (l+1)} } \cdot\\\cdot \exp\( 2 \eta \arctan\( \frac{ 2 \sqrt{\gamma} \sqrt{1/b + 1 - \gamma} }{1-2\gamma} \) + 2 \sqrt{l(l+1)} \sqrt{1 + b - b \gamma} \)
	\end{multline}
	where
	\begin{equation}\label{eqCostanteMoltiplicativaFattoreAstrofisicoWKB}
	C_l = \frac{2 \pi \hbar^2}{m} (2l+1) \frac{ \sqrt{ \mathcal V_l(R_n) \mathcal V_< } }{ \mathcal V_l(R_n) - \mathcal V_< }
	\end{equation}
	For brevity, let $S_l(E) = C_l e^{2 \pi \eta -\mathcal G_l}$, which is %
	the contribution to $S(E)$ from $l$-th partial-wave%
	\footnote{This $S_l$ ha no relation %
		with the scattering $S$-matrix.}.
	Strictly speaking, \cref{eqFattoreAstrofisicoWKBPerEnergiePiccole} %
	is ill-defined for $l=0$, even though it converges to the correct result by continuity. More explicitly, setting $B=0$ in \cref{eqIntegraleIndefinitoLogCoefficienteTrasmissioneWKBBucaSferica}, it is
	\begin{equation}\label{eqSFactorWKBL0}
	S_0(E)
	= C_0 \exp\( 2 \eta \[ \arctan\( \frac{ 2 \sqrt{\gamma} \sqrt{1 - \gamma} }{1-2\gamma} \) + 2 \sqrt{\gamma} \sqrt{1 - \gamma} \] \)
	\end{equation}

	It is now of interest to qualitatively study
	the WKB astrophysical factor for sharp-edge barrier penetration,
	in the limit of very small collision energies, %
	in particular with regard to the role of the effective nuclear radius.
	As computed exactly earlier, see \cref{eqCoefficienteTrasmissioneEsattoPerRaggioNucleareZero}, for $R_n \to 0$ only $S_0(E)$ is non-zero, and it is approximately a constant for small collision energies%
	\footnote{Note that the WKB validity condition in \cref{eqCondizioneValiditaWKB} is not satisfied for $l=0$ and $R_n \to 0$, if the point-like Coulomb potential is adopted. Nonetheless, the WKB formula %
		yields the correct qualitative result, apart from a divergence in the value of $C_0$.}. %
	Both these features disappear for finite values of $R_n$. %

	At zero collision energy \cref{eqFattoreAstrofisicoWKBPerEnergiePiccole} reduces to
	\begin{equation}\label{eqEspansioneInOndeParzialiFattoreAstrofisicoPenetrabilitaWKBBucaEnergiaZero}
	S(E=0) = \sum_l S_l(0)
	= \sum_l C_l \(\frac{ e^{2 \sqrt{1 + b}} }{ 2/b + 1 + 2 \sqrt{1+b}/b }\)^{ 2 \sqrt{l (l+1)} }
	\end{equation}
	where $S_0(0) = C_0 \exp\(4 \sqrt{2 \alpha_e Z_1 Z_2 m c R_n / \hbar} \)$.
	It can be seen that all partial waves contribute. However, $S_l(0)$ monotonically decreases for increasing $l$ (and reasonable choices of $\mathcal V_<$ as a function of $l$). In fact, for typical values of the parameters, only the smallest orbital angular momenta contribute significantly%
	\footnote{For instance, for the $\nuclide[6]{Li}+\nuclide{p}$ system and $R_n \leq \SI{10}{\femto\metre}$, $l(l+1) b$ is $\leq 1.8$. %
		Taking $\mathcal V_l$ and $\mathcal V_<$ as constants for simplicity, $l=2$ contributes by less than \SI{0.5}{\percent} on $S(0)$.}.

	In the following, consider only the $l=0$ term. The astrophysical factor dependence on the collision energy can be explored evaluating its derivate.
	First note from \cref{eqEspressioneGeneraleFattoreGamowGeneralizzatoWKB} that
	$\mathcal G_l / (4 \eta) - \pi/4$ is just the primitive in \cref{eqIntegraleIndefinitoLogCoefficienteTrasmissioneWKBBucaSferica} evaluated at $x=\gamma$. %
	Additionally, by direct inspection it is seen that %
	$\frac{\partial \eta}{\partial E} = -\frac{\eta}{2 E}$ and $\frac{\partial \gamma}{\partial E} = \gamma/E$.
	It is thus
	\begin{multline}
		\frac{\partial}{\partial E} \( 2 \pi \eta - \mathcal G_0 \)
		= - \frac{2 \pi \eta - \mathcal G_0}{2 E} + 4 \eta \sqrt{ \frac{1}{\gamma} - 1 } \cdot \frac{\partial \gamma}{\partial E} = \\
		= \frac{1}{E} \left[ 2 \eta \sqrt{\gamma} \sqrt{1 - \gamma} - \eta \arctan\( \frac{ 2 \sqrt{\gamma} \sqrt{1 - \gamma} }{1-2\gamma} \) \right]
	\end{multline}
	It can be seen (applying de l'Hôpital theorem) that this converges to a finite value in the limit of $E \to 0$.
	Thus, to first order in Taylor expansion around $E=0$ (here, $\mathcal O$ denotes ``big-O'' notation),
	\begin{multline}\label{eqEspansioneAlPrimoOrdineEsponenzialeFattoreAstrofisicoWKBl0}
	2 \pi \eta - \mathcal G_0
	= 8 \eta \sqrt{\gamma} \[1 - \frac{1}{6} \gamma + \mathcal O(E^2) \] = \\
	= 4 \sqrt{2 \alpha_e Z_1 Z_2 \frac{m c}{\hbar} R_n} \[1 - \frac{R_n}{6 \hbar c \alpha_e Z_1 Z_2} E \] + \mathcal O(E^2)
	\end{multline}
	in agreement with \cite[eq.~(4-153)]{Clayton1983}. Bigger values of $R_n$ thus cause a steeper decrease of the astrophysical factor starting from the value at $E=0$.
	This result may be employed as a qualitative guide %
	when studying experimental measurements or theoretically computed cross-sections, especially if it is possible to isolate the contribution due to $s$-waves, %
	by interpreting the first derivative of the logarithm of the corresponding astrophysical factor at very low energies as an assessment of the effective nuclear radius associated to the system.
	
	Given that the first derivative of each partial-wave contribution to the $S$-factor, $\partial_E S_l(E)$, is just $S_l(E) \partial_E \( 2 \pi \eta - \mathcal G_l \)$, an analogous expansion can be applied to $S_l$ itself. %
	In particular, for the $s$-wave component,
	\begin{equation}\label{eqEspansioneAlPrimoOrdineFattoreAstrofisicoWKBl0}
		S_0(E) 
		= S_0(E=0) \[ 1 - \frac{2}{3} \frac{1}{\sqrt{Z_1 Z_2 \alpha_e}} \frac{R_n^{3/2} \sqrt{2 m c^2} E}{(\hbar c)^{3/2}} \] + \mathcal O(E^2)
	\end{equation}
	where the ratio $[\partial_E S_0(E)] / S_0(E)$ %
	at very low collision energies
	bears the information about $R_n$. %

\subsection{Application the \texorpdfstring{$\nuclide[6]{Li}+\nuclide{p}\to\nuclide[3]{He}+\nuclide{\alpha}$}{6Li+p->3He+a} reaction}\label{secPenetrabilityFittingExperimentalData} %

	The expressions derived in the preceding sections, %
	in particular  \cref{eqPenetrabilityFactorRMatrix,eqFattoreAstrofisicoWKBPerEnergiePiccole}, can be employed to phenomenologically describe the non-resonant angle-integrated cross-section
	of a specific nuclear reaction using only few free parameters, which can be fitted on relevant experimental data.
	
	As already shown earlier, the transmission coefficient for a sharp-edge nuclear potential depends %
	on two model-dependent parameters. The first is the radius of the nuclear well, $R_n$, which can be interpreted as the effective interaction size of the system, and is thus expected a priori to roughly scale as in \cref{eqScalingRaggioNucleareLiquidDrop}.
	The second parameter is the depth of the potential in the interior region at $r \approx R_n$, which in general can be different for each partial wave. %
	Even in the most simple case, the corresponding effective radial potential $\mathcal V_<$ is expected to be a constant plus the Coulomb and centrifugal potential (see \cref{eqDefinizionePotenzialeEfficaceARaggiAlti}) evaluated at $r=R_n$, %
	and thus depend also on $R_n$ itself.
	At sufficiently small energies, the dependence on $\mathcal V_<$ can be approximated as an overall scaling factor.
	In addition, in order to obtain the cross-section for a specific reaction, the cross-section for barrier penetrability must be rescaled by another dimensionless factor, to account for the details of the process taking place once the nuclei came into close contact, and in particular selecting the desired final channel.
	Such factor may be estimated through some model (see e.g.~\cite{Kimura2007} for a calculation using the Weisskopf model), but the simplest approach is to treat it as %
	a %
	free parameter, constant with respect to the energy. %
	At small energies, and assuming that any dependence on the nuclear spins can be neglected, the two scaling factors just discussed can be merged into a single set of phenomenological parameters (one for each eigenvalue of the relative orbital angular momentum), $A_l$.
	In particular, it is common to neglect all contributions at $l>0$, so that the cross-section, if expressed in arbitrary units, depends only on $R_n$.

\begin{figure}[tb]%
		\centering
		\includegraphics[keepaspectratio = true, width=\linewidth]{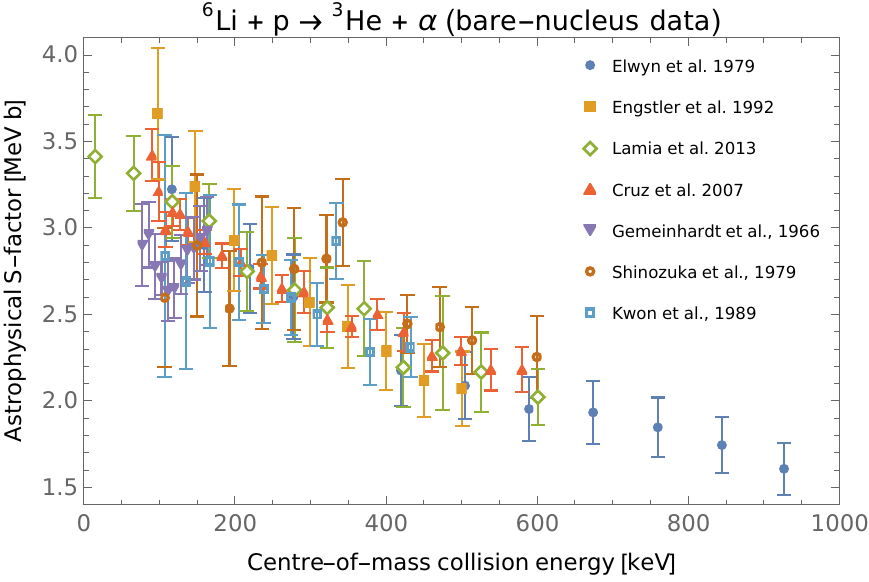}%
		\caption[\texorpdfstring{$\nuclide[6]{Li}+\nuclide{p}\to\nuclide[3]{He}+\nuclide{\alpha}$}{6Li+p->3He+a} bare-nucleus experimental astrophysical factor]{\label{figbareNucleusData}%
			Experimentally measured bare-nucleus astrophysical factor (defined in \cref{eqDefinizioneFattoreAstrofisico}) %
			for the $\nuclide[6]{Li}+\nuclide{p}\to\nuclide[3]{He}+\nuclide{\alpha}$ reaction, as a function of the collision energy. Green open diamonds are Trojan Horse Method data from \cite{Lamia2013}. All other sets are direct measurements at collision energies between \SI{75}{\keV} and \SI{1}{\MeV}, from \cite{Gemeinhardt1966} (violet downward triangles), \cite{Shinozuka1979} (brown open circles), \cite{Elwyn1979} (blue full circles), \cite{Kwon1989} (light blue open squares), \cite{Engstler1992} (orange full squares), \cite{Cruz2007} (red upward triangles).}
	\end{figure}

	\Cref{figbareNucleusData} shows a collection of %
	experimental excitation functions available in literature
	on the $\nuclide[6]{Li}+\nuclide{p}\to\nuclide[3]{He}+\nuclide{\alpha}$ reaction, at centre-of-mass collision energies below \SI{1}{\MeV}.
	The figure includes only bare-nucleus data, that is, measurements where electron screening effects, to be discussed later in \cref{secPhenomenologyScreeningEffects}, do not appear, and reactants can be regarded as isolated.
	Data from \cite{Lamia2013} are an indirect measurement performed using the Trojan Horse Method (see e.g.~\cite{Tribble14} for a review): these data are not affected by electron screening%
	\footnote{%
		Cross-sections measured with the Trojan Horse Method do not %
		include the %
		barrier penetrability contribution, which is added a posteriori to obtain the final result.}.
	All other datasets, from \cite{Gemeinhardt1966,Shinozuka1979,Elwyn1979,Kwon1989,Engstler1992,Cruz2007}, are ``standard'' fixed-target direct measurements on targets which can be considered to bear an atomic structure, %
	at centre-of-mass collision energies such that atomic electrons play a negligible role, in this case above \SI{75}{\keV} (see \cref{secPhenomenologyScreeningEffects}). %
	More data (not shown) can be found in \cite{Marion1956,Fiedler1967,Varnagy1974,Lin1977,Tumino2003}, %
	which for simplicity are currently not included in the analysis.
	Data from \cite{Cruz2005} are not shown here (see instead \cref{figAllDataLowEnergy}) as they involve only collision energies below \SI{75}{\keV} or metallic targets. %
	Data from \cite{He2013} were normalised to points in \cite{Cruz2007} and cover the same energy range, thus are not useful for the present analysis.
	Furthermore, data in \cite{Tumino2004} are not included as they were superseded by data in \cite{Lamia2013}.
	
	One of the datasets in \cite[tab.~1]{Engstler1992} (not shown in \cref{figbareNucleusData}), obtained through a %
	normal kinematics measurement with a \nuclide{LiF} target, was normalized to previous results, and the points at higher energies appear in disagreement %
	with the other datasets in the same work, using an \nuclide{H_2} molecular target, for which absolute values of the astrophysical factor are provided. \cite[tab.~3]{Engstler1992} proposes a normalisation to older measurements also for the molecular-target datasets (scaling them by a factor of $0.93$), which however does not suffice to solve the aforementioned disagreement with normal-kinematics data. %
	The issue is discussed in \cite{Wang2011}, which also proposes a new normalisation for all datasets in \cite{Engstler1992}. Note that a downscaling of inverse-kinematics data in \cite{Engstler1992} would put the points at higher collision energies in tension with data from other experiments, and consequently does not appear to be a fully satisfactory solution.
	In the present study, for simplicity, molecular-target data from \cite{Engstler1992} are employed as originally measured, without any rescaling, while measurements using the \nuclide{LiF} target are not included. %
	
	Data in \cite{Gemeinhardt1966,Shinozuka1979,Kwon1989} show features (peaks or depths) in disagreement with the other datasets (in particular with data in \cite{Cruz2007}, whose quoted errors are relatively small), and which would suggest the presence of resonances, which cannot be described by the penetrability model here under discussion. For simplicity, these datasets were excluded from the following analysis.

\begin{figure}[tbp]%
		\centering
		\includegraphics[keepaspectratio = true, width=\linewidth]{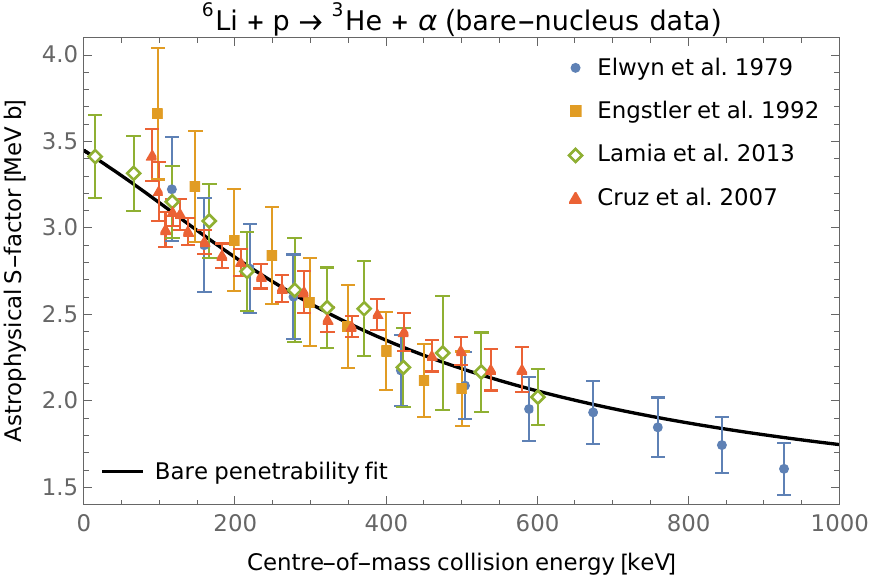}%
		\caption[\texorpdfstring{$\nuclide[6]{Li}+\nuclide{p}\to\nuclide[3]{He}+\nuclide{\alpha}$}{6Li+p->3He+a} bare-nucleus penetrability fit]{\label{figbareNucleusFit}%
			Points are a subset of experimental data in \cref{figbareNucleusData}, from \cite{Lamia2013,Elwyn1979,Engstler1992,Cruz2007}, with identical symbols. Black solid line is a fit on shown data of the model described by \cref{eqDecomposizioneSezioneDurtoPenetrabilitaPotenzialeCentrale,eqPenetrabilityFactorRMatrix} including only the $l=0$ term and adding an overall scaling factor, $A_0$ (see text for details). The fit has 50 degrees of freedom and returns a $\chi^2$ of 19. The best-fitting values for the free parameters, together with the corresponding standard error, are $A_0 = \num{0.39\pm0.03}$ and $R_n = \SI{3.36\pm0.13}{\femto\metre}$.}
	\end{figure}

	The remaining datasets were fitted (for energies below \SI{1}{\MeV} and above \SI{75}{\keV} for direct measurements) adopting the transmission coefficient in \cref{eqPenetrabilityFactorRMatrix} and including only the $s$-wave contribution ($l=0$) in the expansion in \cref{eqDecomposizioneSezioneDurtoPenetrabilitaPotenzialeCentrale}. %
	The results are shown in \cref{figbareNucleusFit}.
	The remarkable agreement between bare-nucleus data below \SI{1}{\MeV} and fitted model is made possible %
	by the absence of significant resonant behaviour in the %
	data in the same energy range, %
	and favours the expectation that the reactants electrostatic repulsion plays a dominant role in the reaction dynamics.
	Another fit was attempted including also the $l=1$ contribution, but the best-fitting value for its weight was \num{0} (and an unconstrained fit suggested a negative value).
	The model, as expected, cannot instead fit the resonance at a centre-of-mass collision energy of about \SI{1.5}{\MeV} found in the data.

\section{Screening effects}\label{secPhenomenologyScreeningEffects}
	Assume now that the collision is not taking place in vacuum but within an interacting environment. %
	In particular, in this work it is of interest to study the effects induced %
	by atomic electrons on the Coulomb barrier penetration process.
	Let $\sigma_b(E)$ be the non-polarised angle-integrated cross-section for a nuclear reaction between isolated reactants, %
	and $\sigma_e(E)$ the corresponding ``screened'' cross-section for the process taking place into the environment. If the environmental interaction in the reactants relative motion %
	frame can be adequately described as an external potential $V_e(\v r)$ (which is not always the case), %
	the reaction can be treated as happening in vacuum with a modified potential. %
	The “screened” barrier penetrability could then be obtained employing the techniques already discussed in \cref{secPhenomenologyBarenucleusnonresonantcrosssection}. %
	
	In the following, the screening potential approach, a widely employed approximation which greatly simplifies the evaluation of screening effects, is introduced.
	Within the context of screening by atomic electrons in fixed-target nuclear reaction experiments, the approach is described in \cite{Assenbaum1987}, and treated in-depth in \cite{Bracci1990} (which also includes some discussion on the approximation accuracy); the same model was employed since much earlier to describe screening effects in plasma environments \cite{Salpeter1954}.
	Some of the concepts %
	presented here have been treated previously in \cite{PerrottaTesiSSC} and, more tangentially, in \cite{Perrotta2019HindranceEffects,PerrottaSantaTecla2019}.

\subsection{The screening potential approach}\label{sezScreeningPotentialDefinizione}

	The interaction between environment and nuclei can be described in terms of energy transfer to (or from) the reactants relative motion, %
	considering %
	an effective collision energy $E_e$ which changes as the reactants approach.
	If it is possible to define an “environmental potential” $V_e$, %
	it is simply $E_e(\v r) = E - V_e(\v r)$, where $E$ is the actual collision energy as before.
	In the screening potential approximation, it is assumed that %
	most of the energy from the environment is transferred when the reactants are still far away. %
	As a consequence,
	$E_e$ can be approximated by a constant in the region relevant for the reaction dynamics, including the barrier penetration if applicable. The total energy gain due to the environmental interaction, %
	$U = E_e(r=0) - E$, is called the \emph{screening potential}
	\footnote{Sometimes $U$ is defined with an opposite sign, to make it consistent with the name “potential”.}.
	With this rationale, %
	the screened and bare reaction cross-section are connected as: %
	\begin{equation}\label{eqSezioneDurtoInPotenzialeScreening}
		\sigma_e(E) = \sigma_b(E+U)
	\end{equation}
	In principle, $U$ is itself a function of $E$, as the environment state can be affected by the reactants motion. In practice, the evaluation of such dynamical effects is a complex problem, and the screening potential is normally approximated to a constant.

	Given a pair of charged reactants, the screening potential approach can be considered valid when %
	the energy transfer from the environment after the tunnelling started is negligible with respect to the effective collision energy possessed by the reactants just before that point. %
	Let $R_e$ be the ``screened'' outer classical turning point for the reactants in-medium interaction. If $U>0$, meaning that the environment favours (and not hinders) the reaction, %
	$R_e$ will be smaller than the corresponding turning point in vacuum.
	For simplicity, assume that the effective collision energy $E_e(\v r)$ is isotropic.
	If $E_e(r)$ is monotonic near the origin, the validity condition for the screening potential approach can then be written as $\m{E_e(0) - E_e(R_e)} \ll E_e(R_e)$.
	In general, the approximation is thus more accurate for greater collision energies, since $R_e$ becomes smaller, increasing the fraction of energy transferred before tunnelling, and a given energy transfer becomes proportionally less important. %
	For the case of screening by atomic electrons, \cite{Bracci1990} discusses how the accuracy of the screening potential approach depends on $U/E$.
	
	The goal when accounting for screening effects is, clearly, to find how the screened reaction cross-section $\sigma_e(E)$ in presence of environmental interaction relates with the bare one $\sigma_b$.
	Usually (see e.g.~\cite[eq.~(3)]{Assenbaum1987}) the results are expressed in terms of the \emph{enhancement factor} $f_e(E)$, defined as the ratio of $\sigma_e(E)$ to $\sigma_b(E)$.
	Adopting the screening potential approach,
	writing the cross-sections in terms of the astrophysical factor in \cref{eqDefinizioneFattoreAstrofisico}, it is:
	\begin{equation}\label{eqFattoreEnhancementPotScreeningCasoGenerale}
		f_e(E,U) %
		= \frac{\sigma_b(E+U)}{\sigma_b(E)}
		= \frac{E}{E + U} \frac{ e^{2 \pi \eta(E)} }{ e^{2 \pi \eta(E+U)} } \frac{S(E+U)}{S(E)}
	\end{equation}
	For the systems of interest, $U$ is of the order of few hundreds of \unit{\eV} at most (see \cref{eqUpperLimitGeneraleScreeningPotentialAtomico}). %
	The astrophysical factor normally varies very slowly through such an energy range, and it is possible to approximate $S(E+U) \approx S(E)$ with good accuracy, making the enhancement factor expression independent of the model adopted for the bare cross-section:
	\begin{equation}\label{eqFattoreEnhancementPotScreeningPerAstrophysicalFactorConstante}
		f_e(E,U) \approx \frac{E}{E + U} \exp\(2 \pi \eta(E) \[ 1 - \sqrt{ \frac{E}{E+U} } \] \)
	\end{equation}

	\Cref{eqFattoreEnhancementPotScreeningPerAstrophysicalFactorConstante} can be employed to predict the range of collision energies where a screening effect is important, provided it can be described within the screening potential approach.
	For instance, for $\nuclide[6]{Li}+\nuclide{p}$ scattering, %
	assuming $U = \SI{186}{\eV}$ for the atomic electron screening potential (the adiabatic limit for a bare proton impinging on a neutral \nuclide[6]{Li} in the ground state, see \cref{secAdiabaticLimitAtomicElectronScreening}), at centre-of-mass energies above \SI{75}{\keV} the net enhancement, $f_e(E)-1$, is approximately 1\% or smaller, which is well below typical experimental errors. At smaller collision energies the correction becomes relevant, for instance at $E = \SI{10}{\keV}$ (approximately the smallest energy at which direct measurements are currently available) %
	and $U$ as before it is $f_e(E)-1 \approx \SI{26}{\percent}$. %

\subsection{Screening by atomic electrons}\label{secScreeningByAtomicElectrons}%
	
	Consider a standard fixed-target experiment with a single projectile nucleus (which may carry some electrons) %
	originated far away and impinging at a given collision energy on an isolated target atom. The atomic electrons screen the nuclear charges and enhance the penetrability.
	Following \cite{Bracci1990}, the problem can be approached assuming that the system total wave-function can be decoupled in the nuclear and electronic parts (as in the Born--Oppenheimer approximation), and treating quantum-mechanically the electronic degrees of freedom. %
	For a sufficiently small $U/E$ ratio, the relative change in the reactants %
	velocity may be neglected, and the motion of the nuclear degrees of freedom approximated as free (\cite[eq.~(3.24)]{Bracci1990} may be adopted in a numerical computation to lift the hypothesis).
	Furthermore, the reaction impact parameter is very small with respect to atomic distances, so it can be approximated with 0 just for the purpose of screening potential evaluation.
	Under these assumptions, %
	the nuclei follow a simple linear trajectory. %
	Once the reactants reach distances sufficiently smaller than typical atomic lengths, %
	the electrons Hamiltonian %
	can be approximated to the time-independent Hamiltonian generated by the compound of the two initial nuclei. %
	If this happens at sufficiently high distances with respect to those relevant for the nuclear process (including the barrier penetration), adopting the screening potential approach is well justified.
	In \cite[eq.~(2.17)]{Bracci1990} it is estimated that the relative error on the screened cross-section induced by the adopted approximations is of the order of $U/E$.
	
	Since the system is isolated, %
	the screening potential $U$ %
	is just the opposite of the change in the total electron energy (namely, the sum of electrons kinetic energy, electrons-ions potential and electrons-electrons potential) between the initial state with two isolated atoms, and the corresponding final state involving the compound system.
	In the initial states of interest here (standard fixed-target experiments), %
	the electrons are always bound to the nuclei, %
	hence, when computing their energy, %
	it is of interest to separate the contribution involving the internal state within each isolated nucleus from the one due to the reactants relative motion. The latter term
	does not include any potential energy,
	as the reactants are initially at infinite distance, and is thus denoted here as $K_e$.
	Let indexes 1 and 2 represent target and projectile in the initial state. Each electron bound to nucleus $i$ is initially moving with the same velocity as the nucleus itself, which, neglecting the role of electrons in fixing the system centre-of-mass, in the centre-of-mass frame has modulus $\hbar k / m_i$, where $m_i$ is nucleus $i$ mass and $k$ is the wave-number connected to the reactants relative motion, as in \cref{secCoulombbarrierpenetrabilityastrophysicalFactor}.
	Then, if there are $N_i$ electrons bound to nucleus $i$, it is
	\begin{equation}\label{eqEspressioneEnergiaCineticaElettroniInComotoConIReagentiRispettoACentroDiMassa}
		K_e = \frac{m_e}{m_1+m_2} \( N_1 \frac{m_2}{m_1} + N_2 \frac{m_1}{m_2} \) E
	\end{equation}
	where $m_e$ is the electron mass.
	The importance of this term depends on the specific configuration under study. For instance, for the scattering of a neutral \nuclide[6]{Li} on a neutral \nuclide[1]{H}, it is $K_e / E \approx \num{5e-4}$,
	while if the hydrogen is a bare proton, it is $K_e / E \approx \num{4e-5}$.

	Finally, let index $1+2$ represent the compound atom, and let $E_i$ be the total energy in $i$ rest frame for the electrons system in the isolated atom $i$ found in the required state for the reaction under study. %
	The screening potential can then be written as
	\begin{equation}\label{eqPotenzialeScreeningAdiabaticLimitAutostatoSingolo}
		U = K_e + E_1 + E_2 - E_{1+2}
	\end{equation}
	Hence, the problem is reduced to correctly model the electrons evolution %
	(using \cite[eq.~(3.15)]{Bracci1990}). %
	
	For any given initial state, %
	a theoretical upper limit for the screening potential can be given fixing the final state to the compound atom ground state (where electrons possess the minimum possible energy), as in \cite[eq.~(4.19)]{Bracci1990}. For the problems of interest here, the initial state of each atom is almost invariably the ground state as well, and the maximum screening potential can be computed in terms of atomic ionization energies, which are generally very well known (tabulated values can be found e.g.~in \cite{NISTasd}). %
	Let $X^{n}$ be the ground state of an ion with charge state $n$ (which can be positive, zero or even negative), %
	and $B(X^{n})$ the corresponding binding energy for ionization of all bound electrons. %
	Then, the maximum screening potential for the scattering of $X^n$ and $Y^m$ with compound $Z^{n+m}$ is
	\begin{equation}\label{eqUpperLimitGeneraleScreeningPotentialAtomico}
		U_{\text{max}} = K_e + B(Z^{n+m}) - B(X^n) - B(Y^m)
	\end{equation}
	If some reactant is in molecular form, %
	one may in the same spirit take into account the atomization energies (namely, the energy to break the molecule into isolated atoms) of the %
	ground states of the initial and final partitions. Since molecular bindings generally involve energies of smaller order of magnitude with respect to atomic ones, the corrections are expected to be relatively small. \cite{NISTcccbdb} reports experimental values for atomization enthalpies, which may be sufficient for the present purpose given that only differences between values for different molecules are of interest. 

	In the general case, the evaluation of the enhancement factor is complicated by dynamical effects in the electrons response to the reactants motion; \cite{Bracci1991,Carraro1988,Kimura2005} are just three examples of several works discussing the problem. %
	However, %
	a simple result is obtained in the limiting cases of very low and high %
	relative velocity between reactants,
	the %
	\emph{adiabatic} and \emph{sudden} limits%
	\footnote{The system is considered isolated in both cases: this excludes, for instance, molecular screening effects. The term adiabatic here %
		refers instead to quasi-static processes.}.
	These are covered in general e.g.~in \cite[sec.~XVII.II]{Messiah2014quantum}, and are applied to the specific problem at hand in \cite{Bracci1990}, which can be consulted for a more detailed treatment.
	In the following, the adiabatic limit formalism is briefly reviewed, as its results are useful for the purposes of the present work.

\subsubsection{Adiabatic limit}\label{secAdiabaticLimitAtomicElectronScreening}
	
	In the adiabatic approximation, valid in the limit of small collision energies, the distance between the nuclei is changing slowly, and so does the Hamiltonian for the electronic degrees of freedom, which evolves in quasi-equilibrium. %
	If the electrons initially are in a non-degenerate eigenstate of the isolated atoms, at any later time they will be found (apart for at most a phase) in the corresponding eigenstate of the new Hamiltonian, where the correspondence is set by continuity with respect to time.
	Note that the notion of non-degenerate state here regards the whole system (not each ion separately). For instance, if the reactants are %
	isotopes %
	and each occupies a distinct atomic state, then the whole system state is approximately degenerate with the configuration obtained swapping the two reactants.
	The set of corresponding states in the final system will in general be a superposition of all degenerate levels in the initial system, and will not be degenerate themselves. %
	\cite[sec.~5]{Bracci1990} discusses %
	a specific example regarding this issue. %
	
	Eigenstates (even excited ones) of the electrons Hamiltonian for isolated atoms %
	are normally rather well known, thus,
	in general, the main difficulty in the calculation lies in %
	the correct identification of the electrons final state.
	If the initial electron state is a superposition of several eigenstates $i$, with coefficients $\alpha_i$, %
	the final state will be a superposition of the corresponding new eigenstates, with all coefficients bearing the same modulus. For each eigenstate the above treatment can be applied, obtaining from \cref{eqPotenzialeScreeningAdiabaticLimitAutostatoSingolo} a different screening potential $U_i$.
	It is then found \cite[eq.~(4.15)]{Bracci1990} that the screened cross-section is
	\begin{equation}
	\sigma_e(E) = \sum_i \m{\alpha_i}^2 \sigma_b(E+U_i)
	\end{equation}

	A particularly simple but widely applicable case is the one where both the initial and final states are the ground-state, meaning that there is no degeneracy issue.
	Then, the adiabatic-limit screening potential is just the upper limit given in \cref{eqUpperLimitGeneraleScreeningPotentialAtomico} with $K_e = 0$ (since here $E \to 0$). %
	For instance, this case %
	can be applied to %
	the scattering of a system initially found in the ground state of a neutral \nuclide{Li} and a neutral \nuclide{H},
	which %
	maps into the ground state of neutral \nuclide{Be}, yielding an adiabatic-limit screening potential of \SI{182}{\eV} \cite[tab.~4]{Bracci1990}.
	Similarly, if the proton is bare, the screening potential in the same limit is \SI{186}{\eV} \cite[tab.~4]{Bracci1990}.
	A \nuclide{Li^+} ion impinging on a neutral \nuclide{H} yields $U = \SI{178}{\eV}$, which is still very similar to the value for neutral atoms, but if the lithium is initially missing two electrons (which leaves its inner electronic shell open), then the adiabatic-limit screening potential is \SI{236}{\eV}, as the ground-state-to-ground-state transition is energetically more convenient in this case.
	This shows that an accurate assessment of the beam particles charge state immediately before they trigger a nuclear reaction can have an impact %
	to correctly evaluate screening effects. Given the very low collision energies of interest, it may be expected that beam particles will pick up electrons from the target with relative ease (and that the initial charge state was not very high to begin with), hence assuming a neutral-on-neutral collision may be reasonable. This is the working hypothesis adopted in this study.
	
	If \nuclide{Li} is neutral and there is a neutral \nuclide{H_2} molecule in place of the hydrogen atom, following the reasoning in the text commenting \cref{eqUpperLimitGeneraleScreeningPotentialAtomico}, and considering a \nuclide{Be H} molecule in the final state
	\footnote{Remind that this is the ``final state'' only with regard to the electronic configuration, in the adiabatic limit, immediately before the quantum tunnelling between the nuclei starts.},
	one would deduce a screening potential of \SI{180}{\eV}, which is only marginally different from the purely atomic case (the correction due to the electrons kinetic energy, computed as in \cref{eqEspressioneEnergiaCineticaElettroniInComotoConIReagentiRispettoACentroDiMassa}, is more important even at $E = \SI{10}{\keV}$). For simplicity, corrections due to the molecular state are consequently ignored in the following.
	
	Finally, %
	it is stressed that, while the adiabatic limit sometimes coincides with the theoretical upper limit at zero collision energies, it is not \emph{only} an upper limit: if the screening potential approach and the approximations employed to treat the atomic case are applicable at all, then $U$ is expected to tend precisely to the adiabatic limit for sufficiently small $E$ (which is also the limit where screening corrections are most important). In particular, the theoretical expectation would be violated both by
	an exceedingly small and an exceedingly high screening enhancement at small energies. %

\subsection[The atomic electron screening problem: application to the \texorpdfstring{$\nuclide[6]{Li}+\nuclide{p}\to\nuclide[3]{He}+\nuclide{\alpha}$}{6Li+p->3He+a} reaction]{The atomic electron screening problem:\\application to the $\nuclide[6]{Li}+\nuclide{p}\to\nuclide[3]{He}+\nuclide{\alpha}$ reaction}\label{secScreeningExperimentalData}
	
\begin{figure}[tb]%
		\centering
		\includegraphics[keepaspectratio = true, width=\linewidth]{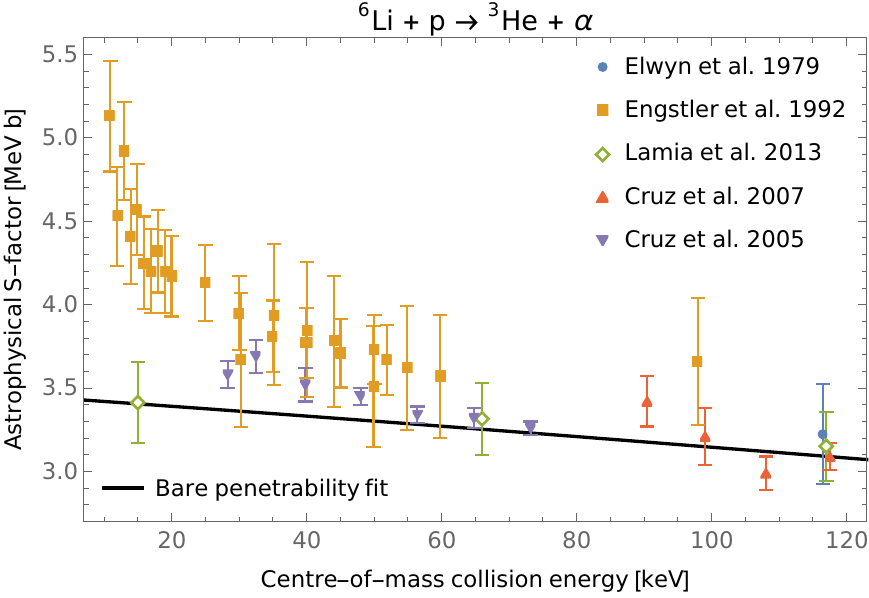}%
		\caption[\texorpdfstring{$\nuclide[6]{Li}+\nuclide{p}\to\nuclide[3]{He}+\nuclide{\alpha}$}{6Li+p->3He+a} low-energy experimental astrophysical factor]{\label{figAllDataLowEnergy}%
			Experimentally measured astrophysical factor (defined in \cref{eqDefinizioneFattoreAstrofisico}) %
			for the $\nuclide[6]{Li}+\nuclide{p}\to\nuclide[3]{He}+\nuclide{\alpha}$ reaction, at collision energies below \SI{120}{\keV}, %
			from \cite{Lamia2013,Elwyn1979,Engstler1992} (same symbols as in \cref{figbareNucleusFit}) and \cite{Cruz2007} (violet downward and red upward triangles for data originally published in 2005 and 2007 respectively).
			The black solid line, shown for comparison, is the same line found in \cref{figbareNucleusFit}.}
	\end{figure}

	\Cref{figAllDataLowEnergy} shows the experimental astrophysical factor for the $\nuclide[6]{Li}+\nuclide{p}\to\nuclide[3]{He}+\nuclide{\alpha}$ reaction %
	from \cite{Lamia2013,Elwyn1979,Engstler1992,Cruz2007} (the same sources of data appearing in \cref{figbareNucleusFit}) and \cite{Cruz2005} (precisely, the updated version %
	shown in \cite{Cruz2007} is employed), at collision energies smaller than \SI{120}{\keV}. More data at higher energies was shown in \cref{figbareNucleusData}. %
	Since the present study is concerned with screening effects due only to atomic electrons, the data shown includes only reactions where reactants can reasonably be considered as atoms (or at most as molecules). In particular, the data in \cite{Cruz2005} regarding metallic targets are not included.
	Apart from the Trojan Horse Method data from \cite{Lamia2013}, %
	the cross-section for all measurements reported in \cref{figAllDataLowEnergy} are expected to be enhanced by the screening due to atomic electrons bound to reactants.
	For comparison, the figure includes the bare-nucleus penetrability fit which reproduces data at higher energies, as shown in \cref{figbareNucleusFit}, but underestimates directly measured data below approximately \SI{50}{\keV}. As expected, a clear disagreement is also found at the lowest energies between the direct measurement from \cite{Engstler1992} and the indirect determination from \cite{Lamia2013}.
	
	The screening effects seen in the data can be studied in several ways, briefly discussed in the following using the $\nuclide[6]{Li}+\nuclide{p}\to\nuclide[3]{He}+\nuclide{\alpha}$ reaction as a concrete example. Some results reported in literature regarding other reactions are also quoted at the end of this section.

\subsubsection{Energy trend of the screening potential}

\begin{figure}[tb]%
		\centering
			\includegraphics[keepaspectratio = true, width=\linewidth]{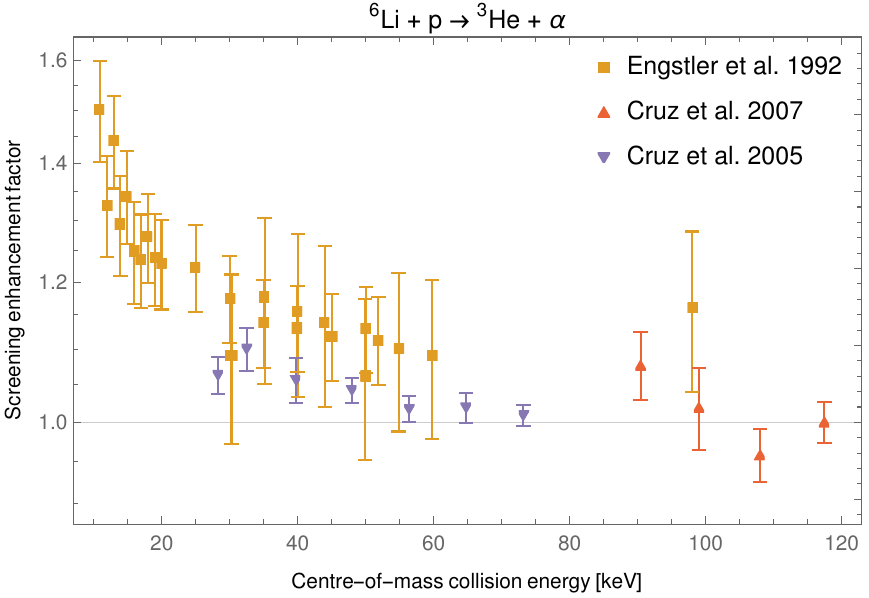}%
			\caption[\texorpdfstring{$\nuclide[6]{Li}+\nuclide{p}\to\nuclide[3]{He}+\nuclide{\alpha}$}{6Li+p->3He+a} screening enhancement factor]{\label{figExperimentalEnhancementFactorAndScreeningPotentialPanelEnhancement}%
				Screening enhancement factor %
				obtained taking the ratio of data points in \cref{figAllDataLowEnergy} to the black solid line in the same figure. The error bars include the propagated uncertainty from the fitted bare-nucleus cross-section.}
		\end{figure}

\begin{figure}[tb]%
			\includegraphics[keepaspectratio = true, width=\linewidth]{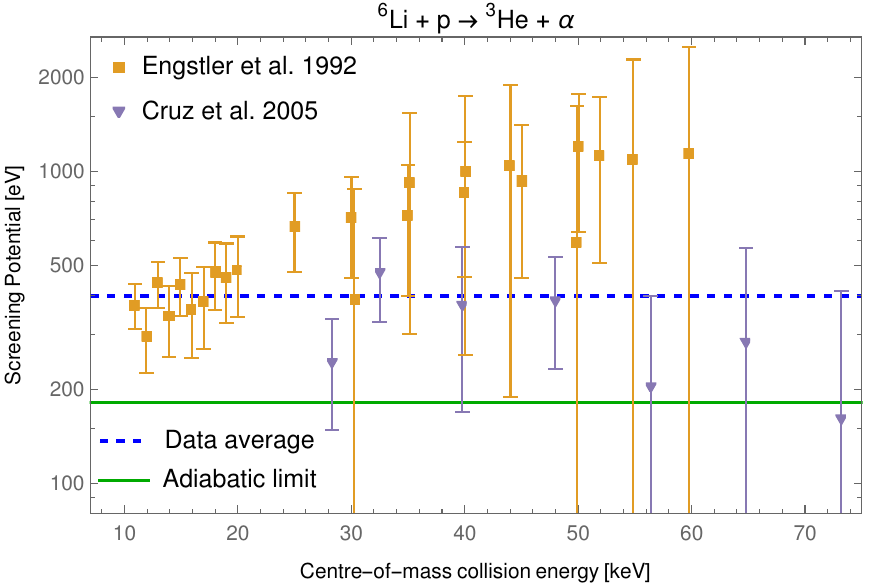}
			\caption[\texorpdfstring{$\nuclide[6]{Li}+\nuclide{p}\to\nuclide[3]{He}+\nuclide{\alpha}$}{6Li+p->3He+a} screening potential]{\label{figExperimentalEnhancementFactorAndScreeningPotentialPanelPotential}%
				Screening potential %
				extracted from %
				the enhancement factor in \cref{figExperimentalEnhancementFactorAndScreeningPotentialPanelEnhancement} %
				using \cref{eqFattoreEnhancementPotScreeningPerAstrophysicalFactorConstante}. %
				The green solid line marks $U = \SI{182}{\eV}$ (adiabatic limit for neutral reactants, see \cref{secAdiabaticLimitAtomicElectronScreening}). The blue solid line is a fit of data at energies below \SI{52}{\keV} on a constant (25 degrees of freedom, $\chi^2$ of 20), whose best fitting value (with its standard error) is \SI{399\pm25}{\eV}.}
	\end{figure}

	If the penetrability fit in \cref{figbareNucleusFit} can be regarded as a faithful estimation of the bare-nucleus cross-section, it can be combined with screened data in \cref{figAllDataLowEnergy} to extract an experimental determination of the screening enhancement factor $f_e$ and the corresponding screening potential $U$, using \cref{eqFattoreEnhancementPotScreeningPerAstrophysicalFactorConstante}. The results are shown in \cref{figExperimentalEnhancementFactorAndScreeningPotentialPanelEnhancement,figExperimentalEnhancementFactorAndScreeningPotentialPanelPotential}.
	The determination of the screening potential is meaningful only at relatively small collision energies, once the deviation between screened and bare cross-section becomes clearly distinguishable within the experimental accuracy. This is partly reflected in the uncertainties obtained for $U$: at higher energies, a given interval of enhancement factors is covered by wider ranges of screening potentials. %
	In the following analysis, only data points at $E < \SI{52}{\keV}$ are taken into account.
	In such region, all data points from \cite{Engstler1992,Cruz2007} suggest a screening potential %
	greater that the adiabatic limit discussed in \cref{secAdiabaticLimitAtomicElectronScreening} (green solid line in \cref{figExperimentalEnhancementFactorAndScreeningPotentialPanelPotential}).
	The datasets are compatible with a constant value for $U$, both separately and combined%
	\footnote{Data from \cite{Engstler1992} seem to suggest a screening potential that increases with energy. %
		However, the uncertainties are too large to draw a conclusive statement.}, %
	even though data from \cite{Engstler1992} favour a higher value.
	The blue dashed line in \cref{figExperimentalEnhancementFactorAndScreeningPotentialPanelPotential} is %
	the weighted average of values extracted from both \cite{Engstler1992} and \cite{Cruz2007}, \SI{399\pm25}{\eV}, which is well beyond the adiabatic limit value. %
	This suggests that the bare-nucleus cross-section shown in \cref{figbareNucleusFit} is incompatible with a screening effect due only to atomic electrons, especially if screened data from \cite{Engstler1992} are taken into account.
	
	The analysis just discussed has the advantage of not requiring to assume a priori any specific functional form for the screening potential %
	with respect to the energy. Its weakness lies in the necessity of first obtaining a trusted bare-nucleus cross-section, which must be derived ignoring all direct data at energies where screening effects are relevant. %

\subsubsection{Fit of direct data on a constant screening potential}
	
	If the screening potential can be approximated to a constant, it is possible to directly fit \cref{eqSezioneDurtoInPotenzialeScreening} with $U$ as free parameter, as commonly done in literature.
	Sometimes, the bare-nucleus cross-section is fitted first, %
	and the %
	enhancement factor over the obtained trend %
	is fitted afterwards using low-energy direct data. This approach is %
	essentially equivalent to computing the weighted average of the observed values of $U$ as in \cref{figExperimentalEnhancementFactorAndScreeningPotentialPanelPotential}%
	\footnote{Since \cref{eqFattoreEnhancementPotScreeningPerAstrophysicalFactorConstante} is not linear, in principle a difference between the fit results could be observed. In practice, the deviations are expected to be small.}. %
	It is adopted %
	especially in older works (e.g.~\cite{Engstler1992}) and when indirect data are taken into account (e.g.~\cite{Lamia2013}), since these are unaffected by screening and thus cannot be fitted together with direct data%
	\footnote{In fact, it is %
		possible, albeit slightly more cumbersome, to make a simultaneous fit of direct and indirect data on different functional forms sharing some parameter, which would allow to implement the sort of fit suggested in \cite{Barker2002} %
		including indirect data. To the author's knowledge this has never been done in literature.}.
	
\begin{figure}[tbp]%
		\centering
		\includegraphics[keepaspectratio = true, width=\linewidth]{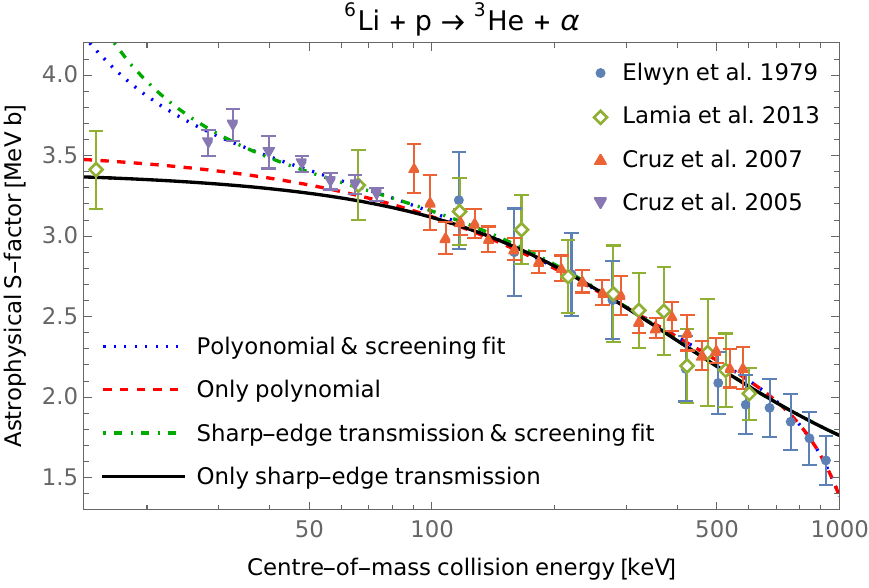}%
		\caption[\texorpdfstring{Fit of $\nuclide[6]{Li}+\nuclide{p}\to\nuclide[3]{He}+\nuclide{\alpha}$}{6Li+p->3He+a} direct data measurements]{\label{figCruzElwynFitBareScreenedInsiemeConfronto}%
			Points are a subset of data in \cref{figbareNucleusFit,figAllDataLowEnergy}, from \cite{Elwyn1979,Cruz2007,Lamia2013}, with identical symbols.
			Blue dotted line is a fit of only data from \cite{Elwyn1979,Cruz2007} (not \cite{Lamia2013}) on a third-order polynomial for the astrophysical factor, $\sum_{n=0}^3 a_n (E/\si{\MeV})^n$, multiplied by the $f_e$ in \cref{eqFattoreEnhancementPotScreeningCasoGenerale}, with the coefficients $a_n$ and the screening potential $U$ as free parameters, whose best-fitting values (with their standard errors) are: $a_0 = \SI{3.54\pm0.08}{\MeV \barn}$, $a_1 = \SI{-4.8\pm0.8}{\MeV\barn}$, $a_2 = \SI{6\pm2}{\MeV\barn}$, $a_3 = \SI{-3\pm2}{\MeV\barn}$, $U = \SI{267\pm129}{\eV}$. %
			Red dashed line is a plot of only the bare-nucleus component of the fitted function.
			Green dot-dashed line is a fit on the same data but using for the bare-nucleus part the same model employed in \cref{figbareNucleusFit}, which has 35 degrees of freedom, returns a $\chi^2$ of 17, and whose best-fitting values and standard errors for the parameters are:
			scaling $A_0 = \num{0.41\pm0.03}$, radius $R_n = \SI{3.27\pm0.13}{\femto\metre}$, screening potential $U = \SI{388\pm100}{\eV}$.
			Black solid line is the bare-nucleus component of this fitted function.
			}
	\end{figure}

	Another possibility, discussed in \cite{Barker2002} and references therein, is to %
	fit both the bare-nucleus cross-section and the screening potential at the same time: the fitted function is the product of a model for the bare-nucleus cross-section (with some free parameters) and the enhancement factor in \cref{eqFattoreEnhancementPotScreeningCasoGenerale} with $U$ as additional free parameter. This was found to %
	yield better fits to direct data and lower values of $U$. Such procedure is followed in %
	\cite{Cruz2007} for the $\nuclide[6]{Li}+\nuclide{p}\to\nuclide[3]{He}+\nuclide{\alpha}$ reaction, using only data from \cite{Elwyn1979,Cruz2007} (including the updated ones from \cite{Cruz2005}) and modelling the bare-nucleus astrophysical factor with a third-order polynomial (4 free parameters), obtaining %
	$U = \SI{273\pm111}{\eV}$,
	which is higher than but compatible with the adiabatic limit. The %
	result could be reproduced in this work %
	under the same conditions.
	However, %
	if the bare-nucleus cross-section is instead fitted using the same model employed in \cref{figbareNucleusFit} (\cref{eqDecomposizioneSezioneDurtoPenetrabilitaPotenzialeCentrale,eqPenetrabilityFactorRMatrix} including only $l=0$ and adding an overall free scaling%
	\footnote{Another fit allowing an $l=1$ contribution was attempted, but as in \cref{secPenetrabilityFittingExperimentalData} the additional component does not contribute to the best-fitting function.}),
	a much higher screening potential is found, \SI{388\pm100}{\eV}, which resembles closely the result shown in \cref{figExperimentalEnhancementFactorAndScreeningPotentialPanelPotential}, albeit which greater uncertainty. The two fits just discussed, using a polynomial or the sharp-edge barrier transmission coefficient, are compared in \cref{figCruzElwynFitBareScreenedInsiemeConfronto}.
	The total (screened) cross-section are essentially the same in both models in the energy range where fitted data is present, %
	but there is a visible difference in the bare cross-section slope %
	at collision energies below about \SI{75}{\keV}, which causes the deviation in the predicted screening potential.%
		\footnote{It is underlined %
		that such difference between the models tends to disappear %
		if bare-nucleus data (from indirect measurements) is supplied for the fit, because in that case the fitted bare-nucleus component %
		is directly constrained. For instance, if data from \cite{Lamia2013} is fitted using \cref{eqPenetrabilityFactorRMatrix}, the excitation function is very similar to the polynomial reported in the same paper.}.
	
	In summary, even taking into account only data from \cite{Elwyn1979} and \cite{Cruz2007} (which reports smaller low-energy cross-sections than \cite{Engstler1992}), and fitting bare cross-section and screening potential simultaneously, the favoured values for the $U$ %
	are significantly higher than the adiabatic limit. %
	The difference can decrease considerably if the bare cross-section is allowed to take a generic functional form (implemented through a polynomial function in the $S$-factor), with respect to the case where the energy trend suggested by Coulomb barrier penetrability is imposed.

	Using the approach just described, %
	excessively high values of the screening potential are %
	similarly found for other reactions. Some examples are listed in \cref{tabPotScreeningAltreReazioni}. As shown above for the $\nuclide[6]{Li}(\nuclide{p},\nuclide{\alpha})\nuclide[3]{He}$ case,
	the screening potential value extracted from data can vary significantly depending on what kind of analysis is performed, and the adiabatic limit value is also subject to small variations depending on the precise reactants state,
	thus a careful comparison of each available determination would be important to obtain a complete picture. On this regard, it can be useful to consult \cite{Lamia2012,Lamia2013,Fang2016,Fang2018}, as %
	they include a list of older measurements for each reaction mentioned in \cref{tabPotScreeningAltreReazioni}. %
	\begin{table}
		\caption[Screening potential determination for several reactions]{\label{tabPotScreeningAltreReazioni}%
		Examples in literature of screening potential experimental determination for several reactions. For each reaction, the table lists (in this order) the reference to a work including a determination %
		for the screening potential, the quoted value and error, %
		and the adiabatic-limit value for the given reaction between neutral atomic reactants. See text for details.}
	\centering
	\begin{tabular}{lcS[table-format=3(3)]S[table-format=3]}
	\multicolumn{1}{c}{Reaction} & Ref. & {$U_e$ exper.\ [\si{\eV}]} & {$U_e$ adiab.~lim.\ [\si{\eV}]} \\ \toprule
	$\nuclide[6]{Li}(\nuclide{p},\nuclide{\alpha})\nuclide[3]{He}$	& \cite{Lamia2013}	& 355 \pm 100	& 182 \\
	$\nuclide[6]{Li}(\nuclide{d},\nuclide{\alpha})\nuclide{\alpha}$	& \cite{Musumarra2001}	& 320 \pm 50	& 182 \\
	$\nuclide[7]{Li}(\nuclide{p},\nuclide{\alpha})\nuclide{\alpha}$	& \cite{Lamia2012}	& 425 \pm 60	& 182 \\ %
	$\nuclide[9]{Be}(\nuclide{p},\nuclide{\alpha})\nuclide[6]{Li}$	& \cite{Fang2018}	& 545 \pm 98	& 258 %
	\end{tabular}
	\end{table}
	The collection of these anomalous observations forms the so-called ``(\emph{atomic}) electron screening problem''. Apart from already quoted papers presenting relevant experimental measurements, %
	this long-standing problem was discussed in a number of data re-analyses and theoretical studies, some examples being %
	\cite{Rolfs1995,Barker2002,Hagino2002,Fiorentini2003,Wang2011,Spitaleri2016}.
	
	As seen, the analysis approach just discussed is rather advantageous, as it %
	allows to employ all available data to fit both the bare cross-section and the screening effects, surpassing the limit of the analysis performed to obtain \cref{figExperimentalEnhancementFactorAndScreeningPotentialPanelPotential}, at the affordable cost of requiring a model for the enhancement factor functional form (in particular, a constant screening potential is usually adopted). %
	Nonetheless, it is relevant to point out that such analysis is normally employed mainly to establish whether the screening potential suggested by experimental data is compatible or not with the theoretical constraints: some care is due when applying the analysis results in this way. %
	In particular, while the standard error associated to the fitted value of $U$ is definitely an useful indicator, it does not correspond to the interval of values of $U$ which keep the total $\chi^2$ within a given interval for any possible choice of the other parameters. %
	A study of the multi-dimensional confidence region of the fit parameters would be required to establish that in a general manner. In this sense, the results reported in \cref{tabPotScreeningAltreReazioni} can risk being misleading. However, a computationally much simpler approach can be adopted to address the problem at hand, as discussed in the following.

\subsubsection{Analysis of direct data rescaled by adiabatic-limit screening}

	First note that if the fit in \cref{figCruzElwynFitBareScreenedInsiemeConfronto} is repeated constraining the maximum value for the screening potential with the adiabatic limit, the fitted value will %
	simply equal the maximum allowed. %
	This was predictable, given that the unconstrained fit suggests values of $U$ which are greater than the upper limit.
	Having acknowledged that data favour higher values of screening potential, it remains to be established whether experiments are at least compatible with the theoretical expectation. To this end, assume in the following that $U$ is constant with respect to the collision energy, $E$, and exactly equals the adiabatic limit. %
	This is though to be a sufficiently good approximation at this level, since the %
	screening potential is in fact expected to reach the adiabatic limit for small $E$, %
	while at higher energies %
	the associated enhancement factor is in any case small.
	Having fixed this %
	``null hypothesis'', it is then possible to examine its likelihood.
	For instance, data can be fitted on a reasonable model for the bare-nucleus cross-section: this is equivalent to repeating the fits in \cref{figCruzElwynFitBareScreenedInsiemeConfronto} after fixing the value of $U$. The quality of the fit %
	can then be employed as an indication of the likelihood of the assumed scenario.
	This sort of statistical analysis may be performed in several ways; here, only one straightforward %
	approach is shown for illustrative purposes. %
	Let $F_n(x)$ %
	be the cumulative distribution function of the chi-square distribution with $n$ degrees of freedom, and $Q_n(x) = 1 - F_n(x)$. %
	Given a fit with $n$ degrees of freedom and a chi-square%
	\footnote{This is %
		the variance of data with respect to the prediction given by the fitted function, weighted on the experimental uncertainties (if $\Delta$ is the uncertainty on a given data point, the associated weight is $1/\Delta^2$).} %
	of $\chi^2$, and assuming that data indeed come from a random distribution with expectation value equal to the best-fitting function, %
	$Q_n(\chi^2)$ %
	is the probability that a random sample extracted from the ``true'' distribution, with the same dimension of the fitted sample, returns a chi-square equal or greater than that found in the fit.
	Often, %
	when $\chi^2 \lesssim n$, which corresponds to $Q_n(\chi^2) \gtrsim 1/2$, %
	the quality of a fit is considered sufficient%
	\footnote{The chosen confidence level can vary depending on the context. If the fitted model is considered reasonable a priori, selection criteria such as $F_n(\chi^2) %
		\leq \SI{90}{\percent}$ can be found.}. %
	Note that, if several different datasets (which may be in disagreement) are fitted together, it is better to analyse separately the quality of the fit on each dataset, to avoid compensations in the aggregated $\chi^2$ which may hide specific issues. %
	
\begin{figure}[tbp]%
		\centering
		\includegraphics[keepaspectratio = true, width=\linewidth]{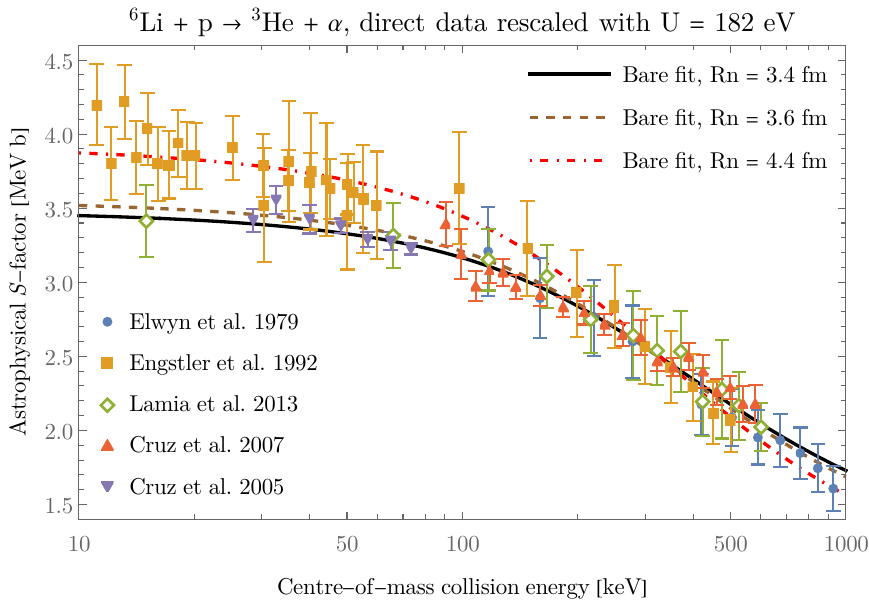}%
		\caption[\texorpdfstring{$\nuclide[6]{Li}+\nuclide{p}\to\nuclide[3]{He}+\nuclide{\alpha}$}{6Li+p->3He+a} data rescaled by adiabatic screening]{\label{figDatiCorrettiAdiabatic}%
			Green open diamonds are the same data in \cref{figbareNucleusFit} from \cite{Lamia2013}. All other points represent data in \cref{figbareNucleusFit,figAllDataLowEnergy} from \cite{Elwyn1979,Engstler1992,Cruz2007}, with identical symbols, after division by an enhancement factor computed using \cref{eqFattoreEnhancementPotScreeningPerAstrophysicalFactorConstante} and $U = \SI{182}{\eV}$.
			Black solid line is a fit of data from \cite{Elwyn1979,Cruz2007,Lamia2013} below \SI{1}{\MeV} on the sharp-edge barrier transmission model employed in \cref{figbareNucleusFit}, which has 48 degrees of freedom, returns a $\chi^2$ of 22, and whose parameter best-fitting values and standard errors are $A_0 = \num{0.37\pm0.02}$ and $R_n = \SI{3.44\pm0.09}{\femto\metre}$.
			Red dot-dashed line is a fit on the same model but using data from \cite{Elwyn1979,Engstler1992} below \SI{1}{\MeV}, which has 42 degrees of freedom, returns a $\chi^2$ of 9, and parameter values and standard errors of $A_0 = \num{0.24\pm0.02}$ and $R_n = \SI{4.36\pm0.17}{\femto\metre}$.
			Brown dashed line is a fit on the same model of all data in the picture, which has 81 degrees of freedom, returns a total $\chi^2$ of 85 (however see text), and whose parameter best-fitting values and standard errors are $A_0 = \num{0.34\pm0.02}$ and $R_n = \SI{3.63\pm0.09}{\femto\metre}$.
		}
	\end{figure}

	\Cref{figDatiCorrettiAdiabatic} shows data in \cref{figAllDataLowEnergy} divided by the enhancement factor connected to $U = \SI{182}{\eV}$.
	If the fit appearing in \cref{figCruzElwynFitBareScreenedInsiemeConfronto} on the sharp-edge barrier transmission model (\cref{eqDecomposizioneSezioneDurtoPenetrabilitaPotenzialeCentrale,eqPenetrabilityFactorRMatrix} including only the $l=0$ term and adding an overall scaling factor $A_0$) is repeated here, precisely on the same data (those from \cite{Elwyn1979,Cruz2007}, including the updated version of data from \cite{Cruz2005}, at energies below \SI{1}{\MeV}), but fixing $U$ to the adiabatic limit, a $\chi^2$ of 21 is returned, corresponding to $Q_n(\chi^2) = \SI{97.5}{\percent}$ for $n = 36$. %
	This is not appreciably worse than the quality obtained allowing $U$ to vary freely ($\chi^2$ of 17, see \cref{figCruzElwynFitBareScreenedInsiemeConfronto}), even though in that case the best-fitting value for the screening potential was $\SI{388}{\eV}$ with standard uncertainty of $\SI{100}{\eV}$. %
	In summary, %
	data from \cite{Elwyn1979,Cruz2007} appear in fact to be compatible with non-anomalous screening effects, even if the sharp-edge barrier transmission model is adopted for the bare-nucleus cross-section (without need to resort to a generic polynomial function).
	
	Once direct data are corrected by the expected screening effects (and provided that such expectation was correct), they represent the bare-nucleus excitation function, just as indirect data unaffected by screening. All datasets can thus be easily fitted together. %
	In particular, the black solid line in \cref{figDatiCorrettiAdiabatic} is a fit on the same sharp-edge barrier transmission model discussed above, on data from both \cite{Elwyn1979,Cruz2007} and \cite{Lamia2013} at collision energies below \SI{1}{\MeV}: the results are very similar to those found without data from \cite{Lamia2013}.

	The situation is different regarding data from \cite{Engstler1992}. If only data from \cite{Elwyn1979,Engstler1992} at collision energies below \SI{1}{\MeV} are fitted using the sharp-edge barrier transmission model, the red dot-dashed line in \cref{figDatiCorrettiAdiabatic} is obtained, which reproduces well the fitted data. However, if a fit of all data from \cite{Elwyn1979,Cruz2007,Engstler1992,Lamia2013} at energies below \SI{1}{\MeV} is attempted, the brown dashed line in \cref{figDatiCorrettiAdiabatic} is obtained. While the total $\chi^2$ of the fit appears acceptable, a visual inspection reveals that the fitted curve is almost identical to the black solid line in the same figure, because the quoted errors in \cite{Cruz2007,Lamia2013} are smaller than those in \cite{Engstler1992}, and that data from \cite{Engstler1992} are consequently not reproduced: quantitatively, the 33 points from \cite{Engstler1992} accrue a $\chi^2$ of 54 (out of the total of 85 for all datasets together), which may be associated to a $Q_{n=33}(\chi^2=54) = \SI{1.3}{\percent}$. Note that the discrepancy is mainly due to the points at lowest energies, approximately below \SI{20}{\keV}, because at higher energies data from \cite{Engstler1992} and \cite{Cruz2007} are mostly compatible, albeit in strong tension. The conclusion is unaltered if indirect data from \cite{Lamia2013} are discarded.
	
	In summary, if anomalous screening effects are excluded, %
	the sharp-edge barrier penetrability model (\cref{eqDecomposizioneSezioneDurtoPenetrabilitaPotenzialeCentrale,eqPenetrabilityFactorRMatrix} with free scaling of each component%
	\footnote{The pictures in this work only show fits restricting to the $l=0$ component. Another fit was performed allowing also an $l=1$ component and excluding indirect data, but the differences in the fitted low-energy cross-section are insignificant.})
	cannot reproduce \emph{all} direct data considered in the present analysis together. %
	It is mentioned that, in this work, an attempt was made %
	to include all known sources of uncertainty in the data (see \cref{appExperimentaldatasourcesandtreatment}; for instance, smaller errors are adopted in the analysis in \cite{Wang2011}): this increases the significance of observed incompatibilities. %

	The analysis approach just discussed does not provide a quantitative estimation of screening effects, but on the contrary suggests whether a predetermined model for the screening is compatible with data, and it does so in a more transparent manner with respect to the results obtained by a fit of the screening potential on data. A similar study may be applied to other reactions for which anomalous screening effects are reported in literature, to understand whether some of the experimental results %
	are in fact compatible with theoretical predictions.

\subsubsection{Non-standard bare-nucleus cross-section}

	The observed discrepancies regarding atomic screening effects may, of course, %
	be %
	due to experimental issues, which would however be beyond the scope of the present work. The rescaling of data in \cite{Engstler1992} considered in \cite{Wang2011} and \cite{Engstler1992} itself (see \cref{secPenetrabilityFittingExperimentalData} for a brief discussion) essentially goes in this direction.
	Another alternative is that screening effects are in fact stronger than expected from the discussion in \cref{secScreeningByAtomicElectrons}. %
	On this regard, note that there are several effects, discussed in \cite{Fiorentini2003} and references therein, %
	that are not directly related to electron screening but may yield variations in the low-energy cross-section. %
	However, none of these effects appears to be sufficiently important to explain the experimental anomalies.
	
	Finally, if the two above scenarios were excluded, it would follow that the bare-nucleus cross-section is affected by some mechanism, related to the reactants bare interaction (and not \emph{exclusively} the surrounding environment), which alters the energy trend with respect to a standard Coulomb-barrier penetrability model. %
	From the phenomenological point of view, it is interesting to check what sort of models may suffice to obtain such result, keeping in mind that no theoretical basis for such models is provided.
	This is done in the following for the $\nuclide[6]{Li}(\nuclide{p},\nuclide{\alpha})\nuclide[3]{He}$ case studied above.

\begin{figure}[tbp]%
		\centering
		\includegraphics[keepaspectratio = true, width=\linewidth]{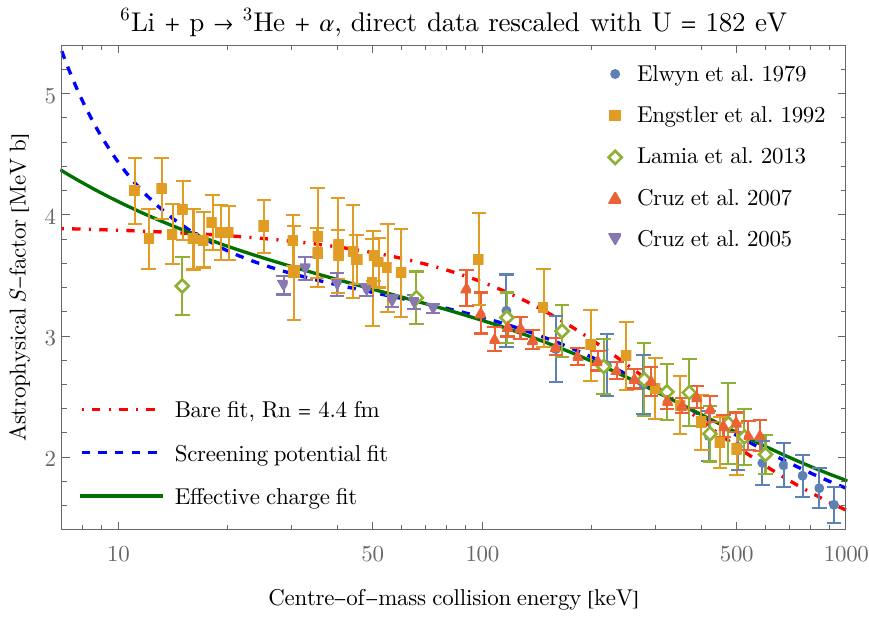}%
		\caption[Non-standard phenomenological models for \texorpdfstring{$\nuclide[6]{Li}+\nuclide{p}\to\nuclide[3]{He}+\nuclide{\alpha}$}{6Li+p->3He+a}]{\label{figDatiCorrettiAdiabaticFitAnomali}%
			Points and red dot-dashed line (shown for comparison) are the same found in \cref{figDatiCorrettiAdiabatic}. %
			Blue dashed line is a fit of screened data from \cite{Elwyn1979,Engstler1992,Cruz2007} below \SI{1}{\MeV} (as shown in \cref{figbareNucleusFit,figAllDataLowEnergy}) on the sharp-edge barrier transmission model employed in \cref{figbareNucleusFit} enhanced according to \cref{eqFattoreEnhancementPotScreeningCasoGenerale} with a screening potential $U$ kept as free parameter. To be %
			comparable, the line is here shown with the same rescaling applied to direct data (see \scref{figDatiCorrettiAdiabatic}). %
			The fit has 68 degrees of freedom and returns a $\chi^2$ of 41; the best-fitting values and standard errors of its parameters are %
			$A_0 = \num{0.39\pm0.02}$, $R_n = \SI{3.34\pm0.10}{\femto\metre}$ and $U = \SI{399\pm32}{\eV}$.
			Green solid line is a fit of data of data from \cite{Elwyn1979,Engstler1992,Cruz2007} in this picture on the ``effective charge'' model discussed in text for the bare-nucleus astrophysical factor, which has 68 degrees of freedom, returns a $\chi^2$ of 40, and whose parameter values and standard errors are $A_0 = \num{0.50\pm0.05}$, $R_n = \SI{2.86\pm0.15}{\femto\metre}$, and $Z_1 Z_2 = \num{2.968\pm0.005}$.
		}
	\end{figure}

	Similarly to what was found when fitting the screening potential value (see \cref{figCruzElwynFitBareScreenedInsiemeConfronto} and commenting text), the incompatibility seen in \cref{figDatiCorrettiAdiabatic} is ameliorated
	if indirect data from \cite{Lamia2013} are excluded %
	and a polynomial form is assumed for the bare-nucleus cross-section, as this is more flexible (and has more free parameters) with respect to a barrier penetrability model. Nonetheless, the qualitative result is similar to the brown line in \cref{figDatiCorrettiAdiabatic}. %
	In order to reproduce satisfactorily both data from \cite{Engstler1992} and from the other direct measurements considered in the present analysis (\cite{Elwyn1979} and \cite{Cruz2007}, including the updated version of data from \cite{Cruz2005}), %
	a rather sharp variation in the bare-nucleus astrophysical factor starting from collision energies of about \SI{20}{\keV} is necessary. %
	One way to achieve the requested behaviour is assigning an anomalous screening potential, as normally done in literature. The blue dashed line in \cref{figDatiCorrettiAdiabaticFitAnomali} is a fit on such model, returning a total screening potential of \SI{399}{\eV} (compare with the similar result in \cref{figExperimentalEnhancementFactorAndScreeningPotentialPanelPotential} obtained trough a different approach). Note that the line shown in the figure is already corrected by the adiabatic-limit enhancement, and thus represents the (anomalous) bare-nucleus cross-section associated to this model.
	
	A different kind of parametrisation was proposed in \cite{Spitaleri2016}:
	the bare-nucleus cross-section is written as the sum of %
	a standard term, based on a barrier-penetrability model (for instance \cref{eqPenetrabilityFactorRMatrix}), and an additional contribution constructed as the standard one but featuring a smaller Coulomb barrier, to phenomenologically account for a dynamical deformation of the system during the reactants approach.
	A fit on such model was attempted here, but the %
	indetermination on the parameters of the %
	non-standard term are too high for the result to be meaningful. %
	The basic idea behind the parametrisation given in \cite{Spitaleri2016} was here employed to formulate a different construction, discussed in the following. %
	Consider again the sharp-edge barrier transmission model employed earlier for the bare-nucleus cross-section ($l=0$ term of \cref{eqDecomposizioneSezioneDurtoPenetrabilitaPotenzialeCentrale,eqPenetrabilityFactorRMatrix} times an overall scaling factor $A_0$), but let the product of the reactants electric charge, $Z_1 Z_2$ in \cref{eqDefinizioneParametroSommerfeld}, be a free parameter: this affects the Coulomb wave-function $H^+$ appearing in \scref{eqPenetrabilityFactorRMatrix}. %
	It is underlined that the corresponding astrophysical factor (defined in \cref{eqDefinizioneFattoreAstrofisico}) is always computed using the bare values of the reactant charges. The same applies when the bare cross-section is multiplied by the standard adiabatic-limit enhancement to compare it with screened (directly measured and ``uncorrected'') data, as those in \cref{figAllDataLowEnergy}. Consequently, small deviations from the standard value of the parameter $Z_1 Z_2$ in the cross-section take here the same role of deviations from the adiabatic-limit screening potential in the ordinary formulation (blue dashed line in \scref{figDatiCorrettiAdiabaticFitAnomali}). A fit on the ``effective charge'' model just discussed of the data under study is shown as the green solid line in \cref{figDatiCorrettiAdiabaticFitAnomali}.

	Both the anomalous-screening-potential model and the effective-charge model (as the two-penetrabilities model in \cite{Spitaleri2016}) %
	predict a positively diverging bare-nucleus astrophysical factor toward zero collision energy.
	However, the ``speed'' of divergence is very different in the two cases. This is most promptly seen considering the trend of the corresponding cross-section $\sigma$. In the effective-charge model, for any $Z_1 Z_2 > 0$ it still is
	$\sigma(E\to0)\to0$, %
	as in the standard case: however small is the effective charge, it still provides a barrier which is penetrated with vanishing probability at small energies.
	The screening potential model, instead, for any screening potential $U$ greater than zero, predicts
	$\sigma(E\to0)\to\sigma_b(U)$ %
	(see \cref{eqSezioneDurtoInPotenzialeScreening}), which is finite.
	This is just connected to a failure of the screening potential approach at vanishing %
	collision energies. %
	As briefly mentioned in \cref{secScreeningByAtomicElectrons}, the relative error committed on the screened cross-section evaluation is of the order of $U/E$: taking into account the adiabatic-limit value for $U$, the adopted formalism is expected to be reasonable only for $E \gtrsim \SI{2}{\keV}$.
	Consequently, the predictions of non-standard models considered here are relevant only within the same region. %
	On this regard, as can be seen also from \cref{figDatiCorrettiAdiabaticFitAnomali}, the effective-charge model %
	yields the %
	the most gradual variation %
	in the astrophysical factor, $S(E)$, in the $E \leq \SI{10}{\keV}$ region (also with respect to the model in \cite{Spitaleri2016}). For comparison, the green solid line in \cref{figDatiCorrettiAdiabaticFitAnomali} returns $S(\SI{2}{\keV}) = \SI{6.0}{\MeV \barn}$, %
	while the blue dashed line in the same picture returns $S(\SI{2}{\keV}) = \SI{48}{\MeV \barn}$.
	In this sense, the proposed model can represent a more attractive alternative (with respect to the others currently present in literature) to phenomenologically study the observed deviations, and in particular the %
	astrophysical impact %
	of the scenario %
	where the %
	anomalies are fully genuine and attributed to bare-nucleus cross-sections. %
\chapter[Ground-state properties of clustered systems]{Ground-state properties\\of clustered systems}\label{chaLinkingCLusterModelToObservables}%
	In the quantum-mechanical formalism, all accessible information regarding the structure of a system can be encoded in its wave-function,
	and in particular as the associated expectation value of appropriate operators. %
	On one hand, the comparison between experimentally observed quantities and the predictions given by a model, attempting to describe such structure, provides an important benchmark to constraint and assess the accuracy of the model itself.
	From the other, the structure information can then be employed as an ingredient to predict the outcome of interactions between the systems under study. %
	While the concepts discussed in this \namecref{chaLinkingCLusterModelToObservables} can in principle be applied to any bound system whose components are each bound in isolation,
	most of the available structure information
	on atomic nuclei
	involves their ground state, which can be %
	examined experimentally with greater ease. %

	This \namecref{chaLinkingCLusterModelToObservables} opens with a succinct review, in \cref{secAntiSymmetrisationFirstQuantisation}, of some general notions regarding anti-symmetrisation which will be useful later. %
	\Cref{secOverlapFunctionSpectroscopicFactors} then %
	covers %
	the formalism of overlap functions. %
	After reviewing the general theory, the %
	discussion is devoted to the explicit derivation of some properties %
	of the overlap function and the value of associated spectroscopic amplitudes %
	within the independent-particle shell model.
	Lastly, \cref{secLegameProprietaCompositoEFunzioneDonda} reports the derivation, from general principles, of formulas to explicitly compute some structure observables
	of interest on a cluster-model wave-function: %
	the results will be employed in \cref{secCalcoliDiTransfer} to check the properties of the overlap functions adopted for the transfer calculations.

\section[Anti-symmetrisation in first-quantisation formalism]{Anti-symmetrisation in first-quantisation\texorpdfstring{\\}{ }formalism}\label{secAntiSymmetrisationFirstQuantisation} %

	The following pages will briefly introduce %
	some basic concepts regarding particle indistinguishability in quantum mechanics and wave-function anti-symmetrisation for systems of identical fermions. It is not of interest here to present a complete survey of %
	the theory behind these concepts: %
	in this \namecref{secAntiSymmetrisationFirstQuantisation}, the results of interest will only be %
	stated without derivation. More in-depth treatments can be found in different books, for instance~\cite[ch.~XIV]{Cohen1977}.

	Consider an orthonormal basis $\{\ket{\nu_k}\}$ of single-particle states for a certain species of particles, where $\nu_k$ represents the quantum number(s) describing the state%
	\footnote{A typical example is a set of states with fixed spin and isospin projection, $\sigma$ and $\tau$, and position, $\v r$, so that each “$\nu_i$” stands for “$\v r_i, \sigma_i, \tau_i$”.}, and let $V$ be the set of values that each single-particle $\nu_k$ can take (i.e.\ it must be $\nu_k \in V$).
	Given a generic single-particle state $\Ket{\psi}$, it is possible to express it as a superposition of the basis states,
	\begin{equation}
	\Ket{\psi} = \sum_{\nu \in V} \psi(\nu) \Ket{\nu}
	\end{equation}
	where $\psi(\nu)$ is a shorthand for $\Braket{\nu|\psi}$, and the symbol $\sum_{\nu \in V}$ is employed to imply a sum over all states in the space $V$, regardless of its cardinality (i.e.\ regardless of whether the sum should actually be a discrete sum or an integral).
	
	Consider now a system of $A$ identical fermions. It is postulated that its only physical states are those anti-symmetric with respect to particle exchange.
	Let
	$\Ket{\psi_1, \dots, \psi_A}$ %
	be a state where
	particle 1 is in the single-particle state $\Ket{\psi_1}$ and so on (thus $\psi_1,\dots,\psi_A$ is an ordered set): this clearly does not respect the required statistics. %
	It is then useful to define an anti-symmetrisation operator $\op W$, a projection which maps any generic $A$-particle state into another respecting the fermion statistics while involving only the single-particle states found in the original state. %
	
	The application of $\op W$ on the generic state $\Ket{\psi_1, \dots, \psi_A}$ %
	can be expressed %
	through the following relation: %
	\begin{equation}\label{eqApplicazioneOperatoreAntisimmetrizzazione}
		\Braket{\nu_1, \dots, \nu_A|\op W|\psi_1, \dots, \psi_A} = \frac{1}{A!} \begin{vmatrix}
			\psi_1(\nu_1) & \psi_1(\nu_2) & \dots & \psi_1(\nu_A) \\
			\psi_2(\nu_1) & \ddots & \vdots & \psi_2(\nu_A) \\
			\vdots & \dots & \ddots & \vdots \\
			\psi_A(\nu_1) & \psi_A(\nu_2) & \dots & \psi_A(\nu_A)
		\end{vmatrix}
	\end{equation}
	where
	$!$ denotes a factorial and %
	$\begin{vmatrix}\dots\end{vmatrix}$ is %
	a matrix determinant. %
	$\op W\Ket{\psi_1, \dots, \psi_A}$ itself is often referred to as a “Slater determinant”. %
	Note that, if the single-particle states $\Ket{\psi_i}$ are distinct and form an orthonormal set, then $\sqrt{A!} \, \op W \Ket{\psi_1, \dots, \psi_A}$ is normalised to~1.
	
	Finally, consider the elements $\Ket{\nu_1,\dots,\nu_A}$ of the orthonormal basis for the complete $A$-particle system: in general, these states do not respect the required statistics and are thus non-physical. %
	A completely generic state $\Ket{\Psi}$ for the $A$-particle system can be certainly written as a superposition such as
	\begin{equation}\label{eqEspansioneStatoFisicoAParticelleBaseNonAntisimmetrizzata}
		\Ket{\Psi} = \sum_{\nu_1 \in V}\dots \sum_{\nu_A \in V} \Psi(\nu_1,\dots,\nu_A) \Ket{\nu_1,\dots,\nu_A}
	\end{equation}
	If $\Ket{\Psi}$ is a physical state, indistinguishability guarantees that $\Psi(\nu_1,\dots,\nu_A)$ changes only by a global sign under exchange of any single pair of particles, e.g.
	\begin{equation}\label{eqEsempioDefinizioneAntisimmetrizzazioneSuCoefficientiStatoCompleto}
	\Psi(\nu_1,\nu_2,\dots,\nu_A) = - \Psi(\nu_2,\nu_1\dots,\nu_A)
	\end{equation}
	This also implies that $\m{\Psi(\nu_1,\dots,\nu_A)}$ is conserved under any permutation of particles. %
	Note that the distinct elements of the set %
	$\{ \op W \Ket{\nu_1,\dots\nu_A} \}_{\nu_1 \in V, \dots, \nu_A \in V}$, once normalised, %
	form a physical (i.e.\ respecting the statistics) basis for the $A$-particles system. For brevity, label $\Ket{\nu_1,\dots\nu_A}_W$ the elements of the orthonormal physical basis. %
	The physical state $\Ket{\Psi}$ can then be expanded as
	\begin{equation}\label{eqEspansioneStatoFisicoAParticelleBaseSlater}
		\Ket{\Psi} = \sum_{\{\nu_1,\dots,\nu_A\}} \Psi_W(\nu_1,\dots,\nu_A) \Ket{\nu_1,\dots,\nu_A}_W
	\end{equation}
	where the sum now runs only over the distinct unordered sets $\{\nu_1,\dots,\nu_A\}$. The coefficients moduli in the two basis are simply connected by the relation
	\begin{equation}\label{eqLegamePesiBaseNonSimmetrizzataEDiSlater}
	\m{\Psi_W(\nu_1,\dots,\nu_A)}^2 = A! \m{\Psi(\nu_1,\dots,\nu_A)}^2
	\end{equation}
	since \cref{eqEspansioneStatoFisicoAParticelleBaseSlater} is constructed from \cref{eqEspansioneStatoFisicoAParticelleBaseNonAntisimmetrizzata} just by grouping together terms differing only by a permutation of particles.
	
	When treating anti-symmetrisation of a state describing a nucleus, it can be convenient to treat proton and neutrons as different isospin projections of an identical particle, a “nucleon”%
	\footnote{It is then still possible to define interactions which depend on the isospin, and possibly do not conserve the isospin modulus.},
	so that there is only a single species to account for. The cost of such approach is to neglect the mass difference between proton and neutron. Also note that, within the non-relativistic formulation here employed, the nucleus mass coincides with the sum of each nucleon mass, and it is consequently proportional to the number of nucleons, $A$.

\section{Overlap functions and spectroscopic factors}\label{secOverlapFunctionSpectroscopicFactors}
	
	In a direct one-step transfer reaction (see \cref{secReactionTheory}) the transferred system is passed from a reactant to the other one without altering its internal state \cite[sec.~4]{Thompson2013},
	and the structure of the involved nuclei, in particular
	the affinity between the reactants wave-functions before and after the transfer,
	consequently plays a relevant role.
	This concept can be formally encoded through \emph{overlap functions}, which constitute the subject of the present \namecref{secOverlapFunctionSpectroscopicFactors}.
	The topic is discussed, for instance, in \cite{Satchler1983Direct} (more detailed references are given later) and \cite[sec.~8.1]{Austern1970direct}.

\subsection{General treatment}\label{secOverlapFunctionSpectroscopicFactorsIntroduzione}

	Let $B$ be any nucleus (“composite”), and $b$ another nucleus composed by an arbitrary subset of $B$ nucleons (“core”). Let $A_i$ be the number of nucleons within nucleus $i$, and $A_x = A_B - A_b$, which is always positive. %
	Further let $\Ket{\Psi^B_{J^\pi_B, M_B, T_B, \tau_B}}$ and $\Ket{\Psi^b_{J^\pi_b, M_b, T_b, \tau_b}}$ be some states of interest for $B$ and $b$ respectively. Each state is anti-symmetric and is an eigenstate of parity and of the modulus and $z$-projection of total spin and %
	isospin, with quantum numbers $\pi_i,J_i,M_i,T_i,\tau_i$ respectively (spin modulus and parity numbers are often compacted into ``$J^\pi$'' for brevity). %
	The optimal reference frame in which to compute each $\Psi$ depends on the application: when the overlap function (see \cref{eqFractionalParentageExpansionProiettata} below) is to be computed explicitly from the $\Psi$, it is usually most convenient to set a single coordinate system for both $\Psi^B$ and $\Psi^b$ (for instance using $B$ centre-of-mass rest-frame). Here, since the overlap is manipulated only at formal level and the goal is to study the relative motion of clusters within $B$,
	it is assumed that each $\Psi$ is computed fixing the nucleus centre-of-mass in the origin, and thus depends only on $A_i-1$ spatial coordinates.

	In the following, consider an orthogonal basis of states, $\{ \Psi^b \}$, spanning the whole $A_b$-nucleons space. For simplicity, assume
	that each state is normalised%
	\footnote{Strictly speaking, this is in fact not possible %
		for continuum states, for which some other convention is to be chosen. However, in practical applications, unbound states are hardly considered within this framework, and for the purposes of the present work it is sufficient to simply ignore the issue, which would complicate the formalism without adding physical insight.}
	to 1, and employ the symbol $\sum_b$ to imply a sum over all elements of the basis $\{ \Psi^b \}$ (the number of nucleons $A_b$ is fixed).

\subsubsection{Fractional parentage expansion and overlap function} %

	Any given state $\Psi_B$
	can be expressed
	through the %
	\emph{fractional parentage expansion} \cite[eq.~(16.12), (16.17), (16.33)]{Satchler1983Direct}:
	the full wave-function is expressed as a superposition of anti-symmetrised products of each possible wave-function for some core nucleus $b$ and a corresponding wave-function for the remaining system $x$ of $A_x$ nucleons (“valence”).
	In a rather compact notation, 
	\begin{equation}\label{eqFractionalParentageExpansionGenerica}
	\Ket{\Psi^B_{J^\pi_B, M_B, T_B, \tau_B}} = \sum_{b,T_x,\transfAngMom} N_{b,T_x,\transfAngMom} \op W \[ \Ket{\Phi^{x}_{T_x, \transfAngMom}} \Ket{\Psi^b_{J^\pi_b, M_b, T_b, \tau_b}} \]
	\end{equation}
	where $\op W$ is the anti-symmetrisation operator defined in \cref{secAntiSymmetrisationFirstQuantisation} and each $N$ is an appropriate coefficient.
	Each $\Phi^{x}_{T_x,\transfAngMom}$ is a wave-function in the space of states at $A_x$ nucleons,
	anti-symmetric under exchange of any pair of its nucleons, %
	and
	accounting for \emph{both} their internal motion and the relative motion between the centre-of-masses of $b$ and $x$. %
	Such wave-function %
	is an eigenstate of total angular momentum and isospin, with modulus quantum numbers $\transfAngMom$ and $T_x$ respectively, and $z$-projections $M_B - M_b$ and $\tau_B - \tau_b$, and is similarly an eigenstate of parity with quantum number $\pi_B \pi_b$.
	It is stressed that $\transfAngMom$ is the total transferred angular momentum, namely it includes both the intrinsic spin of the transferred system, $\intrSpin_x$, and its orbital angular momentum with respect to the core, $l$.
	The wave-function $\Phi$ is normalised following a convention similar to that adopted for $\Psi$ (see \cref{eqEspansioneModelloAClusterGenerica} below), and is such that is satisfies \cref{eqFractionalParentageExpansionGenerica} (which is thus basically its definition).
	The sum over $T_x,\transfAngMom$ runs over all (compatible) values of these quantum numbers. %

	It is of interest to consider the transition between a well-defined pair of states for core nucleus $b$ and composite nucleus $B$. To this end, from the %
	expansion in \cref{eqFractionalParentageExpansionGenerica} only one term at a time is relevant, %
	namely the projection of $B$ state on the components including only the desired $b$ state. In particular, consider the \emph{overlap function}
	\begin{multline}
	\Braket{\Psi^b_{J^\pi_b, M_b, T_b, \tau_b} | \Psi^B_{J^\pi_B, M_B, T_B, \tau_B}} =\\= \sum_{T_x,\transfAngMom} N_{b,T_x,\transfAngMom} \Bra{\Psi^b_{J^\pi_b, M_b, T_b, \tau_b}} \op W \[ \Ket{\Phi^{x}_{T_x,\transfAngMom}} \Ket{\Psi^b_{J^\pi_b, M_b, T_b, \tau_b}} \]
	\end{multline}
	
	When computing the overlap, $\Ket{\Psi^b}$ is assigned to an arbitrary but fixed subspace of the full $A_B$ nucleons space (for instance, the subspace comprising the degrees of freedom of the nucleons labelled from 1 to $A_b$), and a scalar product is performed on that subspace. This selects only $A_b!$ components from $\op W \[ \Ket{\Phi^x} \Ket{\Psi^b} \]$, the ones where the nucleons assigned to $b$ are the same on both sides of the braket: for each component, the result of the overlap is simply %
	$\Ket{\Phi^x}$ (assigned to the sub-space of the remaining $A_x$ single-particle degrees of freedom).
	Defining conveniently the numerical factors, it is %
	\begin{multline}\label{eqFractionalParentageExpansionProiettata}
	\Braket{\Psi^b_{\intrSpin^\pi_b, M_b, T_b, \tau_b} | \Psi^B_{\intrSpin^\pi_B, M_B, T_B, \tau_B}} = \\
	= \frac{1}{\sqrt{\binom{A_B}{A_b}}} \sum_{T_x, \transfAngMom} \Braket{(\transfAngMom,M_B-M_b),(\intrSpin_b,M_b)|\intrSpin_B, M_B} \mathcal{A}^{x}_{T_x, \transfAngMom} \Ket{\Phi^{x}_{T_x,\transfAngMom}}
	\end{multline}
	where $\Braket{(\transfAngMom,m),(J_b,M_b)|J_B, M_B}$ is a Clebsch-Gordan coefficient (see \cref{eqDecomposizioneClebschGordan})
	and
	$\binom{A_B}{A_b} = A_B! / [ A_b! (A_B - A_b)! ]$ (a binomial coefficient).

	$\mathcal{A}$, which is the \emph{spectroscopic amplitude}, is a number such that \cref{eqFractionalParentageExpansionProiettata} is satisfied%
	\footnote{In \cite{Satchler1983Direct}, the spectroscopic amplitude is absorbed in the wave-function $\Phi^{x}$, whose norm consequently bears physical significance. %
	}.
	The modulus square of $\mathcal{A}$ is the \emph{spectroscopic factor} $\mathcal{S}$. If the global phase of all $\Psi$ and $\Phi$ is chosen consistently, the phase of $\mathcal{A}$ itself can be significant for interference effects. If, for a given component, $\Psi^B, \Psi^b$ and $\Phi^x$ are all bound states, they can all be taken purely real with no loss of generality, %
	so that $\mathcal{A}$ is real as well. %
	
	The coefficient $\binom{A_B}{A_b}$ %
	in \cref{eqFractionalParentageExpansionProiettata} %
	takes into account %
	the desired normalisation of the anti-symmetrised states. %
	From the discussion in \cref{secAntiSymmetrisationFirstQuantisation}, it can be seen that each anti-symmetrised state of $A$ particles carries a factor%
	\footnote{Even if the state is not a simple Slater determinant, it can be written as a superposition of several of them, each carrying the same factor.}
	$1/\sqrt{A!}$.
	Furthermore, the scalar product $\Braket{\Psi^b \Phi^x | \Psi^B}$ consists of $A_b! A_x!$ identical terms (one for each possible ordering of $A_b$ nucleons within the core nucleus and $A_x$ nucleons within the valence system). This two contributions combined yield $1/\sqrt{\binom{A_B}{A_b}}$. %
	Also
	note that %
	$\binom{A_B}{A_b}$ %
	is the number of distinct ways in which the $A_B$ nucleons can be arranged as a group of $A_b$ indistinguishable nucleons in the core nucleus, and another distinct group of $A_B - A_b$ indistinguishable nucleons.
	If $\Psi^B$ could be written as a single Slater determinant in some single-particle basis, then given a basis of core states $\{\Psi^b\}$ constructed from the same single-particle basis, there would thus be precisely $\binom{A_B}{A_b}$ core states giving non-zero overlap, each equal to $1/\sqrt{\binom{A_B}{A_b}}$ times a normalised valence state, in agreement with the hypothesis that $\Psi^B$ is normalised.

	If an explicit expression is provided for the $A$-body wave-functions $\Psi^B$ and $\Psi^b$, %
	\cref{eqFractionalParentageExpansionProiettata} can be employed directly to derive all $\Phi^{x}$ and their respective spectroscopic amplitudes.
	For example, if the expansion of each $\Psi$ %
	in the Slater-determinant basis is known, \cref{eqOverlapEspansoInBaseDeterminantiSlater} may be applied (see \cref{secOverlapFunctionExpansionintheSlaterdeterminantbasis}). %
	The wave-functions may be found numerically from ab-initio methods (for instance, Quantum Monte Carlo methods are discussed in \cite{Forest1996,Brida2011}), or an analytical form may be devised phenomenologically (as done for instance in \cite{Typel2020}).
	Another possible approach (see e.g.~\cite[sec.~16.4.1.3]{Satchler1983Direct}) is to disregard %
	the detailed internal structure of the involved particles through the use of a cluster model, %
	and directly provide a wave-function for the clusters %
	relative motion, usually constructed to reproduce experimental structure or scattering properties.
	In any case, the computed overlap function, including its norm, %
	will clearly depend on the assumptions involved in the adopted model. In particular, as pointed out also in \cite[sec.~16.4.1.3]{Satchler1983Direct}, the value of the spectroscopic factor is not independent from the choice of the normalised wave-function for the core-valence motion.
	In fact, spectroscopic factors are not observables, as they are not conserved under unitary transformations that leave unaltered all physical predictions \cite[sec.~3.6]{Mukhamedzhanov2022}. Notwithstanding this, in practical cases it is often %
	possible to observe a limited %
	consistency between spectroscopic factors given by different models (an explicit example is considered in \cite[sec.~11.4.1]{Mukhamedzhanov2022}). %

\subsubsection{Expansion of the valence states}\label{secGeneraltreatmentExpansionvalencestates} %

	In the common occurrence of $x$ being a single nucleon (this is the case treated explicitly in \cite[sec.~16.4.1]{Satchler1983Direct} and \cite[sec.~8.1]{Austern1970direct}), $\Phi^x$ reduces to the $x$-$b$ relative motion appropriately coupled to the nucleon spin.
	In general, consider an orthogonal basis of anti-symmetrised states for the internal motion of $A_x$ nucleons, with each element $\Ket{\Psi^x_{\nu,\pi_x,\intrSpin_x,M_x,T_x,\tau_x}}$ having definite modulus and $z$-projection of total isospin (quantum numbers $T_x$, $\tau_x$), total spin (quantum numbers $\intrSpin_x$, $M_x$) and of parity (quantum number $\pi_x$), and being normalised as the other $\Psi$. An additional quantum number $\nu$ is also included to allow for several distinct states of system $x$ with equal spin and isospin. Then, $\Phi$ can be expanded in a basis of states where the internal motion within the valence cluster $x$ and the relative motion between $x$ and $b$ are factorised \cite[eq.~(16.35)]{Satchler1983Direct},
	\begin{multline}\label{eqEspansioneModelloAClusterGenerica}
	\Ket{\Phi^x_{T_x,\transfAngMom}} = \sum_{\nu,l,\intrSpin_x} A_{\nu, l, \intrSpin_x, T_x, \transfAngMom} \sum_{m_l} \Braket{(l,m_l),(\intrSpin_x,M_B-M_b-m_l)|\transfAngMom,M_B-M_b} \\ \Ket{\Psi^x_{\nu, (-1)^l \pi_b\pi_B, \intrSpin_x,M_B-M_b-m_l, T_x,\tau_B-\tau_b}} \Ket{\phi_{\nu,\intrSpin_x,l,m_l}}
	\end{multline}
	where $\phi$ is a function describing solely the relative motion between $x$ and $b$ centre-of-masses, with normalisation set analogously to the one adopted for the $\Psi$ (for bound states, it is simply $\Braket{\phi|\phi} = 1$), and $A_{\nu, l, \intrSpin_x, T_x, \transfAngMom}$ is the appropriate weight%
	\footnote{In \cite{Satchler1983Direct}, as done for the spectroscopic amplitude $\mathcal A$, the weight $A$ is absorbed into $\phi$.}
	to satisfy \cref{eqEspansioneModelloAClusterGenerica}, so that
	\begin{equation}
	\sum_{\nu,l,\intrSpin_x} \m{A_{\nu, l, \intrSpin_x, T_x, \transfAngMom}}^2 = 1
	\end{equation}
	Note that relative-motion states $\phi$ with different orbital angular momentum $(l,m_l)$ are certainly orthogonal (because spherical harmonics are), while internal-motion states $\Psi^x$ with different quantum numbers $(\nu,\intrSpin_x,T_x)$ are orthogonal by construction.

	In \cref{eqEspansioneModelloAClusterGenerica}, the sum over $m_l$ enforces the appropriate coupling of $\Psi^x \phi$ to the desired total transferred angular momentum $\transfAngMom$. This also guarantees that the set of $\Phi^x$ with different $\transfAngMom$ %
	are orthogonal.
	It may be useful to define a set of orthogonal functions $\tilde\Phi^x_{\nu,\pi_x,\intrSpin_x,T_x,\tau,l,\transfAngMom,M}$ as
	\begin{multline}\label{eqDefinizioneAlternativaStatiValence}
	\Ket{\tilde\Phi^x_{\nu,\pi,\intrSpin_x,T_x,\tau,l,\transfAngMom,M}} =\\
	= \sum_{m_l} \Braket{(l,m_l),(\intrSpin_x,M-m_l)|\transfAngMom,M} \Ket{\Psi^x_{\nu,\pi,\intrSpin_x,M-m_l,T_x,\tau}} \Ket{\phi_{\nu,\intrSpin_x,l,m_l}}
	\end{multline}
	with
	\begin{equation}
	\Ket{\Phi^x_{T_x,\transfAngMom}} = \sum_{\nu,l,\intrSpin_x} A_{\nu, l, \intrSpin_x, T_x, \transfAngMom} \Ket{\tilde\Phi^x_{\nu,(-1)^l\pi_b\pi_B,\intrSpin_x,T_x,\tau_B-\tau_b,l,\transfAngMom,M_B-M_b}}
	\end{equation}
	where, in analogy with $\Phi^x_{T_x,\transfAngMom}$, the indexes $\pi, M, \tau$ in $\tilde\Phi^x$ will be often dropped for brevity. %

	Note that the treatment given here is %
	independent from the precise framework employed to describe the internal motion of the valence system: $\Psi^x$ is here in principle the full $A_x$-body wave-function, but it may as well be approximated as just the spin and isospin state of a structureless particle $x$ %
	(which is useful to perform a one-particle transfer calculation) or as a cluster-model wave-function (for multi-particle transfer).
	
\paragraph{} %
	The overlap function defined in \cref{eqFractionalParentageExpansionProiettata} includes the appropriate superposition of all states of the valence system $x$ which can couple with the given core state to yield the desired composite state, possibly with several allowed values of $T_x$ and $\transfAngMom$, each bearing a separate spectroscopic amplitude.
	It can %
	be interesting to project on a precise state $\Phi^x$, obtaining a number directly proportional to the spectroscopic amplitude for the given configuration:
	\begin{multline}\label{eqOverlapFunctionProiettataSuSingoloTrasferito}
	\Braket{\Psi^b_{J^\pi_b, M_b, T_b, \tau_b} \Phi^{x}_{T_x,\transfAngMom} | \Psi^B_{J^\pi_B, M_B, T_B, \tau_B}} = \\
	= \frac{1}{\sqrt{\binom{A_B}{A_b}}} \Braket{(\transfAngMom,M_B-M_b),(J_b,M_b)|J_B, M_B} \mathcal{A}^{x}_{T_x,\transfAngMom}
	\end{multline}
	It is also %
	possible to select a single component for the internal motion of $x$ and its relative motion with $b$: using the definition in \cref{eqDefinizioneAlternativaStatiValence}, %
	\begin{multline}\label{eqAmpiezzaSpettroscpicaSingolaComponenteMotoValenza}
	\Braket{\Psi^b_{J^\pi_b, M_b, T_b, \tau_b} \tilde\Phi^x_{\nu,\intrSpin_x,T_x,l,\transfAngMom} | \Psi^B_{J^\pi_B, M_B, T_B, \tau_B}} = \\
	= \frac{1}{\sqrt{\binom{A_B}{A_b}}} \Braket{(\transfAngMom,M_B-M_b),(J_b,M_b)|J_B, M_B} \mathcal{A}^{x}_{T_x,\transfAngMom} A_{\nu,l,\intrSpin_x,T_x,\transfAngMom}
	\end{multline}
	where now the relevant amplitude is $\mathcal{A}^{x}_{T_x,\transfAngMom} A_{\nu,l,\intrSpin_x,T_x,\transfAngMom}$.
	
	The scalar product $\Braket{\Psi^b_{J^\pi_b, M_b, T_b, \tau_b} \Phi^x_{T_x,\transfAngMom} | \Psi^B_{J^\pi_B, M_B, T_B, \tau_B}}$ (as the one in \cref{eqAmpiezzaSpettroscpicaSingolaComponenteMotoValenza}) %
	embodies the coupling of angular momenta $\transfAngMom$ and $J_b$ to yield $J_B$, which produces the coefficient $\Braket{(\transfAngMom,M_B-M_b),(J_b,M_b)|J_B, M_B}$ appearing in \cref{eqFractionalParentageExpansionProiettata}. More formally, $\Phi^x_{T_x,\transfAngMom}$ can be seen as a $(\transfAngMom,M_B-M_b)$ spherical tensor, thus, by Wigner-Eckart theorem,
	\begin{multline}\label{eqDimostrazioneFattoreSpettroscopicoIndipendenteDaProiezioniSpinTotali}
	\Braket{ \Psi^B_{J^\pi_B, M_B, T_B, \tau_B} | \Phi^x_{T_x,\tau_B-\tau_b,\transfAngMom,M_B-M_b} \Psi^b_{J^\pi_b, M_b, T_b, \tau_b} } = \\
	= \Braket{(\transfAngMom,m),(J_b,M_b)|J_B, M_B} \ \Braket{\Psi^B_{J^\pi_B,T_B,\tau_B} || \Phi^x_{T_x,\tau_B-\tau_b,\transfAngMom} || \Psi^b_{J^\pi_b, T_b, \tau_b} }
	\end{multline}
	where $\Braket{\Psi^B || \Phi^x || \Psi^b }$ (the reduced matrix element) does not depend on the projections $M_b, M_B$. This implies that the spectroscopic amplitude is the same for any %
	spin $z$-projection of core and composite nuclei (provided they can actually couple, i.e.\ the Clebsch-Gordan coefficient is not zero), and can thus be computed for the most convenient pair of projections.
	On this regard, it is relevant to point out that the set of admissible components $\Phi^x_{T_x,\transfAngMom}$ of a given overlap can depend on $M_b$ and $M_B$. As a trivial example, valence states coupling to $\transfAngMom=0$ are admissible only for $M_b = M_B$ (in all other cases, their contribution is cancelled by the Clebsch-Gordan coefficient in \cref{eqDimostrazioneFattoreSpettroscopicoIndipendenteDaProiezioniSpinTotali}). %
	
	A similar discussion could be repeated for isospin. This is the reason why sometimes an extra factor $\Braket{(T_b,\tau_b),(T_x,\tau_B-\tau_b)|T_B, \tau_B}$ appears in \cref{eqFractionalParentageExpansionGenerica}, changing the definition of $\mathcal A$ \cite[sec.~16.4.1]{Satchler1983Direct}.

\subsubsection{Extreme cluster model and Wildermuth connection}
	
	In practice, it is impossible to consider a complete basis of states, and even an approximation including bound states and low-lying continuum is challenging. The extreme cluster model prescribes to approximate the overlap function $\Braket{\Psi^b|\Psi^B}$ in \cref{eqFractionalParentageExpansionProiettata} including a single component %
	$\Phi^x$ which in turn comprises only a single component
	$\tilde\Phi^x$ (defined in \cref{eqDefinizioneAlternativaStatiValence}), %
	\begin{multline}\label{eqExtremeClusterModelApproximation}
	\Braket{\Psi^b_{J^\pi_b, M_b, T_b, \tau_b} | \Psi^B_{J^\pi_B, M_B, T_B, \tau_B}} \approx \\
	\approx \frac{1}{\sqrt{\binom{A_B}{A_b}}} \Braket{(\transfAngMom,M_B-M_b),(J_b,M_b)|J_B, M_B} \mathcal{A}^{x} \Ket{\tilde\Phi^x_{\nu,\intrSpin_x,T_x,l,\transfAngMom}} %
	\end{multline}
	More explicitly, using \cref{eqDefinizioneAlternativaStatiValence},
	\begin{multline}
	\Braket{\Psi^b_{J^\pi_b, M_b, T_b, \tau_b} | \Psi^B_{J^\pi_B, M_B, T_B, \tau_B}}
	\approx \frac{1}{\sqrt{\binom{A_B}{A_b}}} \Braket{(\transfAngMom,M_B-M_b),(J_b,M_b)|J_B, M_B} \cdot\\\cdot \mathcal{A}^{x} \sum_{m_l} \Braket{(l,m_l),(\intrSpin_x,M-m_l)|\transfAngMom,M} \Ket{\Psi^x_{\nu,\intrSpin_x,M-m_l,T_x}} \Ket{\phi_{\nu,\intrSpin_x,l,m_l}}
	\end{multline}
	Note that, even if the stated approximation were exact, %
	the spectroscopic amplitude could still take non-trivial values, depending on the microscopic structure of core and composite nuclei. In addition, a phenomenological value for $\mathcal{A}^{x}$ may partially correct the error introduced by the approximation.
	
	The %
	internal motion of the valence system $x$, $\Psi^x$, is normally chosen to match some relevant state of system $x$ when observed in isolation, often the most bound among its allowed states, such that the total spin and isospin modulus quantum numbers, $s$ and $t$, can correctly couple the desired $\Psi^b$ and $\Psi^B$. %
	As mentioned earlier, the precise model employed for $\Psi^x$ %
	depends on what process or features are being described.
	
	The core-valence relative motion state, $\phi$, can be modelled as an eigenfunction of a phenomenological potential for the interaction between $b$ and $x$. Its energy eigenvalue is most commonly set in terms of the experimental mass (and excitation energy) of the states of interest for $B$, $b$ and $x$. %
	If $\phi$ is a bound state,
	then its orbital angular momentum modulus quantum number, $l$, and the number of nodes of its radial part, labelled $n$, are often set from shell-model considerations, adopting the so-called \emph{Wildermuth connection}, as follows.
	Consider an independent-particle shell-model state for the isolated systems $b$, $B$ and $x$, where each nucleon $i$ is assigned to a specific shell-model single-particle state with orbital angular momentum $l_i$ and number of radial nodes $n_i$ (the minimum possible value for $n_i$ is zero). %
	The number of “energy quanta” $Q_j$ in system $j$ is defined as
	\begin{equation}
	Q_j = \sum_{i=1}^{A_j} 2 n_i + l_i
	\end{equation}
	and, similarly, the $b$-$x$ relative motion is assigned an amount $Q_{bx}$ of energy quanta equal to $2 n + l$. $Q_{bx}$ is then found requiring that the total amount of quanta in $B$ matches the sum of quanta in $b$, $x$ and their relative motion, \cite[eq.~(16.32)]{Satchler1983Direct}
	\begin{equation}\label{eqWildermuthconnection}
	Q_B = Q_b + Q_x + Q_{bx}
	\end{equation}
	In general, there may be several pairs of $n,l$ satisfying the relation, in which case the smallest permitted $l$ is often adopted. It is also possible that none of the values of $l$ allowed by the criterion matches the required one to describe a particular state of interest: this can be taken as a suggestion that the extreme-shell-model configuration which has been chosen %
	is not the favoured one to describe the system of interest.
	
	Finally, note that setting the binding energy for a bound state $\phi$ amounts to put a constraint on the $b$-$x$ potential. %
	Once the orbital angular momentum and number of nodes of the state are provided, the constraint is sufficiently detailed to completely fix one degree of freedom in the expression for the potential (for example a global scaling factor), which is useful to construct the potential phenomenologically.

\subsubsection{Expansion in the Slater-determinant basis}\label{secOverlapFunctionExpansionintheSlaterdeterminantbasis} %
	
	Differently from what was done above, %
	consider now a fixed coordinate system, and express all wave-functions $\Psi^B$ and $\Psi^b$ in such coordinates (instead of computing each of them in the respective centre-of-mass rest frame).
	As a consequence, %
	$\Phi^x_{T_x,\transfAngMom}$ is not cleanly factorised between internal motion of $x$ and $x$-$b$ relative motion any more, and it makes more sense to just refer to it as a valence system state %
	computed in the same coordinate system as the other $\Psi$ functions.
	
	Consider an orthonormal single-particle basis (for instance the one provided by a shell model), and let $\chi^B_i$ be elements of the orthonormal basis of the $A_B$-body system constructed as normalised Slater determinants (defined as in \cref{eqApplicazioneOperatoreAntisimmetrizzazione}) %
	of the aforementioned single-particle states ($\chi^B_i$ was labelled $\Ket{\nu_1,\dots\nu_A}_W$ in \cref{secAntiSymmetrisationFirstQuantisation}). In general, the composite nucleus state $\Psi^B$ of interest can be expanded in such basis, obtaining $\Ket{\Psi^B} = \sum_i c^B_i \Ket{\chi^B_i}$, where the $c$ are numerical coefficients such that $\sum_i \m{c_i}^2 = 1$
	and one can proceed identically for the core nucleus state $\Psi^b$.
	The overlap function can then be computed explicitly in the shell-model basis:
	\begin{equation}
		\Braket{\Psi^b|\Psi^B} = \sum_{ik} \coniugato{c^b_i} c^B_k \Braket{\chi^b_i|\chi^B_k}
	\end{equation}
	the overlap %
	$\Braket{\chi^b_i|\chi^B_k}$ is either zero, if %
	any single-particle state is occupied in $\chi_i^b$ but empty in $\chi_k^B$, or proportional to a single Slater determinant with $A_x$ nucleons filling the states occupied in $\chi_k^B$ and empty in $\chi_i^b$: let $\chi^x_{ik}$ be this Slater determinant, normalised as the other $\chi$ functions. %
	Using the same reasoning described in \cref{secOverlapFunctionSpectroscopicFactorsIntroduzione} when discussing the binomial factor in \cref{eqFractionalParentageExpansionProiettata}, the proportionality factor is such that (for a pair $i,k$ whose overlap is not zero)
	\begin{equation}\label{eqProdottoScalareDeterminantiDiSlater}
		\m{ \Braket{\chi^b_i \chi^x_{ik} | \chi^B_k } } = \frac{1}{\sqrt{\binom{A_B}{A_b}}}
	\end{equation}
	The phase (which for bound states can always be defined to be just a sign) depends on the adopted conventions (it is necessary to fix an ordering for the single-particle basis elements) and on what states are occupied in $\chi^b$ or $\chi^x$. For simplicity, the present discussion on phases will be limited to the few relative signs required to deduce the spectroscopic factors of interest within the in\-de\-pen\-dent-particle shell-model.
	
	It can be useful to note that, if the overlap of a given pair of components, $\Braket{\chi^b_i|\chi^B_k}$, is $\sqrt{\frac{A_b! A_x!}{A_B!}} \Ket{\chi^x_{ik}}$,
	then the corresponding overlap between the Slater determinants of composite and valence systems, $\Braket{\chi^x_{ik}|\chi^B_k}$, is similarly $\sqrt{\frac{A_x! A_b!}{A_B!}} \Ket{\chi^b_i}$,
	where the coefficient is the same. %
	
	To summarise,
	let $f_{ij}$ be 0 if $\Braket{\chi^b_i|\chi^B_j} = 0$, or the phase of $\Braket{\chi^b_i \chi^x_{ij} | \chi^B_j }$ otherwise. Once the phase convention has been fixed, $f_{ij}$ can be computed %
	for any pair of Slater determinants by inspection of the involved single-particle states. Then,
	\begin{equation}\label{eqOverlapEspansoInBaseDeterminantiSlater}
		\Braket{\Psi^b_{J^\pi_b, M_b, T_b, \tau_b} | \Psi^B_{J^\pi_B, M_B, T_B, \tau_B}} = \frac{1}{\sqrt{\binom{A_B}{A_b}}} \sum_{ij} \coniugato{c^b_i} c^B_j f_{ij} \Ket{\chi^x_{ij}}
	\end{equation}
	The last remaining step is to regroup the different components into valence states $\Ket{\Phi^x_{T_x,\tau_B - \tau_b, \transfAngMom, M_B-M_b}}$ with the desired quantum numbers. The details of this step depend on the precise single-particle basis in use, but it essentially requires to couple the single-particle states %
	to the desired total spin and isospin. For instance, in a shell-model basis state all nucleons in closed shells couple to angular momentum 0, simplifying the problem, and the remaining nucleons are in states with their intrinsic spin and orbital angular momentum already coupled.

\subsection{Spectroscopic factors within the independent-particle shell model}\label{secOverlapSpectroscopicfactorswithinExtremeshellmodel} %
	
	In many cases, experimental or ab-initio evaluations of spectroscopic factors (together with the corresponding overlap functions) can be found in literature. Even in these cases, evaluating %
	the spectroscopic amplitudes phenomenologically, %
	within a sufficiently simple model, can be very useful, as it allows to pinpoint the most salient features affecting the result, and %
	to gauge the corrections introduced by the more sophisticated determinations. %
	In the following, the predictions given by the in\-de\-pen\-dent-particle shell-model shell model will be discussed.

	Start from \cref{eqOverlapEspansoInBaseDeterminantiSlater}, and approximate %
	the states for core and composite nucleus, $\Psi^b$ and $\Psi^B$, as in\-de\-pen\-dent-particle shell-model wave-functions.
	This means that %
	each nucleon occupies a well-defined shell among the elements of the orthogonal single-particle basis, %
	but may not have defined angular momentum $z$-projection.
	In the simplest case, the state is just a single Slater determinant:
	for instance, an \nuclide{\alpha}-particle in its ground-state is described with its four nucleons in the $0s1/2$ shell (0 nodes, orbital angular momentum $l=0$, total angular momentum $1/2$), with anti-symmetrisation requiring that precisely one proton and one neutron have positive spin $z$-projection. %
	However, %
	if there are several nucleons in partially-empty shells, whose angular momentum $z$-projections can be assigned in more than one way to obtain a nucleus with the desired spin projection,
	then it necessary to describe the state as a superposition of several Slater determinants. The weight and relative phase of each component can be found by explicitly coupling spin and isospin of all nucleons.
	For instance, a “deuteron” in the singlet state (total spin modulus 0, total isospin modulus 1) may have its proton with either spin up or down, and both components have the same absolute weight. %
	As a consequence, the overlap function $\Braket{\Psi^b_{J^\pi_b, M_b, T_b, \tau_b} | \Psi^B_{J^\pi_B, M_B, T_B, \tau_B}}$ into a given valence state $\Phi^x_{T_x,\transfAngMom}$ %
	can in turn be a superposition of several Slater determinants, in which the occupied shells are always the same, but the angular momentum projections of occupied states may not, even though the total spin projection of the system is defined.

	\paragraph{} There are several specific cases where the spectroscopic factors can be computed with particular ease. Such cases are often applicable when both $\Psi^b_{J^\pi_b, M_b, T_b, \tau_b}$ and $\Psi^B_{J^\pi_B, M_B, T_B, \tau_B}$ represent the ground state of the respective nuclei (which is also the situation where an in\-de\-pen\-dent-particle shell-model state can be expected to work best). %
	
	For instance, when both $\Psi^B$ and $\Psi^b$ comprise only a single Slater determinant (all nucleon angular momentum $z$-projections are assigned unambiguously), their overlap is a single Slater determinant as well. If the overlap is not zero, from \cref{eqOverlapEspansoInBaseDeterminantiSlater} %
	it is %
	\begin{equation}\label{eqOverlapEspansoInBaseDeterminantiSlaterCasoSingoloDeterminante}
	\Braket{\Psi^b_{J_b^\pi, M_b, T_b, \tau_b} | \Psi^B_{J_B^\pi, M_B, T_B, \tau_B}} = \frac{1}{\sqrt{\binom{A_B}{A_b}}} \Ket{\chi^x}
	\end{equation}
	where the phase factor prescribed in \cref{eqOverlapEspansoInBaseDeterminantiSlater} was neglected, because in this case it would be just a global phase, irrelevant here. %
	The only possible complication is that $\Ket{\chi^x}$ may not be a state with definite total spin and isospin (this is the “regrouping” issue mentioned when commenting \cref{eqOverlapEspansoInBaseDeterminantiSlater}). For instance, a proton with spin up and a neutron with spin down in the $0s1/2$ shell have definite total spin $z$-projection (equal to 0), but are in a superposition of a state with total spin modulus $j=1$ and another one with $j=0$.
	In general, $\Ket{\chi^x}$ will thus be written as some superposition $\sum_{T_x,\transfAngMom} X_{T_x,\transfAngMom} \Ket{\Phi^x_{T_x,\transfAngMom}}$. %
	From the definition in \cref{eqFractionalParentageExpansionProiettata} it follows that the spectroscopic factor for each component, $\mathcal S^x_{T_x,\transfAngMom} = \m{\mathcal A^x_{T_x,\transfAngMom}}^2$, is
	\begin{equation}\label{eqFattoreSpettroscopicoExtremeShellModelCasoUnaSolaComponente}
	\mathcal{S}^{x}_{T_x,\transfAngMom}
	= \frac{ \m{ X_{T_x,\transfAngMom} }^2 }{ \m{ \Braket{(\transfAngMom,M_B-M_b),(J_b,M_b)|J_B, M_B} }^2 }
	\end{equation}
	As was discussed earlier (see text commenting \cref{eqDimostrazioneFattoreSpettroscopicoIndipendenteDaProiezioniSpinTotali}), the spectroscopic amplitude for a given state $\Phi^x_{\transfAngMom, M_B-M_b, T_x,\tau_B - \tau_b}$ does not depend on the core and composite total spin projections $M_B$ and $M_b$. In several cases, it is possible to find an appropriate pair of projections such that the hypothesis for \cref{eqFattoreSpettroscopicoExtremeShellModelCasoUnaSolaComponente} are satisfied.
	
	As a simple but often useful corollary of \cref{eqFattoreSpettroscopicoExtremeShellModelCasoUnaSolaComponente} is drawn in the case where %
	$B$ is a nucleus with only closed shells, and thus has spin 0, and $x$ is a nucleon in a shell with total angular momentum $\transfAngMom$. The hypothesis guarantee that $\Psi^B$ and $\Psi^b$ are Slater determinants %
	and $\Phi^x$ has only one allowed value for $\transfAngMom$ and $T_x$. It follows that the spectroscopic factor for the only appearing component is $\mathcal{S}^{x} = 2\transfAngMom+1$. %
	
	\paragraph{} Another simple case is found when the core nucleus has total spin 0 and total isospin 0, which typically happens if the state has no partially filled shells. %
	If the overlap is to be non-zero, the composite nucleus will necessarily have the completely occupied shells found in the core, plus some more nucleons.
	These nucleons need not to have definite angular momentum $z$-projection, %
	($\Psi^B$ may not be a single Slater determinant), but they must be coupled to total spin and isospin equal to the composite nucleus quantum numbers. Since there is only one possible valence state, which is $\Ket{\Phi^x_{T_x=T_B,\tau=\tau_B, \transfAngMom=J_B, m=M_B}}$ (regardless of how many or what Slater determinants it comprises), %
	it is necessarily
	\begin{equation}\label{eqOverlapExtremeShellModelPerCoreInerte}
	\Braket{\Psi^b_{J^\pi_b, M_b, T_b, \tau_b} | \Psi^B_{J^\pi_B, M_B, T_B, \tau_B}} = \frac{1}{\sqrt{\binom{A_B}{A_b}}} \Ket{\Phi^x_{T_B,J_B}} %
	\end{equation}
	Furthermore, $\Braket{(J_B,M_B),(0,0)|J_B, M_B} = 1$, thus, from \cref{eqFractionalParentageExpansionProiettata}, the associated spectroscopic factor is just $\mathcal S^x = 1$.
	More in general, shells which are completely occupied both in core and composite system can be %
	ignored, as they will play no actual role.
	
	\paragraph{} As an application of the concepts just discussed, %
	in the following paragraphs
	the overlaps and %
	spectroscopic factors of interest for the nuclear reaction discussed in this thesis are %
	computed within the in\-de\-pen\-dent-particle shell-model.

\subsubsection{Overlaps for the $\nuclide[6]{Li} + \nuclide{p} \to \nuclide[4]{He} + \nuclide[3]{He}$ reaction as a deuteron transfer} %
	
	All reactants are assumed to be in their ground state, with definite spin $J$, parity $\pi$ and isospin $T$. \nuclide[6]{Li} has $J^\pi=1^+$ and $T=0$, while \nuclide{p} and \nuclide[3]{He} have $J^\pi=\frac{1}{2}^+$ and $T=1/2$, and finally \nuclide[4]{He} has $J^\pi=0^+$ and $T=0$.
	The process is described as the direct one-step transfer of a system with $A_x = 2$ (“deuteron”; the only possible alternative description would have been $A_x=3$), thus the required overlaps are $\Braket{\nuclide[4]{He}|\nuclide[6]{Li}}$ and $\Braket{\nuclide{p}|\nuclide[3]{He}}$.  All quantum numbers regarding the internal motion of the valence system are not altered during the reaction \cite[sec.~4]{Thompson2013}.
	It follows that the only valence \emph{internal motion} states relevant for the process are those compatible with \emph{both} overlaps. In particular, in $\Braket{\nuclide[4]{He}|\nuclide[6]{Li}}$ the valence system can only have isospin modulus $T_x = 0$, consequently the $T_x=1$ states, despite being allowed in $\Braket{\nuclide{p}|\nuclide[3]{He}}$, will not contribute to the one-step transfer. %
	Furthermore, the transferred particle parity, $\pi_x$, can be either even or odd (provided the core-valence orbital angular momentum has the same parity), but only configurations where $\pi_x$ is identical for both projectile and target will couple.
	On the contrary, the total transferred angular momentum, $\transfAngMom$, %
	needs not in general to be identical in both overlaps, as the core-valence relative orbital angular momentum can differ. In $\Braket{\nuclide[4]{He}|\nuclide[6]{Li}}$, it is necessarily $\transfAngMom=1$. %
	In $\Braket{\nuclide{p}|\nuclide[3]{He}}$ both $\transfAngMom=0$ and 1 are in general possible (even after fixing $T_x=0$), but the shell-model states considered here are such that only $\transfAngMom=1$ contributes, as discussed below. %
	
	The in\-de\-pen\-dent-particle shell-model state of the \nuclide[4]{He} core, as commented earlier, comprises four nucleons completely filling the $1s1/2$ shell.
	As was just discussed explicitly, or directly from %
	\cref{eqOverlapExtremeShellModelPerCoreInerte}, %
	$\Braket{\nuclide[4]{He}|\nuclide[6]{Li}}$ is thus just a single state $\Ket{\Phi^x_{T_x=0,\tau=0,\transfAngMom=1,m=M_B}}$, and the associated spectroscopic factor is 1.

	The three nucleons of \nuclide[3]{He} ground state all occupy, within the in\-de\-pen\-dent-particle shell-model, the $1s1/2$ shell (that is, the most bound shell with orbital angular momentum zero). The angular momentum $z$-projection of the lone neutron coincides with the total projection $M_B$, thus each distinct $\Psi^B$ state is a single Slater determinant. The core is requested to be %
	a single proton, %
	and must %
	be assigned to the $1s1/2$ shell as well for the overlap to be non-zero, but it can have either positive or negative spin projection.
	Let $\Ket{\nuclide{p}(m)}$ be the single-particle state for a proton (positive isospin projection) in the $1s1/2$ shell with spin $z$-projection $m$, and similarly $\Ket{\nuclide{n}(m)}$ for neutrons in the same shell.
	For definiteness, consider only the \nuclide[3]{He} state with $M_B = +1/2$ (the reasoning can be repeated for the other projection), labelled $\Ket{\nuclide[3]{He}(+1/2)}$.
	As mentioned, it would be of interest to consider both $\Braket{\nuclide{p}(+1/2)|\nuclide[3]{He}}$ and $\Braket{\nuclide{p}(-1/2)|\nuclide[3]{He}}$, %
	but since %
	spectroscopic amplitudes %
	are independent from the core and composite spin projections (see text commenting \cref{eqDimostrazioneFattoreSpettroscopicoIndipendenteDaProiezioniSpinTotali}), it is %
	sufficient to treat the overlap $\Braket{\nuclide{p}(+1/2)|\nuclide[3]{He}(+1/2)}$, %
	where all possible components %
	appear
	\footnote{$\Braket{\nuclide{p}(-1/2)|\nuclide[3]{He}(+1/2)}$ instead includes only a $\transfAngMom=1, T_x=0$ component.}.
	\Cref{eqOverlapEspansoInBaseDeterminantiSlaterCasoSingoloDeterminante} applies, with $\Ket{\chi^x}$ being here the normalised Slater determinant constructed from the only possible assignment of single-particle levels, namely
	\begin{equation}
	\Ket{\chi^x} = \frac{1}{\sqrt{2}} \( \Ket{\nuclide{p}(-1/2), \nuclide{n}(+1/2)} - \Ket{\nuclide{n}(+1/2), \nuclide{p}(-1/2)} \)
	\end{equation}
	(the notation is as in \cref{secAntiSymmetrisationFirstQuantisation}, particle 1 is in the first listed state and so on), %
	where the global phase is conventional.
	Each component of $\Ket{\chi^x}$ %
	can be expanded in %
	the valence total spin and total isospin basis. %
	Let $\tau_i$ and $m_i$ be the isospin and spin projection of each nucleon (protons are assigned $\tau_i>0$ as usual), and $T_x$ and $\transfAngMom$ the modulus quantum numbers for total isospin and spin (the total projections are both fixed to 0 by the choice of core and composite states). Then,
	\begin{multline}\label{eqEspansioneStanoNonSimmetrizatoDiDueNucleoni}
	\Ket{(\tau_1 m_1), (\tau_2 m_2)} =\\= \sum_{T_x,\transfAngMom} \Braket{(1/2, m_1),(1/2,m_2)|\transfAngMom,0} \Braket{(1/2, \tau_1),(1/2, \tau_2)|t,0} \Ket{\Phi^x_{T_x,\transfAngMom}}
	\end{multline}
	revealing that
	\begin{equation}
	\Ket{\chi^x} = \frac{1}{\sqrt{2}} \( \Ket{\Phi^x_{T_x=0,\transfAngMom=1}} - \Ket{\Phi^x_{T_x=1,\transfAngMom=0}} \)
	\end{equation}
	The result could have been anticipated noting that both spin and isospin are, in this specific case, being coupled exactly in the same manner (in both cases two $1/2$ spins are being added) %
	consequently states with $\transfAngMom=T_x$ are certainly symmetric under particle exchange and thus inadmissible.
	Finally, from \cref{eqFattoreSpettroscopicoExtremeShellModelCasoUnaSolaComponente}, the spectroscopic factors %
	are $\mathcal{S}^{x}_{T_x=0,\transfAngMom=1} = 3/2$ and %
	$\mathcal{S}^{x}_{T_x=1,\transfAngMom=0} = 1/2$.

\subsubsection{Overlaps for the $\nuclide[6]{Li} + \nuclide{p} \to \nuclide[5]{Li} + \nuclide{d} \to \nuclide[4]{He} + \nuclide[3]{He}$ two-step transfer}
	
	Instead of the one-step transfer of two nucleons treated in the preceding example, consider now two separate transfers of a neutron and a proton. %
	The internal state of the transferred system in each step is now trivial, in particular it must be $T_x=1/2$ and $J^\pi_x = \frac{1}{2}^+$, thus, differently than in the deuteron-transfer case, imposing the conservation of such state during each step separately yields no restriction.
	
	It is %
	of interest to generalise the allowed initial state, in particular let \nuclide[6]{Li} be composed by an inert \nuclide[4]{He} core of four nucleons occupying the $1s1/2$ shell, and two valence nucleons placed in any unoccupied pair of shells capable of coupling to the required total spin-parity, $1^+$.
	Let $j_{\nuclide{p}}$ and $j_{\nuclide{n}}$ be the total angular momentum modulus of the two selected shells, and $\Ket{\nuclide[6]{Li}(j_{\nuclide{p}},j_{\nuclide{n}})}$ and $\Ket{\nuclide[5]{Li}(j_{\nuclide{p}})}$ the associated states of \nuclide[6]{Li} and \nuclide[5]{Li}. It must be $\m{j_{\nuclide{p}}-j_{\nuclide{n}}} \leq 1 \leq j_{\nuclide{p}}+j_{\nuclide{n}}$, and the shells orbital angular momentum must bear the same parity.
	In general, the complete \nuclide[6]{Li} wave-function will be a superposition of several such configurations, $\Ket{\nuclide[6]{Li}} = \sum_{j_{\nuclide{p}},j_{\nuclide{n}}} C_{j_{\nuclide{p}},j_{\nuclide{n}}} \Ket{\nuclide[6]{Li}(j_{\nuclide{p}},j_{\nuclide{n}})}$, but it is possible to consider only one configuration at a time and then recombine the complete overlap function. %

	Reasoning as before, the spectroscopic factor for $\Braket{\nuclide[4]{He}|\nuclide[5]{Li}(j_{\nuclide{p}})}$ is 1, because the \nuclide[4]{He} core, which has total spin and isospin 0, completely fills a shell and its nucleons do not couple in any way with the additional neutron. %
	Similarly, the $\Braket{\nuclide[5]{Li}(j_{\nuclide{p}})|\nuclide[6]{Li}(j_{\nuclide{p}},j_{\nuclide{n}})}$ case can, within the present model, be treated as the $\Braket{\nuclide{p}|\nuclide{d}}$ overlap, %
	with the two nucleons placed in arbitrary shells with angular momentum modulus $j_{\nuclide{p}}$ and $j_{\nuclide{n}}$ as described above.
	The \nuclide{d} state cannot, in general, be written as a single Slater determinant: %
	rather, it will comprise a combination of states with different spin projections, coupled to the desired total spin and isospin (it would be possible to write an explicit expansion in analogy with \cref{eqEspansioneStanoNonSimmetrizatoDiDueNucleoni}). %
	Let $\chi^B_{m_{\nuclide{p}},m_{\nuclide{n}}}$ be a Slater determinant with proton and neutron angular momentum projection fixed to %
	$m_{\nuclide{p}}$ and %
	$m_{\nuclide{n}}$ respectively. The complete composite nucleus state, with fixed spin projection $M_B$, is then
	\begin{equation}\label{eqEspansioneDeuterioInDeterminantiDiSlater}
	\Ket{d(M_B)} = \sum_{m_{\nuclide{p}}} \Braket{(j_{\nuclide{p}},m_{\nuclide{p}}),(j_{\nuclide{n}},M_B-m_{\nuclide{p}})|1,M_B} \Ket{\chi^B_{m_{\nuclide{p}},M_B-m_{\nuclide{p}}}}
	\end{equation}
	Note that the isospin coupling is included in the anti-symmetrisation enforced within the Slater determinant. The \nuclide{p} state is instead a single ``Slater determinant'' (a single state, in fact) with the desired quantum numbers $j_{\nuclide{p}}$, $m_{\nuclide{p}}$ %
	for the corresponding \nuclide[5]{Li} system. The projection $\Braket{\nuclide{p}|\nuclide{d}}$ is thus proportional to the only overlapping configuration between the two systems, namely a neutron in a shell with angular momentum modulus $j_{\nuclide{n}}$ and projection $M_B-m_{\nuclide{p}}$. By combining \cref{eqOverlapFunctionProiettataSuSingoloTrasferito,eqOverlapEspansoInBaseDeterminantiSlater,eqEspansioneDeuterioInDeterminantiDiSlater}, it follows that the spectroscopic factor is always 1.
	In passing, note that the situation would have been identical for $\Braket{\nuclide{n}|\nuclide{d}}$.
	
	The last overlap of interest for this application is $\Braket{\nuclide{d}|\nuclide[3]{He}}$. Restrict to the case in which \nuclide[3]{He} is in the independent-particle shell-model ground state. %
	The system is then the same already discussed when considering the $\Braket{\nuclide{p}|\nuclide[3]{He}}$ overlap, just with the role of core and valence inverted. The associated spectroscopic factors are thus the same, since in e.g.~\cref{eqOverlapFunctionProiettataSuSingoloTrasferito}, once all states (including the transferred system one) are fixed, exchanging $x$ and $b$ can at most invert the sign of the overlap%
	\footnote{It is of course possible to perform the calculation explicitly. If the core \nuclide{d} has spin 1 and spin projection 1, its state is a %
		single Slater determinant and the result follows from \cref{eqFattoreSpettroscopicoExtremeShellModelCasoUnaSolaComponente}. %
		In the $J=0$ case, the \nuclide{d} state comprises two Slater determinants, each with amplitude $1/\sqrt{2}$, but only one of them overlaps with a \nuclide[3]{He} state with fixed spin projection, and the result follows using \cref{eqOverlapEspansoInBaseDeterminantiSlater}.}.
	
	For completeness, finally consider the %
	$\Braket{\nuclide[2]{He}|\nuclide[3]{He}}$ overlap, which would be relevant if the two-step transfer from $\nuclide[6]{Li} + \nuclide{p}$ to $\nuclide[4]{He} + \nuclide[3]{He}$ were accomplished by first transferring a proton. The core state is now a single Slater determinant with nucleons coupled to $T=1$ and $J=0$, thus by using \cref{eqFattoreSpettroscopicoExtremeShellModelCasoUnaSolaComponente} it follows that the spectroscopic factor is $\mathcal S = 1$.
	Note how the three core states $\nuclide{d}(T=0)$, $\nuclide{d}(T=1)$ and \nuclide[2]{He} suffice to complete the fractional parentage expansion of \nuclide[3]{He} in \cref{eqFractionalParentageExpansionGenerica} for $A_b=2$. %

\subsection{Asymptotic normalisation coefficient}\label{secANCdefinition}

	Consider a bound state of two charged nuclei. %
	At distances such that %
	the interaction between the two particles is just the point-like Coulomb potential,
	namely in the ``external'' region,
	the wave-function is certainly a combination of Coulomb functions, defined in~\cite[sec.~33]{NISTdlmf}. These are the same functions employed in \cref{secCoulombbarrierpenetrabilityastrophysicalFactor}, but with a negative value for the energy (since the state is bound), and thus imaginary wave-number $k$ and Sommerfeld parameter $\eta$ (defined in \cref{eqDefinizioneParametroSommerfeld}). In fact, under this condition there is only one class of %
	functions which %
	do %
	not diverge at high distances: adopting the convention that $i k < 0$ (and thus $i \eta > 0$), these are the spherical outgoing Coulomb functions. %

	In general, the system bound state may comprise several components, each with definite relative orbital angular momentum modulus $l$. For each component, let $u_l(r)$ be the associated reduced radial wave-function%
	\footnote{In general, there may be several distinct components with equal $l$, differing by some other quantum number (e.g.~in the coupling with the spins). For brevity, such possibility is not reflected by the notation.}.
	As discussed above, at high radii $u_l(r)$ will be proportional to $H^+(k r)$.
	The proportionality constant %
	depends
	on the features of the bound state at smaller radii.
	In principle, since the bound state is normalised to some known value, %
	it would be sufficient to additionally fix the norm of the wave-function in the external region (or equivalently in the internal one) to deduce the scaling factor. %
	In practice,
	once the complete wave-function has been computed under some model, the proportionality constant is simply deduced by comparison with the expected trend.
	For convention, instead of the Coulomb function, $u_l$ is normally compared to the corresponding Whittaker function, $W_{l+1/2}(- 2 i k r)$
	(see e.g.~\cite[box 4.3, eq.~(4.5.24)]{ThompsonNunes2009})%
	\footnote{The Whittaker function also depends on the Sommerfeld parameter, but the associated index is here omitted for brevity.},
	which differs from $H_l^+(k r)$ by a multiplicative factor independent of $r$ (see~\cite[eq.~33.2.7]{NISTdlmf}). More explicitly:
	\begin{equation}\label{eqDefinizioneANC}
	u_l(r \gg 1) \to C_l \, W_{l+1/2}(- 2 i k r)
	\end{equation}
	where $C_l$ is the \emph{asymptotic normalisation coefficient} (ANC).
	Sometimes, a dimensionless version of the ANC, here labelled $\tilde{C}_l$ for definiteness, is %
	employed in literature. The two quantities are connected by the relation:
	\begin{equation}
	C_l = \sqrt{2 \m{k}} \, \tilde{C}_l
	\end{equation}
	(as can be found comparing e.g.~\cite[eq.~29]{Kukulin1984} with \scref{eqDefinizioneANC}). %

	Let $\mathcal{S}_l$ be the norm of $u_l$ (which, if $u_l$ is an overlap function, is given by the spectroscopic factor).
	Note that $C_l^2$ is proportional to $\mathcal{S}_l$: %
	it can thus be convenient to define the ``single-particle'' asymptotic normalisation coefficient, $b_l$, as:
	\begin{equation}
	b_l = C_l / \sqrt{\mathcal{S}_l} %
	\end{equation}
	
	Being an asymptotic property, $C_l$ lends itself to be probed experimentally, and plays a major role in peripheral processes (see e.g.~\cite[sec.~14.1.4]{ThompsonNunes2009}). %
	An in-depth review on %
	asymptotic normalisation coefficients can be found in %
	\cite{Mukhamedzhanov2022}.

	Differently than for spectroscopic amplitudes, the concept of asymptotic normalisation coefficient is not trivially generalised to bound states of more than two particles, %
	because the wave-function asymptotic trend becomes more complicated, especially if the system comprises several charged particles.
	Some results for a bound state of three particles can be found for instance in \cite{Grigorenko2020,Yarmukhamedov2018}.
	Note that, if it is of interest to study a peripheral %
	transfer reaction, the relevant asymptotic behaviour is the one at high core-transferred displacement and small distances between the internal components of both core and transferred nucleus.

\section[Expectation value of one-body observables]{Expectation value of one-body observables:\texorpdfstring{\\}{ }radii and quadrupole moments}%
\label{secLegameProprietaCompositoEFunzioneDonda}%

	Let $\Ket{\Psi}$ be the full anti-symmetrised wave-function for a given state %
	of a nucleus consisting of $A$ nucleons. %
	$\Ket{\Psi}$ describes all the system degrees of freedom (i.e.\ includes the coordinates of all nucleons). %
	The goal of the present \namecref{secLegameProprietaCompositoEFunzioneDonda} is to evaluate analytically the expectation value on $\Psi$ of some observables, %
	under the assumption that the system can in fact be described as a bound state of few inert clusters (in particular two or three). %
	The case of a two-cluster system is widely treated in literature. Several %
	concepts relevant for the present work appear %
	for instance in \cite{Nishioka1984,Merchant1985,Walliser1985,Fortunato2005,Mason2008} and references therein. %
	The evaluation of structure properties of systems of three or more clusters is instead less commonly covered. Some relevant discussion regarding \nuclide[6]{Li} was found in \cite{Bang1979}.
	
	The quantities considered in this work are, specifically: the binding energy, the root-mean-square radius, and the electric quadrupole moment. %
	The problem of finding the binding energy of a state is intimately connected to the problem of constructing the state itself, so it is not %
	discussed in this \namecref{secLegameProprietaCompositoEFunzioneDonda}. %
	All other mentioned observables are constructed as %
	one-body operators, and bear a particularly convenient form when expressed in the spatial coordinate representation, on which the discussion will thus be focused.
	
\subsection{Introduction}\label{secOperatoreUnCorpoDensitaUnCorpo}
	
	Consider an arbitrary labelling, by index $i$, of the $A$ nucleons in a system. Let $\op o_i$ be a \emph{single-particle} operator which acts only on the $i$-th nucleon degrees of freedom (possibly including its isospin projection). A \emph{one-body} observable $\op O$ is an operator with a structure such as %
	\begin{equation}\label{eqDefinizioneOperatoreUnCorpo}
		\op O = \sum_{i=1}^A \op o_i
	\end{equation}
	Due to indistinguishability, all $\op o_i$ must possess precisely the same form, and differ only by the subspace on which they act.
	More explicitly, consider a basis $\{\ket{\nu}\}$ of single-particle states, as in \cref{secAntiSymmetrisationFirstQuantisation}, chosen %
	so that each of its elements, $\ket{\nu}$, is an eigenstate of $\op o_i$ with eigenvalue $o(\nu)$. The operator $\op o_i$ can then be written as %
	\begin{equation}\label{eqDefinizioneOperatoreDiParticellaSingolaSecValAspettOsservabili}
		\op o_i = \sum_{\nu \in V} \Ket{i:\nu} o(\nu) \Bra{i:\nu}
	\end{equation}
	where the index $i$ is meant to specify that the states belong to the space of the $i$-th nucleon. %
	Consequently, the application of operator $\op O$ on a given element $\Ket{\nu_1,\dots,\nu_A}$ of the “non-anti-symmetric” $A$-particle system basis is
	\begin{equation}
		\op O \Ket{\nu_1,\dots,\nu_A} = \sum_{i=1}^A o(\nu_i) \Ket{\nu_1,\dots,\nu_A}
	\end{equation}
	$\op O$ is thus diagonal in this basis, with the diagonal matrix elements being precisely $\sum_i o(\nu_i)$.
	It is pointed out that no extra coefficient was added to \cref{eqDefinizioneOperatoreUnCorpo}%
	\footnote{%
	In a sense, the complete observable $\op O$ is here being treated as “extensive” with respect to the single-particle operator $\op o$.}. %
	However, the definition of some observables includes a scaling which depends on the number of particles involved. In that case, the discussion presented here can be repeated identically, carrying the extra factor. The slight modifications appearing in the results will be discussed explicitly for the case of root-mean-square radii in \cref{secRootMeanSquareRadius}. %
	
	In order to express the expectation value of $\op O$ on the state $\Psi$ of interest, $\media{O}$, it is convenient to use the expansion in the “non-anti-symmetric” basis in \cref{eqEspansioneStatoFisicoAParticelleBaseNonAntisimmetrizzata}. One obtains
	\begin{equation}%
		\media{O} = \frac{\Braket{\Psi|\op O|\Psi}}{\Braket{\Psi|\Psi}}
		= \frac{ \sum\limits_i \sum\limits_{\nu_1 \in V}\dots \sum\limits_{\nu_A \in V} o(\nu_i) \m{\Psi(\nu_1,\dots,\nu_A)}^2 }{ \sum\limits_{\nu_1 \in V}\dots \sum\limits_{\nu_A \in V} \m{\Psi(\nu_1,\dots,\nu_A)}^2 }
	\end{equation}
	The expectation value $\media{O}$ can in general depend on the specific reference frame and/or coordinate system employed (just as in classical physics). Most observables of interest acquire or retain the intended meaning when the composite system centre-of-mass is fixed in the origin. Here, such condition will be enforced when providing an explicit cluster-model expression for $\Ket{\Psi}$, see e.g.\ \cref{eqCentroDiMassaClustersNellOrigine}. %
	
	Due to the %
	invariance of $\m{\Psi(\nu_1,\dots,\nu_N)}$ under particle exchange (see \cref{eqEsempioDefinizioneAntisimmetrizzazioneSuCoefficientiStatoCompleto}), all elements of the sum over $i$ (running over the $A$ particles) are in fact equal. Arbitrarily choosing a convenient value for $i$, e.g.\ $i=1$, it is
	\begin{equation}\label{eqEspressioneValoreAspettazionePsi}
		\media{O}
		= \frac{A}{\Braket{\Psi|\Psi}} \sum\limits_{\nu_1 \in V} o(\nu_1) \sum\limits_{\nu_2 \in V} \dots \sum\limits_{\nu_A \in V} \m{\Psi(\nu_1,\dots,\nu_A)}^2
	\end{equation}
	The function %
	\begin{equation}\label{eqDefinizioneOneBodyDensity}
	\rho(\nu) = A \sum\limits_{\nu_2 \in V} \dots \sum\limits_{\nu_A \in V} \m{\Psi(\nu,\nu_2,\dots,\nu_A)}^2
	\end{equation}
	is the one-body “density” for the single-particle configuration $\nu$, i.e.\ the probability for nucleon 1 to be found in state $\Ket{\nu}$ regardless of the state of all other nucleons
	(see %
	\cite[eq~(5.1.21)]{ThompsonNunes2009} for a more general definition, not requiring $\Psi$ to be anti-symmetrised or computed in the centre-of-mass frame.
	\cite[eq~(5.1.20)]{ThompsonNunes2009} is instead the analogous of the definition given here).
	The normalisation of $\rho$ is here set such that $\sum_{\nu \in V} \rho(\nu) = A \Braket{\Psi|\Psi}$.
	While $\rho$ was defined as related specifically to nucleon 1, %
	it is %
	identical for any other one. %
	In fact, since $\m{\Psi(\nu_1,\dots,\nu_A)}$ is invariant under particle permutation, it is
	\begin{equation}
	\rho(\nu) = A! \sum_{\{\nu_2,\dots,\nu_A\}} \m{\Psi(\nu,\nu_2,\dots,\nu_A)}^2 = \sum_{\{\nu_2,\dots,\nu_A\}} \m{\Psi_W(\nu,\nu_2,\dots,\nu_A)}^2
	\end{equation}
	where the sum runs only over the distinct unordered sets $\{\nu_2,\dots,\nu_A\}$, and $\Psi_W$ is the same appearing in \cref{eqLegamePesiBaseNonSimmetrizzataEDiSlater}.
	It is here relevant to highlight the difference between a sum $\sum_{\{\nu_1,\nu_2,\dots,\nu_A\}}$ over the unordered set $\{\nu_1,\nu_2,\dots,\nu_A\}$, and the double sum $\sum_{\mu \in V} \sum_{\{\mu_2,\dots,\mu_A\}}$: for each term appearing in the former, there are $A$ distinct terms appearing in the latter (one for each possible choice of $\mu \in \{\nu_1,\nu_2,\dots,\nu_A\}$). %
	
	The expectation value of $\op O$ can then be rewritten in terms of the one-body density $\rho$:
	\begin{equation}
		\media{O} %
		= \frac{1}{\Braket{\Psi|\Psi}} \sum_{\nu \in V} o(\nu) \rho(\nu)
		= A \frac{ \sum\limits_{\nu \in V} o(\nu) \rho(\nu) }{ \sum\limits_{\nu \in V} \rho(\nu) }
	\end{equation}
	Let now, in general, $O(F)$ be a functional of the function $F(\nu)$, defined as:
	\begin{equation}
		O(F) = \sum_{\nu \in V} o(\nu) F(\nu)
	\end{equation}
	For convenience, assume in the following that $\Braket{\Psi|\Psi} = 1$, and thus
	\begin{equation}\label{eqEspressioneValoreAspettazioneRho}
		\sum\limits_{\nu \in V} \rho(\nu) = A
		\quad , \quad
		\media{O} = O(\rho) %
	\end{equation}
	$\media{O}$ would of course be the same for any other choice for $\Braket{\Psi|\Psi}$, but the given one simplifies the expressions in terms of $\rho$.

	The formalism involving the wave-function at the individual-nucleons level, employed here, was useful to derive \cref{eqEspressioneValoreAspettazioneRho}.
	In the following paragraphs, %
	instead, \cref{eqEspressioneValoreAspettazioneRho} will be taken as starting point to derive explicit expressions adopting a cluster model for $\Ket{\Psi}$.

	As a final note, %
	remind that \cref{eqEspressioneValoreAspettazioneRho} %
	neglects the nucleons internal structure, and may consequently be unsuited for computing an expectation value directly on a $A$-nucleons wave-function, unless the observable of interest is corrected appropriately. For instance, a nucleus root-mean-square radius would be greater than predicted by \cref{eqEspressioneValoreAspettazioneRho} because nucleons are not point-like (see e.g.\ \cite[eq.~(4.40)]{Satchler1990Introduction} \cite[eq.~(5.1.25)]{ThompsonNunes2009} for a correction on radii). %
	This is not an issue %
	within this work, %
	as \cref{eqEspressioneValoreAspettazioneRho} %
	is never employed directly, %
	in favour of cluster-model expressions where the contribution from each cluster internal structure is explicitly accounted for.

\subsection{General treatment of observables quadratic in the spatial position}\label{secObservablesdependingonlyonpositions} %

	Consider an observable $\op O$ whose single-particle %
	operator acts only on the nucleon position and is diagonal in the position basis. For the purpose of deriving the relations of interest here, all other degrees of freedom can consequently be ignored, as they are integrated away in \cref{eqEspressioneValoreAspettazionePsi}, and the $\nu$ in \cref{secOperatoreUnCorpoDensitaUnCorpo} can represent just the position $\v r$. The one-body density defined in \cref{eqDefinizioneOneBodyDensity} is then just the spatial particle number density of the nucleus (probability to find indifferently any nucleon in a given position $\v r$).
	Note %
	that this does \emph{not} imply that the other possible degrees of freedom (as the spin) do not play any role in determining the expectation values %
	(see for instance the discussion in \cref{secElectricQuadrupoleMoment,secLiDeformation}). %
	The point is, rather, that these extra degrees of freedom can only contribute to fix the nucleus structure, i.e.\ to derive a complete expression for $\rho(\v r)$, which is here taken as external input, but do not affect its connection with the observables of interest. %

	With this assumption,
	\cref{eqEspressioneValoreAspettazioneRho} becomes
	\begin{equation}\label{eqValAspettazioneOsservabileDipendenteDaPosizioneEspressioneGenerale}
		\int_{\campo{R}^3} \rho(\v r) \d^3 \v r = A
		\quad , \quad
		\media{O} = \int_{\campo{R}^3} o(\v r) \, \rho(\v r) \d^3 \v r %
	\end{equation}
	It will be convenient to define, in general, the functional
	\begin{equation}\label{eqDefinizioneFunzionaleValoreAtteso}
		O(F) \equiv \int_{\campo{R}^3} o(\v r) \, F(\v r) \d^3 \v r
	\end{equation}
	
	Suppose now that the composite system can be described as a bound state of $n$ inert clusters, usually distinguishable, which will be typically expressed through a wave-function in some Jacobi coordinates related to the relative motion between the clusters. From such wave-function it is possible to extract the probability density function for each cluster position, $\tilde\Phi_i(\v x)$, i.e.\ the probability of finding the centre-of-mass of cluster $i$ in position $\v x$, regardless of the position of all other clusters.
	The procedure will be discussed explicitly in \cref{secObservablesPositionTransformationtoJacobicoordinates}. %
	It is convenient to fix the normalisation of each $\tilde\Phi_i$ to 1, and consider two other integrals of interest:
	\begin{equation}\label{eqValAspettazioneOsservabileDipendenteDaPosizioneDefinizioneClusterCoMProbabilityDensity}\begin{aligned}
		\int_{\campo{R}^3} &\tilde\Phi_i(\v x) \d^3\v x = 1 \\
		\int_{\campo{R}^3} \v x \, &\tilde\Phi_i(\v x) \d^3 \v x = \v X_i \\
		\int_{\campo{R}^3} o(\v x) \, &\tilde\Phi_i(\v x) \d^3 \v x = O({\tilde\Phi_i})
	\end{aligned}\end{equation}
	where $\v X_i$ is thus the average position of cluster $i$ centre-of-mass in the system of coordinates of choice, and $O({\tilde\Phi_i})$ is the expected value of $o$ on $\tilde\Phi_i$ (note that the expression follows the definition in \cref{eqDefinizioneFunzionaleValoreAtteso}). %
	Each cluster, in turn, has a defined and immutable one-body density for its internal structure. Let $\tilde\rho_i(\v r)$ be the $i$-th cluster one-body density, defined with respect to the cluster own centre-of-mass (that is, placing the centre-of-mass in the origin), so that
	\begin{equation}\label{eqValAspettazioneOsservabileDipendenteDaPosizioneDefinizioneIntrinsicClusterOneBodyDensity}\begin{aligned}
		&\int_{\campo R^3} \tilde\rho_i(\v r) \d^3 \v r = A_i \\
		&\int_{\campo R^3} \v r \, \tilde\rho_i(\v r) \d^3\v r = \v 0 \\
		 A_i \frac{ \int_{\campo{R}^3} o(\v r) \, \tilde\rho_i \d^3 \v r }{ \int_{\campo{R}^3} \tilde\rho_i(\v r) \d^3 \v r } = &\int_{\campo R^3} o(\v r) \, \tilde\rho_i(\v r) \d^3\v r = O(\tilde\rho_i)
	\end{aligned}\end{equation}
	where $A_i$ is the $i$-th cluster mass number, so that $\sum_i A_i = A$, while $O(\tilde\rho_i)$ is the expectation value of $\op O$ on the isolated cluster $i$, defined precisely as $\media{O}$ in \cref{eqValAspettazioneOsservabileDipendenteDaPosizioneEspressioneGenerale}. %
	If the cluster centre-of-mass is in position $\v x$, its one-body density at position $\v r$ will be $\tilde\rho_i(\v r - \v x)$.
	Therefore, within the inert-cluster-model approximation, the one-body density $D_i$ for each cluster within the composite nucleus, taking into account the distribution for the cluster centre-of-mass position, is defined in terms of the “intrinsic” one-body density $\tilde\rho_i$ as the folding (see e.g.\ \cite[eq.~(4.40)]{Satchler1990Introduction} \cite[eq.~(5.1.25)]{ThompsonNunes2009})
	\begin{equation}
	D_i(\v r) = \int_{\campo{R}^3} \tilde\Phi_i(\v x) \, \tilde\rho_i(\v r - \v x) \d^3 \v x
	\end{equation}
	and the one-body density of the composite nucleus is then
	\begin{equation}\label{eqDefinizioneDensitaOneBodyComplessivaInModelloACluster}
	\rho(\v r) = \sum_{i=1}^n D_i(\v r)
	\end{equation}
	Note that, even within an \emph{inert}-cluster model, it may be necessary to take into account several possible states $\tilde\rho_i$ for some cluster (for instance, the different possible spin projections), each of them possibly corresponding to a different $O(\tilde\rho_i)$ and $\tilde\Phi_i$. If that is the case, each component has to be taken into account separately. The total one-body density for the composite nucleus would then be, using index $\nu$ to enumerate the cluster configurations,
	\begin{equation}
	\rho(\v r) = \sum_{i=1}^n \sum_{\nu} c_\nu D_{i,\nu}(\v r)
	\end{equation}
	where $c_\nu$ is the appropriate weight for each configuration.
	 As discussed in \cref{secOverlapFunctionSpectroscopicFactors}, the formalism would in principle become exact when a sufficient (infinite) amount of possible states %
	is included.
	Here, for brevity, %
	it will be assumed that each cluster has only one possible $\tilde\rho_i$ and $\tilde\Phi_i$. %
	
	Fix the composite nucleus centre-of-mass into the origin, so that
	\begin{equation}\label{eqValAspettazioneOsservabileDipendenteDaPosizioneEquazionePosizioneCentroDiMassaNucleoComposito}
	\v 0 = \int_{\campo{R}^3} \v r \rho(\v r) \d^3\v r %
	= \sum_{i=1}^n \int_{\campo{R}^6} \v r \, \tilde\Phi_i(\v x) \, \tilde\rho_i(\v r - \v x) \d^3\v r \d^3 \v x
	\end{equation}
	Performing first the integral over $\v r$, applying the variable change $\v y = \v r - \v x$, and then using \cref{eqValAspettazioneOsservabileDipendenteDaPosizioneDefinizioneIntrinsicClusterOneBodyDensity,eqValAspettazioneOsservabileDipendenteDaPosizioneDefinizioneClusterCoMProbabilityDensity}, \cref{eqValAspettazioneOsservabileDipendenteDaPosizioneEquazionePosizioneCentroDiMassaNucleoComposito} becomes
	\begin{equation}\label{eqCentroDiMassaClustersNellOrigine}
	\v 0 = \sum_{i=1}^n A_i \v X_i
	\end{equation}
	It is worth reminding that, as mentioned in \cref{secAntiSymmetrisationFirstQuantisation}, the formalism employed (non-relativistic treatment, nucleons share the same mass) causes each composite system (including each cluster separately, seen as a collection of nucleons) to have a mass proportional to their mass number.
	
	The expectation value $O(\rho)$ %
	of interest for the composite nucleus, as defined in \cref{eqValAspettazioneOsservabileDipendenteDaPosizioneEspressioneGenerale}, can then be computed adopting  %
	\cref{eqDefinizioneDensitaOneBodyComplessivaInModelloACluster} for the total one-body density, with the goal of removing the explicit dependence on the specific form of each cluster internal density $\tilde \rho_i$. %
	Applying again the variable change $\v y = \v r - \v x$, %
	it is
	\begin{equation}
		O(\rho) %
		= \sum_{i=1}^n \int_{\campo{R}^3} \d^3 \v x \ \tilde\Phi_i(\v x) \int_{\campo{R}^3} \d^3 \v y \ o(\v y + \v x) \, \tilde\rho_i(\v y)
	\end{equation}
	The result thus depends on the specific form of $o(\v r)$. Given that, here, it is of interest to study the root-mean-square radius and the electric quadrupole moment, it is sufficient to specialise to the case of operators with quadratic dependence on the position, meaning that
	\begin{equation}\label{eqDefinizioneFunzioneQuadratica}
	o(\v r) = %
	\sum_{k = \{x,y,z\}} o_k \, r_k^2 %
	\end{equation}
	where the sum runs over all spatial dimensions%
	\footnote{In particular, the “$x$” and “$y$” displayed here bear a completely different meaning than the vectors $\v x$ and $\v y$.},
	$r_k$ is the component of $\v r$ along the $k$-th dimension, and the $o_k$ %
	are arbitrary constants%
	\footnote{In general, the term “quadratic function” denotes a function which can also include linear and constant terms. Note %
		that linear terms give no contribution to $\media{O}$, as is it calculated in the centre-of-mass coordinate system, see %
		\cref{eqValAspettazioneOsservabileDipendenteDaPosizioneDefinizioneIntrinsicClusterOneBodyDensity,eqCentroDiMassaClustersNellOrigine}. Constant terms do contribute to $\media{O}$, but only as an overall offset which would add little to the discussion.}.
	Using this assumption, %
	it is
	\begin{equation}\label{eqValAspettazioneInserimentoIpotesiOsservabileQuadraticaNellaPosizione}
	O(\rho) %
	= \sum_{i=1}^n \int_{\campo{R}^3} \d^3 \v x \ \tilde\Phi_i(\v x) \int_{\campo{R}^3} \d^3 \v y \, \sum_{k} o_k \(y_k + x_k\)^2 \, \tilde\rho_i(\v y)
	\end{equation}
	The mixed term $2 y_k x_k$ appearing from $\(y_k + x_k\)^2$ %
	cancels out %
	by virtue of the second line in \cref{eqValAspettazioneOsservabileDipendenteDaPosizioneDefinizioneIntrinsicClusterOneBodyDensity}, i.e.\ because each $\rho_i(\v y)$ is defined placing the cluster centre-of-mass in the origin.
	The two remaining terms %
	are decoupled, and the three spatial components can be regrouped:
	\begin{equation}
	O(\rho) %
	= \sum_{i=1}^n \int_{\campo{R}^3} \d^3\v x \ \tilde\Phi_i(\v x) \int_{\campo{R}^3} \d^3 \v y \ \[ o(\v y) + o(\v x) \] \, \tilde\rho_i(\v y)
	\end{equation}
	Substituting the definitions given in \cref{eqValAspettazioneOsservabileDipendenteDaPosizioneDefinizioneClusterCoMProbabilityDensity,eqValAspettazioneOsservabileDipendenteDaPosizioneDefinizioneIntrinsicClusterOneBodyDensity},  one finds
	\begin{equation}\label{eqValAspOsservQuadraticiPosizioneRisultato}
	O(\rho) %
	= \sum_{i=1}^n \[ O(\tilde\rho_i) + A_i O({\tilde\Phi_i}) \]
	\end{equation}
	In \cref{eqValAspOsservQuadraticiPosizioneRisultato} there is no explicit dependence on the clusters internal structure, but only on the observables $O(\tilde\rho_i)$, which can be obtained from an experimental measurement or a dedicated model. In this way, it is possible to obtain the desired result $O(\rho)$ %
	for the composite nucleus considering only the bound state between the clusters, treated as structureless particles.

\subsubsection{Isospin-dependent one-body operators -- “Charge” observables} %

	A relevant generalization to the hypothesis employed in \cref{secObservablesdependingonlyonpositions}, of an operator acting solely on the nucleons position, is that of an operator distinguishing between protons and neutrons, or in other words, depending also on the nucleon isospin projection. The generic index $\nu$ defined in \cref{secAntiSymmetrisationFirstQuantisation} consequently must now include the isospin projection, %
	so the system one-body density, defined in \cref{eqDefinizioneOneBodyDensity}, can be written as $\rho_\tau(\v r)$,
	where $\rho_{\nuclide{p}}(\v r)$ and $\rho_{\nuclide{n}}(\v r)$ are the densities for protons and neutrons respectively.
	Similarly, the single-particle operator $\op o_i$ defined in \cref{eqDefinizioneOperatoreDiParticellaSingolaSecValAspettOsservabili} is now
	\begin{equation}
	\op o_i = \sum_\tau \int_{\campo{R}^3} \d^3 \v r \ \Ket{i: \v r, \tau} o_\tau(\v r) \Bra{i: \v r, \tau} %
	\end{equation}
	where $o_{\nuclide{p}}(\v r)$ and $o_{\nuclide{n}}(\v r)$ are the eigenvalues for the single-particle observable for protons and neutrons respectively. %
	\Cref{eqValAspettazioneOsservabileDipendenteDaPosizioneEspressioneGenerale} is thus generalised to
	\begin{equation}\label{eqValAspettazioneOsservPosizioneIsospinEspressioneGenerale}\begin{gathered}
		\int_{\campo{R}^3} \left[ \rho_{\nuclide{p}}(\v r) + \rho_{\nuclide{n}}(\v r) \right] \d^3 \v r = A
		\quad , \quad
		o_\tau(\v r) = \sum_{k} o_{\tau,k} \, r_k^2
	\\
		\media{O} = O(\rho) = A \frac{ \sum_\tau \int_{\campo{R}^3} o_{\tau}(\v r) \, \rho_{\tau}(\v r) \d^3 \v r }{ \sum_\tau  \int_{\campo{R}^3} \rho_{\tau}(\v r) \d^3 \v r } = \sum_\tau \int_{\campo{R}^3} o_{\tau}(\v r) \, \rho_{\tau}(\v r) \d^3 \v r
	\end{gathered}\end{equation}
	The contribution for each $\tau$ can be treated separately precisely as seen earlier.
	To this end, it is convenient to define
	\begin{equation}\label{eqDefinizioneValAspettazioneSingoloIsospinSuFGenerica}
	O_\tau(F) = \int_{\campo{R}^3} o_{\tau}(\v r) \, F(\v r) \d^3 \v r
	\end{equation}
	Furthermore, for each cluster, let $Z_i$ and $N_i$ be the number of protons and neutrons in cluster $i$, so that $Z_i + N_i = A_i$, and $\tilde\rho_{\tau,i}$ the one-body density for the internal structure of cluster $i$ regarding species $s$, normalised to the appropriate number of particles, e.g.\ $\int_{\campo R^3} \tilde\rho_{\nuclide{p},i}(\v r) \d^3 \v r = Z_i$, so that
	\begin{equation}\label{eqDefinizioneDensitaComplessivaIsospinDefinitoDaFoldingDeiCluster}
		\rho_\tau(\v r) = \sum_{i=1}^n \int_{\campo{R}^3} \tilde\Phi_i(\v x) \, \tilde\rho_{\tau,i}(\v r - \v x) \d^3 \v x
	\end{equation}
	where $n$ is again the total number of clusters.
	For observables obeying \cref{eqDefinizioneFunzioneQuadratica} (“quadratic”), one then finds, in analogy to \cref{eqValAspOsservQuadraticiPosizioneRisultato},
	\begin{equation}\label{eqValAspOsservQuadraticiPosizioneDistinzioneProtoniNeutroniRisultato}
	O(\rho) = \sum_{i=1}^n \[ O_{\nuclide{p}}(\tilde\rho_{\nuclide{p},i}) + O_{\nuclide{n}}(\tilde\rho_{\nuclide{n},i}) + Z_i O_{\nuclide{p}}({\tilde\Phi_i}) + N_i O_{\nuclide{n}}({\tilde\Phi_i}) \]
	\end{equation}
	As before, the values for $O_{\tau}(\tilde\rho_{\tau,i})$ are normally supplied independently, thus the observable definition and the cluster distributions $\tilde\Phi_i$ are again sufficient to determine the expectation value for the composite nucleus.

	A common occurrence, of interest for this work, is to distinguish between the “matter” and “charge” versions of an observable. “Matter” quantities arise by treating protons and neutrons equally, so that $o_{\nuclide{p}} = o_{\nuclide{n}}$ as in the derivation leading to \cref{eqValAspOsservQuadraticiPosizioneRisultato}.
	“Charge” quantities are often more readily accessible experimentally, and as such also of greater practical interest. They are measured by probing nuclei electromagnetically, for instance through electron scattering (see e.g.~\cite{Angeli2013} and references therein for a list of charge radius measurement techniques). %
	To evaluate charge observables theoretically at the nucleus level, in treatments where nucleons are assumed to be elementary, the standard recipe (see e.g.\ \cite[app.~A]{Mason2008}, \cite[eq.~(6.120), (6.132)]{Greiner1996}) %
	is to weight the nucleons densities on their global electric charge: this amounts to consider an operator acting exclusively on the protons, putting $o_{\nuclide{n}}(\v r) = 0$ and reducing \cref{eqValAspOsservQuadraticiPosizioneDistinzioneProtoniNeutroniRisultato} to %
	\begin{equation}\label{eqValAspOsservChargeQuadraticiPosizioneRisultato}
	\media{O}_{\text{ch}} = O_{\nuclide{p}}(\rho_{\nuclide{p}}) = \sum_{i=1}^n \[ O_{\nuclide{p}}(\tilde\rho_{\nuclide{p},i}) + Z_i O_{\nuclide{p}}({\tilde\Phi_i}) \]
	\end{equation}
	Such recipe is adopted in the present work as well, and it is expected to represent an acceptable approximation. It is relevant to point out, however, that neutrons, despite being electrically globally neutral, are in fact not insensible to electromagnetic probes due to their internal structure%
	\footnote{For instance, the neutron has a negative mean-square charge radius %
		\cite{Angeli2013,Atac2021}. This can be reconnected to its positively and negatively charged internal constituents not sharing the same spatial distribution.}.
		
	It is also underlined that, %
	when using \cref{eqValAspOsservChargeQuadraticiPosizioneRisultato}, it would \emph{not} be equivalent to consider a system of clusters composed exclusively of protons with density $\tilde\rho_{\nuclide{p},i}$, completely neglecting the neutrons.
	As can be seen from \cref{eqCentroDiMassaClustersNellOrigine,eqDefinizioneDensitaComplessivaIsospinDefinitoDaFoldingDeiCluster}, the centre-of-mass position of each cluster, which affects the total proton density, still depends on the total mass of each cluster%
	\footnote{The calculations could be repeated %
		considering the protons centre-of-mass instead of the total one, %
		but %
		this is normally not the quantity of interest. As an extreme example, imagine a system composed of a purely neutronic core and a single bound proton. The experimental charge radius of such system would not be just the proton internal radius, as the proton motion around the core would for instance make the scattering target appear bigger.}.

\subsubsection{Transformation to Jacobi coordinates}\label{secObservablesPositionTransformationtoJacobicoordinates}

	As anticipated, the last step is to connect the cluster-model wave-function of the composite nucleus
	with the distribution of each cluster %
	position, $\tilde\Phi_i$, and in particular their expected value $O({\tilde\Phi_i})$ as defined in \cref{eqDefinizioneFunzionaleValoreAtteso}, or equivalently the isospin-dependent $O_\tau({\tilde\Phi_i})$ defined in \cref{eqDefinizioneValAspettazioneSingoloIsospinSuFGenerica}.

	Consider a system of $n$ clusters, and let $\v x_i$ be the centre-of-mass position of the $i$-th cluster. Further let $\v d_i$ be a set of Jacobi coordinates: $\v d_1$ is the distance between clusters 1 and 2, $\v d_2$ is the distance between the centre-of-mass of clusters 1 and 2 and cluster 3, and so on, while $\v d_n$ is the whole system centre-of-mass position. The orientation of each vector is set as in \cite[eq.~(3.43)]{Wildermuth1977}~%
	\footnote{In \cite{Wildermuth1977} the equations refer to a set of particles with identical masses, thus some factors appear different. Also note that the opposite choice on the vectors orientation is often adopted, see e.g.~\cite{Nielsen2001}.}, so that
	\begin{equation}\label{eqDefinizioneCoordinateJacobi}\begin{aligned}
	\v d_1(\{\v x_i\}) &= \v x_1 - \v x_2 \\
	\v d_j(\{\v x_i\}) &= \frac{ \sum_{\alpha=1}^{j} A_\alpha \v x_\alpha }{ \sum_{\alpha=1}^{j} A_\alpha } - \v x_{j+1} \\
	\v d_n(\{\v x_i\}) &= \frac{ \sum_{\alpha=1}^{n} A_j \v x_j}{ \sum_{\alpha=1}^{n} A_j }
	\end{aligned}\end{equation}
	where the notation $\v d_j(\{\v x_i\})$ is a shorthand to signal that $\v d_j$ in the equation is expressed as a function of the set of single-particle coordinates $\v x_1, \dots, \v x_n$.
	The inverse transformation is
	\begin{equation}\label{eqDefinizioneCoordinateJacobiInversa}\begin{aligned}
	\v x_n(\{\v d_i\}) &= \v d_n - \frac{\sum_{\alpha=1}^{n-1} A_\alpha}{\sum_{\alpha=1}^{n} A_\alpha} \v d_{n-1} \\
	\v x_{n-1}(\{\v d_i\}) &= \v d_n + \frac{A_n}{\sum_{i=1}^{n} A_i} \v d_{n-1} - \frac{\sum_{i=1}^{n-2} A_i}{\sum_{i=1}^{n-1} A_i} \v d_{n-2} \\
	\v x_{j}(\{\v d_i\}) &= \v d_n + \sum_{\beta=j}^{n-1} \frac{A_{\beta+1}}{\sum_{\alpha=1}^{\beta+1} A_\alpha} \v d_{\beta} - \frac{\sum_{\alpha=1}^{j-1} A_\alpha}{\sum_{\alpha=1}^{j} A_\alpha} \v d_{j-1} \\
	\v x_1(\{\v d_i\}) &= \v d_n + \sum_{j=2}^n \frac{A_j}{\sum_{i=1}^{j} A_i} \v d_{j-1}
	\end{aligned}\end{equation}

	The cluster-model wave-function of the composite nucleus will be usually expressed in such a coordinate system, further setting the origin in the system centre-of-mass, that is, $\v d_n = \v 0$. Since in this \namecref{secLegameProprietaCompositoEFunzioneDonda} only observables diagonal in the position are of interest, %
	it is sufficient to consider the wave-function modulus-square, labelled $\Phi(\v d_1,\dots,\v d_{n-1})$, which is the probability density function to find the system in the configuration corresponding to the specified Jacobi coordinates%
	\footnote{Note that $\Phi$ and $\tilde \Phi_i$ have nothing to do with the $\Phi$ in \cref{secOverlapFunctionSpectroscopicFactors}.}.
	Let $f$ be an auxiliary function defined as follows:
	\begin{equation}\label{eqValAspettModelloAClusterDiagonaliPosizioneFunzioneAusiliaraMotoRelativo}
	f(\{\v d_i\}) = \Phi( \v d_1, \dots, \v d_{n-1} ) \ \delta( \v d_n )
	\end{equation}
	The one-cluster density $\tilde\Phi_j$ can then be written as
	\begin{equation}
	\tilde\Phi_j(\v x_j) = \int f(\{\v d_i(\{\v x_i\})\}) \d^3\v x_1\dots \d^3\v x_{j-1} \d^3\v x_{j+1} \dots \d^3\v x_n
	\end{equation}
	Its expected value $O(\tilde\Phi_j)$ is thus
	\begin{equation}
	O(\tilde\Phi_j) = \int_{\campo{R}^3} \tilde\Phi_j(\v r) o(\v r) \d^3\v r
	= \int o(\v x_j) f(\{\v d_i(\{\v x_i\})\}) \d^3\v x_1 \dots \d^3\v x_n
	\end{equation}
	the integral will be computed by changing variables to the Jacobi coordinates. In general, given two vectors of variables $\v y$ and $\v x(\v y)$ (of arbitrary length) and a function $f(\v x)$, it is %
	\begin{equation}
		\int f(\v x) \d\v x = \int f(\v x(\v y)) \m{J(\v y)} \d\v y
		\quad , \quad
		J(\v y) = \begin{vmatrix}
			\frac{\partial x_1}{\partial y_1}(\v y) & \frac{\partial x_1}{\partial y_2}(\v y) & \dots \\
			\frac{\partial x_2}{\partial y_1}(\v y) & \ddots & \vdots \\
			\dots & \dots & \dots
		\end{vmatrix}
	\end{equation}
	meaning that the measure $\m{J(\v y)}$ is the absolute value of the determinant $J(\v y)$.
	For the case of interest, it is convenient to compute the determinant for the variable change $\{ \v d_i \} \to \{ \v x_i \}$. For brevity, consider only the variable change for one component of each spatial vector (say, the $z$-projection), as the result is identical for all components. %
	The matrix is easily found from \cref{eqDefinizioneCoordinateJacobi}, yielding
	\begin{equation}\label{eqMatriceJacobianoMotoRelativoOsservabiliPosizione}
	J = \begin{vmatrix}
			1 & -1 & 0 & 0 & \dots \\
			\frac{A_1}{A_1+A_2} & \frac{A_2}{A_1+A_2} & -1 & 0 & \dots \\
			\vdots & \vdots & \vdots & \ddots & \vdots \\
			\frac{A_1}{\sum_i A_i} & \frac{A_2}{\sum_i A_i} & \dots & \dots & \frac{A_n}{\sum_i A_i}
		\end{vmatrix}
	\end{equation}
	Any matrix obtained by combining different rows, or different columns, of the original matrix has the same determinant \cite{EncyclopediaOfMathDeterminant}. By summing each column to the next one, starting from the first one, in order, then appropriately combining the rows, the matrix in \cref{eqMatriceJacobianoMotoRelativoOsservabiliPosizione} can be reduced to the identity matrix, implying that $J=1$.
	As a consequence,
	\begin{multline}\label{eqCollegamentoTermineMotoRelativoCasoNClusterFormaGenerica}
	O(\tilde\Phi_j) = \int o(\v x_j(\{\v d_i\})) f(\{\v d_i\}) \d^3\v d_1 \dots \d^3\v d_n = \\
	= \int o(\v x_j(\{\v d_i\}, \v d_n=\v 0)) \Phi(\v d_1, \dots, \v d_{n-1}) \d^3\v d_1 \dots \d^3\v d_{n-1}
	\end{multline}
	where the functions $\v x_j(\{\v d_i\}, \v d_n=\v 0)$, given in \cref{eqDefinizioneCoordinateJacobiInversa}, are to be computed in the coordinate system where the centre-of-mass position $\v d_n$ is the origin, due to the Dirac $\delta$ in appearing in \cref{eqValAspettModelloAClusterDiagonaliPosizioneFunzioneAusiliaraMotoRelativo}. %
	In the following, for brevity, let $\v x_j(\{\v d_i\})$ be a shorthand for $\v x_j(\{\v d_i\}, \v d_n=\v 0)$, where it is understood that calculations are performed in the system centre-of-mass.
	
	Specialise now to the case of a function $o(\v x)$ defined as in \cref{eqDefinizioneFunzioneQuadratica} (“quadratic” and not depending on isospin).
	The expectation value of interest is then given by \cref{eqValAspOsservQuadraticiPosizioneRisultato}, where the relative motion contribution is given by $\sum_j A_j O(\tilde\Phi_j)$ (sum over all clusters). As shown in the following, this takes %
	a notably simple form, for any $\Phi$.
	Let $R_o$ be the relevant term within the integrand in \cref{eqCollegamentoTermineMotoRelativoCasoNClusterFormaGenerica}, defined so that %
	\begin{equation}\label{eqDefinizioneAusiliareRoEspressioneMotoRelativoCoordinateJacobi}
	\sum_j A_j O(\tilde\Phi_j) = \int R_o(\v d_1, \dots, \v d_{n-1}) \, \Phi(\v d_1, \dots, \v d_{n-1}) \d^3\v d_1 \dots \d^3\v d_{n-1}
	\end{equation}
	More explicitly:
	\begin{equation}
	R_o(\v d_1, \dots, \v d_{n-1}) %
	= \sum_j A_j o(\v x_j(\{\v d_i\})) %
	= \sum_{k = \{x,y,z\}} o_k \sum_j A_j x_{j,k}^2(\{\v d_i\}) %
	\end{equation}
	Further let $A = \sum_{i=1}^{n} A_i$.
	First, note that all “mixed” terms appearing in $R_o$ when squaring each $x_{j,k}$, %
	i.e.\ terms depending on two distinct Jacobi coordinates, %
	cancel out when performing the complete sum. %
	As an explicit example, consider all the terms in $d_{n-1,k} d_{n-2,k}$, which are easily extracted by inspection of \cref{eqDefinizioneCoordinateJacobiInversa}:
	\begin{equation}
	\frac{2 o_k A_n d_{n-1,k} d_{n-2,k}}{A (A-A_n)} \[ A_{n-1} (A-A_n-A_{n-1}) - (A-A_n-A_{n-1}) A_{n-1} \] %
	\end{equation}
	the same cancellation takes place for all other “mixed” terms as well. As a result, $R_o$ is a sum of only terms which depend on a single Jacobi coordinate $\v d_i$: this simplifies the integral in \cref{eqCollegamentoTermineMotoRelativoCasoNClusterFormaGenerica}. %
	These terms also %
	take a simple form. %
	Let $\mu_{i}$ be the reduced mass number between cluster $i+1$ and the composition of clusters from 1 to $i$. Then, as can be found again by inspection of \cref{eqDefinizioneCoordinateJacobiInversa}, it is
	\begin{equation}\label{eqDefinizioneNumeroDiMassaRidotto}
	R_o = \sum_{k = \{x,y,z\}} o_k \sum_i \mu_i d_{i,k}^2 = \sum_{i=1}^{n-1} \mu_i o(\v d_i)
	\quad , \quad
	\mu_i \equiv \frac{A_{i+1} \sum_{j=1}^{i} A_j }{\sum_{j=1}^{i+1} A_j}
	\end{equation}
	Finally, for brevity, let $\phi_i$ be the “one-coordinate” density derived by integrating $\Phi$ on all Jacobi coordinates except $\v d_i$,
	\begin{equation}\label{eqMatterValAspInCoordJacobiDefinizioneOneCoordinateDensity}
	\phi_i(\v d_i) = \int \Phi(\v d_1, \dots, \v d_{n-1}) \d^3\v d_1 \dots \d^3\v d_{i-1} \d^3\v d_{i+1} \dots \d^3\v d_{n-1}
	\end{equation}
	Using such definition, the expectation value of any one-body operator whose single-particle operator acts only on the spatial position (in particular it is isospin-independent) and is diagonal in the spatial position basis with “quadratic” eigenvalues (as in \cref{eqDefinizioneFunzioneQuadratica}), given in \cref{eqValAspOsservQuadraticiPosizioneRisultato}, can be written as %
	\begin{equation}\label{eqValAspOsservQuadraticiPosizioneNoIsospinRisultatoCasoNCluster}
	\media{O} = O(\rho) = \sum_{i=1}^n O(\tilde\rho_i) + \sum_{i=1}^{n-1} \mu_i O(\phi_i)
	\end{equation}
	where $O(F)$ and $\tilde\rho_i$ are defined in \cref{eqDefinizioneFunzionaleValoreAtteso,eqValAspettazioneOsservabileDipendenteDaPosizioneDefinizioneIntrinsicClusterOneBodyDensity}, respectively, and $n$ is the number of clusters.
	
	For a two-cluster system there is only one relevant Jacobi coordinate, the distance between the two clusters (the centre-of-mass position plays no role, as mentioned), thus $\phi_1$ is just the full probability density $\Phi$. It is
	\begin{equation}\label{eqValAspOsservQuadraticiPosizioneRisultatoCasoDueCluster}
	O(\rho) = \sum_i O(\tilde\rho_i) + \frac{A_1 A_2}{A} O(\Phi)
	\end{equation}
	For a three-cluster system, \cref{eqValAspOsservQuadraticiPosizioneNoIsospinRisultatoCasoNCluster} is more explicitly written as
	\begin{equation}\label{eqValAspOsservQuadraticiPosizioneRisultatoCasoTreCluster}
	O(\rho) = \sum_i O(\tilde\rho_i) + \frac{A_1 A_2}{A_1+A_2} O(\phi_1) + \frac{(A_1+A_2) A_3}{A_1+A_2+A_3} O(\phi_2)
	\end{equation}

\subsubsection{Transformation to Jacobi coordinates for “charge” observables}
	
	If the expectation value for a “charge” observable, given in \cref{eqValAspOsservChargeQuadraticiPosizioneRisultato}, is of interest, the reasoning described in the preceding section is still valid, but the quantity of interest is now $\sum_j Z_j O_{\nuclide{p}}(\tilde\Phi_j)$. The unbalance between $Z_i$ and the factors appearing in the coordinate transformation, related to the clusters mass, cause the equivalent of \cref{eqValAspOsservQuadraticiPosizioneNoIsospinRisultatoCasoNCluster} to be more cumbersome in the general case, mainly due to the appearance of “mixed” terms involving the product of two Jacobi coordinates. Depending on the specific situation, in order to compute an expectation value it may be more convenient to first deduce the distribution of positions for each cluster, $\tilde\Phi_i$ (possibly converting from the probability density in Jacobi coordinates $\Phi$), and then use \cref{eqValAspOsservChargeQuadraticiPosizioneRisultato} directly. Here, instead, the expectation value in terms of $\Phi$ is derived.
	In analogy with \cref{eqDefinizioneAusiliareRoEspressioneMotoRelativoCoordinateJacobi}, let $R_c$ be the function such that
	\begin{equation}\label{eqDefinizioneAusiliareRcEspressioneMotoRelativoCoordinateJacobiOsservabiliCarica}
	\sum_j Z_j O(\tilde\Phi_j) = \int R_c(\v d_1, \dots, \v d_{n-1}) \, \Phi(\v d_1, \dots, \v d_{n-1}) \d^3\v d_1 \dots \d^3\v d_{n-1}
	\end{equation}
	It is found that %
	\begin{multline}\label{eqEspressioneFunzioneAusliariaRValoreAspettazioneOsservabileCaricaJacobiCoordinates}
	R_{c}(\v d_1, \dots, \v d_{n-1}) %
	= \sum_j Z_j o_{\nuclide{p}}(\v x_j(\{\v d_i\}))
	= \sum_{k = \{x,y,z\}} o_{\nuclide{p},k} \sum_j Z_j x_{j,k}^2(\{\v d_i\}) = \\
	\shoveleft{= \sum_{j=1}^{n-1} \frac{ \( \sum_{\alpha=1}^{j} A_\alpha \)^2  Z_{j+1} + A_{j+1}^2 \( \sum_{\alpha=1}^{j} Z_\alpha \) }{ \( \sum_{\alpha=1}^{j+1} A_\alpha \)^2 } o_{\nuclide{p}}(\v d_{j}) +}\\+ 2 \sum_{i=1}^{n-2} \sum_{j=i+1}^{n-1} A_{j+1} \frac{ A_{i+1} \( \sum_{\alpha=1}^{i} Z_\alpha \) - \( \sum_{\alpha=1}^{i} A_\alpha \) Z_{i+1} }{ \( \sum_{\alpha=1}^{i+1} A_\alpha \) \( \sum_{\alpha=1}^{j+1} A_\alpha \) } \sum_k o_{\nuclide{p},k} d_{i,k} d_{j,k}
	\end{multline}
	For instance, for a three-cluster system, %
	\begin{multline}\label{eqEspressioneFunzioneAusliariaRperTreClusterValoreAspettazioneOsservabileCaricaJacobiCoordinates}
	R_{c}(\v d_1, \v d_2) %
	= \frac{ A_1^2  Z_2 + A_2^2 Z_1 }{ \( A_1 + A_2 \)^2 } o_{\nuclide{p}}(\v d_1) + \frac{ \( A_1 + A_2 \)^2  Z_3 + A_3^2 \( Z_1 + Z_2 \) }{ (A_1+A_2+A_3)^2 } o_{\nuclide{p}}(\v d_2) +\\+ 2 \frac{A_1 A_2 A_3}{A_1 + A_2 +A_3} \frac{ Z_1/A_1 -Z_2/A_2 }{ A_1 + A_2 } \sum_k o_{\nuclide{p},k} d_{1,k} d_{2,k}
	\end{multline}
	
	The mixed terms $d_{i,k} d_{j,k}$ %
	do not, of course, appear for a two-cluster system, since there is only one relevant Jacobi coordinate in that case. Thus, by combining \cref{eqValAspettazioneOsservPosizioneIsospinEspressioneGenerale,eqDefinizioneCoordinateJacobiInversa,eqCollegamentoTermineMotoRelativoCasoNClusterFormaGenerica}, one finds directly that, for any two-particle system with relative-motion probability density $\Phi$ and any “charge” operator $\op O_{\nuclide{p}}$ (obeying the same hypothesis required for \cref{eqValAspOsservQuadraticiPosizioneNoIsospinRisultatoCasoNCluster} apart for isospin-independence), %
	the expectation value in \cref{eqValAspOsservChargeQuadraticiPosizioneRisultato} can be rewritten as
	\begin{equation}\label{eqValAspOsservDiCaricaQuadraticiPosizioneRisultatoCasoDueCluster}
	\media{O}_{\text{ch}} = O_{\nuclide{p}}(\rho_{\nuclide{p}}) = \sum_i O_{\nuclide{p}}(\tilde\rho_{\nuclide{p},i}) + \frac{Z_1 A_2^2 + Z_2 A_1^2}{A^2} O_{\nuclide{p}}(\Phi)
	\end{equation}
	
	Going back to the general case, %
	in analogy with \cref{eqMatterValAspInCoordJacobiDefinizioneOneCoordinateDensity} let $\phi_{ij}$ be the “two-coordinate” density of $\v d_i$ and $\v d_j$,
	\begin{equation}\label{eqMatterValAspInCoordJacobiDefinizioneTwoCoordinateDensity}
	\phi_{ij}(\v d_i,\v d_j) = \int \Phi(\v d_1, \dots) \d^3\v d_1 \dots \d^3\v d_{i-1} \d^3\v d_{i+1} \dots \d^3\v d_{j-1} \d^3\v d_{j+1} \dots
	\end{equation}
	The expectation value of interest may then %
	be computed from \cref{eqDefinizioneAusiliareRcEspressioneMotoRelativoCoordinateJacobiOsservabiliCarica,eqEspressioneFunzioneAusliariaRValoreAspettazioneOsservabileCaricaJacobiCoordinates} in terms of all $\phi_i$ and $\phi_{ij}$.
	In the following, the result is specialised employing a decomposition in the spherical-harmonics basis. %
	
	Note that each $d_{\alpha,k}$, being the $x$, $y$ or $z$ component of $\v d_\alpha$, can be written as $d_{\alpha,k} = d_\alpha \sum_{m} c_{k,m} Y_{1,m}(\Omega_\alpha)$, where $Y_{l,m}$ is a spherical harmonic, %
	$\Omega_\alpha$ (defined as in \cref{secAppendiceSphericalHarmonics}) is the direction of $\v d_\alpha$ with respect to $Y_{l,m}$ quantisation axis, %
	$d_\alpha = \sqrt{\v d_\alpha \cdot \v d_\alpha}$, and $c_{k,m}$ is a set of appropriate coefficients.
	The combination $\sum_k o_{\nuclide{p},k} d_{i,k} d_{j,k}$ can in this way be written separating radial and angular coordinates of both Jacobi coordinates:
	\begin{equation}
		\sum_k o_{\nuclide{p},k} d_{i,k} d_{j,k}
		= d_i d_j \sum_{k,m,m'} o_{\nuclide{p},k} c_{k,m} c_{k,m'} Y_{1,m}(\Omega_i) Y_{1,m'}(\Omega_j) %
	\end{equation}

	The precise result depends on $o_{\nuclide{p},k}$, and thus on the observable under study. Two cases in particular are of interest here. First, if $o_{\nuclide{p}}(\v r) = o_{\nuclide{p}} r^2$, so that $o_{\nuclide{p},k} = o_{\nuclide{p}}$ is independent of the direction, then %
	\begin{multline}\label{eqEspressioneOperatoreDiRaggioQuadraticoDecompostoInArmonicheSferiche}
	\sum_k o_{\nuclide{p},k} d_{i,k} d_{j,k} = o_{\nuclide{p}} \v d_i \cdot \v d_j = o_{\nuclide{p}} d_i d_j \sqrt{\frac{4 \pi}{3}} Y_{10}(\Omega_{ij}) = \\
	= o_{\nuclide{p}} d_i d_j \frac{4 \pi}{3} \sum_m \coniugato{Y}_{1m}(\Omega_{j}) Y_{1m}(\Omega_{i}) = \\
	= o_{\nuclide{p}} d_i d_j \frac{4 \pi}{3} \sum_m (-1)^m Y_{1,-m}(\Omega_{j}) Y_{1m}(\Omega_{i})
	\end{multline}
	where $\Omega_{ij}$ is the direction of $\v d_i$ with respect to $\v d_j$, and \cref{eqAdditionTheoremSphericalHarmonics,eqComplexConjugateSphericalHarmonics} were employed.

	The second useful case is when $o_{\nuclide{p}}(\v r) = o_{\nuclide{p}} z^2$, so that $o_{\nuclide{p},k} = 0$ except for the $z$ component. Then,
	\begin{equation}
	\sum_k o_{\nuclide{p},k} d_{i,k} d_{j,k} = o_{\nuclide{p}} d_{i,z} d_{j,z}
	= o_{\nuclide{p}} d_i d_j \frac{4 \pi}{3} Y_{10}(\Omega_{j}) Y_{10}(\Omega_{i})
	\end{equation}
	For convenience, let
	$\xi_{ij}(\v d_i, \v d_j)$ be a function whose square-modulus is the “two-coordinate” probability density $\phi_{ij}(\v d_i, \v d_j)$ in \cref{eqMatterValAspInCoordJacobiDefinizioneTwoCoordinateDensity}.
	$\xi$ can be decomposed in the basis of the spherical harmonics of each Jacobi coordinate, giving
	\begin{equation}\label{eqScritturaCompattaXiPerTrasformazioneJacobiOsservabiliCarica}
	\xi(\v d_i, \v d_j) = \sum_{\lambda_i,\mu_i,\lambda_j,\mu_j} \frac{ F_{\lambda_i,\mu_i,\lambda_j,\mu_j}(d_i,d_j) }{ d_i d_j } Y_{\lambda_i,\mu_i}(\Omega_i) Y_{\lambda_j,\mu_j}(\Omega_j)
	\end{equation}
	where $F_{\lambda_i,\mu_i,\lambda_j,\mu_j}$ is a reduced radial function giving the appropriate weight to each term: in the following, the subscripts may be omitted for brevity. In $\m{\xi}^2$ there are several “mixed” terms involving the product of two different spherical harmonics for some Jacobi coordinate. These terms can contribute in the mixed terms in \cref{eqEspressioneFunzioneAusliariaRValoreAspettazioneOsservabileCaricaJacobiCoordinates}, if the $Y_{1m}$ appearing there can couple them. More explicitly, consider
	\begin{multline}
	\sum_k o_{\nuclide{p},k} \int d_{i,k} d_{j,k} \, \Phi(\v d_1, \dots, \v d_{n-1}) \d^3\v d_1 \dots \d^3\v d_{n-1} = \\
	= \sum_{k,m,m'} o_{\nuclide{p},k} c_{k,m} c_{k,m'} \int d_{i} d_{j} Y_{1m}(\Omega_i) Y_{1m'}(\Omega_j) \, \phi_{ij}(\v d_i, \v d_j) \d^3\v d_i \d^3\v d_{j} = \\
	\shoveleft{= \sum_{k,m,m'} o_{\nuclide{p},k} c_{k,m} c_{k,m'} \sum_{\substack{\lambda_i,\mu_i,\lambda_j,\mu_j,\\\lambda'_i,\mu'_i,\lambda'_j,\mu'_j}} \int \coniugato{F'}(d_i,d_j) d_i d_j F(d_i,d_j) \d d_i \d d_j \cdot}\\\cdot \int \coniugato{Y}_{\lambda'_i,\mu'_i}(\Omega_i) Y_{1m}(\Omega_i) Y_{\lambda_i,\mu_i}(\Omega_i) \d\Omega_i \int \coniugato{Y}_{\lambda'_j,\mu'_j}(\Omega_j) Y_{1m'}(\Omega_j) Y_{\lambda_j,\mu_j}(\Omega_j) \d\Omega_{j} = \\
	\shoveleft{= \sum_{k,m,m'} o_{\nuclide{p},k} c_{k,m} c_{k,m'} \sum_{\substack{\lambda_i,\mu_i,\lambda_j,\mu_j,\\\lambda'_i,\mu'_i,\lambda'_j,\mu'_j}} \Braket{F_{\lambda'_i,\mu'_i,\lambda'_j,\mu'_j}|d_i d_j|F_{\lambda_i,\mu_i,\lambda_j,\mu_j}} \cdot}\\\cdot \Braket{ Y_{\lambda'_i,\mu'_i} | Y_{1m} | Y_{\lambda_i,\mu_i} } \Braket{ Y_{\lambda'_j,\mu'_j} | Y_{1m'} | Y_{\lambda_j,\mu_j} }
	\end{multline}
	The result of the angular integration is reported in \cref{eqBraketTreArmonicheSferiche}. %
	As required by Wigner-Eckart theorem, %
	only terms with $\m{\lambda-\lambda'}=1$ and $\m{\mu-\mu'}\leq 1$ can contribute. %
	This result can then be substituted into \cref{eqDefinizioneAusiliareRcEspressioneMotoRelativoCoordinateJacobiOsservabiliCarica}. For clarity, the resulting expression for $\media{O}_{\text{ch}}$ is reported:
	\begin{multline}\label{eqValAspettazioneOsservabileCaricanCluster}
	\media{O}_{\text{ch}}
	= \sum_i O_{\nuclide{p}}(\tilde\rho_{\nuclide{p},i})
		+ \sum_{j=1}^{n-1} \frac{ \( \sum_{\alpha=1}^{j} A_\alpha \)^2  Z_{j+1} + A_{j+1}^2 \( \sum_{\alpha=1}^{j} Z_\alpha \) }{ \( \sum_{\alpha=1}^{j+1} A_\alpha \)^2 } O_{\nuclide{p}}(\phi_j) +\\
		+ 2 \sum_{i=1}^{n-2} \sum_{j=i+1}^{n-1} A_{j+1} \frac{ A_{i+1} \( \sum_{\alpha=1}^{i} Z_\alpha \) - \( \sum_{\alpha=1}^{i} A_\alpha \) Z_{i+1} }{ \( \sum_{\alpha=1}^{i+1} A_\alpha \) \( \sum_{\alpha=1}^{j+1} A_\alpha \) } \sum_{k,m,m'} o_{\nuclide{p},k} c_{k,m} c_{k,m'} \cdot \\
		\cdot \sum_{\substack{\lambda_i,\mu_i,\lambda_j,\mu_j,\\\lambda'_i,\mu'_i,\lambda'_j,\mu'_j}} \Braket{F'|d_i d_j|F} \Braket{ Y_{\lambda'_i,\mu'_i} | Y_{1m} | Y_{\lambda_i,\mu_i} } \Braket{ Y_{\lambda'_j,\mu'_j} | Y_{1m'} | Y_{\lambda_j,\mu_j} }
	\end{multline}
	Regarding the determination of the weights $F_{\lambda_i,\mu_i,\lambda_j,\mu_j}$, in practice it is often useful to recast them trough some angular momentum coupling scheme.

\subsection{Root-mean-square radius}\label{secRootMeanSquareRadius}

\subsubsection{Root-mean-square matter radius}

	The root-mean-square matter radius operator for a nucleons composed of $A$ nucleons, $\op r_{\text{m}}^2$, is defined analogously to a classical root-mean-square, see e.g.\ \cite[eq.~(5.1.24)]{ThompsonNunes2009} or \cite[app.~A]{Mason2008}:
	\begin{equation}\label{eqDefinizioneMatterrootmeansquareradius}
	\op r_{\text{m}}^2 = \frac{1}{A} \sum_{i=1}^A \op r_i^2
	\end{equation}
	where $\op r_i$ is nucleon $i$ position operator, in the composite-nucleus centre-of-mass coordinates.
	Apart from the square, note the factor $1/A$. This grants, for instance, that two systems having precisely the same profile for the one-body density but composed of a different number of particles will share the same radius.
	
	An expression for $\media{r_{\text{m}}^2}$ can be readily found using the results shown earlier. Start from an operator $\op O$ including no scaling factor, as in \cref{eqDefinizioneOperatoreUnCorpo}: here, it would be $\op O = \sum_{i=1}^A \op r_i^2$.
	\Cref{eqValAspOsservQuadraticiPosizioneRisultatoCasoDueCluster} consequently holds for $\media{O}$. %
	However, the quantities of interest are now instead the expectation values of the “rescaled” operator
	\begin{equation}
	\op O^* = \frac{1}{A} \op O %
	\end{equation}
	Provided that each cluster one-body density (and thus also the nucleus one) is normalised to the corresponding number of particles, the expectation value of $\op O^*$ on a distribution $F(\v r)$, labelled $O^*(F)$, can be defined, considering \cref{eqValAspettazioneOsservabileDipendenteDaPosizioneEspressioneGenerale}, as %
	\begin{equation}\label{eqDefinizioneValoreAspettazioneOperatoreRiscalato}
	O^*(F) = \frac{ \int_{\campo{R}^3} o(\v r) \, F(\v r) \d^3 \v r }{ \int_{\campo{R}^3} F(\v r) \d^3 \v r } = \frac{ O(F) }{ \int_{\campo{R}^3} F(\v r) \d^3 \v r }
	\end{equation}
	where $O(F)$ is the expectation value of $\op O$. %
	Substituting \cref{eqDefinizioneValoreAspettazioneOperatoreRiscalato} into \cref{eqValAspOsservQuadraticiPosizioneRisultato} immediately yields
	\begin{equation}\label{eqValAspOsservQuadraticiPosizioneRiscalatiRisultato}
	A \, O^*(\rho) = \sum_i A_i \[ O^*(\tilde\rho_i) + O^*({\tilde\Phi_i}) \]
	\end{equation}
	Note that the normalisations and definitions were chosen so that, conveniently, $O(\tilde\Phi_i) = O^*(\tilde\Phi_i)$, thus the relative-motion contribution is precisely the same found in \cref{eqValAspOsservQuadraticiPosizioneNoIsospinRisultatoCasoNCluster}, %
	hence,
	\begin{equation}
	A \, O^*(\rho) = \sum_{i=1}^{n} A_i O^*(\tilde\rho_i) + \sum_{i=1}^{n-1} \mu_i O^*(\phi_i)
	\end{equation}
	with $\mu_i$ and $\phi_i$ defined as in \cref{eqDefinizioneNumeroDiMassaRidotto,eqMatterValAspInCoordJacobiDefinizioneOneCoordinateDensity}.
	
	In particular, the root-mean-square matter radius of a two-cluster system is, %
	in analogy with \cref{eqValAspOsservQuadraticiPosizioneRisultatoCasoDueCluster} and as in \cite[eq.~(A.4)]{Mason2008},
	\begin{equation}\label{eqRisultatoEsplicitoRaggioDiMassaDueCluster}
	\media{r_{\text{m},\text{composite}}^2} = \frac{1}{A} \[ A_1 \media{r_{\text{m},1}^2} + A_2 \media{r_{\text{m},2}^2} + \frac{A_1 A_2}{A} \, \media{d^2}_\Phi \]
	\end{equation}
	where a more explicit notation was adopted for clarity: $\media{r_{\text{m},\text{composite}}^2}$ is the expectation value of $\op r_{\text{m}}^2$ for the composite nucleus, $\media{r_{\text{m},i}^2}$  the expectation value for the $i$-th cluster, composed of $A_i$ nucleons, such that $\sum A_i = A$, and $\media{d^2}_\Phi$ %
	is the expected value of $d^2$ for the relative-motion probability density $\Phi(\v d)$, namely $\media{d^2}_\Phi = \int_{\campo{R}^3} d^2 \Phi(\v d) \d^3 \v d$.
	
	For completeness, in the limit where one cluster is much lighter than the other one, $A_1 \gg A_2$, and possibly also smaller, $\media{r_{\text{m},1}^2} \gg \media{r_{\text{m},2}^2}$, \cref{eqRisultatoEsplicitoRaggioDiMassaDueCluster} can be approximated to
	\begin{equation}
	\media{r_{\text{m},\text{composite}}^2} = \frac{A_1}{A} \media{r_{\text{m},1}^2} + \frac{A_2}{A} \media{d^2}_\Phi
	\end{equation}
	which is sometimes employed %
	for practical calculations. In this work, the full expression \cref{eqRisultatoEsplicitoRaggioDiMassaDueCluster} is always adopted.
	
	For a three-cluster system, in analogy with \cref{eqValAspOsservQuadraticiPosizioneRisultatoCasoTreCluster}, %
	\begin{multline}
	\media{r_{\text{m},\text{comp}}^2} = \frac{A_1}{A} \media{r_{\text{m},1}^2} + \frac{A_2}{A} \media{r_{\text{m},2}^2} + \frac{A_3}{A} \media{r_{\text{m},3}^2} +\\+ \frac{A_1 A_2}{A (A_1+A_2)} \media{d^2}_{\phi_1} + \frac{(A_1+A_2) A_3}{A^2} \media{d^2}_{\phi_2}
	\end{multline}

\subsubsection{Root-mean-square charge radius}

	In analogy with \cref{eqDefinizioneMatterrootmeansquareradius}, the root-mean-square charge radius operator is defined as (see \cite[eq.~(4.40)]{Satchler1990Introduction} or \cite[app.~A]{Mason2008})
	\begin{equation}\label{eqDefinizioneChargerootmeansquareradius}
	\op r_{\text{ch}}^2 = \frac{1}{Z} \sum_{i=1}^Z \op r_i^2
	\end{equation}
	where $Z$ is the nucleus charge number, and the sum runs only over the protons in the nucleus (as in \cref{eqValAspOsservChargeQuadraticiPosizioneRisultato}). Note how the rescaling for the operator is not computed considering the normalisation of the complete density, $A$, but solely that of the protons fraction. With this definition, for instance, two systems sharing the same proton density but including a different number of neutrons would have the same radius.
	
	The reasoning applied to the matter radius %
	can be repeated identically here, keeping into account the different scaling required (essentially, in \cref{eqValAspOsservQuadraticiPosizioneRiscalatiRisultato} $A_i$ is substituted by $Z_i$): \cref{eqDefinizioneValoreAspettazioneOperatoreRiscalato} is defined so that it still holds here.
	
	For a two-cluster system, \cref{eqValAspOsservDiCaricaQuadraticiPosizioneRisultatoCasoDueCluster} becomes
	\begin{equation}\label{eqRaggioDiCaricaDueCluster}
	\media{r_{\text{ch},\text{comp}}^2} = \frac{1}{Z} \[ Z_1 \media{r_{\text{ch},1}^2} + Z_2 \media{r_{\text{ch},2}^2} + \frac{Z_1 A_2^2 + Z_2 A_1^2}{A^2} \, \media{d^2}_\Phi \]
	\end{equation}
	where $\sum Z_i = Z$. This is the same as in \cite[eq.~(A.7)]{Mason2008}.
	
	For a three-cluster system, combining \cref{eqEspressioneFunzioneAusliariaRperTreClusterValoreAspettazioneOsservabileCaricaJacobiCoordinates,eqEspressioneOperatoreDiRaggioQuadraticoDecompostoInArmonicheSferiche,eqValAspettazioneOsservabileCaricanCluster,eqValAspOsservQuadraticiPosizioneRiscalatiRisultato},
	\begin{multline}\label{eqRaggioDiCaricaTreCluster}
		\media{r_{\text{ch},\text{comp}}^2}
		= \sum_i \frac{Z_i}{Z} \media{r_{\text{ch},i}^2} + 2 \frac{A_3}{A} \frac{ Z_1 A_2 - Z_2 A_1 }{ Z (A_1 + A_2) } \media{\v d_1 \cdot \v d_2}_{\Phi} +\\
		+ \frac{ Z_1 A_2^2 + Z_2 A_1^2 }{ Z \( A_1 + A_2 \)^2 } \media{d^2}_{\phi_1} + \frac{ \( Z_1 + Z_2 \) A_3^2 + Z_3 \( A_1 + A_2 \)^2 }{ Z A^2 } \media{d^2}_{\phi_2}
	\end{multline}
	where $\media{\v d_1 \cdot \v d_2}_{\Phi} = \int_{\campo{R}^3} \v d_1 \cdot \v d_2 \Phi(\v d) \d^3\v d_1 \d^3\v d_2$.

\subsection{Electric quadrupole moment}\label{secElectricQuadrupoleMoment}

\subsubsection{Matter quadrupole moment}

	Arbitrarily choose a quantisation axis $z$. The matter quadrupole moment operator, $\op Q_{\text{m}}$ can be defined through \cref{eqDefinizioneOperatoreUnCorpo,eqDefinizioneOperatoreDiParticellaSingolaSecValAspettOsservabili}, adopting as single-particle eigenvalues (“$o(\nu)$” in \cref{secLegameProprietaCompositoEFunzioneDonda})
	\begin{equation}\label{eqEspressioneAutovaloriMomentoQuadrupoloElettricoDiMateria}
	Q(\v r) %
	= 3 z^2 - r^2 = 4 \sqrt{\frac{\pi}{5}} r^2 Y_{20}(\theta,\phi)
	\end{equation}
	(see \cite[app.~A]{Mason2008} or \cite[eq.~(5.1.36)]{ThompsonNunes2009}), %
	where the spherical harmonic $Y_{20}$ (see \cref{secAppendiceSphericalHarmonics}) is %
	defined using the %
	$z$-axis as polar axis.

	Differently than with the radius, due to the tensorial nature of the quadrupole operator, if a nucleus has non-zero total spin $J$, the associated quadrupole moment depends on the total spin $z$-projection (which in turn depends on the nucleus orientation with respect to the arbitrarily chosen coordinates).
	Expectation values for multipole moments are customarily quoted for %
	eigenstates of the total spin $z$-projection, with eigenvalue $M$ equal to the maximum possible, $M = J$ (the nucleus is chosen to be “maximally aligned” to the quantisation axis).
	The expectation value for a state in a different spin projection, $\Ket{J,M}$, can then be found using the Wigner-Eckart theorem. Given \cref{eqEspressioneAutovaloriMomentoQuadrupoloElettricoDiMateria}, it is manifest that the quadrupole operator is a $(2,0)$ tensor (meaning that it transforms as $Y_{20}(\theta,\phi)$ under spatial rotations), hence
	\begin{equation}\label{eqTeoremaWignerEckartMomentoQuadrupoloElettrico}
	\Braket{J, M |\op Q_{\text{m}}| J, M}
	= \frac{\Braket{(J,M), (2,0)|J,M}}{\Braket{(J,J), (2,0)|J,J}} \Braket{J, J |\op Q_{\text{m}}| J, J}
	\end{equation}
	where $\Braket{(j_1,m_2), (j_2,m_2)|J,M}$ is a Clebsch-Gordan coefficient. %
	
	Assume now that a nucleus %
	can be described as a bound-state of $n$ clusters. Such bound state can be split into several components. In each of them, %
	each cluster $i$ is in turn defined as a bound state of $A_i$ nucleons coupling to a definite total spin modulus and $z$-projection, denoted by the quantum numbers $j_i$ and $j^{(z)}_i$.
	The clusters orbital angular momenta (with respect to the composite nucleus centre-of-mass) can be coupled to form the relative orbital angular momenta of interest in the Jacobi coordinates of choice, so it is possible to consider a set of configurations with definite values for the relative orbital angular momentum moduli and $z$-projections, denoted by the quantum numbers $\lambda_i, \mu_i$ %
	(with $i$ running from $1$ to $n-1$).
	The clusters total spins and their total orbital angular momentum (with respect to the composite nucleus centre-of-mass) %
	further couple to form the total spin of the composite nucleus, with modulus $J$ and projection $M$.
	Even within an inert-cluster model, there normally will be several choices for the angular momenta $j^{(z)}_i$, $\lambda_i$ and $\mu_i$ %
	which couple to the desired $M$. Since the quadrupole moment does not couple the nucleons spin (as all observables treated in this \namecref{secObservablesdependingonlyonpositions}), it is possible to consider only a single possibility at a time for the set of $j^{(z)}_i$: the complete result can be obtained combining each component with the appropriate weight (usually a Clebsch-Gordan coefficient). An explicit example will be discussed in \cref{secLiDeformation}. Even after this simplification, there can be several possible values for the set of $\lambda_i$ %
	which can be coupled by the spherical harmonic in \cref{eqEspressioneAutovaloriMomentoQuadrupoloElettricoDiMateria}.
	There may also be several choices for the set of $\mu_i$, %
	but these are not coupled by a $Y_{\lambda0}$, and can thus be taken into account separately, as the spins.
	
	Using \cref{eqValAspOsservQuadraticiPosizioneRisultatoCasoDueCluster}, it is immediately, for a two-cluster system,
	\begin{equation}\label{eqValAspMatterQuadrupoleMomentDueCluster}
	\media{Q_{\text{m},\text{comp}}} = \sum_i \media{Q_{\text{m},i}} + 4 \sqrt{\frac{\pi}{5}} \frac{A_1 A_2}{A} \int_{\campo{R}^3} r^2 Y_{20}(\theta,\phi) \, \Phi(\v r) \d^3 \v r
	\end{equation}
	in agreement with \cite[eq.~(A.12)]{Mason2008}%
	\footnote{More precisely, in \cite[eq.~(A.12)]{Mason2008} the factor $4 \sqrt{\frac{\pi}{5}}$ does not appear, %
		as this is included only later in the expectation value ``$Q^{mo}$'' in \cite[eq.~(A.13)]{Mason2008}. The factor is not present in %
		\cite[eq.~(A.14), (A.15)]{Mason2008} as well.}. %
	This holds for a generic $\Phi$, but given the coupling scheme mentioned above, it is of interest to specialise it to the case where the relative-motion wave-function is an eigenstate of the $z$-projection of the relative orbital angular momentum, $\mu$, but not necessarily an eigenstate of the angular momentum modulus. As in \cref{eqDefinizioneFunzioneRadialeRidotta}, the probability density $\Phi$ can thus be written as the modulus square of a composition of several components $\xi_{\lambda\mu}$ defined through the following equation:
	\begin{equation}\label{eqDecomposizioneDensitaMotoRelativoPerMomentoQuadrupolo}
	\Phi(\v r) = \m{ \sum_\lambda c_\lambda \xi_{\lambda,\mu}(\v r) }^2 = \m{ \sum_\lambda c_\lambda \frac{\chi_\lambda(r)}{r} Y_{\lambda,\mu}(\Omega) }^2
	\end{equation}
	where the $c_\lambda$ are the coefficients of the expansion, such that $\sum_\lambda \m{c_\lambda}^2 = 1$. %
	Using bra-ket notation, the integral of each possible term is thus
	\begin{multline}
	\Braket{\xi_{\lambda_2,\mu}|r^2 Y_{20}|\xi_{\lambda_1,\mu}} \equiv \int_{\campo{R}^3} \coniugato{\xi}_{\lambda_2,\mu}(\v r) \, r^2 Y_{20}(\theta,\phi) \, \xi_{\lambda_1,\mu}(\v r) \d^3 \v r = \\
	= \int_{4 \pi} \coniugato{Y}_{\lambda_2,\mu}(\Omega) Y_{20}(\Omega) Y_{\lambda_1,\mu}(\Omega) \d\Omega \int_0^{+\infty} \coniugato{\chi}_{\lambda_2}(r) r^2 \chi_{\lambda_1}(r) \d r \equiv \\
	\equiv \Braket{ Y_{\lambda_2,\mu} | Y_{20} | Y_{\lambda_1,\mu} } \Braket{ \chi_{\lambda_2} | r^2 | \chi_{\lambda_1} }
	\end{multline}
	where the radial wave-functions %
	are normalised so that $\Braket{\chi_\lambda|\chi_\lambda} = 1$.
	The result of the angular integration is %
	reported in \cref{eqBraketTreArmonicheSferiche},
	and is non-zero only when either $\lambda_1 = \lambda_2 \neq 0$ or $\lambda_1 = \lambda_2 \pm 2$, %
	which may reduce the number of components which have to be taken into account together in \cref{eqDecomposizioneDensitaMotoRelativoPerMomentoQuadrupolo}.
	
	In summary, the relative-motion contribution to $\media{Q_{\text{m},\text{comp}}}$ in \cref{eqValAspMatterQuadrupoleMomentDueCluster} is
	\begin{equation}
	4 \sqrt{\frac{\pi}{5}} \frac{A_1 A_2}{A} \sum_{\lambda_1, \lambda_2} \coniugato{c}_{\lambda_2} c_{\lambda_1} \Braket{ Y_{\lambda_2,\mu} | Y_{20} | Y_{\lambda_1,\mu} } \Braket{ \chi_{\lambda_2} | r^2 | \chi_{\lambda_1} }
	\end{equation}
	
	For a three-cluster system, using \cref{eqValAspOsservQuadraticiPosizioneRisultatoCasoTreCluster},
	\begin{multline}
		O(\rho) = \sum_i \media{Q_{\text{m},i}} +\\+ 4 \sqrt{\frac{\pi}{5}} \[ \frac{A_1 A_2}{A_1+A_2} \media{r^2 Y_{20} }_{\phi_1} + \frac{(A_1+A_2) A_3}{A} \media{r^2 Y_{20} }_{\phi_2} \]
	\end{multline}
	where each $\media{r^2 Y_{20} }_{\phi_i}$ can be treated identically to the analogous term in \cref{eqValAspMatterQuadrupoleMomentDueCluster}.

\subsubsection{Charge quadrupole moment} %

	In analogy with its matter counterpart, the charge electric quadrupole moment can be defined through \cref{eqValAspettazioneOsservPosizioneIsospinEspressioneGenerale}, adopting
	as single-particle eigenvalues
	\begin{equation}
	Q_{\tau}(\v r) = Z_\tau \, 4 \sqrt{\frac{\pi}{5}} r^2 Y_{20}(\theta,\phi)
	\end{equation}
	(see \cite[ex.~1.10]{Satchler1990Introduction} or \cite[app.~A]{Mason2008}) %
	where $Z_\tau$ is the charge number of species $\tau$ (that is, $1$ for protons and $0$ for neutrons)%
	\footnote{Consequently, the charge quadrupole moments, in the present work, bear the dimensions of an area.
	It is also possible to use the particle charge ($Z_\tau$ times the proton charge) instead of the charge number: this would be equivalent to replace the system number density with the charge density.}. %
	Apart from the charge dependence, all considerations made for the matter quadrupole moment hold unaltered. Using \cref{eqValAspOsservDiCaricaQuadraticiPosizioneRisultatoCasoDueCluster}, for a two-cluster system it is
	\begin{multline}\label{eqValAspChargeQuadrupoleMomentDueCluster}
	\media{Q_{\text{ch},\text{comp}}} = \sum_i \media{Q_{\text{ch},i}} +\\+ 4 \sqrt{\frac{\pi}{5}} \frac{Z_1 A_2^2 + Z_2 A_1^2}{A^2} \int_{\campo{R}^3} r^2 Y_{20}(\theta,\phi) \, \Phi(\v r) \d^3 \v r
	\end{multline}
	(again similarly to \cite[eq.~(A.12)]{Mason2008}), where the integral involving the inter-cluster relative motion is identical to the one discussed for the matter quadrupole moment.
	
	For a three-cluster system,
	combining \cref{eqEspressioneFunzioneAusliariaRperTreClusterValoreAspettazioneOsservabileCaricaJacobiCoordinates,eqEspressioneOperatoreDiRaggioQuadraticoDecompostoInArmonicheSferiche,eqValAspettazioneOsservabileCaricanCluster},
	\begin{multline}
		\media{Q_{\text{ch},\text{comp}}}
		= \sum_i \media{Q_{\text{ch},i}} + 2 \frac{A_3}{A} \frac{ Z_1 A_2 - Z_2 A_1 }{ A_1 + A_2 } \media{3 z_1 z_2 - \v r_1 \cdot \v r_2}_{\Phi} +\\
		+ 4 \sqrt{\frac{\pi}{5}} \frac{ Z_1 A_2^2 + Z_2 A_1^2 }{ A^2 } \media{r^2 Y_{20}}_{\phi_1} +\\+ 4 \sqrt{\frac{\pi}{5}} \frac{ \( Z_1 + Z_2 \) A_3^2 + Z_3 \( A_1 + A_2 \)^2 }{ A^2 } \media{r^2 Y_{20}}_{\phi_2}
	\end{multline}
\chapter{Direct transfer nuclear reactions}\label{secReactionTheory}

	Consider once more the scattering of two nuclei, $A$ and $b$, in vacuum. The system final state, in general, can comprise a different number of particles, and of different species. In a \emph{transfer nuclear reaction}, the final state is composed again by two particles $a$ and $B$ (thus the process is labelled as $A + b \to a + B$), and it is possible to identify a specific collection of nucleons, $\nu$, such that $A$ comprises the nucleons in $a$ and $\nu$ together, and similarly $B$ is the aggregate of systems $b$ and $\nu$ (this does not necessarily imply any cluster model). In general, any possible rearrangement of the nucleons comprising the whole system is called a \emph{partition}.

	This \namecref{secReactionTheory} is concerned in particular with the description of processes that can be modelled as \emph{direct} reactions, %
	where %
	the system initial and final states can be connected explicitly through the properties of the interaction (see e.g.~\cite[sec.~2.18]{Satchler1990Introduction}).
	More specifically, the goal of this \namecref{secReactionTheory} is to review some approximate approaches %
	which allow to explicitly evaluate, in a fully quantum framework, the cross-section of a direct transfer reaction as a function of the aforementioned physical ingredients.
	First of all, %
	different formal %
	expressions of the cross-section in terms of the scattering problem exact solution are presented in \cref{secGeneralReactionTheoryApplicataAlTransfer}. These are then employed to draw explicit formulas in the following sections. \Cref{secTeoriaScatteringOneParticleTransfer} presents two different frameworks for single-particle transfers: %
	the well-known %
	first-order Distorted-Wave Born Approximation (DWBA), %
	and a selection of continuum-discretised coupled-channels schemes.
	Finally, \cref{secReactionTheoryTwoParticleTransfer} discusses the application of second-order DWBA to the transfer of two particles whose relative motion is explicitly taken into account.

\section{Formal exact expressions for the reaction cross-section}\label{secGeneralReactionTheoryApplicataAlTransfer} %

	For definiteness, the specific case of transfer reactions is considered in the following, but note that the very same formalism can be applied to any direct process. Furthermore,
	some details of the derivations
	are skipped in this section for simplicity.
	Sources as \cite[sec.~2]{Satchler1983Direct}, \cite[ch.~3]{Austern1970direct}, \cite[ch.~3, 5, 6]{Glendenning2004direct} and \cite[sec.~2, 9]{MoroVarenna} may be consulted for a more complete treatment. %
	Furthermore, bra-ket notation is widely adopted here for compactness: %
	the quoted sources often provide more explicit expressions in coordinate representation.

\subsection{Introduction}\label{secIntroductionGeneralReactionTheoryApplicataAlTransfer}
	
	The system Hamiltonian, $\mathcal H$,
	can be expressed as $H_a + H_b + H_\nu + V_{ab} + V_{\nu a} + V_{\nu b}$, where each $H_i$ (with $i = a,b,\nu$) includes all kinetic and potentials terms for the isolated $i$ system, and each $V_{ij}$ includes all interactions between systems $i$ and $j$ (i.e.~between their internal components, if $i$ and $j$ are not elementary). Note that splitting the interaction in isolated pieces essentially amounts to assume that, for instance, the interaction between the nucleons in $a$ and $\nu$ is independent of the state of nucleons in $b$: this is strictly true only if the total interaction involves just two-body forces. %
	It is possible to lift the aforementioned assumption in the general treatment, but this is not done for simplicity, given that all practical calculations adopt this framework.
	
	Let $K_i$ be the kinetic energy of particle $i$ centre-of-mass. As is customary, the coordinates are chosen so to decouple the irrelevant motion of the centre-of-mass of the whole system ($a+b+\nu$). %
	For instance, in the initial partition, consider the three Jacobi coordinates (defined as in \cref{eqDefinizioneCoordinateJacobi}) constructed from the position of particles $\nu,a,b$ (in this order): $\nu$--$a$ distance, $\v r_{\nu a}$, then $A$--$b$ distance, $\v R_{Ab}$, and centre-of-mass position, $\v{\mathcal R}_{\text{cm}}$. Let $K_A$ be the kinetic energy of the centre-of-mass of $\nu$ and $a$, similarly $K_{\text{cm}}$ the kinetic energy of the whole system centre-of-mass, then $K_{\nu a} = K_\nu + K_a - K_A$ and similarly $K_{Ab} = K_A + K_b - K_{\text{cm}}$. Since all Jacobi coordinates except the last one (here $\mathcal R_{\text{cm}}$)
	are frame-invariant, each $K_{ij}$ is invariant as well, and in particular coincides with the total kinetic energy of $i$ and $j$ computed in $i+j$ centre-of-mass frame (this is essentially König's decomposition theorem).
	From the standard treatment of the classical two-body problem, or explicitly for the quantum case %
	by applying the coordinates transformation to the Laplacian operator, it can be seen that $K_{ij}$ is the kinetic energy of a fictitious particle having the $i$--$j$ distance (i.e.~the associated Jacobi coordinate) as position and the reduced mass of $i$ and $j$ %
	as mass.
	$\v r_{\nu b}$ and $\v R_{Ba}$ are defined analogously in the final partition.
	
	Let $H_\alpha = H_a + H_b + H_\nu + V_{a\nu} - K_{Ab} - K_{\text{cm}}$, which is is just the sum of the Hamiltonians for the internal motion of $A$ and $b$ (each isolated and in its own centre-of-mass frame).
	Analogous consideration can be made regarding the final partition defining $H_\beta = H_a + H_b + H_\nu + V_{b\nu} - K_{aB} - K_{\text{cm}}$.
	Symbols $\alpha$ and $\beta$ will be thoroughly employed to denote the initial and final partition, respectively.
	The total Hamiltonian can be written as
	\begin{equation}\label{eqRearrangementCompleteHamiltonian}
		\mathcal{H} = K_{\text{cm}} + K_{Ab} + H_\alpha + V_{ab} + V_{b\nu} = K_{\text{cm}} + K_{aB} + H_\beta + V_{ab} + V_{a\nu}
	\end{equation}
	In the following, set the reference frame to the system centre-of-mass rest frame, %
	where $K_{\text{cm}}$ gives no contribution. Such term is thus omitted in the following.
	Also note that $V_{ab} + V_{b\nu}$ is just the total interaction between $b$ and the composite system $A$, thus the expression could be made slightly more general by writing such term as a generic $V_{Ab}$, which in principle could even include the distortions in the $a$--$\nu$ interaction caused by the presence of $b$. %
	
	The symbols $\ket{\psi_\alpha^j}$ and $\bra{\psi_\beta^j}$ will instead denote eigenvectors of $H_\alpha$ and $H_\beta$, i.e.\ the internal motion states for, respectively, projectile and target, and ejectile and residue. The index $j$ labels the eigenbasis elements.
	The eigenvalue of $\Ket{\psi_\alpha^j}$ for $H_\alpha$ is denoted by $E - E_\alpha^j$: thus, $E_\alpha^j - E$ is the total binding energy of $A$ and $b$ in $\Ket{\psi_\alpha^j}$, while $E_\alpha^j = E - (E - E_\alpha)$ is just the %
	kinetic energy at infinite distance for $A$--$b$ relative motion when the nuclei are in state $\Ket{\psi_\alpha^j}$. Let $E^j_\beta$ be defined similarly, in regard to the final states. In summary:
	\begin{equation}\label{eqProblemaAutovaloriStatiMotoInternoPartIniziale}
	\(E - E_\alpha^j - \op H_\alpha\) \Ket{\psi_\alpha^j} = 0
	\end{equation}
	and similarly for partition $\beta$.
	If the index is omitted, as in ``$\Ket{\psi_\alpha}$'', %
	the precise initial ($A$ and $b$) or final ($a$ and $B$) state involved in the reaction of interest is being referenced, unless otherwise noted. Note that no requirement is made on the structure of such states (in particular with regard to the division in core, $a$ or $b$, and valence system $\nu$).
	
	In addition, let $\ket{\v k_\alpha^j}$ and $\ket{\v k_\beta^j}$ be plane-wave states with momentum $\v k_\alpha^j$ and $\v k_\beta^j$, pertaining, respectively, to the initial-partition projectile-target motion, and to the final-partition ejectile-residue motion. Unless otherwise stated, the module $k_\gamma^j$ will be such that $\hbar^2 {k_\gamma^j}^2 / 2 m_\gamma = E_\gamma^j$, where $m_\gamma$ is the reduced mass of reactants in the $\gamma$ partition (where $\gamma=\alpha$ or $\beta$). %
	Further let “$\v r_\gamma$” be a multi-dimensional coordinate %
	representing all internal coordinates for nuclei in $\gamma$ partition
	\footnote{In a simple cluster model where $a,b,\nu$ are inert and structureless, %
	and their spins are not coupled,
	$\v r_\alpha$ would be just the three-dimensional vector $\v r_{\nu a}$ defined previously.}. %
	Finally, let $\mathds{R}_\alpha$ be a shorthand for $(\v R_{Ab}, \v r_\alpha)$, and similarly $\mathds{R}_\beta = (\v R_{aB}, \v r_\beta)$. %
	Both $\mathds{R}$ can describe any configuration in the full space under consideration, but either one can be more practical depending on the specific situation.
	In some occasions it will also be useful to consider the position eigenvectors for a given coordinate, as $\Ket{\v R_{Ab}}$ etc.
	
	Finally, let $\Ket{\Psi_+}$ be an exact eigenfunction of the operator associated to the complete system Hamiltonian, $\op{\mathcal H}$, for a given eigenvalue $E$ (this is the total energy, including the binding energy of all subsystems). %
	According to the physical problem of interest, the boundary conditions are chosen to fix the desired initial state for the system as the product of a wave-function $\Ket{\psi_\alpha}$, describing the reactants internal state, and a plane-wave $\ket{\v k_\alpha}$ directed %
	as the beam (and oriented toward the target) for the reactants relative motion. %
	$\Ket{\Psi_+}$ can thus be written as the sum of $\ket{\psi_\alpha \v k_\alpha}$ and, asymptotically at large distances, of a spherical scattered outgoing wave for each open exit channel. Reminding that $\Psi_+(\mathds{R})$ depends on several coordinates, each scattered wave can be observed when approaching $\m{\mathds{R}} \to \infty$ from a specific ``direction''. For instance, a scattered wave involving a bound state of ejectile and residue can be observed if all internal coordinates of partition $\beta$, ``$\v r_\beta$'', are kept finite (otherwise the reactants internal-motion wave-function vanishes), and only $R_{aB}$ is sent to infinity. This statement can be written in formulas as in \cite[eq.~(3.34)]{Glendenning2004direct}.
	The normalisation adopted for $\ket{\psi_\alpha \v k_\alpha}$ fixes proportionally the norm of all other terms.
	It is also possible to consider the time-reversed version of the same problem (see \cite[sec.~(2.7)]{Satchler1983Direct} and referenced sections for a more complete discussion), drawing an eigenfunction $\Ket{\Psi_-}$ of $\op{\mathcal H}$ with boundary conditions that fix the final state: $\Ket{\Psi_-}$ is written as $\ket{\psi_\beta \v k_\beta}$, where the plane-wave is now directed from the target toward the detector, plus (asymptotically at infinity) ingoing waves in all initial channels compatible with the chosen final state.
	If the distinction is relevant (in particular if complex potentials are admitted), $\Ket{\Psi_+}$ and $\Ket{\Psi_-}$ are taken to be, respectively, a right and a left eigenfunction, meaning that
	\begin{equation}\label{eqProblemaAutovaloriCompletoFormaGenerica}
		0 = \( \op{\mathcal H} - E \) \Ket{\Psi_+} = \Bra{\Psi_-} \( \op{\mathcal H} - E \)
	\end{equation}

	The reaction cross-section for a given process can be found, in \emph{post form}, through the projection of the complete solution $\Psi_+$ on the desired final state $\psi_\beta$ for the internal motion, precisely the asymptotic limit of such projection for big ejectile-residue distance in the detector direction. Analogously, in \emph{prior} form, $\Psi_-$ can be projected on the initial state of interest.

\subsection{Plane-wave formulation}\label{secTeoriaScatteringGeneralePlaneWaves}

	For definiteness, consider here the post form (similar considerations may be employed in prior form).
	A formal solution for the $\Psi_+$ eigenvalue problem can be expressed through the Green's functions formalism.
	Start from \cref{eqProblemaAutovaloriCompletoFormaGenerica}, splitting $\mathcal H$ in $(\v R_{aB}, \v r_\beta)$ coordinates:
	\begin{equation}\label{eqProblemaAutovaloriCompletoPost}
		(E - \op K_{a B} - \op H_\beta) \Ket{\Psi_+} = (\op V_{ab} + \op V_{a\nu}) \Ket{\Psi_+}
	\end{equation}
	Seek now a Green's function $\op{\mathcal G}$ for the operator ($E - \op K_{a B} - \op H_\beta$), namely a distribution such that, for any two coordinates $\mathds{R}$ and $\mathds{X}$, it is
	\begin{equation}\label{eqDefinizioneFunzioneDiGreenCompletaPlaneWave}
		\Braket{\mathds{X} | \[ E - \op K_{a B}(\v R_{aB}) - \op H_\beta(\v r_\beta) \] \op{\mathcal G}(\mathds{R}, \mathds{X}) | \mathds{R} } = \Braket{\mathds{X} | \mathds{R}}
	\end{equation}
	where the symbols “$\op K_{a B}(\v R_{aB})$” etc.\ %
	were employed here just to underline that each operator acts only on the space specified by the coordinate.
	Given a solution $\op{\mathcal G}$, %
	it can be seen by direct substitution that $\op G [\op V_{ab} + \op V_{a\nu}] \Ket{\Psi_+}$ (i.e.~$\op G$ times the source term in \scref{eqProblemaAutovaloriCompletoPost}) is a solution for $\Ket{\Psi_+}$ in \cref{eqProblemaAutovaloriCompletoPost}. In explicit notation, this is
	\begin{equation}\label{eqFormaEsplicitaSoluzioneFunzioneDiGreenGenericaConGradiDiLiberta}
	\Psi_+(\mathds{R}) = \int \mathcal G(\mathds{R}, \mathds{X}) [V_{ab}(\mathds{X}) + V_{a\nu}(\mathds{X})] \Psi_+(\mathds{X}) \d\mathds{X}
	\end{equation}

	There are in general several solutions for $\op{\mathcal G}$, each generating a different $\Psi$ which solves \cref{eqProblemaAutovaloriCompletoPost}%
	\footnote{Furthermore, as for any eigenvalue problem, a linear combination of solutions $\Psi$ is still a solution.}. %
	The set of %
	solutions depends on the structure of the operator under study, and the correct one must thus be chosen to satisfy the required boundary conditions. For the problem under study, %
	some freedom is generated because the operator $E - \op K_{a B} - \op H_\beta$ is not invertible.
	Consider any function $v(\mathds{R})$ belonging to the operator eigenspace %
	with eigenvalue 0: given a solution $\mathcal G$, clearly $\mathcal G + v(\mathds{R}) g(\mathds{X})$, with $g$ an arbitrary function, still satisfies \cref{eqDefinizioneFunzioneDiGreenCompletaPlaneWave}. %
	By substitution in \cref{eqFormaEsplicitaSoluzioneFunzioneDiGreenGenericaConGradiDiLiberta}, it is seen that the additional term in the Green's function generates an additional term $v(\mathds{R})$ in the wave-function (the scaling constant is irrelevant, since $v$ has arbitrary norm anyway). For the present purposes, it is then convenient to add $v$ explicitly to the expression for $\Psi^+$, and remove the discussed degree of freedom from $\mathcal G$ in a way that satisfies the required boundary conditions.
	In particular, it is customary to change the problem into %
	\begin{equation}\label{eqDefinizioneFunzioneDiGreenCompletaPlaneWaveConIEspilon}
		\Braket{\mathds{X} | \[ E - \op K_{a B} - \op H_\beta + i \epsilon \] \op{\mathcal G}_+ | \mathds{R} } = \Braket{\mathds{X} | \mathds{R}}
	\end{equation}
	where $\epsilon$ is an arbitrary positive number, and later take the limit of $\epsilon \to 0^+$ (see e.g.~\cite[sec.~2.8]{Satchler1983Direct}), drawing the solution
	\begin{equation}\label{eqSoluzioneGeneraleFunzioneDondaDaFunzGreenPlaneWave}
		\Ket{\Psi_+} = \Ket{v} + \op{\mathcal G}_+ (\op V_{ab} + \op V_{a\nu}) \Ket{\Psi_+}
	\end{equation}
	Regarding the explicit form of $\Ket{v}$, note that $\op K_{aB}(\v R_{aB})$ and $\op H_\beta(\v r_\beta)$ act on different subspaces and are not coupled. Consequently, $\Ket{v}$ is any linear combination of decoupled states, each %
	written as %
	$\Ket{\psi_\beta^j \v k_\beta^j}$, where $\Ket{\psi_\beta^j}$ is any eigenvector of $H_\beta$, and $\Ket{\v k_\beta^j}$ is a plane wave in $\v R_{aB}$ whose momentum is such that the total energy of $\Ket{\psi_\beta^j \v k_\beta^j}$ is $E$.
	As mentioned earlier, asymptotically $\Psi_+$ is requested to be composed by $\Ket{\psi_\alpha \v k_\alpha}$ %
	plus scattered outgoing waves (not plane waves) in all channels. $\Ket{\psi_\alpha \v k_\alpha}$ does \emph{not} belong to the space spanned by the set of $\Ket{\psi_\beta^j \v k_\beta^j}$ unless $\alpha$ and $\beta$ are actually the same partition (i.e.\ this is not a transfer reaction). When $\alpha \neq \beta$, the boundary conditions are such that the $\Ket{v}$ term in \cref{eqSoluzioneGeneraleFunzioneDondaDaFunzGreenPlaneWave} is just the null vector (see e.g.~\cite[eq.~(2.29)]{Satchler1983Direct} or \cite[eq.~(3.39)]{Austern1970direct}): %
	\begin{equation}\label{eqSoluzioneGeneraleFunzioneDondaDaFunzGreenPlaneWaveConBoundaryConditionApplicata}
		\Ket{\Psi_+} = \op{\mathcal G}_+ (\op V_{ab} + \op V_{a\nu}) \Ket{\Psi_+}
	\end{equation}

\subsubsection{Explicit expression of the Green's function}

	Since, as already mentioned, %
	$\op K_{aB}(\v R_{aB})$ and $\op H_\beta(\v r_\beta)$ are decoupled, %
	it is %
	possible to split \cref{eqDefinizioneFunzioneDiGreenCompletaPlaneWaveConIEspilon} into two independent equations:
	\begin{equation}\label{eqEquazioneFunzGreenSpazioCompletoDisaccoppiata}\begin{aligned}
			& \Braket{\v X_{aB} | \[ E_\beta^j - \op K_{a B} + i \epsilon \] G^j_{\text{PW}} | \v R_{aB} } = \Braket{\v X_{aB} | \v R_{aB}} \\
			& \Braket{\v x_\beta | \[ (E - E_\beta^j) - \op H_\beta \] \op G^j_{\text{int}} | \v r_\beta } = \Braket{\v x_\beta | \v r_\beta}
	\end{aligned}\end{equation}
	These generate a set of solutions $\{ G_{\text{PW}}^{j}(\v X_{aB},\v R_{aB}) G_{\text{int}}^{j}(\v x_\beta,\v r_\beta) \}_{j}$, one  %
	for each $E_\beta^j$ such that $E-E_\beta^j$ is an eigenvalue of $H_\beta$. %
	Any linear combination of the associated wave-functions is then a solution for $\Psi_+$, and the correct one is chosen to satisfy the boundary conditions.
	
	Provided that the eigenvalue problem of $H_\beta$ can be solved (as it is normally assumed), $\op{\mathcal G}_+$ can thus be found as well, because
	the Green's function associated to the free-particle problem %
	can be calculated explicitly and analytically. %
	For definiteness, set the normalisation of the position eigenvectors so that $\Braket{\v X_{aB} | \v R_{aB}} = \delta^3\(\v R_{aB} - \v X_{aB}\)$.
	For brevity, let $\op L = E_\beta^j - \op K_{a B} + i \epsilon$, and drop the index $j$ in the following. First note that $\op L$ is translational invariant, meaning that if $\op L(\v r) f(\v r) = h(\v r)$ then $\op L(\v r) f(\v r - \v r') = h(\v r - \v r')$. In particular, if $G_{\text{PW}}(\v R, \v X)$ is a solution, then %
	$G_{\text{PW}}(\v R - \v X, \v 0)$ %
	must be a solution of the same equation.
	Let %
	$g(\v R, \v X) = G_{\text{PW}}(\v R, \v X) - G_{\text{PW}}(\v R - \v X, \v 0)$. By construction, $g$ belongs to the eigenspace of $\op L$ with eigenvalue 0, but $\op L$ is %
	invertible, thus $g = 0$. %
	In summary, the solution must obey $G_{\text{PW}}(\v R, \v X) = G_{\text{PW}}(\v R - \v X, \v 0)$. For brevity, let $G_+(\v R - \v X) = G_{\text{PW}}(\v R - \v X, \v 0)$.
	The equation to be solved is reduced to
	\begin{equation}\label{eqEquazioneFunzGreenMotoReagentiSemplificata}
	\frac{\hbar^2}{2 m_{\beta}} \[ \nabla^2 + k_\beta^2 + i \tilde\epsilon\] G_+(\v r) = \delta^3\(\v r\)
	\end{equation}
	where the explicit expressions of $\op K_{a B}$ and $k_\beta^j$ were employed, and similarly $\tilde\epsilon$ was defined as $2 m_{\beta} \epsilon / \hbar^2$.
	Define the three-dimensional Fourier transform, $\Fourier(f(\v x), \v q)$, as $\int_{\campo{R}^3} f(\v x) \exp\(- i \v q \cdot \v x\) \d^3 \v x$, which is linear in $f$.
	It is found that $\Fourier\( \delta^{3}(\v x), \v q \) = 1$, while, in general, $\Fourier\(\nabla^2 f(\v x), \v q\) = - q^2 \Fourier\(f(\v x), \v q\)$.
	Transforming both sides of \cref{eqEquazioneFunzGreenMotoReagentiSemplificata}, an expression for $\Fourier(G_+(\v r), \v q)$ is deduced. The sought solution can then be computed applying the inverse transform, $G_+(\v r) = \int_{\campo{R}^3} \Fourier(G_+(\v r), \v q) \exp\(+ i \v q \cdot \v x\) \d^3 \v q$.
	Given that $\Fourier(G_+(\v r), \v q)$ actually depends only on the modulus of $\v q$, the integral is conveniently computed in spherical coordinates, and reduces to a one-dimensional Fourier transform (see e.g.~\cite[eq.~(6.16)]{Glendenning2004direct}):
	\begin{equation}
	G_+(\v r) = - \frac{2 m_\beta}{\hbar^2} \frac{1}{(2 \pi)^2} \int_{\campo{R}} \frac{q}{i r} \frac{e^{- i q r}}{k_\beta^2 + i \tilde\epsilon - q^2} \d q
	= - \frac{m_\beta}{2 \pi \hbar^2} \frac{1}{r} e^{- \sqrt{- (k_\beta^2 + i \tilde\epsilon)} \, r }
	\end{equation}
	where it was taken into account that $r > 0$ for the physical problem of interest.
	The sign of the exponential depends on the adopted convention on the value to adopt as square root of a complex number, $\sqrt{z}$. If this is defined as $\sqrt{\m{z}} \exp(\frac{i}{2} \arg z)$, with $\arg z \in [-\pi, \pi]$, in the limit of $\tilde\epsilon \to 0^+$ the desired sign for the spherical waves is found, and %
	\begin{equation}\label{eqDerivazioneAmpiezzaTransferScritturaFunzioneGreenEsplicita}
		G^j_{\text{PW},+}(\v R_{aB}, \v X_{aB}) = - \frac{m_\beta}{2 \pi \hbar^2} \frac{ e^{i k^j_{\beta} \left| \v R_{aB} - \v X_{aB} \right| } }{ \left|\v R_{aB} - \v X_{aB} \right| }
	\end{equation}

\subsubsection{Reaction amplitude}

	As mentioned, in order to evaluate (in post form) the reaction cross-section for a specific process, it is sufficient to find $\Braket{\psi_\beta | \Psi_+}$, the projection of $\Psi_+$ on the desired final state for the internal motion of ejectile and residue%
	\footnote{When writing such projections, it is understood that the integration is performed only on the degrees of freedom appearing in both Bra and Ket, so that e.g.~$\Braket{\psi_\beta | \Psi_+}$ is still a Ket in the $a$--$B$ relative motion space.},
	and not the full solution.
	As a consequence, in this case it is more convenient to not proceed through \cref{eqEquazioneFunzGreenSpazioCompletoDisaccoppiata}. Instead, start from \cref{eqProblemaAutovaloriCompletoPost} and project on $\Bra{\psi_\beta}$, using \cref{eqProblemaAutovaloriStatiMotoInternoPartIniziale} and the separation of variables between $\op K_{aB}$ and $\op H_\beta$:
	\begin{equation}\label{eqEquazioneProiettataOndePiane}
		(E_\beta - \op K_{a B}) \Braket{\psi_\beta|\Psi_+} = \Braket{\psi_\beta | \op V_{ab} + \op V_{a\nu} | \Psi_+}
	\end{equation}
	Both $\Braket{\psi_\beta|\Psi_+}$ and $\Braket{\psi_\beta | \op V_{ab} + \op V_{a\nu} | \Psi_+}$ are functions of the $\v R_{aB}$ coordinate only, and the equation can be solved directly for $\Braket{\psi_\beta|\Psi_+}$. This is helpful because the associated Green's function can then be restricted to the $\v R_{aB}$ coordinate.
	will depend only on that degree of freedom as well.
	Note how the simplification of the internal-motion Hamiltonian was made possible by the choice, in \cref{eqProblemaAutovaloriCompletoPost}, to split the complete Hamiltonian in the final-partition expression, even though the boundary conditions of $\Psi_+$ would have suggested the opposite choice.
	Incidentally, the analogous equation in prior form would be
	\begin{equation}\label{eqEquazioneProiettataOndePianePrior}
		\Braket{\Psi_- | \psi_\alpha } \(E_\alpha - \op K_{Ab}\) = \Braket{\Psi_- | \op V_{ab} + \op V_{b\nu} | \psi_\alpha }
	\end{equation}

	Resuming with the post form, the analogous of \cref{eqDefinizioneFunzioneDiGreenCompletaPlaneWaveConIEspilon} is %
	now just the first line of \cref{eqEquazioneFunzGreenSpazioCompletoDisaccoppiata}, thus the solution is the same $G_{\text{PW},+}$ shown in \cref{eqDerivazioneAmpiezzaTransferScritturaFunzioneGreenEsplicita}
	with $j$ corresponding to the specific state $\Ket{\psi_\beta}$ of interest.
	Note that $G$ is the identity operator in the $\v r_\beta$ space, thus, in operator notation, one could just write $\op G = \op G_{\text{PW},+}$.
	The solution for $\Braket{\psi_\beta|\Psi_+}$, in analogy with \cref{eqSoluzioneGeneraleFunzioneDondaDaFunzGreenPlaneWaveConBoundaryConditionApplicata}, is then
	\begin{equation}\label{eqSoluzioneOndePianeNotazioneBraKetConFunzioneGreen}
		\Braket{\psi_\beta|\Psi_+} = \Braket{\psi_\beta | \op G_{\text{PW},+} \(\op V_{ab} + \op V_{a\nu}\) | \Psi_+}
	\end{equation}
	or, in explicit notation, defining $\xi_\beta(\v R_{aB}) = \Braket{\psi_\beta \v R_{aB}|\Psi_+}$,
	\begin{equation}\label{eqSoluzioneOndePianeNotazioneEsplicitaConFunzioneGreen}
		\xi_\beta(\v R_{aB}) = \int \coniugato{\psi}_\beta(\v x_\beta) G_{\text{PW},+}(\v R_{aB}, \v X_{aB}) [V_{ab}(\mathds{X}) + V_{a\nu}(\mathds{X})] \Psi_+(\mathds{X}) \d\mathds{X} %
	\end{equation}
	
	To evaluate the reaction cross-section, it is sufficient to study the form of $\xi_\beta$ in the limit of $R_{aB} \to +\infty$ and $\v R_{aB}$ directed toward the detector.
	If the potentials appearing in \cref{eqSoluzioneOndePianeNotazioneEsplicitaConFunzioneGreen} are short-ranged, it is possible to simply consider the asymptotic form of $G_{\text{PW},+}$. Note that such assumption excludes the relevant case of a bare Coulomb potential between reactants \cite[sec.~2.4.1]{Satchler1983Direct}, %
	which may be treated separately \cite[sec.~4.4]{MoroVarenna}. For simplicity, this step is not reviewed here.
	Asymptotically, \cite[eq.~(2.30)]{Satchler1983Direct} \cite[eq.~(18)]{MoroVarenna}
	\begin{equation}
		G_{\text{PW},+}(\v R_{aB}, \v X_{aB}) \xrightarrow{R_{aB} \to +\infty} - \frac{m_\beta}{2 \pi \hbar^2} \frac{ e^{i k_{\beta} R_{aB} } }{ R_{aB} } e^{- i \v k_{\beta} \cdot \v X_{aB} }
	\end{equation}
	where $\v k_\beta = k_\beta \v R_{aB} / R_{aB}$, the ejectile-residue momentum directed toward the detector. The relation is verified explicitly taking the limit of the ratio of $G_{\text{PW},+}$ to its asymptotic form.
	By construction, $G_{\text{PW},+}$ is the only term in $\xi_\beta$ depending on $\v R_{aB}$, all other factor being constants. It is thus convenient to define the \emph{transition amplitude}: %
	\begin{equation}
		\mathcal T^{\text{post}}_{\alpha\to\beta}(\v k_\alpha, \v k_\beta) = \int \coniugato{\psi}_\beta(\v x_\beta) e^{- i \v k_{\beta} \cdot \v X_{aB} } [V_{ab}(\mathds{X}) + V_{a\nu}(\mathds{X})] \Psi_+(\mathds{X}) \d\mathds{X}
	\end{equation}
	so that $\xi_\beta \to - \frac{ e^{i k_{\beta} R_{aB} } }{ R_{aB} } \frac{m_\beta}{2 \pi \hbar^2} \mathcal T_{\alpha\to\beta}$.
	
	In general, in the time-independent picture employed here, the differential cross-section per unit angle for detection of desired final-state particle in the direction of $\v k_\beta$ (in the centre-of-mass frame) is the modulus of the total scattered flux through an infinitesimal surface centred in $\v k_\beta$, per unit angle, divided by the incoming beam flux modulus (see e.g.~\cite[eq.~(7.1.35)]{SakuraiModern1994}):
	\begin{equation}
	\frac{\d \sigma_{\alpha\to\beta}}{\d\Omega}(\v k_\alpha, \v k_\beta) = \frac{R_{aB}^2 \m{\v{\mathcal J}_{\beta}(\v R_{aB})}}{\m{\v{\mathcal J}_{\text{in}}}}
	\end{equation}
	where each $\v{\mathcal J}$ is defined as in \cite[eq.~(2.4.16)]{SakuraiModern1994}.
	If the initial internal state $\Ket{\psi_\alpha}$ has square-norm of 1, and
	the initial plane-wave $\Braket{\v R_{Ab}|\v k_\alpha}$ is set to $B \exp\(i \v k_\alpha \v R_{Ab}\)$, the incoming current is just $\m{B}^2 \hbar k_\alpha / m_\alpha$. %
	Similarly,
	$R_{Ab}^2 \m{\v{\mathcal J}_{\beta}} = \frac{\hbar k_\beta m_\beta}{(2 \pi \hbar^2)^2} \m{\mathcal T^{\text{post}}_{\alpha\to\beta}}^2$, %
	hence
	\begin{equation}\label{eqScritturaSezioneDurtoDifferenzialeReazioneInTerminiDiAmpiezzaDiScattering}
	\frac{\d \sigma_{\alpha\to\beta}}{\d\Omega}(\v k_\alpha, \v k_\beta) = \frac{m_\alpha m_\beta}{(2 \pi \hbar^2)^2} \frac{k_\beta}{k_\alpha} \frac{\m{\mathcal T_{\alpha\to\beta}(\v k_\alpha, \v k_\beta)}^2}{\m{B}^2}
	\end{equation}
	It is convenient to choose $B = 1$ to remove the irrelevant factor (in any case, $\mathcal T$ is linear in $\Psi$ and thus proportional to $B$), %
	even though with this choice $\Psi_+$ does not bear the correct dimensions.
	Furthermore, if the non-polarised cross-section is of interest, \cref{eqScritturaSezioneDurtoDifferenzialeReazioneInTerminiDiAmpiezzaDiScattering} is to be averaged over all possible spin projections of projectile and target and summed over all possible spin projections of ejectile and residue, as usual (see e.g.~\cite[eq.~(3.40)]{Glendenning2004direct}).
	
	A similar derivation may be performed in prior form, employing the time-reversal properties of the solutions \cite[sec.~2.7]{Satchler1983Direct}. The differential cross-section can be written again as in \cref{eqScritturaSezioneDurtoDifferenzialeReazioneInTerminiDiAmpiezzaDiScattering}, with a different expression for the transition amplitude. In bra-ket notation,
	\begin{equation}\label{eqRisultatoTransitionAmplitudePWPriorPost}\begin{aligned}
			& \mathcal T_{\alpha\to\beta}^{\text{post}}(\v k_\alpha, \v k_\beta) = \Braket{ \psi_\beta \v k_{\beta} | \op V_{ab} + \op V_{a\nu} | \Psi_+} \\
			& \mathcal T_{\alpha\to\beta}^{\text{prior}}(\v k_\alpha, \v k_\beta) = \Braket{\Psi_- | \op V_{ab} + \op V_{b\nu} | \psi_\alpha \v k_{\alpha} }
	\end{aligned}\end{equation}
	where, as mentioned, $\Ket{\psi_\alpha}$ and $\Bra{\psi_\beta}$ are the internal motion wave-functions for the initial and final states of interest, normalised to 1, $\v k_\alpha$ is the wave-vector in the beam direction for the initial-state collision energy, $\v k_\beta$ is the wave-vector in the detector direction for the reactants final-state asymptotic kinetic energy, and $\Braket{\v r|\v k} = \exp\(i \v k \v r\)$.
	The expressions in prior and post form necessarily coincide, because they are derived from the same model without applying approximations.
	
	\Cref{eqRisultatoTransitionAmplitudePWPriorPost} only represents a formal solution to the problem, given that the full wave-function $\Psi$ appears in it. This equation, and the corresponding more elaborate forms shown in the following, are useful as starting points for approximate methods, as will be discussed later.

\subsection{Distorted-wave formulation}\label{sezDistortedWaves}
	
	The problem in \cref{eqProblemaAutovaloriCompletoPost} is now addressed again in a slightly different manner. %
	The exact expressions for $\Psi$ and $\mathcal T$ to be found will of course be equivalent to those in the plane-wave formalism, as long as the unattainable exact solutions are employed, but will differ once approximations are introduced. %
	A more complete treatment of the topic can be found in \cite[ch.~5]{Glendenning2004direct} or \cite[sec.~2.5]{Satchler1983Direct} and references therein.

	Consider, as in \cite[eq.~(2.37)]{Satchler1983Direct} the following variation %
	of \cref{eqEquazioneProiettataOndePiane} (and \emph{not} of \scref{eqEquazioneProiettataOndePianePrior}): %
	\begin{equation}\label{eqEquazioneOndaDistortaStandard}
		\Bra{\chi_{\beta^-}} \(E_\beta - \op K_{aB}\) = \Bra{\chi_{\beta^-}} \op U_{aB}
	\end{equation}
	where $U_{aB}$ is an arbitrary “auxiliary potential” (it needs not to be real or local) acting only on the $a$--$B$ relative motion space: this limitation is very useful, as it allows to formulate the equation as a 1-body problem. $\Bra{\chi_{\beta^-}}$ is a vector of the same space of the adjoint of $\Braket{\psi_\beta|\Psi_+}$. %
	The boundary conditions %
	match those holding for $\Braket{\Psi_-|\psi_\beta}$, namely, %
	at large $R_{aB}$ distances $\Bra{\chi_{\beta^-}}$ is written as $\bra{\v k_\beta}$ plus
	spherical outgoing scattered waves.
	The plane wave is the same appearing in \cref{eqRisultatoTransitionAmplitudePWPriorPost}.
	This is formally identical %
	to a standard elastic scattering problem (with ``initial'' wave-vector equal to $-\v k_\beta$ and the adjoint of $U_{aB}$ as potential%
	\footnote{It is also possible to consider the time-reversed solution $\ket{\chi_{\beta^+}}$, to the same end, see e.g.~\cite[eq.~(2.36)]{Satchler1983Direct}.}),
	and can be easily solved numerically. %

	Consider now the formal solution of \cref{eqEquazioneOndaDistortaStandard} obtained trough the Green's function method. %
	The operator on the left-hand side, %
	$E_\beta - \op K_{aB}$, is identical to that in \cref{eqEquazioneProiettataOndePiane}. Consider the same procedure employed for \cref{eqDefinizioneFunzioneDiGreenCompletaPlaneWaveConIEspilon}, where an imaginary term is added to the equation to fix the Green's function. Given that outgoing scattered waves are requested both in \cref{eqEquazioneProiettataOndePiane,eqEquazioneOndaDistortaStandard},
	the associated Green's function operator will thus be the same for both cases: %
	its explicit expression was shown in \cref{eqDerivazioneAmpiezzaTransferScritturaFunzioneGreenEsplicita}.
	Instead, the ``$v$'' term appearing in \cref{eqSoluzioneGeneraleFunzioneDondaDaFunzGreenPlaneWave} in this case is $\bra{\v k_\beta}$, which is indeed an eigenfunction of the operator $E_\beta - \op K_{aB}$ in \cref{eqEquazioneOndaDistortaStandard}
	\footnote{To compare with the case of \scref{eqEquazioneProiettataOndePianePrior}, note that, even though $\braket{\Psi_-|\psi_\alpha}$ and $\bra{\chi_{\beta^-}}$ have the same boundary conditions, the operator defining the respective differential equations is very different. This causes the difference between \scref{eqSoluzioneOndePianeNotazioneBraKetConFunzioneGreen,eqSoluzioneGreenFunctionOndaDistortaStandardPost} regarding the ``$v$'' term.}.
	In summary,
	\begin{equation}\label{eqSoluzioneGreenFunctionOndaDistortaStandardPost}
		\Bra{\chi_{\beta^-}} = \Bra{\v k_\beta} + \Bra{\chi_{\beta^-}} \op U_{aB} \op G_{\text{PW},+}
	\end{equation}
	Substitute now $\bra{\v k_\beta}$ from \scref{eqSoluzioneGreenFunctionOndaDistortaStandardPost} in \cref{eqRisultatoTransitionAmplitudePWPriorPost}, finding: %
	\begin{equation}
		\mathcal T_{\alpha\to\beta}^{\text{post}} = \Braket{\psi_\beta \chi_{\beta^-} | (V_{ab} + V_{a\nu}) - \op U_{aB} \op G_{\text{PW},+} (V_{ab} + V_{a\nu}) | \Psi_+}
	\end{equation}
	Using \cref{eqSoluzioneOndePianeNotazioneBraKetConFunzioneGreen},
	and assuming (as usual) that the order in which the integrals are performed can be changed without altering the result,
	this is simplified into
	\begin{equation}\label{eqAmpiezzaTransizionePostEsattaInOndeDistorte}
		\mathcal T_{\alpha\to\beta}^{\text{post}}(\v k_\alpha, \v k_\beta) = \Braket{\psi_\beta \chi_{\beta^-} | \op V_{ab} + \op V_{a\nu} - \op U_{aB} | \Psi_+}
	\end{equation}
	Similarly, in prior form one finds: %
	\begin{equation}\label{eqAmpiezzaTransizionePostEsattaInOndeDistortePrior}%
		\mathcal T_{\alpha\to\beta}^{\text{prior}}(\v k_\alpha, \v k_\beta) = \Braket{\Psi_- | \op V_{ab} + \op V_{b\nu} - \op U_{Ab} | \psi_\alpha \chi_{\alpha^+} }
\end{equation}
	where $\Ket{\chi_{\alpha^+}}$ is a solution of $(E_\alpha - \op K_{Ab}) \Ket{\chi_{\alpha^+}} = \op U_{Ab} \Ket{\chi_{\alpha^+}}$ with boundary conditions corresponding to those for $\Braket{\psi_\alpha|\Psi_+}$ (plane-wave directed as the beam plus outgoing scattered waves) and $U_{Ab}$ is an auxiliary potential acting only on coordinate $\v R_{Ab}$.
	
	The advantage of such expressions, with respect to \cref{eqRisultatoTransitionAmplitudePWPriorPost}, is a simplification on the operator appearing in the transition amplitude. For instance, $U_{aB}$ is normally constructed to include at least the long-range Coulomb potential between $a$ and $B$, which at high distances matches the analogous contribution appearing from $V_{ab} + V_{a\nu}$. Hence, the total potential is short-ranged in $R_{aB}$, %
	which causes the necessary approximations on $\Psi_+$ to be less critical. %

\subsection{Generalized distorted-wave formulation}\label{sezGeneralizedDistortedWaves}

	The procedure in \cref{sezDistortedWaves} can be extended to allow for a more general auxiliary potential, which depends on all coordinates (and not only on the reactants relative motion degree of freedom).
	Here, the results are only reported without derivation. For a more complete treatment, see \cite[sec.~6.F]{Glendenning2004direct}, \cite[sec.~11.6]{Goldberger1964collision}, or \cite{Timofeyuk1999,Moro2009} and references therein.
	In post-form, start back from the non-projected \cref{eqProblemaAutovaloriCompletoPost}, and consider the following auxiliary equation (similarly to e.g.~\cite[eq.~(2.41)]{Satchler1983Direct} or, more explicitly, \cite[eq.~(10)]{Moro2009}):
	\begin{equation}\label{eqAusiliariaOndeGeneralizzate}
	\Bra{\Phi_{\beta^-}} \[ E - \op K_{aB}(\v R_{aB}) - \op H_\beta(\v r_\beta) \] = \Bra{\Phi_{\beta^-}} \op U_\beta(\mathds{R})
	\end{equation}
	where $\bra{\Phi_{\beta^-}}$ is the ``generalised distorted-wave'', a vector belonging to the same complete space of $\bra{\Psi_-}$, and $\op U_\beta$ is any “potential” operator acting on it. %
	The boundary conditions match those %
	required for $\bra{\Psi_-}$ (mentioned in \cref{secIntroductionGeneralReactionTheoryApplicataAlTransfer}):
	asymptotically $\bra{\Phi_{\beta^-}}$ %
	is composed by $\bra{\psi_\beta \v k_\beta}$ plus spherical outgoing scattered waves in any channel which $\op U_\beta$ couples to the initial one.
	The adopted differential equation instead matches the one found in \cref{eqProblemaAutovaloriCompletoPost}, so that the distorted-wave can then be employed to re-express the post-form transition amplitude.

	For the prior form, the auxiliary equation is:
	\begin{equation}\label{eqAusiliariaOndeGeneralizzatePrior}
	\[ E - \op K_{Ab}(\v R_{Ab}) - \op H_\alpha(\v r_\alpha) \] \Ket{\Phi_{\alpha^+}} = \op U_\alpha(\mathds{R}) \Ket{\Phi_{\alpha^+}}
	\end{equation}
	with analogous meaning of all symbols and similar boundary conditions.
	The case in which $U$ depends only on the projectile-target coordinate, as in \cref{sezDistortedWaves}, could %
	be seen as a sub-case of the present situation (thus the name “generalized distorted waves”). %
	
	Appropriately manipulating the relation between the full solutions and the distorted-waves, it is found that the transition amplitude can, in the most general case, be written as in %
	\cite[eq.~(6.39), (6.46)]{Glendenning2004direct}%
	\footnote{The trick exploited in \cref{sezDistortedWaves} cannot be applied here, because, as mentioned when discussing \cref{eqEquazioneFunzGreenSpazioCompletoDisaccoppiata}, there is some freedom in the choice of the Green's function $\op{\mathcal G}$ solving \cref{eqDefinizioneFunzioneDiGreenCompletaPlaneWaveConIEspilon}, and the correct choice, in general, will be different for %
		$\ket{\Psi_+}$ and $\bra{\Phi_{\beta^-}}$.}:
	\begin{equation}\label{eqFormulaADuePotenzialiGeneralizzata}\begin{gathered}
	\mathcal T_{\alpha\to\beta}^{\text{post}} = \Braket{\Phi_{\beta^-} | \op V_{b\nu} - \op V_{a\nu} + \op U_\beta | \psi_\alpha \v k_{\alpha} } + \Braket{\Phi_{\beta^-} | \op V_{a\nu} + \op V_{ab} - \op U_\beta | \Psi_+} \\
	\mathcal T_{\alpha\to\beta}^{\text{prior}} = \Braket{\psi_\beta \v k_{\beta}| \op V_{a\nu} - \op V_{b\nu} + \op U_\alpha |\Phi_{\alpha^+}} + \Braket{\Psi_- | \op V_{b\nu} + \op V_{ab} - \op U_\alpha | \Phi_{\alpha^+} }
	\end{gathered}\end{equation}
	As noted in \cite[sec.~6.F]{Glendenning2004direct}, by employing the definitions in \cref{secIntroductionGeneralReactionTheoryApplicataAlTransfer}, the potential $V_{a\nu} - V_{b\nu} + U_\alpha$ can be rewritten as $K_{Ab} + H_\alpha + U_\alpha - \( K_{aB} + H_\beta \)$. In favourable cases, the operator written in this second form can be treated as self-adjoint, and then the first term of the prior transition amplitude vanishes using \cref{eqProblemaAutovaloriStatiMotoInternoPartIniziale,eqAusiliariaOndeGeneralizzatePrior}. The same consideration applies to the post-form amplitude. %
	For instance, this is the case when $\op U_\beta$ acts only on the $a$--$B$ relative motion space \cite[sec.~6.F]{Glendenning2004direct}, causing \cref{eqFormulaADuePotenzialiGeneralizzata} to reduce to \cref{eqAmpiezzaTransizionePostEsattaInOndeDistorte} as expected.
	However, the simplification does not hold in general%
	\footnote{Similar counter-intuitive results can be also found in simpler expressions (for instance the transition amplitude for an optical-model elastic scattering) when inserting equal and opposite non-self-adjoint terms (often kinetic terms) and attempting to evaluate them separately on non-normalisable states. The discussion of such issue is beyond the scope of this work, see \cite[sec.~6.K]{Glendenning2004direct} or \cite[ch.~1]{BurrelloTesiTriennale} for some additional information.}.
	For instance, choosing $\op U_\beta = \op V_{a\nu} + \op V_{ab}$ (the full potential of the exact problem), the distorted wave $\bra{\Phi_{\beta^-}}$ coincides with $\bra{\Psi_-}$, the second term of the transition amplitude in \cref{eqFormulaADuePotenzialiGeneralizzata} vanishes, and the first term becomes the prior-form plane-wave amplitude in \cref{eqRisultatoTransitionAmplitudePWPriorPost}. %

	The generalised auxiliary potential $U_\gamma$ is often explicitly set to induce only elastic and inelastic scattering between a finite number of excited states within partition $\gamma$, see e.g.~\cite[sec.~2.6]{Satchler1983Direct}. %
	A different possible choice %
	is to set $U_\alpha = U_\beta = V_{ab}$.
	It can be found that with this choice the the first term in the transition amplitude in \cref{eqFormulaADuePotenzialiGeneralizzata} still vanishes \cite[eq.~(561)]{Goldberger1964collision}.
	The exact amplitude in \cref{eqFormulaADuePotenzialiGeneralizzata} is thus
	\begin{equation}\label{eqAmpiezzaTransizioneGGW}\begin{gathered}
			\mathcal T_{\alpha\to\beta}^{\text{post}}(\v k_\alpha, \v k_\beta) = \Braket{ \Phi_{\beta^-} | \op V_{a\nu} | \Psi_+} \\
			\mathcal T_{\alpha\to\beta}^{\text{prior}}(\v k_\alpha, \v k_\beta) = \Braket{\Psi_- | \op V_{b\nu} | \Phi_{\alpha^+} }
	\end{gathered}\end{equation}
	In the standard distorted-wave formalism, the difference $V_{ab} - U$, which can never be strictly zero in that case, is called ``remnant'' term. %
	Such term is sometimes neglected to simplify the practical computation, especially in less recent %
	works. It is underlined that in \cref{eqAmpiezzaTransizioneGGW} the remnant term is, on the contrary, %
	cancelled exactly. %
	This approach %
	was
	first proposed (to the author's knowledge) by Greider in \cite{Greider1959}, then studied by Goldberger and Watson \cite[sec.~11.6]{Goldberger1964collision}, and later employed for example by Timofeyuk and Johnson in \cite{Timofeyuk1999}. %
	For definiteness, in this work %
	\cref{eqAmpiezzaTransizioneGGW} is referred to as ``Greider-Goldberger-Watson'' (GGW) amplitude.
	
	Note that, if both $b$ and $\nu$ are charged particles, their mutual interaction $V_{b\nu}$ will include a Coulomb term, which is long-ranged (and similarly for the post form). This could potentially degrade the accuracy of approximated expressions derived from this scheme, and it is conceivable that a modified choice for $U$, cancelling at least the long-range part of the Coulomb interaction, may turn out to be more accurate.
	In literature, \cref{eqAmpiezzaTransizioneGGW} was normally applied to reactions of deuteron stripping or pickup \cite{Greider1959,Goldberger1964collision,Timofeyuk1999,Ogata2003,Moro2009}, so that $V_{a\nu}$ (or $V_{b\nu}$) is the proton-neutron potential, and the issue is avoided.

\subsubsection{Explicit calculation of the generalised distorted wave}
	
	Differently than in the standard distorted-waves formalism in \cref{sezDistortedWaves}, the accuracy on the generalised distorted-wave can in itself be an issue affecting the overall accuracy of the reaction amplitude.
	In fact, depending on the precise $U_\beta$ chosen, finding $\Phi_\beta$ in \cref{eqAusiliariaOndeGeneralizzate} can be as complicated as finding the complete solution $\Psi$.

	In general, $\Bra{\Phi_{\beta^-}}$ can be expressed by expanding it into a basis %
	of final-partition states, i.e.\ a complete orthogonal set $\{\psi_\beta^j\}$ of eigenstates of $H_\beta$, each with eigenvalue $E - E_\beta^j$: %
	\begin{equation}\label{eqScomposizioneCDCCOndaDistortaGeneralizzata}
	\Bra{\Phi_{\beta^-}} = \sum_j C_j \Bra{\psi_\beta^j \chi_\beta^j}
	\end{equation}
	where each $C_j \chi_\beta^j$ is the “coefficient” of the associated term in the expansion. The $\sum$ can actually be an integral on $E - E_\beta^j$ if the $\phi_\beta^j$ are unbound, and in general will be a combination of a discrete sum and an integral, if the spectrum is mixed. $C_j$ is a numeric coefficient, whose modulus is determined if each $\chi_\beta^j$ and $\psi_\beta^j$ is normalized (to 1 or to an appropriate $\delta$ distribution). The phases are similarly fixed given a convention on the phases of each wave-function.
	
	The choice of expanding $\Phi_\beta$ into the final-partition basis is motivated by the form of the operator in \cref{eqAusiliariaOndeGeneralizzate}. Projecting on any basis state and reasoning as for \cref{eqEquazioneProiettataOndePiane}, a coupled set of differential equations for the set of $C_j \Bra{\chi_\beta^j}$ is derived. For each $j$,
	\begin{equation}\label{eqProblemaCDCCAccoppiatoSpazioCompleto}
		C_j \Bra{\chi_\beta^j} \( E_\beta^j - \op K_{aB}(\v R_{aB}) \) = \sum_k C_k \Bra{\chi_\beta^k} \Braket{\psi_\beta^k | \op U_\beta(\v R_{aB}, \v r_\beta) | \psi_\beta^j }
	\end{equation}
	Such coupled system can be practically solved through a continuum-discretized coupled channels (CDCC)
	numerical method. Essentially, in order for the problem to become numerically solvable, the number of coupled equations, namely the number of eigenfunctions $\psi_\beta^j$ to include in the spectrum, must be first reduced to a finite number, thus at least the continuum part of the spectrum must be somehow discretised and truncated. Several approaches exist, each with its strengths, and their discussion is beyond the scope of this work.
	The approximate solution for $\Bra{\Phi_{\beta^-}}$ can be expected to be accurate in the limit where the couplings induced by $U_\beta$ are strong only within the space that is well represented by the discretised and truncated basis.

\section{One-particle transfer}\label{secTeoriaScatteringOneParticleTransfer}
	
	Consider now a truncated model space where particles $a$, $b$ and $\nu$ are ``elementary'' (inert and structureless). All microscopic potentials ($V_{ab}$, $V_{a\nu}$, $V_{b\nu}$) are substituted with effective ones involving only the position of each particle centre-of-mass. %

	Any wave-function $\ket{\psi_\alpha^j}$ for the internal motion of projectile and target in the initial partition can be factorised as the product of the fixed intrinsic states for $a$, $b$ and $\nu$ (those referred to as $\Psi_b$ etc.~in \cref{secOverlapFunctionSpectroscopicFactorsIntroduzione}), and an appropriate one-body bound state, $\ket{\phi_\alpha^j}$, for the $\nu$--$a$ relative motion; similar considerations apply to $\ket{\psi_\beta^j}$. The elementary particles fixed state then play no role, because in all transition amplitudes (consider e.g.~\cref{eqRisultatoTransitionAmplitudePWPriorPost}) it is simplified in the scalar product, %
	as the effective potential does not act on it.
	Formally, each $\ket{\phi_\alpha^j}$ %
	is deduced from %
	the overlap function (see the discussion in \cref{secOverlapFunctionSpectroscopicFactorsIntroduzione}) between the desired microscopic states for $a$, $\nu$, and the composite system $A$.
	The anti-symmetrised wave-functions %
	for the composite systems consist of several components essentially differing just for the assignment of labelled coordinates to each nucleon. %
	At the same time, the precise identity %
	of the nucleons removed from $A$ (and added to $B$) during the transfer process %
	is irrelevant for the result. The reaction cross-section is then the sum of several %
	contributions, which counter the binomial coefficients appearing in the overlap functions (which are defined projecting the composite-system wave-function on a precise repartition of nucleons between core and valence), see for instance \cref{eqFractionalParentageExpansionProiettata}. Similarly, the Clebsh-Gordan coefficient in the overlap expression is employed when coupling the angular momenta to form a composite-system state with defined total spin. %
	The spectroscopic amplitudes appearing in the overlaps instead do not cancel out, and the first-order reaction amplitude is proportional to these.
	In summary, it is possible to compute the cross-section as a single term, adopting as $\Ket{\phi_\alpha^j}$ the spectroscopic amplitude for the overlap times a one-body bound state normalised to 1, constructed from the functions $\Ket{\phi_{\nu,l,s,m_l}}$ in \cref{eqEspansioneModelloAClusterGenerica} by appropriately coupling the angular momenta.
	In the expressions shown in the following, the symbol $\psi_\gamma$ for the internal-motion states is retained for conciseness.
	
	Having specified a model for the reactants structure, the last missing ingredient is to choose an approximation for the full scattering solution $\Psi_+$ to obtain an explicit expression for the transition amplitude. Several approaches, parallel to the schemes %
	employed for generating the distorted waves in \cref{secGeneralReactionTheoryApplicataAlTransfer}, are applicable, and some of them are discussed in the following. %
	Many other schemes have been devised in literature, which are not covered in this work. See for instance \cite{ThompsonNunes2009} and references therein for the adiabatic and eikonal approximation, and \cite{Broglia2004} for semi-classical methods. %

\subsection{Distorted-wave Born approximation (DWBA)}\label{sezDWBA}
	
	Consider %
	the prior-form standard-distorted-wave transition amplitude in \cref{eqAmpiezzaTransizionePostEsattaInOndeDistortePrior}. Let $U_{aB}$ be a potential depending only on the $a$--$B$ relative motion coordinates, and find a solution $\Bra{\chi_{\beta^-}}$ for that potential, in complete analogy with \cref{eqEquazioneOndaDistortaStandard}. The first-order distorted-wave Born approximation (DWBA) prescribes%
	\footnote{See \cite[sec.~2.8.7]{Satchler1983Direct} and referenced section for a more in-depth discussion on the interpretation of the approximation. The simple recipe reported here is sufficient within the scope of this text.} %
	to approximate $\Bra{\Psi_-} \approx \Bra{\psi_\beta \chi_{\beta^-}}$. %
	The same approach is followed in post form for $\Psi_+$. In summary:
	\begin{equation}\label{eqAmpiezzaDiTransizioneDWBAPrimoOrdine}\begin{gathered}
			\mathcal T_{\text{DWBA}}^{\text{post}} = \Braket{\psi_\beta \chi_{\beta^-} | \op V_{ab} + \op V_{a\nu} - \op U_{aB} | \psi_\alpha \chi_{\alpha^+} } \\
			\mathcal T_{\text{DWBA}}^{\text{prior}} = \Braket{\psi_\beta \chi_{\beta^-} | \op V_{ab} + \op V_{b\nu} - \op U_{Ab} | \psi_\alpha \chi_{\alpha^+} }
	\end{gathered}\end{equation}

	In prior form, if $U_{aB}$ properly describes the elastic scattering between two inert particles $a$ and $B$, the approximation basically amounts to claiming that the exact solution is dominated by the elastic component, and all other channels are a small perturbation.
	The potential can also phenomenologically contain some effects related to non-elastic channels, for instance through imaginary components. The approximation could be expected to be more accurate for a smaller residual interaction (in prior, $V_{ab} + V_{a\nu} - U_{aB}$), so that $\chi_{\beta^-}$ resembles more the exact solution. %
	
	Incidentally, if both $U_{aB}$ and $U_{Ab}$ are set to zero, the exact transition amplitude reduces to the plane-wave expression in \cref{eqRisultatoTransitionAmplitudePWPriorPost}, and the exact wave-function is approximated to just its non-scattered component, $\Ket{\Psi_+} \approx \Ket{\psi_\alpha \v k_{\alpha}}$ etc. The result is know as ``plane-wave Born approximation'' (PWBA). As long as a numerical solution to the problem as posed is sought, the PWBA scheme is rarely employed in recent works, as the practical computation of the distorted waves is not particularly expensive or difficult, compared to current computational capabilities.

	DWBA prior and post forms are equivalent, meaning that, if the same pair of potentials is adopted in both forms, and %
	no further approximation is made, it is $\mathcal T_{\text{DWBA}}^{\text{post}} = \mathcal T_{\text{DWBA}}^{\text{prior}}$, as shown in \cite[sec.~2.8.8]{Satchler1983Direct}. %
	Such result is not trivial because, while the exact expressions in \cref{eqAmpiezzaTransizionePostEsattaInOndeDistorte,eqAmpiezzaTransizionePostEsattaInOndeDistortePrior} necessarily coincide, they are supplied with two different approximations in order to obtain \cref{eqAmpiezzaDiTransizioneDWBAPrimoOrdine}. %
	The prior-post equivalence has two relevant consequences in practical calculations. First,
	for any given calculation it is possible to freely choose the form giving rise to the simplest, and thus most accurate, calculation from the numerical point of view.
	Second, %
	the difference between prior and post transition amplitudes is a %
	measure of %
	numerical inaccuracies, %
	disentangled from the impact of physical approximations. %
	In particular, if the computation is accurate in both forms, the results must coincide.
	Note that %
	if the results do not coincide, it is still possible that the calculation in one form is accurate.
	
	There is an apparent asymmetry %
	between the role of $U_{aB}$ and $U_{Ab}$ in each (prior or post) DWBA scheme. One of the potentials %
	is an auxiliary, in principle arbitrary, operator employed to reformulate an exact expression: note that this does not exclude that the precise choice of the auxiliary potential can have an impact on the computed amplitude, since an approximation is then being performed. The other $U$, instead, is introduced while attempting to generate a physically sound %
	approximation for the complete scattering wave-function.
	However, the aforementioned property of equivalence between prior and post forms (in which the role of the two auxiliary potentials is opposite) %
	actually suggest that, once the approximation of the full solution is introduced, both $U_{aB}$ and $U_{Ab}$ equally contribute %
	to determine the final result.

\subsection{Coupled-channels approaches}\label{secOneParticleTransferCCApproaches}

	Another relevant class of approximate methods %
	includes all the cases where %
	a wave-function is computed using a continuum-discretised coupled-channels scheme. This is the case when a generalised-distorted-waves formalism is adopted, computing the distorted wave as described in \cref{sezGeneralizedDistortedWaves}, or when the full solution of the scattering problem is approximated by a CDCC wave-function. The possible combinations are listed below, followed by a general discussion on the features of such methods.

\subsubsection{Standard distorted-waves scheme} %

	Start from the standard-distorted-wave reaction amplitude in \cref{eqAmpiezzaTransizionePostEsattaInOndeDistorte,eqAmpiezzaTransizionePostEsattaInOndeDistortePrior}, but now
	approximate $\Psi_+$ with a continuum-discretized coupled channels solution of the full problem \cref{eqProblemaAutovaloriCompletoFormaGenerica}. %
	Such scheme is essentially a simplified version %
	of the %
	coupled-channels Born Approximation discussed in \cite[sec.~3.6]{Satchler1983Direct}.
	The %
	differential equation for the solution $\Ket{\Psi_+}$ (required in post-form), equipped with the desired boundary condition,
	is expressed using the prior-form coordinates and the corresponding rearrangement of the complete Hamiltonian, $\mathcal H$: %
	\begin{equation}\label{eqEquazioneDifferenzialeSoluzioneCompletaPsiPiuPerCDCC}\begin{aligned}
			& \[ E - \op K_{Ab}(\v R_{Ab}) - \op H_\alpha(\v r_\alpha) - \op V_{ab}(\v R_{Ab}, \v r_\alpha) - \op V_{b\nu}(\v R_{Ab}, \v r_\alpha) \] \Ket{\Psi_+} = 0 \\
			& \Ket{\Psi_+} = \Ket{\psi_\alpha \v k_\alpha} + \text{outgoing scattered waves}
	\end{aligned}\end{equation}
	analogously, in prior:
	\begin{equation}\begin{aligned}
			& \Bra{\Psi_-} \[ E - \op K_{aB}(\v R_{aB}) - \op H_\beta(\v r_\beta) - \op V_{ab}(\v R_{aB}, \v r_\beta) - \op V_{a\nu}(\v R_{aB}, \v r_\beta) \] = 0 \\
			& \Bra{\Psi_-} = \Bra{\psi_\beta \v k_\beta} + \text{outgoing scattered waves}
	\end{aligned}\end{equation}
	It is also possible to supply a simplified version of the full problem, including only simpler couplings within the initial partition states.
	The differential equation is then solved approximately trough a scheme entirely analogous to the one employed for $\Phi$ in \cref{eqProblemaCDCCAccoppiatoSpazioCompleto},
	and the result is inserted in \cref{eqAmpiezzaTransizionePostEsattaInOndeDistorte} or \eqref{eqAmpiezzaTransizionePostEsattaInOndeDistortePrior}.

	It is underlined that, in principle, any set of coordinates and any form for $\mathcal H$ would be acceptable, and the choice is dictated by practical matters. %
	When deriving the exact transition amplitude, the goal was %
	to analytically evaluate the projection of the unknown full solution on the desired final (or initial) state, thus the Hamiltonian was expressed so to simplify such task. Here, where an explicit solution for $\Psi$ %
	is sought, the given choice for the coordinates removes any difficulty in properly enforcing the boundary condition, and allows to describe with good accuracy elastic and inelastic couplings from the initial (or final) state of interest, which presumably give rise to the dominant components in the full wave-function.
	The disadvantage is that %
	the two wave-functions appearing in the reaction amplitude %
	are originally computed in different coordinates, thus a coordinate change will be required to obtain the final result, with the main effect of complicating the explicit form of the angular-momentum expansions. Note that this sort of issue already arises in \cref{eqAmpiezzaDiTransizioneDWBAPrimoOrdine}.

\subsubsection{Generalised-distorted-waves schemes}

	Consider now the expression in \cref{eqFormulaADuePotenzialiGeneralizzata} for the exact reaction amplitude, together with the CDCC solution for the generalised distorted wave, computed from \cref{eqProblemaCDCCAccoppiatoSpazioCompleto}.
	In order to approximate the full solution, $\Psi$,
	one possible approach is to apply again the CDCC method %
	(see text commenting \cref{eqEquazioneDifferenzialeSoluzioneCompletaPsiPiuPerCDCC}).

	Another %
	possibility is to employ %
	the same approximation adopted when deriving the DWBA, illustrated in \cref{sezDWBA}: $\Bra{\Psi_-}$ is set to $\Bra{\psi_\beta \chi_{\beta^-}}$, where $\chi_{\beta^-}$ is the solution of a one-body problem with the form of \cref{eqEquazioneOndaDistortaStandard}.
	Such construction is simpler to implement and less demanding computationally, %
	and was thus employed in the coupled-channels calculation discussed %
	in \cref{secCalcoliDiTransfer}. %

\subsubsection{Common features of coupled-channels schemes}
	
	The CDCC solution for either $\Psi$ or $\Phi$ %
	is not limited to the elastic scattering component.
	This framework thus allows to take into account some ``dynamical'' effects, namely features that cannot be explained solely in terms of ground-state properties of the reactants, and which manifest quantum-mechanically as %
	coupling to excited states. %
	If the spectrum of $\psi_\alpha^j$ is approximated with acceptable accuracy, $\Psi_+$ (or $\Phi_{\alpha}$) may %
	provide a good account of inelastic excitations of nucleus $A$ induced by the interaction with nucleus $b$.
	This approach may thus be expected to be superior to \cref{eqAmpiezzaDiTransizioneDWBAPrimoOrdine}. %
	However, as mentioned when discussing \cref{eqProblemaCDCCAccoppiatoSpazioCompleto}, the practical execution of the CDCC recipe %
	can be significantly %
	more complex with respect to the standard first-order DWBA prescription. %
	On this regard, remind that the reactants structure assumed in this section (associated to the transfer of an inert cluster) causes $\psi_\alpha$ and $\psi_\beta$ to reduce to just the $\nu$--$a$ and $\nu$--$b$ relative-motion wave-functions, greatly simplifying the problem.
	Even in the case where both the full solution and the generalised distorted-wave are computed in CDCC, prior and post forms are in general not equivalent.
	Of course, in the limit where the approximated $\Psi_{\pm}$ approach the exact solutions, the calculated amplitudes would tend to the same, exact value. %
	Consequently, in this framework, prior-post invariance cannot be used to test specifically the formal consistency of the calculation and the appropriate implementation of the numerical algorithm, but can only give a suggestion on %
	the “global” accuracy of the calculation, being sensitive also (but not only) to the adopted physical approximations. %
	At the same time,
	achieving satisfactory numerical accuracy can be difficult, %
	thus, %
	even more than in DWBA,
	for each specific calculation it is useful to identify the most convenient form regarding the amount of required computational resources.
	Some information from this point of view can be deduced checking convergence with respect to the numerical parameters of the calculation (see e.g.\ \cite{Moro2009}).

\section{Two-particle transfer}\label{secReactionTheoryTwoParticleTransfer}

	If the transferred system %
	is a composite nucleus, especially a loosely bound one, it can be interesting to take into account its internal structure and the associated impact on the reaction.
	Applications in literature almost invariably involve the transfer of two nucleons, very often two neutrons. %
	A discussion on the associated formalism, from different points of view, can be found in \cite[sec.~3.7]{Satchler1983Direct}, \cite{Catara1984,Bang1985,Oertzen2001,Thompson2013,Potel2013}.
	Such calculations are customarily performed in second-order distorted-waves Born approximation (DWBA).
	In \cref{secCalcoliDiTransfer}, the formalism will be applied to the transfer of a proton and a neutron. %
	As in \cref{secTeoriaScatteringOneParticleTransfer}, the core particles of both reactants %
	are treated as inert and their internal structure is neglected.
	Notwithstanding the formal similarities with the two-neutron-transfer case, the application of the same %
	scheme to \nuclide{p}+\nuclide{n} transfer reactions is %
	quite new, with the very first published calculation apparently dating back to 2017 \cite{Ayyad2017}.

	In the second-order DWBA formalism,
	(see \cite[sec.~3.7]{Satchler1983Direct}), %
	the transition amplitude can be decomposed into %
	the coherent sum of a first-order term, connected to the one-step (or ``simultaneous'') transfer of both particles, and a second-order contribution, associated with two-step processes, and in particular the ``sequential'' transfer of each particle separately.
	It is found that the equivalence between prior and post forms of the distorted-wave Born approximation holds order-by-order (provided that non-orthogonality corrections are taken into account) \cite[sec.~3.7.2]{Satchler1983Direct}. The considerations made in \cref{sezDWBA} on this regard can thus be applied here to the simultaneous and sequential contribution separately.

	In principle, the calculation of each contribution to the two-particle-transfer transition amplitude involves addressing a four-body problem (the two core nuclei and the two transferred particles), where the reactants internal-motion wave-functions depend on two independent coordinates (e.g.~the position of each transferred particle).
	Here, the full problem is instead approximated so that all internal-motion states can be constructed in terms of two-body problems, which can be solved with greater ease.
	The following parts of this \namecref{secReactionTheoryTwoParticleTransfer} are principally dedicated to %
	the formal development of such approximation. %
	The discussion will be focused in particular on approaches that can be implemented in the coupled reaction channel calculations code \textsc{Fresco} \cite{Thompson2004}, which was employed to perform the practical calculations in \cref{secCalcoliDiTransfer}.

\paragraph{The system Hamiltonian}	

	The notation employed in this \namecref{secReactionTheoryTwoParticleTransfer} mirrors %
	the one found in \cref{secIntroductionGeneralReactionTheoryApplicataAlTransfer}, with some modification to allow explicit discussion of the additional degree of freedom, discussed in the following. %
	Consider the transfer reaction $\mathcal{A}+b \to a + \mathcal{B}$, %
	where nucleus $\mathcal{A}$ is a bound system of three elementary clusters, $a+\nu+\mu$, and similarly $\mathcal{B} = b+\mu+\nu$. The transferred system is $N = \nu+\mu$. Also let $A = a + \mu$, $A' = a + \nu$, $B = b + \nu$, and $B' = b + \mu$. %
	
	As before, given any pair of systems $i,j$ and their composition $J = i+j$, let $K_{ij} = K_i + K_j - K_J$, where $K_i$ is the kinetic energy of the centre-of-mass of $i$.
	Assume all interactions can be described by two-body forces, %
	with the potential energy between any pair of particles $i,j$ being $V_{ij}$.
	Let $\v r_{ij}$ be the distance between particles $i$ and $j$. %
	
	The system Hamiltonian, is the same in %
	\cref{secIntroductionGeneralReactionTheoryApplicataAlTransfer}: updating the notation and explicitly splitting the terms involving $\nu$ or $\mu$ (rather than their composition $N$), it is
	$\mathcal H = \sum_{i\in\lbrace a,b,\mu,\nu\rbrace} K_{i} + V_{ab} + V_{a\nu} + V_{b\nu} + V_{a\mu} + V_{b\mu} + V_{\nu\mu}$.
	All calculations are performed in the system centre-of-mass rest frame. $\mathcal H$ can be rewritten in several ways, for instance
\begin{subequations}\label{eqTwoNucTranHamiltExplicitSubEqPriorPost}
	\begin{equation}\label{eqTwoNucTranHamiltExplicitPrior}
	\mathcal H = \left(K_{a\mu} + V_{a\mu}\right) + \left(K_{A\nu} + V_{a\nu} + V_{\mu\nu}\right) + \left(K_{\mathcal{A}b} + V_{ab} + V_{b \mu} + V_{b \nu} \right)
	\end{equation}
	which is useful to perform a calculation in prior form, or analogously
	\begin{equation}\label{eqTwoNucTranHamiltExplicit}
		\mathcal H = \left( K_{b\nu} + V_{b\nu} \right) + \left( K_{B\mu} + V_{b\mu} + V_{\mu \nu} \right) + \left(K_{\mathcal{B}a} + V_{ab} + V_{a\nu} + V_{a\mu} \right)
	\end{equation}
\end{subequations}
	which is the post-form rearrangement.
	Similar expressions, involving $A'$ and $B'$, can be similarly obtained (note that the role of $\mu$ and $\nu$, and thus $A$ and $A'$, is entirely symmetrical).

\subsection{First-order (``simultaneous'') contribution}\label{secTNTSimultaneoTeoria}
	
	From the formal point of view, the DWBA simultaneous contribution to the reaction amplitude has precisely the same form in \cref{eqAmpiezzaDiTransizioneDWBAPrimoOrdine}, see \cite[eq.~(3.61)]{Satchler1983Direct}, with the relevant difference that now the Hamiltonian depends also on $\v r_{\mu\nu}$, and full internal-motion states $\psi_\alpha$ and $\psi_\beta$ reduce to three-particle wave-functions explicitly describing the $a+\nu+\mu$ and $b+\mu+\nu$ relative motion, scaled, as in \cref{secTeoriaScatteringOneParticleTransfer}, by the spectroscopic amplitude for the associated overlaps (see \cref{secOverlapFunctionSpectroscopicFactors}). %
	In the following, the symbol $\psi_\gamma$ will still denote the wave-function for reactants internal motion in partition $\gamma$, as in \cref{secIntroductionGeneralReactionTheoryApplicataAlTransfer}, but ignoring the trivial factors related to the internal state of $a,b,\mu$, and $\nu$.

	Consider for instance the ``post-form'' Hamiltonian in \cref{eqTwoNucTranHamiltExplicit}. The grouping of terms enclosed in braces suggests a formal procedure to construct all required ingredient for the simultaneous transfer calculation. First, the spectrum for the $b$--$\nu$ relative-motion is constructed as the solution of a standard two-body problem under the Hamiltonian $K_{b\nu} + V_{b\nu}$, obtaining an orthogonal set of wave-functions $\phi^j_{b\nu}$ indexed by $j$.
	Then, $\psi_\beta$ %
	is found expanding it in the $b$--$\nu$ basis and solving for the $B$--$\mu$ motion under the potential $V_{\mu b} + V_{\mu \nu}$, trough a procedure analogous to \cref{eqProblemaCDCCAccoppiatoSpazioCompleto}%
	\footnote{In \cref{sezGeneralizedDistortedWaves}, the $a$--$\mathcal B$ motion was studied expanding in a basis of reactant internal-motion states $\psi_\beta^j$. Here, a single state $\psi_\beta$ is studied expanding in a basis for the $b$--$\nu$ motion.}.
	The internal state for $\mathcal A$ is constructed analogously. Finally, the first-order DWBA amplitude is derived precisely as in \cref{eqAmpiezzaDiTransizioneDWBAPrimoOrdine}, using $V_{ab} + V_{a\nu} + V_{a\mu} - U_{a\mathcal{B}}$ as transition potential.
	Note that %
	explicit three-body structure calculations are usually not performed in this manner%
	\footnote{One difficulty is %
		that a good account of $b$--$\mu$ and $B$--$\nu$ motion separately as described %
		would normally require inclusion of continuum states.} %
	(see instead e.g.~\cite{Bang1979} an actual example): %
	the %
	purpose of the formal construction just illustrated is rather to clarify the meaning of approximations shown later, %
	which are aimed at decoupling %
	the two degrees of freedom appearing in both $\psi_\alpha$ and $\psi_\beta$. %
	
	Regardless of how $\psi_\beta$ %
	is originally computed, to perform a practical calculation it is convenient to express it in the $\(\v r_{\mu\nu},\v r_{Nb}\)$ Jacobi coordinates (sometimes known as ``t'' coordinates), and similarly $\psi_\alpha$ in $\(\v r_{\mu\nu},\v r_{Na}\)$ coordinates \cite[sec.~4]{Thompson2013}: note that $\v r_{\mu\nu}$ is the same vector in both cases.
	In this way, taking into account that it is possible to express $\v r_{\mathcal A b}$ and $\v r_{Na}$ as functions of $\v r_{\mathcal B a}$ and $\v r_{Nb}$, the first-order DWBA transition amplitude could be written as
	\begin{equation}\label{eqScritturaAmpiezzaSimultaneoPerSpiegareCoordinatet}
		\mathcal T = \Braket{\psi_\beta\(\v r_{\mu\nu},\v r_{Nb}\) \chi_{\beta^-}(\v r_{\mathcal B a}) | V %
			| \psi_\alpha\( \v r_{\mu\nu}, \v r_{Nb}, \v r_{\mathcal B a} \) \chi_{\alpha^+}(\v r_{Nb}, \v r_{\mathcal B a}) }
	\end{equation}
	where $V$ is a shorthand for the appropriate transition potential appearing in the scalar product, which depends on all coordinates, $\v r_{\mu\nu}, \v r_{Nb}, \v r_{\mathcal B a}$. For any fixed value of $\v R_{Nb}$ and $\v r_{\mathcal B a}$, perform first the integration on $\v r_{\mu\nu}$, defining
	\begin{multline}
	\coniugato{\widetilde\psi}_\beta\(\v r_{Nb}\) \widetilde V(\v r_{Nb}, \v r_{\mathcal B a}) \widetilde\psi_\alpha\( \v r_{Nb}, \v r_{\mathcal B a} \) =\\= \int \coniugato{\psi}_\beta\(\v r_{\mu\nu},\v r_{Nb}\) V\( \v r_{\mu\nu}, \v r_{Nb}, \v r_{\mathcal B a} \) \psi_\alpha\( \v r_{\mu\nu}, \v r_{Nb}, \v r_{\mathcal B a} \) \d^3\v r_{\mu\nu}
	\end{multline}
	so that $\mathcal T = \Braket{\widetilde\psi_\beta\(\v r_{Nb}\)|\widetilde V(\v r_{Nb}, \v r_{\mathcal B a})|\widetilde\psi_\alpha\( \v r_{Nb}, \v r_{\mathcal B a} \)}$, which is formally identical to a standard one-particle-transfer transition amplitude, %
	and can thus be computed using the same numerical approaches. %
	Hence, it is relevant to point out that %
	a function expressed in $(\v r_{a\mu}, \v r_{a\nu})$ or $(\v r_{b\nu}, \v r_{b\mu})$ coordinates (also known as ``v'' coordinates) can be mapped to the corresponding one in ``t'' coordinates
	using the Moshinsky transformation \cite{Moshinsky1959}.
	The change from $(\v r_{a\mu}, \v r_{A\nu})$ and $(\v r_{b\nu}, \v r_{B\mu})$ (known as ``y'' coordinates) to ``t'' coordinates is instead governed by the Raynal-Revai transformation \cite{Raynal1970}.

\subsubsection{The “heavy-ion” approximation}\label{secHeavyIonScheme}

	The Hamiltonian in \cref{eqTwoNucTranHamiltExplicit} may be rewritten, in post-form rearrangement, as %
	\begin{multline}
		\mathcal H = \[ \( K_{b\nu} + V_{b\nu} \) + \( K_{b\mu} + V_{b\mu} \) + V_{\mu\nu} + K_{B} + K_{B'} - K_b - K_{\mathcal{B}} \] +\\+ K_{\mathcal{B}a} + V_{ab} + V_{a\nu} + V_{a\mu}
	\end{multline}
	or
	\begin{multline}
		\mathcal H = \[ \( K_{B' \nu} + V_{b\nu} + V_{\mu \nu} \) + \( K_{B \mu} + V_{b\mu} + V_{\mu \nu} \) - V_{\mu\nu} +\right.\\\left.+ K_{\mathcal{B}} + K_b - K_{B} - K_{B'} \] + K_{\mathcal{B}a} + V_{ab} + V_{a\nu} + V_{a\mu}
	\end{multline}
	These exact expressions by themselves bring %
	no profit. %
	However, suppose now that the core nuclei $a$ and $b$ are much heavier than the transferred system $N$. %
	Then, the centre-of-mass position of each composite system, $B$, $B'$ and $\mathcal{B}$, approximately coincides with $b$ position, so that %
	$\v r_{b\mu} \approx \v r_{B\mu}$ and so on. This also implies that
	the expectation value of $\op K_{\mathcal{B}} + \op K_b - \op K_{B} - \op K_{B'}$ on any physical state is approximately zero
	(and similarly for $a$).
	Furthermore, assume that the interaction between the two transferred particles, $V_{\mu\nu}$, is negligible compared to their interaction with the cores, for instance $V_{b\mu}$, so that $V_{b\mu}(\v r_{b\mu}) + V_{\mu \nu}(\v r_{\mu\nu}) \approx V_{b\mu}(\v r_{b\mu})$, which, if convenient, may also be seen as a potential $V_{B \mu}(\v r_{B \mu})$ depending only on the distance between $\mu$ and the centre-of-mass of $B$. %
	The Hamiltonians are as a result approximated to: %
	\begin{equation}\label{eqHamiltonianaApprossimataPostSchemiHeavyIon}\begin{aligned}
			& \mathcal H_{\text{HI-cc}}^{\text{post}} = \[ \( K_{b\nu} + V_{b\nu} \) + \(K_{b\mu} + V_{b \mu} \) \] + K_{a\mathcal{B}} + V_{ab} +  V_{a\nu} + V_{a\mu} %
			\\
			& \mathcal H_{\text{HI-CC}}^{\text{post}} = \[ \( K_{B' \nu} + V_{b\nu} \) + \(K_{B \mu} + V_{b \mu} \) \] + K_{a\mathcal{B}} + V_{ab} + V_{a\nu} + V_{a\mu} %
	\end{aligned}\end{equation}
	which could be labelled, respectively, as %
	``core-core'' %
	and
	``composite-composite''	%
	schemes of the ``heavy-ion'' approximation.
	In this limit, %
	the problems for the $b$--$\nu$ and $B$--$\mu$ motions are decoupled, so that any state of $\mathcal{B}$ can be obtained as a linear combination of products of solutions for each independent problem.
	Note that, as a result of the adopted approximations, %
	the solutions $\psi_\alpha$ and $\psi_\beta$ deduced from the model will not coincide with the ``exact'' ones. This is not a serious concern in itself, given that the ``exact'' $\mathcal H$ in \cref{eqTwoNucTranHamiltExplicit} is of course a model Hamiltonian as well, and if phenomenological forms for $V_{b\nu}$, $V_{b\mu}$ and so on are chosen appropriately, it is possible to partially correct for the neglected terms. %
	For instance, the potentials are usually selected so that the sum of the binding energies of each product of single-particle solutions matches the experimental binding energy of the composite nucleus.
	In order to mimic phenomenologically an interaction between $\mu$ and $\nu$, the potential $V_{b\nu}+V_{b \mu}$ may be set to depend in a non-standard manner on the quantum numbers characterising the solutions%
	\footnote{For example, the depth of the nuclear potential may depend on %
	the angular momentum in a way that could not be described as a spin-orbit potential.}. %
	In studies aimed at investigating correlations %
	between the two transferred particles, a framework as the one just described may be employed to generate a starting basis of solutions, which is then coupled adding an explicit $V_{\mu\nu}$ interaction, %
	often focused on pairing effects (a lot of work exists in this direction, see e.g.~\cite{Singh2019} just for a recent example). %

	In the ``core-core'' scheme, states for the $B$--$\mu$ problem can be approximated as solutions of $K_{b\mu} + V_{b\mu}$, thus the complete state is constructed using only bound-state wave-functions of the core nucleus $b$ with each valence particle independently, labelled $\phi^j_{b\nu}$ %
	and $\phi^j_{b\mu}$, %
	which are most conveniently %
	described using ``v'' coordinates (see comment to \cref{eqScritturaAmpiezzaSimultaneoPerSpiegareCoordinatet}). %
	In the same way, %
	the ``composite-composite'' scheme instead allows to describe the complete state in terms of removal of one valence particle from the given composite state $\mathcal B$.
	This sort of %
	approach
	is particularly convenient when
	$\mu$ and $\nu$ are the same nuclide, so that $V_{b\mu}(\v r)=V_{b\nu}(\v r)$ and $\phi^j_{b\nu}(\v r) = \phi^j_{b\mu}(\v r)$, %
	further simplifying the practical computation.
	
	Furthermore, %
	in the heavy-ion approximation, the transition potential $V_{ab} +  V_{a\nu} + V_{a\mu}$ can be obtained as the sum of $V_{ab}$ %
	and of the two binding potentials required to construct $\psi_\alpha$ %
	in the same approximate scheme, $V_{a\nu}$ and $V_{a\mu}$.
	Such condition, %
	which simplifies the explicit computation of the transition amplitude,
	arises naturally in
	one-particle transfers (see \cref{eqAmpiezzaDiTransizioneDWBAPrimoOrdine}), %
	but
	is not found in the full four-body treatment of the two-particle transfer (see \cref{eqTwoNucTranHamiltExplicitSubEqPriorPost}), where $V_{\mu\nu}$ %
	appears in the equations defining $\psi_\alpha$ and $\psi_\beta$ %
	but not in the transition operator. %

	With the same reasoning employed for \cref{eqHamiltonianaApprossimataPostSchemiHeavyIon}, %
	the prior form of the ``core-core'' scheme %
	is found to be %
	\begin{equation}\label{eqHamiltonianaApprossimataPostSchemaHeavyIonCoreCorePrior}
	\mathcal H_{\text{HI-cc}}^{\text{prior}} = \[ \(K_{a\mu} + V_{a \mu} \) + \( K_{a\nu} + V_{a\nu} \) \] + K_{\mathcal{A}b} + V_{ab} +  V_{b\nu} + V_{b\mu} %
	\end{equation}
	this is different than the corresponding post form %
	shown above. %
	The discrepancy in the Hamiltonians causes a difference in the calculated transition amplitudes, thus the heavy-ion approximation %
	breaks the DWBA prior-post invariance.
	In particular, %
	\begin{equation}
	\mathcal H_{\text{HI-cc}}^{\text{post}} - \mathcal H_{\text{HI-cc}}^{\text{prior}}
	= \left( K_{b\mu} - K_{B\mu} \right) - \left( K_{a\nu} - K_{A\nu} \right)
	\end{equation}
	thus the difference lies only in the kinetic terms and becomes negligible in the limit of heavy core nuclei. If the difference is instead significant, it is expected that the most accurate form will be the one related to the partition where %
	the heaviest core particle is bound to the transferred system%
	\footnote{More explicitly, if the target is the heaviest reactant, the smallest error related to this issue %
		is expected in prior for a pick-up reaction and in post for a stripping one.}.
	Regarding the $\nuclide[6]{Li} + \nuclide{p} \to \nuclide[3]{He} + \nuclide{\alpha}$ reaction studied in \cref{secCalcoliDiTransfer}, the approximation appears to be unsuitable in both the initial and final partition. %

\subsubsection{The “core-composite” approximation}\label{secTNTSimultaneoTeoriaSchemaCoreComposite}

	It is here noted that the shortcomings of %
	\cref{eqHamiltonianaApprossimataPostSchemiHeavyIon} %
	can be mitigated %
	by correctly computing the kinetic terms.
	In the exact expression for the total Hamiltonian, in \cref{eqTwoNucTranHamiltExplicitSubEqPriorPost},
	let $V_{A\nu}(\v r_{a\nu}, \v r_{\mu\nu}) = V_{a\nu}(\v r_{a\nu}) + V_{\mu\nu}(\v r_{\mu\nu})$, and similarly define $V_{B\mu} = V_{b\mu} + V_{\mu\nu}$, %
	obtaining
	\begin{equation}\label{eqHamiltonianaSimultaneoTNTIntuitivo}\begin{aligned}
		& \mathcal H = \left[\left(K_{a\mu} + V_{a\mu}\right) + \left(K_{A\nu} + V_{A\nu} \right) \right] + \left(K_{\mathcal{A}b} + V_{ab} + V_{b\nu} + V_{B\mu} - V_{\mu\nu} \right) \\
		& \mathcal H = \left[ \left( K_{b\nu} + V_{b\nu} \right) + \left( K_{B\mu} + V_{B\mu} \right) \right] + \left(K_{\mathcal{B}a} + V_{ab} + V_{a\mu} + V_{A\nu} - V_{\mu\nu} \right)
	\end{aligned}\end{equation}
	which is still exact. If now $V_{A\nu}$ is approximated %
	by a potential depending only on the distance between $A$ and $\nu$, and similarly
	$V_{B\mu}$ is simplified %
	to a function of only $\v r_{B\mu}$,
	the problem for $A$--$\nu$ and $B$--$\mu$ relative motion is again %
	decoupled from the equations for $A$ and $B$ internal state. However, in this case, the Hamiltonian is the same for both prior and post form, and the scheme can be applied indifferently to heavy and light systems.
	Regarding the choice of the approximate potentials, from the formal point of view %
	one may for instance set $V_{A\nu}$ to %
	the ``diagonal'' component of the full potential on the solution $\phi_{a\mu}$ for the $a$--$\mu$ relative motion, $\Braket{\phi_{a\mu}(\v r_{\mu\nu})|V_{a\nu}(\v r_{A\nu},\v r_{\mu\nu}) + V_{\mu\nu}(\v r_{\mu\nu})|\phi_{a\mu}(\v r_{\mu\nu})}$. %
	In practice, the potentials are usually constructed phenomenologically.
	
	The proposed approach presents two issues from the practical point of view. %
	The first one is that,
	differently than in \cref{eqHamiltonianaApprossimataPostSchemiHeavyIon}, the transition operator, e.g.~$V_{a\mu} + V_{A\nu} - V_{\mu\nu}$ in post form, does not coincide
	with the sum of the binding potentials employed to construct $\psi_\alpha$ (here $V_{a\mu} + V_{A\nu}$)%
	\footnote{This is required, for instance, in the \textsc{Fresco} code.}. %
	The problem disappears in the limit where $V_{\mu\nu}$ is negligible, as assumed when writing %
	\cref{eqHamiltonianaApprossimataPostSchemiHeavyIon}: then \cref{eqHamiltonianaSimultaneoTNTIntuitivo,eqHamiltonianaSimultaneoTNTColonna} coincide, and $V_{A\nu}(\v r_{A\nu})$ can be formally thought as $\Braket{\phi_{a\mu}(\v r_{\mu\nu})|V_{a\nu}(\v r_{A\nu},\v r_{\mu\nu})|\phi_{a\mu}(\v r_{\mu\nu})}$, for instance%
	\footnote{Such prescription %
		is expected to be superior to simply setting $V_{A\nu}(\v r_{A\nu})$ to $V_{a\nu}$ computed at the same value for the function variable, $V_{a\nu}(\v r_{a\nu}\equiv\v r_{A\nu})$.}.
	In general,
	it is possible to overcome the limitation, at least partially, %
	adopting a suitable phenomenological form for $V_{A\nu}$ and $V_{B\mu}$. Formally, %
	define $\widetilde V_{A\nu}(\v r_{A\nu}, \v r_{\mu\nu}) = V_{a\nu}(\v r_{A\nu}, \v r_{\mu\nu}) + \frac{1}{2} V_{\mu\nu}(\v r_{\mu\nu})$ and $\widetilde V_{B\mu} = V_{b\nu} + \frac{1}{2} V_{\mu\nu}$, %
	and rewrite the total Hamiltonian as %
	\begin{equation}\label{eqHamiltonianaSimultaneoTNTColonna}\begin{aligned}
			& \mathcal H = \left[\left(K_{a\mu} + V_{a\mu}\right) + \left(K_{A\nu} + \widetilde V_{A\nu} \right) \right] + \left(K_{\mathcal{A}b} + V_{ab} + V_{b\nu} + \widetilde V_{B\mu} \right) \\
			& \mathcal H = \left[ \left( K_{b\nu} + V_{b\nu} \right) + \left( K_{B\mu} + \widetilde V_{B\mu} \right) \right] + \left(K_{\mathcal{B}a} + V_{ab} + V_{a\mu} + \widetilde V_{A\nu} \right)
	\end{aligned}\end{equation}
	which, if the complete expression of all potentials is employed, is still exact: in principle, the calculation scheme suggested by this form could be employed to address the full four-body problem%
	\footnote{Given a basis of eigenfunctions of %
	the modified ``internal'' Hamiltonians (the terms within square brackets in \cref{eqHamiltonianaSimultaneoTNTColonna}), the physical initial and final states for the system can be expressed in such basis, and subsequently the transition amplitude for these states can be computed, using $V_{ab} + V_{b\nu} + \widetilde V_{B\mu}$ as prior-form transition operator (and similarly in post form). Of course, there is no advantage in such approach with respect to the standard one, as long as no further approximations are introduced.}, %
	just as %
	\cref{eqTwoNucTranHamiltExplicitSubEqPriorPost} or \eqref{eqHamiltonianaSimultaneoTNTIntuitivo}.
	If $\widetilde V_{A\nu}$ and $\widetilde V_{B\mu}$ are approximated to potentials depending only on, respectively, $\v r_{A\nu}$ and $\v r_{B\nu}$, the resulting calculation scheme has the same features found from \cref{eqHamiltonianaSimultaneoTNTIntuitivo} and additionally the desired simplified form for the transition potential. %
	Once again, the corresponding solutions for $\psi_\alpha$ and $\psi_\beta$ %
	do not coincide with the ``exact'' ones (namely, the solutions of e.g.~the Hamiltonian $K_{a\mu} + V_{a\mu} + K_{A\nu} + V_{a\nu} + V_{\mu\nu}$), %
	but this is true for all approximate schemes discussed here, and the deviation may in fact be less severe in \cref{eqHamiltonianaSimultaneoTNTColonna} than in \cref{eqHamiltonianaApprossimataPostSchemiHeavyIon} (without adding further corrections), depending on the choice of the approximate potentials.
	
	The second issue involves the coordinates adopted to express the wave-functions. As mentioned earlier, in the core-core heavy-ion scheme, in \cref{eqHamiltonianaApprossimataPostSchemaHeavyIonCoreCorePrior}, %
	$\psi_\alpha$ and $\psi_\beta$ are first computed using ``v'' coordinates, %
	while from \cref{eqHamiltonianaSimultaneoTNTColonna} a solution in ``y'' coordinates %
	is drawn. %
	In order to convert the wave-functions to ``t'' coordinates (which is very useful in a practical calculation, see the comment to \cref{eqScritturaAmpiezzaSimultaneoPerSpiegareCoordinatet}), a different transformation is thus required depending on which scheme is employed. %
	However, most publicly available codes (including \textsc{Fresco}) %
	implement only the Moshinsky transformation (the one changing from ``v'' to ``t'', see the comment to \cref{eqScritturaAmpiezzaSimultaneoPerSpiegareCoordinatet}).
	Even in case it is necessary to employ ``v'' coordinates to construct $\psi_\alpha$ and $\psi_\beta$, %
	it still can be useful to substitute \cref{eqHamiltonianaApprossimataPostSchemaHeavyIonCoreCorePrior} with an adapted ``core-composite-like'' Hamiltonian, to preserve at least part of its features. %
	For instance, starting from \cref{eqHamiltonianaSimultaneoTNTColonna} consider %
	\begin{subequations}\label{eqHamiltonianaSimultaneoSchemaColonnaConCoordinateCoreCore}
	\begin{multline}
	\mathcal H_{cC/cc}^{\text{prior}} = \left[ K_{a\mu}(\v r_{a\mu}) + V_{a\mu}(\v r_{a\mu}) + K_{A\nu}(\v r_{a\nu}) + \widetilde V_{A\nu}(\v r_{a\nu}) \right] +\\+ K_{\mathcal{A}b}(\v r_{\mathcal{A}b}) + V_{ab}(\v r_{ab}) + V_{b\nu}(\v r_{b\nu}) + \widetilde V_{B\mu}(\v r_{b\mu})
	\end{multline}
	\begin{multline}
	\mathcal H_{cC/cc}^{\text{post}} = \left[ K_{b\nu}(\v r_{b\nu}) + V_{b\nu}(\v r_{b\nu}) + K_{B\mu}(\v r_{b\mu}) + \widetilde V_{B\mu}(\v r_{b\mu}) \right] +\\+ K_{\mathcal{B}a}(\v r_{\mathcal{B}a}) + V_{ab}(\v r_{ab}) + V_{a\mu}(\v r_{a\mu}) + \widetilde V_{A\nu}(\v r_{a\nu})
	\end{multline}
	\end{subequations}
	where
	$\widetilde V_{A\nu}$ is here a phenomenological potential depending only on $\v r_{a\nu}$ (and analogously for $\widetilde V_{B\mu}$),
	while $K_{A\nu}(\v r_{a\nu}) = - \frac{\hbar^2}{2 m_{A\nu}} \nabla_{a\nu}^2$, with $\v \nabla_{a\nu}$ being the gradient with respect to coordinate $\v r_{a\nu}$, and $m_{A\nu}$ being the $A$--$\nu$ reduced mass.
	Then, similarly to what is done in the calculation scheme suggested by \cref{eqHamiltonianaApprossimataPostSchemaHeavyIonCoreCorePrior}, two sets of single-particle states, $\phi^j_{a\mu}(\v r_{a\mu})$ and $\phi^j_{A\nu}(\v r_{a\nu})$, are constructed as the solutions of, respectively, $K_{a\mu}(\v r_{a\mu}) + V_{a\mu}(\v r_{a\mu})$ and $K_{A\nu}(\v r_{a\nu}) + \widetilde V_{A\nu}(\v r_{a\nu})$. The index $j$ is set to that each product $\phi^j_{a\mu} \phi^j_{A\nu}$ has the same energy eigenvalue. Finally, $\psi_\alpha$ is written in ``v'' coordinates as a linear combination of products $\phi^j_{a\mu}(\v r_{a\mu}) \phi^j_{A\nu}(\v r_{a\nu})$, and converted using the Moshinsky transformation.
	In this way, the single-particle state %
	for %
	the $A$--$\nu$ motion, $\phi_{A\nu}(\v r_{a\nu})$, is computed using the correct kinetic term (that is, the correct reduced mass), even though later, when constructing $\psi_\alpha$, the wave-function is associated to coordinate $\v r_{a\nu}$ instead of $\v r_{A\nu}$. %
	Such approximation is fully justified %
	when $a$ is much heavier than $\mu$, in which case \cref{eqHamiltonianaApprossimataPostSchemaHeavyIonCoreCorePrior,eqHamiltonianaSimultaneoSchemaColonnaConCoordinateCoreCore} %
	coincide
	if the same potentials are employed. For light ions, both calculation schemes are approximate, but the one derived from \cref{eqHamiltonianaSimultaneoSchemaColonnaConCoordinateCoreCore} is thought to be more accurate. %
	The approximation on the potentials is not the same as before, but is similar in spirit. %
	The prior- and post-form Hamiltonians are not identical due to the choice on the coordinates. Their difference is
	\begin{equation}\label{eqDifferenzaHamiltonianaPriorPostPerSchemaAdaptedCoreCompositeLike}
	\mathcal H_{cC/cc}^{\text{post}} - \mathcal H_{cC/cc}^{\text{prior}} = K_{B\mu}(\v r_{b\mu}) - K_{B\mu}(\v r_{B\mu}) - K_{A\mu}(\v r_{a\mu}) + K_{A\mu}(\v r_{A\mu})
	\end{equation}
	An explicit comparison, for a simple choice of potentials, between the simultaneous cross-section computed in %
	prior or post form, using either the ``core-composite-like'' or the ``heavy-ion'' scheme, %
	is given later in \cref{figpnTransferSimPriorPostOtticiCentrali}.

\subsection{Second-order (``sequential'') contribution}\label{secTeoriaTransferContributoSequenziale}

	The two-step contribution to a reaction is connected %
	to processes where the system, starting from the given initial configuration, %
	is first excited to an intermediate state through a one-step (i.e.~first-order) process, %
	and then reaches the desired final state through another one-step process.
	For each intermediate state appearing in the model space, %
	the associated transition amplitude is given by \cite[eq.~(3.76), (3.77), (3.78)]{Satchler1983Direct}, and the total two-step amplitude is the coherent sum of all such terms, as in \cite[sec.~3.7.3]{Satchler1983Direct}.
	
	In principle, any state %
	within the initial or final partition (but different from the precise initial and final states under study) would be an eligible intermediate state for the reaction. %
	However, %
	these ``inelastic'' excitations would be better treated %
	through a coupled-channels approach analogous to those in \cref{secOneParticleTransferCCApproaches}, where their coupling can be evaluated to all orders.
	The most interesting applications of this formalism regard instead intermediate states which belong to a third partition, $\gamma$, different from both the initial and final one ($\alpha$ and $\beta$). %

	The presence of the propagator (the Green's function in \cite[eq.~(3.64)]{Satchler1983Direct}) %
	complicates the computation of \cite[eq.~(3.77)]{Satchler1983Direct}, %
	therefore, from the practical point of view, it may be more convenient to address the problem %
	solving explicitly the coupled set of differential equations in \cite[eq.~(3.59)]{Satchler1983Direct} (one of the terms in \cite[eq.~(3.59c)]{Satchler1983Direct} is the contribution due to the simultaneous process, which can be calculated separately)%
	\footnote{This is the approach adopted in the \textsc{Fresco} code.}.
	This implies the following steps. %
	First, find the distorted wave $\chi_\alpha$ %
	of the auxiliary potential $U_{\mathcal A + b}$ for the elastic scattering in the initial channel. %
	Second, use $\chi_\alpha$ %
	to solve the problem for the first step of the reaction, for instance %
	the $\mathcal{A}+b \to A + B$ process as a direct transfer of $\nu$,
	in first-order DWBA, finding explicitly the corresponding wave-function $\chi_\gamma$ %
	for the projectile-target motion in the intermediate channel: this is not simply a distorted wave of an optical potential, as it includes the coupling from the initial state.
	Finally, repeat again the same computation scheme and use $\chi_\gamma$ %
	as source term to solve the problem for the second step, for instance $A + B \to a + \mathcal{B}$ as transfer of $\mu$, again in first-order DWBA.
	In summary, the calculation of each contribution to the two-step process is reduced to solving two one-step problems analogous %
	to the one discussed in \cref{sezDWBA}, again with the complication that both the potentials and the internal-motion states depend also on the internal coordinate within the transferred system, $\v r_{\mu \nu}$. %
	
	Each step can be solved either in prior or post form (which are distinct only if initial and final partition in the given step are different), giving rise to a total of four possible forms: prior-post (i.e.~first step in prior, second step in post), prior-prior, post-post, post-prior. As commented in \cite[sec.~3.7]{Satchler1983Direct}, all forms except prior-post involve a correction, \cite[eq.~(3.78)]{Satchler1983Direct}, not found in a standard one-particle transfer calculation, and proportional to the overlap between the internal-motion states in different partitions, e.g.~$\Braket{\psi_\gamma|\psi_\alpha}$, which is not zero because such states are not orthogonal (they belong to two different basis representing different degrees of freedom), hence the name of ``non-orthogonality'' terms for such corrections.

\subsubsection{Two-body approximation}

	As in \cref{secTNTSimultaneoTeoria}, %
	it is of interest %
	to approximate the evaluation of the sequential contribution to a computation including only two-body problems. %
	For definiteness, let $\mathcal{A}+b \to A + B$ and $A + B \to a + \mathcal{B}$ be the first and second step under study.
	Regarding the first step, it may be solved in prior form employing %
	the exact Hamiltonian given in \cref{eqTwoNucTranHamiltExplicitPrior}, or in post form through the equivalent rearrangement
	\begin{equation}\label{eqHamiltonianaSequenziale}
		\mathcal H = \[ K_{a\mu} + V_{a\mu} \] + \( K_{AB} + V_{ab} + V_{b\mu} + V_{a\nu} + V_{\mu \nu} \) + \[ K_{b\nu} + V_{b\nu} \]
	\end{equation}
	Similarly, the prior form for the second step is deduced again from \cref{eqHamiltonianaSequenziale}, and the post form using \cref{eqTwoNucTranHamiltExplicit}.

	Given that the first step involves only the transfer of $\nu$, one may approximate \cref{eqTwoNucTranHamiltExplicitPrior,eqHamiltonianaSequenziale} substituting $V_{ab}(\v r_{ab}) + V_{b\mu}(\v r_{b\mu})$ with a potential $V_{Ab}$ depending only on $\v r_{Ab}$, and similarly $V_{a\nu} + V_{\mu\nu}$ with a function $V_{A\nu}(\v r_{A\nu})$, obtaining, for the prior and post forms respectively,
	\begin{equation}\begin{aligned}
	& \mathcal H_{\text{1st}} = \left[K_{a\mu} + V_{a\mu}\right] + \left[K_{A\nu} + V_{A\nu} \right] + \left(K_{\mathcal{A}b} + V_{Ab} + V_{b\nu} \right) \\
	& \mathcal H_{\text{1st}} = \[ K_{a\mu} + V_{a\mu} \] + \( K_{AB} + V_{Ab} + V_{A\nu} \) + \[ K_{b\nu} + V_{b\nu} \]
	\end{aligned}\end{equation}
	The Hamiltonian is the same in both forms. %
	Note that, in both forms, the degree of freedom concerning the $a$--$\mu$ motion is now completely decoupled, and can be simply ignored. The problem is now formally identical to a single-particle transfer.
	
	Using the same reasoning for the second step, substitute $V_{ab} + V_{a \nu}$ with a $V_{aB}$ depending only on $\v r_{aB}$, and similarly $V_{b\mu} + V_{\mu\nu}$ with a function $V_{B\mu}(\v r_{B\mu})$, finding, %
	for the prior and post forms respectively,
	\begin{equation}\begin{aligned}
	& \mathcal H_{\text{2nd}} = \[ K_{a\mu} + V_{a\mu} \] + \( K_{AB} + V_{aB} + V_{B \mu} \) + \[ K_{b\nu} + V_{b\nu} \] \\
	& \mathcal H_{\text{2nd}} = \left[ K_{b\nu} + V_{b\nu} \right] + \left[ K_{B\mu} + V_{B \mu} \right] + \left(K_{\mathcal{B}a} + V_{aB} + V_{a\mu} \right)
	\end{aligned}\end{equation}
	where the $b$--$\nu$ relative-motion state is irrelevant. Post and prior forms are equivalent also in this second step, but note that $\mathcal H_{\text{1st}} \neq \mathcal H_{\text{2nd}}$, as the approximation applied in each step is different. This can be sufficient to %
	cause discrepancies between results computed using each form. In particular, the potential $V_{ab} + V_{b\mu} + V_{a\nu} + V_{\mu \nu}$ appearing in the ``transition part'' of \cref{eqHamiltonianaSequenziale} (in \cite[sec.~3.7]{Satchler1983Direct} this quantity is written as $W_\gamma - U_{\gamma}$) is approximated differently, as $V_{Ab} + V_{A\nu}$ or $V_{aB} + V_{B\mu}$, depending on whether it appears in the first or second step, thus %
	the expressions in \cite[eq.~(3.77), (3.78)]{Satchler1983Direct} may cease to be %
	equivalent
	after applying the approximations. %
	
	It is also relevant to underline that neither of the approximated Hamiltonians considered here coincides with any of the approximated Hamiltonians shown in \cref{secTNTSimultaneoTeoria} (at least the $V_{ab}$ %
	potential is always treated differently), %
	thus the approximated computation schemes employed for the two contributions are in fact not perfectly %
	consistent. %
	Nevertheless, for appropriate choices of the phenomenological potentials, such inconsistency is %
	expected to be confined within the accuracy of the adopted approximations, %
	and is thus not critical in the limit where the treatment of each contribution separately is valid.%
\chapter{The \texorpdfstring{$\nuclide[6]{Li} + \nuclide{p} \to \nuclide[3]{He} + \nuclide{\alpha}$}{6Li+p->3He+4He} transfer reaction}\label{secCalcoliDiTransfer}
	
	Distorted-wave Born approximation (DWBA) calculations and coupled reaction channels (CRC) methods are the most commonly adopted approaches in the study of transfer reactions. %
	Here, %
	the concepts introduced in \cref{chaLinkingCLusterModelToObservables,secReactionTheory} are %
	employed to investigate their application to the $\nuclide[6]{Li} + \nuclide{p} \to \nuclide[3]{He} + \nuclide{\alpha}$ reaction.
	
	\Cref{secCalcoliDWBADeuteronTransfer} is concerned with the DWBA one-step transfer of the deuteron, described as an elementary particle, using the approximated expression in \cref{eqAmpiezzaDiTransizioneDWBAPrimoOrdine} for the transition amplitude. This sort of calculations is very well established in literature, and has provided results in reasonable agreement with experimental data in a rather wide range of regimes. %
	It can thus provide a solid basis to test the appropriateness of the adopted physical ingredients %
	and the computational limitations, %
	and serve as benchmark for the other calculations.
	
	In \cref{secCalcoliDWBApnTransfer},
	the DWBA calculation is %
	repeated adopting %
	a more complex structure for the reactants, and in the particular for the internal motion of the transferred system, which is now described as a generic state of a proton and a neutron.
	Such framework %
	allows to exert a greater control over the clustering strength in reactants, in terms of correlations between the transferred particles. This was investigated in particular for the \nuclide[6]{Li} nucleus, whose ground state within the three-cluster model was studied in greater detail using the results of existing three-body calculations.
	The transfer cross-section, including its energy trend at astrophysical energies, was then explicitly computed as a function of the adopted \nuclide[6]{Li} wave-function, to study the impact of clustering phenomena.
	A preliminary version of the calculations discussed in \cref{secCalcoliDWBADeuteronTransfer,secCalcoliDWBApnTransfer} %
	previously appeared in \cite{PerrottaLNSReport201920}.
	
	Finally, \cref{secCDCCCalcoloPratico} presents an attempt to apply %
	the ``Greider-Goldberger-Watson'' generalised-distorted-waves scheme (see \cref{eqAmpiezzaTransizioneGGW}) on the deuteron-transfer process.
	This sort of approach can be a useful tool in the study of reaction dynamics at low energies, given its capability to include %
	virtual dynamical excitations to unbound states of the reactants.
	The application to the $\nuclide[6]{Li} + \nuclide{p} \to \nuclide[3]{He} + \nuclide{\alpha}$ reaction is also interesting from the technical perspective, %
	as most existing works employing this method
	were restricted to deuteron stripping or pickup reactions (see the discussion in \cref{sezGeneralizedDistortedWaves}), and the current study aims to overcome this boundary.
	The calculation allowed to identify an improvement over the standard formalism, that appears to be required in order to obtain an accurate result. %

	All cross-sections and phase shifts %
	shown in this \namecref{secCalcoliDiTransfer} were computed using the \textsc{Fresco} code \cite{Thompson2004}. The practical computation of the transfer cross-sections
	is performed solving directly the relevant coupled differential equations, instead of numerically evaluating the expressions for the reaction amplitude introduced in \cref{secReactionTheory}.
	It was shown that the two approaches are formally equivalent \cite{Ascuitto1969}.

\section{DWBA deuteron transfer}\label{secCalcoliDWBADeuteronTransfer}

	The first part of this section, up to \cref{sec3HepPotential}, presents and comments all the physical ingredients required to compute the cross-section of interest, together with some details regarding how %
	they were derived. Some of these ingredients are later employed also in the other transfer calculations reported in this \namecref{secCalcoliDiTransfer}, and in the $\nuclide[6]{Li}+\nuclide{p}$ barrier penetrability estimation in \cref{secLiDeformationQuantumModel}.
	The results on the transfer channel are then shown in \cref{secTransferDeuterioTransferCrossSection}.

\subsection{Optical potentials}\label{secDWBAOpticalPotentials} %
	
	The following paragraphs discuss %
	the form of %
	some potentials mainly aimed at describing the elastic scattering between the particles involved in the transfer reaction.
	The projectile-target interactions are employed to construct both the distorted waves, as discussed in \cref{sezDistortedWaves}, and to approximate the full scattering solution, using the approach reported in \cref{sezDWBA}.
	The ``core-core'' potential is the $V_{ab}$ appearing in the transition amplitudes in \cref{eqAmpiezzaDiTransizioneDWBAPrimoOrdine}, and refers to the interaction between reactants deprived of the transferred system.

\subsubsection{\texorpdfstring{$\nuclide[6]{Li}+\nuclide{p}$}{6Li+p} projectile-target potential}\label{sec6LipOpticalPotential} %
	
	Most optical potentials existing in literature for $\nuclide[6]{Li}+\nuclide{p}$ elastic scattering include at most spin-orbit couplings. However, it is not possible to reproduce the experimental phase-shifts %
	without a component coupling the spins of both reactants together. %
	\begin{figure}[tbp]%
		\centering
		\includegraphics[keepaspectratio = true, width=\linewidth]{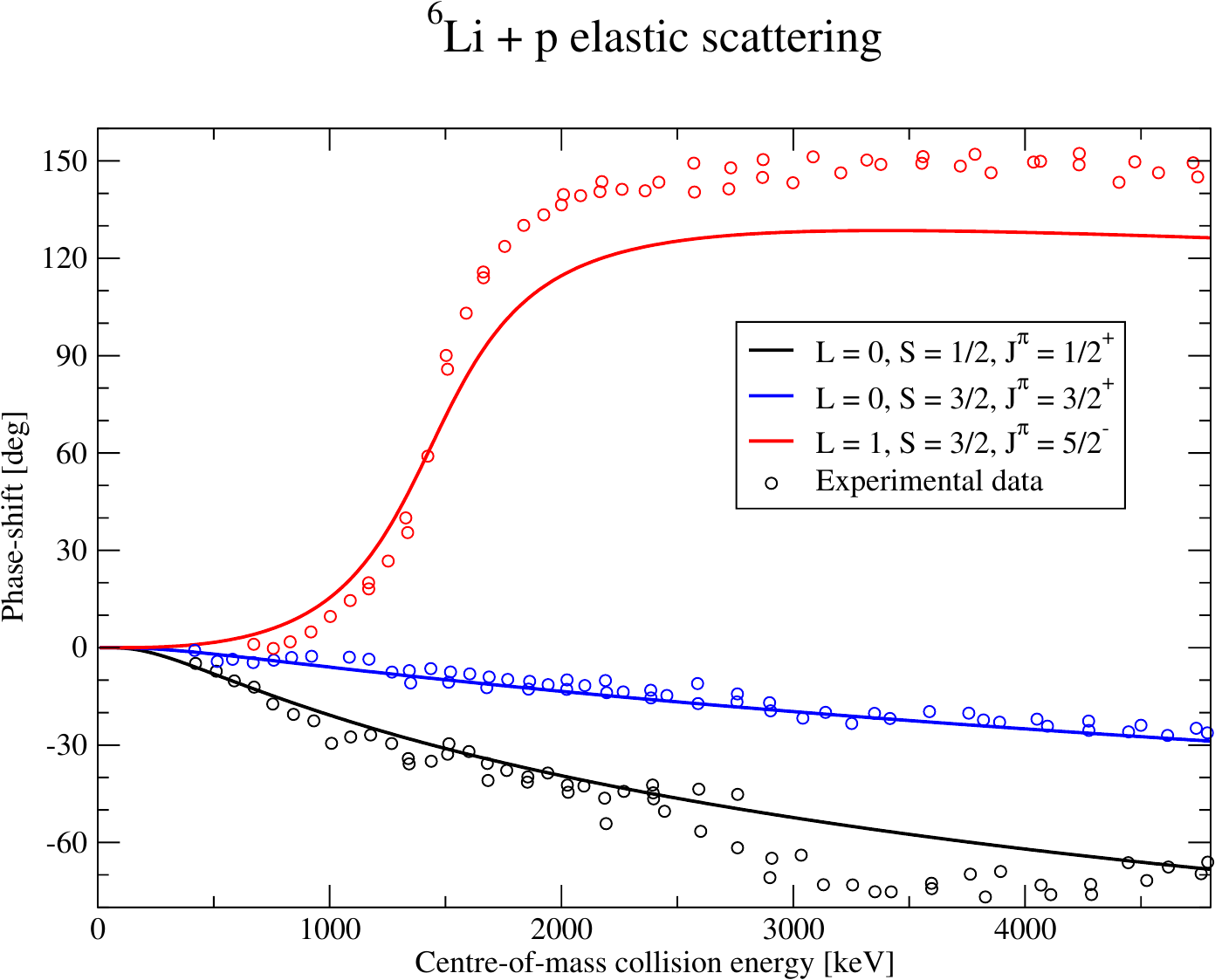}%
		\caption[\texorpdfstring{$\nuclide[6]{Li}+\nuclide{p}$}{6Li+p} elastic-scattering phase-shifts]{\label{figPhaseShiftPotenziale6LipFittato}%
			Points are experimental $\nuclide[6]{Li}+\nuclide{p}$ elastic-scattering phase-shifts, %
			collected in \cite{Petitjean1969} from earlier experiments (see refs.\ therein). %
			Lines in corresponding colours are the prediction of the optical potential discussed in \cref{sec6LipOpticalPotential}. %
			In the figure legend, $L$ is the $\nuclide[6]{Li}$--$\nuclide{p}$ relative orbital angular momentum modulus quantum number, $S$ the modulus quantum number for the sum of $\nuclide[6]{Li}$ and $\nuclide{p}$ spins, $J$ %
			the modulus quantum number for the sum of $\v L$ and $\v S$, and $\pi$ the parity of the state.
		}
	\end{figure}
	Furthermore, no published energy-independent potential known to the author compares acceptably with elastic scattering cross-section in the energy range of interest here.
	For this work, an energy-independent %
	potential was adjusted to reproduce the most relevant partial waves in the experimental phase-shifts, namely the $s$-waves and the $p_{5/2}$ wave %
	(projectile-target orbital angular momentum of 1, total angular momentum of $\frac{5}{2}$), the latter showing a resonant trend, %
	attributed %
	to a \nuclide[7]{Be} level, which is visible also in the transfer channel (see the discussion later in \cref{secTransferDeuterioTransferCrossSection}).
	The result is shown in \cref{figPhaseShiftPotenziale6LipFittato}. Note that the experimental phase-shifts for other waves are %
	not reproduced, but the adjusted potential introduces no spurious resonances (namely, resonances with incorrect energy or quantum numbers) %
	within the region of interest: this is a relevant advantage of this potential with respect to others tested before.
	An imaginary component was then added to the potential to improve the agreement with elastic-scattering cross-sections from~\cite{McCray1963,Dubovichenko2011} at the relevant energies. %
	The precise form and parameters of this potential are reported in \cref{secPotentialsDefinition}.

\subsubsection{\texorpdfstring{$\nuclide[3]{He}+\nuclide{\alpha}$}{3He+a} projectile-target potential}\label{sec3HeaOpticalPotential} %

	Different \nuclide{\alpha} -- \nuclide[3]{He} potentials are available in literature, either related to scattering \cite{Dunnill1967} or bound state properties \cite{Kubo1972,Mason2008}. Generic parametrizations, fitted on the elastic scattering of \nuclide[3]{He} (or \nuclide{\alpha}) on heavy-ion targets at energies above the Coulomb barrier, %
	are also available. However, all potentials considered %
	within the present project %
	poorly reproduced the experimental elastic scattering at the energies of interest, and yielded an unsatisfactory description of the transfer cross-section as well.
	
\begin{figure}[tbp]%
		\centering
		\includegraphics[keepaspectratio = true, width=\linewidth]{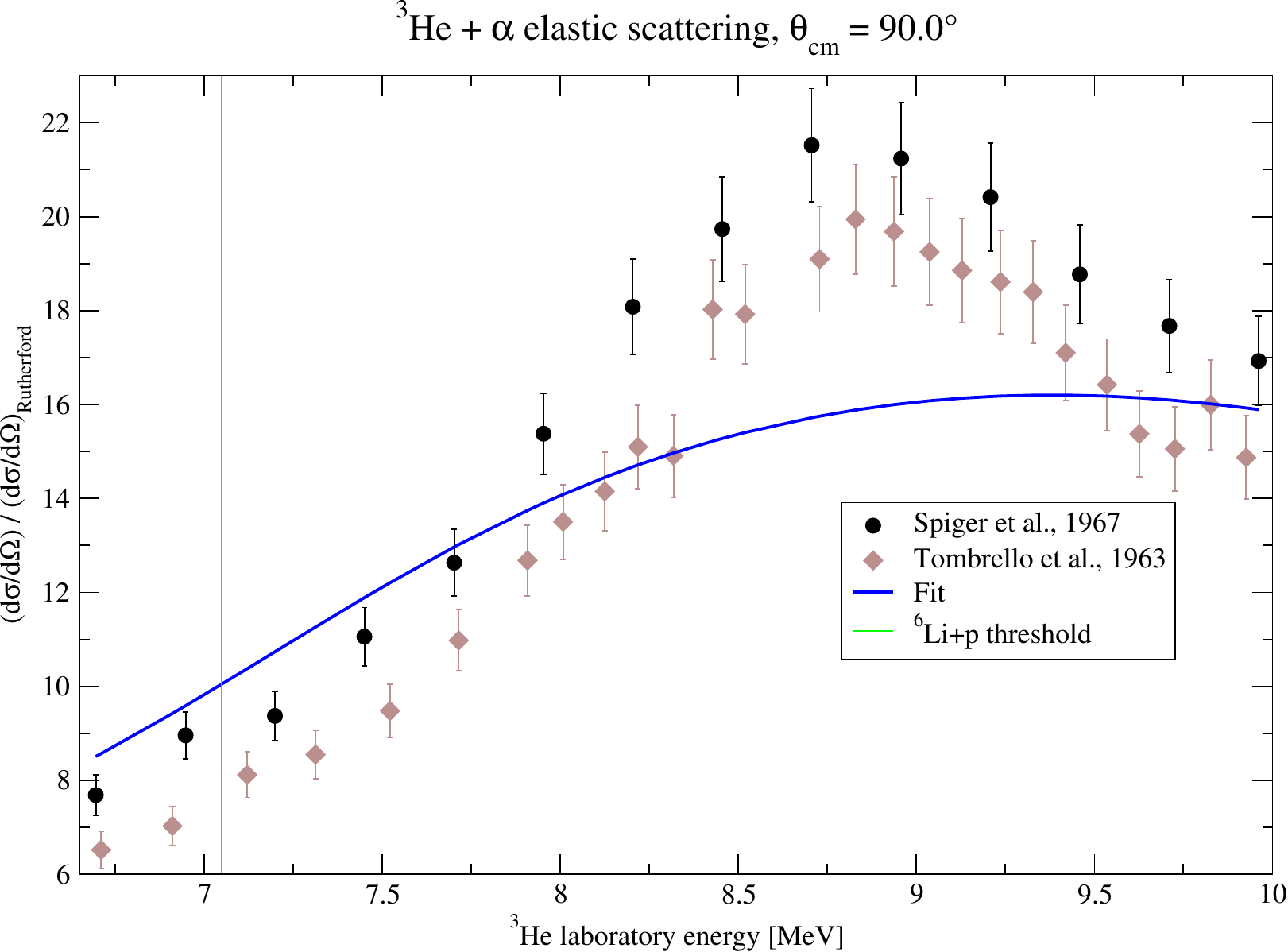}%
		\caption[\texorpdfstring{$\nuclide[3]{He}+\nuclide{\alpha}$}{3He+a} elastic-scattering cross-section sample]{\label{figElastico90gradiPotenziale3He4HeFittato}%
			Points are the experimental $\nuclide[3]{He}+\nuclide{\alpha}$ elastic-scattering differential cross-section at a centre-of-mass scattering angle of \ang{90}, rescaled to the %
			Rutherford cross-section, as a function of \nuclide[3]{He} laboratory energy, from \cite{Tombrello1963,Spiger1967}. Blue line is the prediction of the fitted optical potential discussed in \cref{sec3HeaOpticalPotential}. Green vertical line is the threshold for the $\nuclide[6]{Li}+\nuclide{p}$ channel.
		}
	\end{figure}

	The projectile-target optical potential for the final partition %
	adopted in the present work was created by fitting the experimental elastic scattering differential cross-sections %
	found in \cite{Tombrello1963,Spiger1967}, using \textsc{Sfresco} \cite{Thompson2004} (based on \textsc{Minuit} \cite{Minuit1975}).
	The potential included a spin-orbit component and an imaginary surface term. The complete set of parameters is reported in column ``\nuclide{\alpha} -- \nuclide[3]{He}'' of \cref{tabParametriNumericiPotenzialiOttici}. %
	For illustration, a sample of the results
	is shown in \cref{figElastico90gradiPotenziale3He4HeFittato}.
	Attempts to fit the data using a purely real potential, with or without using phase-shifts in \cite{Spiger1967} as guide, yielded inferior results. %
	While a visual comparison of datasets in \cite{Tombrello1963} and \cite{Spiger1967} suggests that some experimental errors are underestimated, it is apparent that the fitted potential does not describe perfectly the data (a similar result is found when attempting to fit phase-shifts, as only few waves at a time can be reasonably adjusted).
	
	Given the fit complexity and the high number of free parameters, more than one local minimum in $\chi^2$ can be found through the fitting %
	procedure, %
	depending on the computation specific details.
	Some of the potentials obtained in this manner were discarded as they displayed extreme parameter values%
	\footnote{For instance, a vanishing diffuseness for the spin-orbit component.}. %
	The remaining candidate potentials %
	yielded similar results regarding the transfer cross-section,
	which suggests that the specific choice adopted here does not influence the conclusions significantly. %

	The real part of the adopted potential was also employed to check the \nuclide[7]{Be} bound states it generates, as shown in \cref{tabStatiLegatiDelPotenziale3HeaFittato}.
	It is stressed that %
	no explicit adjusting was performed on these quantities.
	\begin{table}[tbp]
		\caption[Properties of \texorpdfstring{\nuclide[7]{Be}}{7Be} bound states]{\label{tabStatiLegatiDelPotenziale3HeaFittato}%
			Experimental values for the binding energies of \nuclide[7]{Be} bound states and the ground-state root-mean-square radius are compared with the predictions of the real part of the $\nuclide[3]{He}+\nuclide{\alpha}$ optical potential discussed in \cref{sec3HeaOpticalPotential}.} %
		\centering
		\begin{tabular}{lcc}
			{Quantity} & {Experimental} & {$\nuclide[3]{He}+\nuclide{\alpha}$ potential} \\ \toprule	%
			\nuclide[7]{Be} g.s.\ $3/2^-$ BE	& \SI{1587}{\keV}	& \SI{2022}{\keV}	\\
			\nuclide[7]{Be} 1st $1/2^-$ BE	& \SI{1158}{\keV}	& \, \SI{951}{\keV}	\\
			\nuclide[7]{Be} g.s.\ rms radius & \SI{2.48}{\femto\metre} & \SI{2.64}{\femto\metre}	%
		\end{tabular}
	\end{table}

\subsubsection{\texorpdfstring{\nuclide{\alpha} -- \nuclide{p}}{4He-p} core-core potential}\label{secapPotential}

	\cite[eq.~4.3]{Bang1979} quotes a real energy-independent \nuclide{\alpha}-nucleon potential, found by fitting scattering phase-shifts.
	Such potential was also tested here, finding good qualitative agreement with experimental elastic scattering cross-sections from \cite{Freier1949,Miller1958,Barnard1964,Kraus1974,Nurmela1997,Nurmela1998,Lu2009,Cai2010,Godinho2016} (in this regard, it should be taken into account that the different datasets do not fully agree with each other).
	No information about the Coulomb repulsion term is given in \cite{Bang1979}, thus, in the present work, the shape in \cref{eqCoulombWoodsSaxonDefinition} was assumed, fitting the radius of the uniformly charged sphere, $R_C$, on experimental cross-sections, using the \textsc{SFresco} code.
	The result is reported in \cref{tabParametriNumericiPotenziali6Li}.

\subsection{Overlap functions}\label{secCalcoloDWBAOverlapFunctions}

	The description of the ingredients for the DWBA transfer calculation is now completed by a discussion on the model Hamiltonians for the internal motion within each nucleus, and the overlap functions for both projectile and target systems. %
	In fact, within the inert di-cluster model employed here for the structure, the required physical input is reduced to the potentials between the transferred system and each core nucleus, since all inter-cluster relative-motion wave-functions are %
	constructed as appropriate eigenstates of Hamiltonians involving these potentials. %

\subsubsection{\texorpdfstring{$\Braket{\nuclide{\alpha}|\nuclide[6]{Li}}$}{<a|6Li>} overlap}\label{secadPotential}

	The $\nuclide[6]{Li}$ nucleus is being described here as a bound state between two structureless clusters, an \nuclide{\alpha} particle and a deuteron, %
	hence the $\Braket{\nuclide{\alpha}|\nuclide[6]{Li}}$ overlap function depends only on the distance between the clusters centre-of-mass:
	the normalised wave-function %
	was constructed as detailed in \cref{secCostruzioneGroundState6LiDeformato}, using the binding potential given in \cite[eq.~(13)]{Kubo1972} and including both the $2s$ and $1d$ components.
	Most calculations in this \namecref{secCalcoliDiTransfer} adopt a value of $-\sqrt{0.05}$ for the amplitude of the $L=2$ component, $c_2$ in \cref{eqFunzione6LiGroundStateDeformataGenerica}: this is similar to the value which reproduces the experimental magnetic dipole moment, see the discussion in \cref{secCostruzioneGroundState6LiDeformato}. Some calculations employing $c_2 = \num{-0.0909}$ (the value reproducing the electric quadrupole moment) are reported in \cref{figdTransferPotenziale3HeaAlternativo}.

	Regarding the %
	spectroscopic factor for the overlap, $\mathcal S$, %
	the independent-particle shell model (in which $c_2=0$) %
	would %
	suggest to assign a value of 1 %
	(see \cref{secOverlapSpectroscopicfactorswithinExtremeshellmodel}).
	The variational Monte Carlo (VMC) %
	calculation in \cite[sec.~V.C]{Forest1996} yields a value of $\mathcal S = \num{0.841}$ %
	including both $L=0$ and 2 components%
	\footnote{There had been attempts to experimentally evaluate the spectroscopic factor through the analysis of experimental cross-sections of the $\nuclide{d}+\nuclide{\alpha} \to \nuclide[6]{Li} + \gamma$ radiate-capture reaction within a two-cluster model, see e.g.~\cite{Robertson1981} and, importantly, the comment in \cite{Langanke1986}. However, in \cite[sec.~12.4]{Mukhamedzhanov2022} it is shown that the $\nuclide{d}+\nuclide{\alpha} \to \nuclide[6]{Li} + \gamma$ reaction, being strongly peripheral, can be described equally well employing as \nuclide[6]{Li} ground state any ``reasonable'' $\nuclide{d}+\nuclide{\alpha}$ wave-function reproducing the experimental binding energy and asymptotic normalisation coefficient, and thus having arbitrary spectroscopic factor.}
	In the present %
	calculation, an overall spectroscopic factor $\mathcal S$ of \num{0.85} %
	was adopted (meaning that the spectroscopic factor for each specific component is $c_L^2 \mathcal S$). %
	With this choice (and using $c_2 = -\sqrt{0.05}$), the $L=0$ component of the wave-function has an asymptotic normalisation coefficient (defined in \cref{secANCdefinition}) of \SI[per-mode=power]{2.28}{\femto\metre\tothe{-1/2}}, %
	which is very similar to the value of \SI[per-mode=power]{2.29}{\femto\metre\tothe{-1/2}} quoted in \cite[sec.~4.2]{Mukhamedzhanov2022}.

\subsubsection{\texorpdfstring{$\Braket{\nuclide{p}|\nuclide[3]{He}}$}{<p|3He>} overlap}\label{sec3HepPotential} %

	Using the same reasoning and notation as in \cref{secadPotential}, the \nuclide[3]{He} ground state is described as a $\nuclide{p}+\nuclide{d}$ bound state with orbital angular momentum $L = 0$ and zero nodes (``$1s$'').
	An $L=2$ component (a $1d$ state) is also admissible. Both components were constructed using the binding potentials published in \cite[tab.~VIII]{Brida2011} and reported in \cref{tabParametriNumericiPotenziali3He}, under the columns \mbox{``\nuclide{p} -- \nuclide{d} ($L=0$)''} and ``($L=2$)'' respectively. These potentials were fitted in order to reproduce the $\Braket{\nuclide{p}|\nuclide[3]{He}}$ overlap functions calculated through the Green's function Monte Carlo method, but keeping the binding energy of the eigenfunctions of interest %
	fixed to the experimental value. %
	This %
	procedure ensures %
	that the solutions bear the correct asymptotic form: as reported in %
	\cite[tab.~VI]{Brida2011}, the eigenfunctions of the fitted potentials have asymptotic normalisation coefficients in agreement with experimental data.
	
	\cite[tab.~IV]{Brida2011} also lists the spectroscopic factors found within the same calculation for each component, \num{1.31} and \num{0.0221} respectively, %
	which are adopted here%
		\footnote{The relative phase between the $L=0$ and 2 components can be deduced from the asymptotic normalisation coefficients in \cite[tab.~VI]{Brida2011}.}
	The total spectroscopic factor is slightly smaller than the value found within the independent-particle shell model, $1.5$, and is similar to the values found in past literature using different models, see for instance \mbox{\cite[sec.~IV.A]{Werby1973}} and references therein.
	For comparison, %
	very similar results are found for the $\nuclide[6]{Li}+\nuclide{p} \to \nuclide[3]{He}+\nuclide{\alpha}$ cross-section employing instead the overlap function (including the interaction potential and spectroscopic factor) from \cite{Werby1973}.
	
	It is pointed out %
	that these overlaps functions have slightly higher root-mean-square extensions $d_{\text{rms}}$ than would be expected within the cluster model. Restricting to the $L=0$ component for simplicity, %
	the overlap from \cite{Brida2011} has $d_{\text{rms}} = \SI{2.79}{\femto\metre}$, which, adopting experimental values in \cref{tabExperimentalGroundStateData}
	for each cluster rms charge radius and using \cref{eqRaggioDiCaricaDueCluster}, corresponds to an \nuclide[3]{He} rms charge radius of
	\SI{2.20}{\femto\metre}, %
	against the experimental value in \cite{Angeli2013} of \SI{1.966 \pm 0.003}{\femto\metre}. For comparison, an rms inter-cluster distance of $d_{\text{rms}} = \SI{2.07}{\femto\metre}$ would be required in order to reproduce the experimental \nuclide[3]{He} radius.

\subsection{Transfer cross-section}\label{secTransferDeuterioTransferCrossSection}
	
\begin{figure}[tbp]%
		\centering
		\includegraphics[width=\textwidth,keepaspectratio=true]{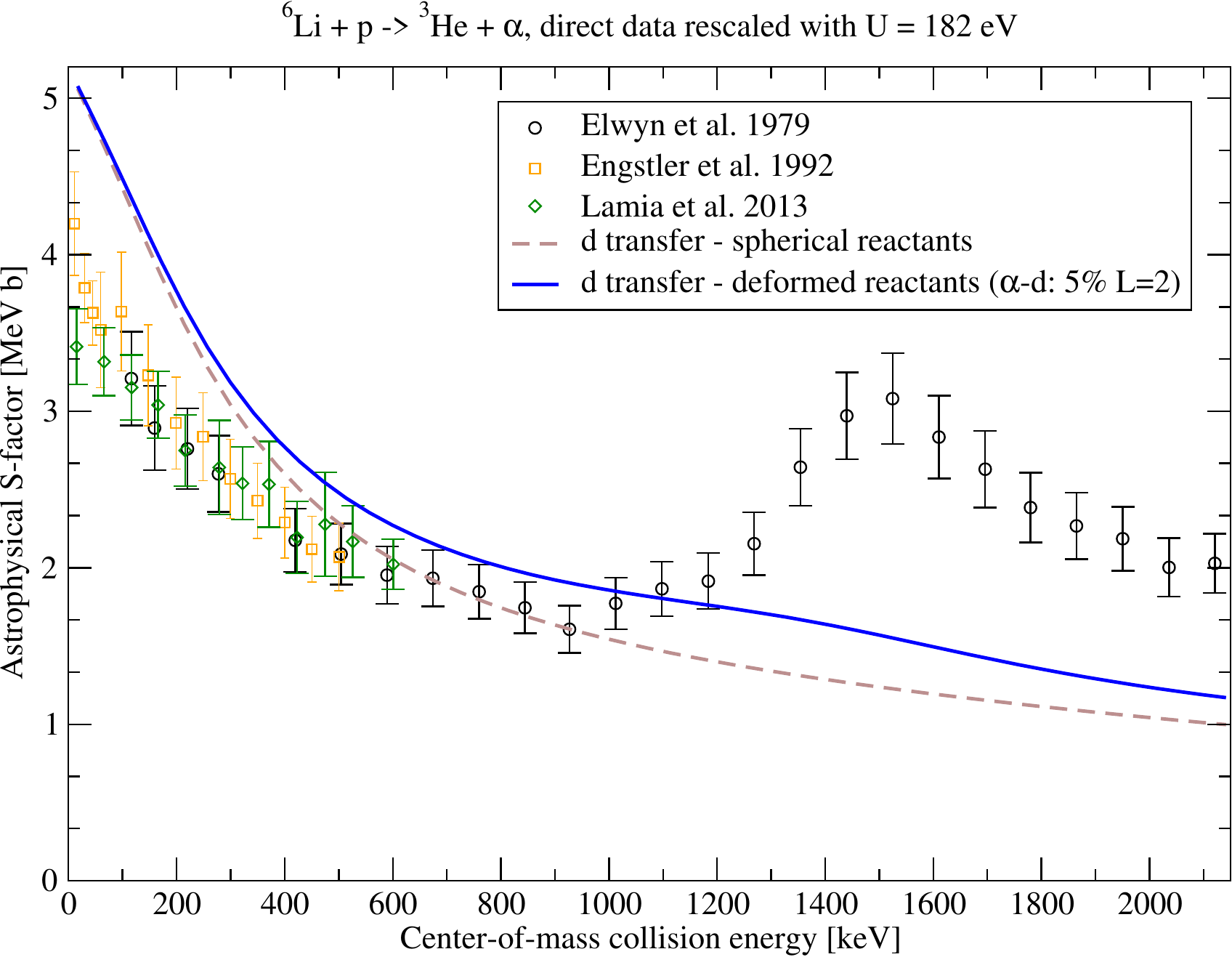}%
		\caption[\texorpdfstring{$\nuclide[6]{Li} + \nuclide{p} \to \nuclide[3]{He} + \nuclide{\alpha}$}{6Li+p->3He+4He} deuteron-transfer calculation]{\label{figdTransferConOSenzaL2}%
			Points are experimental $\nuclide[6]{Li} + \nuclide{p} \to \nuclide[3]{He} + \nuclide{\alpha}$ bare-nucleus astrophysical factors, from \cite{Elwyn1979} (black circles), \cite{Engstler1992} (orange squares), \cite{Lamia2013} (green diamonds). Data from direct measurements was rescaled by the atomic adiabatic-limit screening enhancement factor as in \cref{figDatiCorrettiAdiabatic}. For readability, only a selection of data is shown. Brown dashed line is DWBA deuteron transfer calculation where both \nuclide[6]{Li} and \nuclide[3]{He} are spherical. Blue solid line is the same calculation including a deformed component for both nuclei, as detailed in \cref{secadPotential,sec3HepPotential}.
		}
	\end{figure}

	The ingredients discussed earlier in this \namecref{secCalcoliDWBADeuteronTransfer} are here employed to estimate the non-polarised angle-integrated cross-section of the $\nuclide[6]{Li} + \nuclide{p} \to \nuclide[3]{He} + \nuclide{\alpha}$ reaction described as the transfer of a structureless deuteron in DWBA. %
	\Cref{figdTransferConOSenzaL2} shows both the computed astrophysical factors and a selection of experimental data for comparison.
	The results may be contrasted, for instance, with more microscopic calculations %
	in \cite{Arai2002,Vasilevsky2009,Solovyev2018}. %
	To this end, note that, in this \namecref{secCalcoliDiTransfer}, cross-sections obtained from direct fixed-target experiments are always shown corrected by the expected adiabatic-limit atomic screening (see \cref{secPhenomenologyScreeningEffects}). %
	As in the quoted earlier work, %
	the predicted cross-section %
	overestimates bare-nucleus data at low energies, while (see in particular \cite[fig.~8]{Arai2002}) the resonance at about \SI{1.5}{\MeV} above the $\nuclide[6]{Li}+\nuclide{p}$ threshold is not reproduced.
	However, the calculation predicts the correct order of magnitude for the astrophysical factor trough the whole energy range, suggesting that the physical ingredients of the calculation are at least qualitatively appropriate. %

\subsubsection{The \texorpdfstring{\nuclide[7]{Be} $5/2^-$}{7Be 5/2-} resonance}\label{secCalcolidDWBARisonanza7Be}
	
\begin{figure}[tbp]%
		\centering
		\includegraphics[width=0.95\textwidth,keepaspectratio=true]{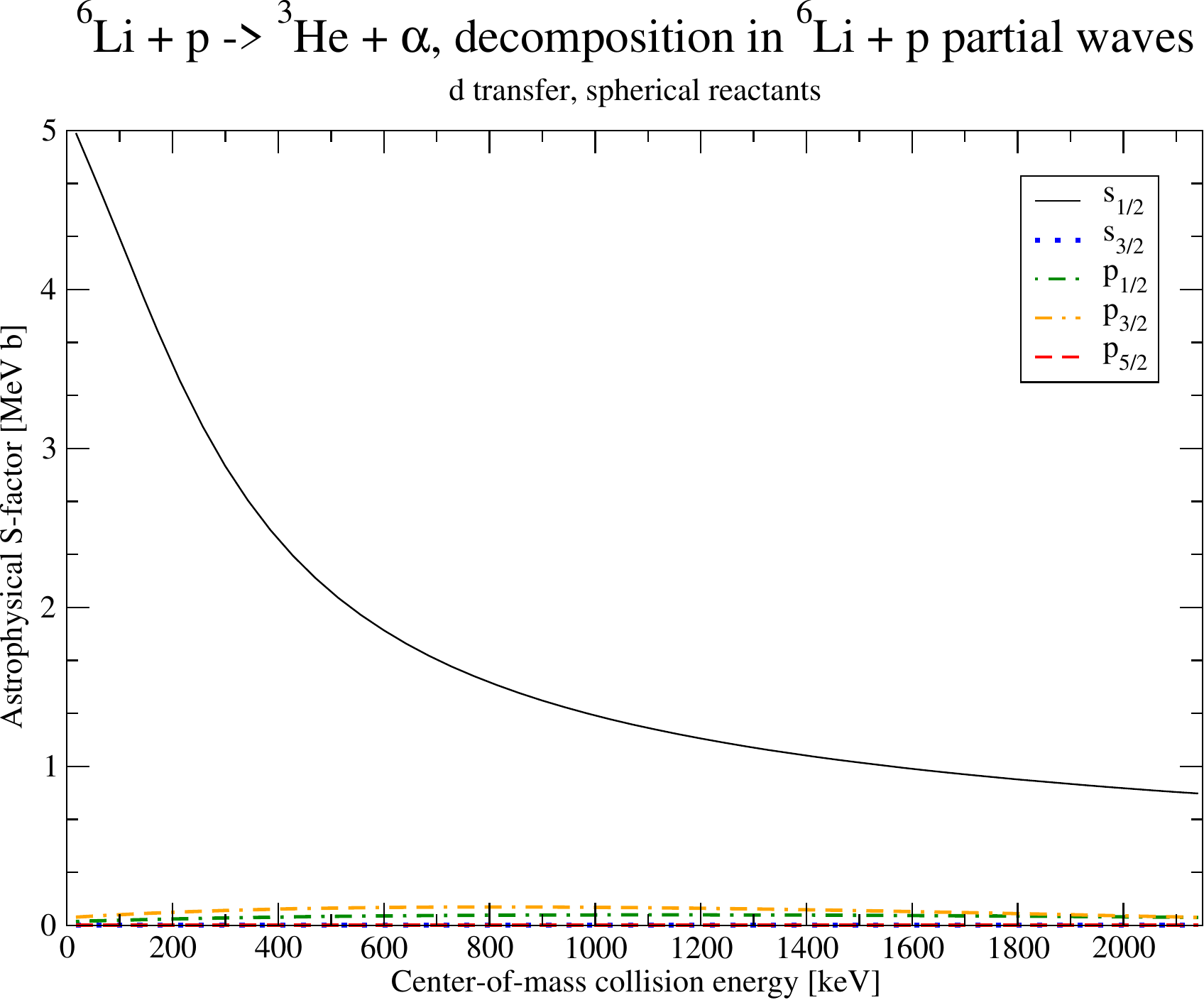} %
		
	\bigskip %
	
		\includegraphics[width=0.95\textwidth,keepaspectratio=true]{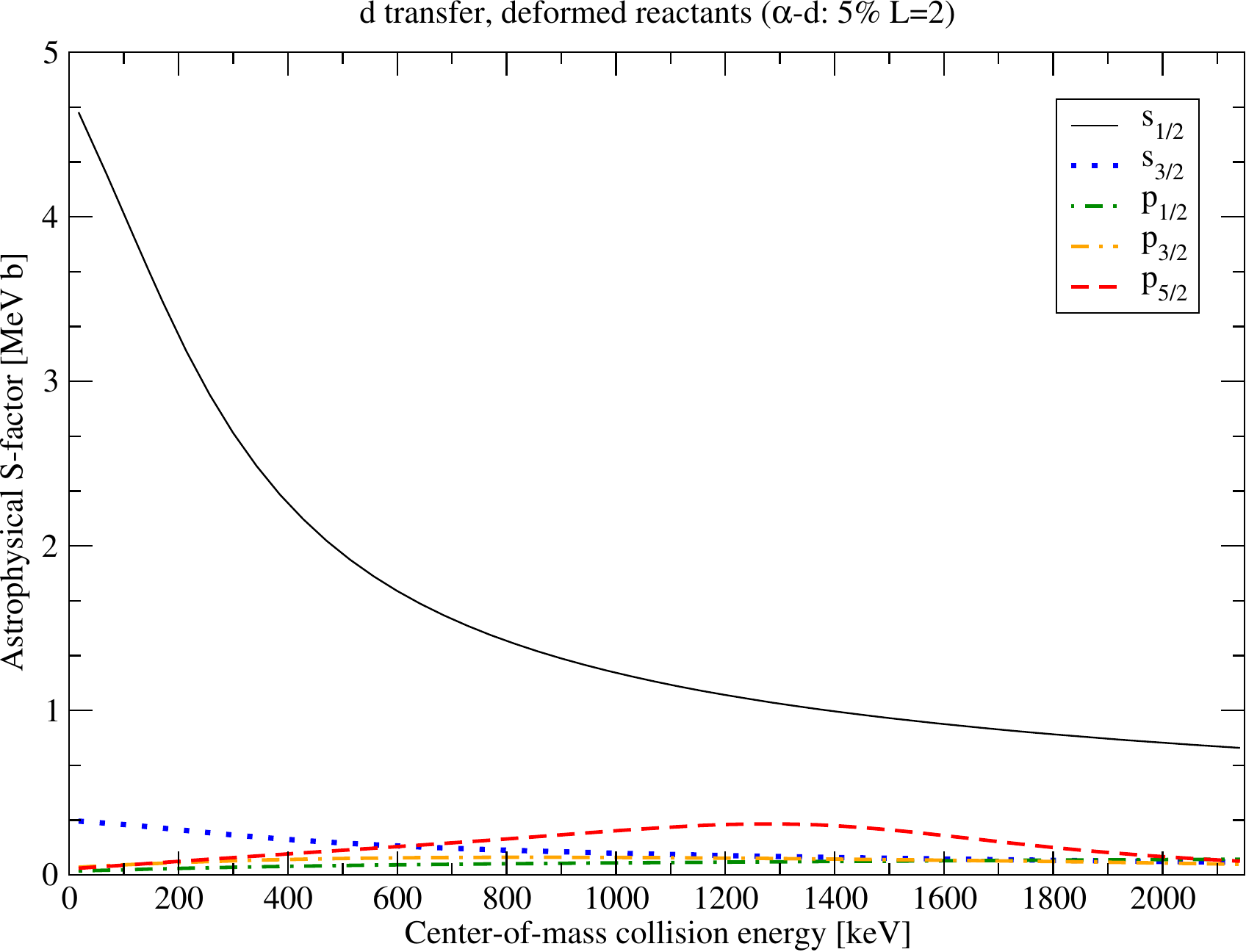}%
		\caption[\texorpdfstring{$\nuclide[6]{Li} + \nuclide{p} \to \nuclide[3]{He} + \nuclide{\alpha}$}{6Li+p->3He+4He} deuteron-transfer partial-wave expansion]{\label{figdTransferPartialWaveExpansions}%
			Upper panel: partial-wave decomposition for the astrophysical factor for spherical reactants in \cref{figdTransferConOSenzaL2}. Each line represents the contribution of a partial wave in the \nuclide[6]{Li}+\nuclide{p} channel as per the legend (``$s$'' and ``$p$'' refer to a projectile-target orbital angular momentum of 0 or 1, the subscript number %
			is the total angular momentum). Bottom panel: same for the case of deformed reactants (brown dashed line in \cref{figdTransferConOSenzaL2}).
		}
	\end{figure}

	The peak in the data at about \SI{1.5}{\MeV} is associated in literature (see e.g.~\cite{Tumino2003}) %
	to the \nuclide[7]{Be} $5/2^-$ level at an excitation energy of about \SI{7.2}{\MeV}, %
	manifesting in the $p$-wave (i.e.~projectile-target angular momentum of $1$) $\nuclide[6]{Li}+\nuclide{p}$ scattering%
	\footnote{There is another, much broader $5/2^-$ level in the \nuclide[7]{Be} spectrum at about \SI{6.7}{MeV} of excitation energy, %
		which is instead visible in the \nuclide[3]{He}--\nuclide{\alpha} elastic scattering (see e.g.~\cref{figElastico90gradiPotenziale3He4HeFittato}), together with a $7/2^-$ level which is however about \SI{1}{\MeV} below the \nuclide[6]{Li}+\nuclide{p} threshold.}.
	Note that, in the \nuclide[3]{He}--\nuclide{\alpha} channel, a $5/2^-$ level necessarily correspond to a projectile-target orbital angular momentum of 3. To observe such resonance in a calculation, it is thus necessary that the model includes a mechanism not conserving projectile-target orbital angular momentum.
	All calculations shown here are such that the total orbital angular momentum (sum of the projectile-target and inter-cluster orbital momenta) is conserved during the reaction. As a consequence, if both projectile and target have a spherical structure (inter-cluster orbital momentum zero) the sought resonance %
	cannot be populated.
	\Cref{figdTransferConOSenzaL2} shows two calculations, considering respectively spherical (brown crosses) and deformed (blue solid line) reactants. Each of these astrophysical factors is expanded in partial waves in the two panels in \cref{figdTransferPartialWaveExpansions}. %
	In accordance with the considerations made above, the calculation involving spherical nuclei gives no outgoing cross-section through the $p_{5/2}$ wave in the \nuclide[6]{Li}--\nuclide{p} channel, while the other case displays such contribution, albeit small, approximately at the correct energy. Another component which is allowed when reactants are deformed is the $s_{3/2}$ wave in the \nuclide[6]{Li}--\nuclide{p} channel (which translates to an orbital angular momentum of 2 between \nuclide[3]{He} and \nuclide{\alpha}). However, it is interesting to note that the total $s$-wave cross-section, which is the dominant component at low energies, remains essentially unaltered in both calculations, thus the astrophysical region is not affected by this difference.
	
\begin{figure}[tbp]%
		\centering
		\includegraphics[width=\textwidth,keepaspectratio=true]{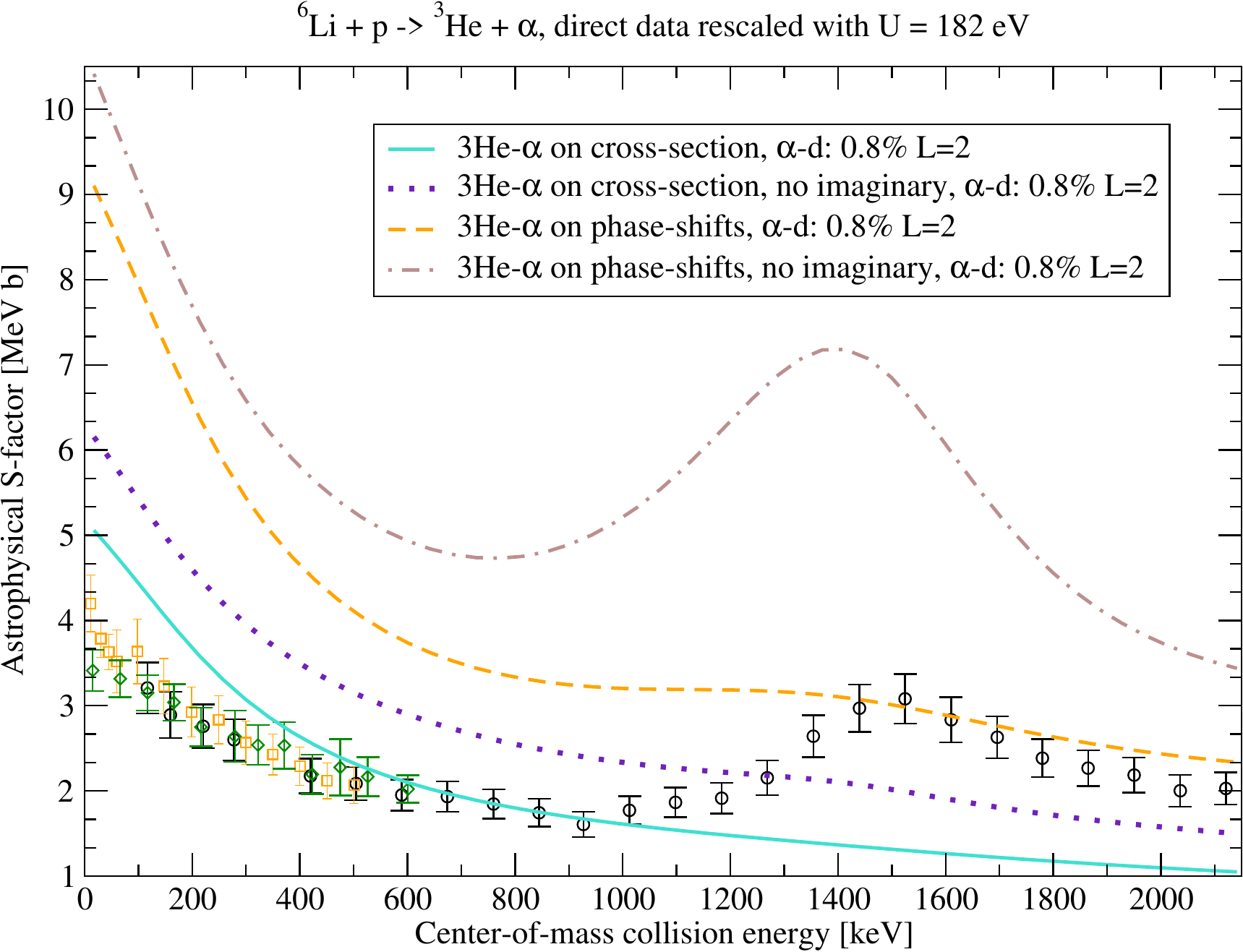}%
		\caption[Deuteron-transfer calculation with different \texorpdfstring{\nuclide[3]{He}--\nuclide{\alpha}}{3He-4He} potentials]{\label{figdTransferPotenziale3HeaAlternativo}%
			Points are the same in \cref{figdTransferConOSenzaL2} (legend suppressed for readability): direct data was rescaled as in \cref{figDatiCorrettiAdiabatic}. Turquoise solid line is the same calculation as the blue solid line in \cref{figdTransferConOSenzaL2} but with a smaller deformed component for \nuclide[6]{Li}.
			Orange dashed line is the same calculation as the turquoise solid line but with a different \nuclide[3]{He}--\nuclide{\alpha} potential, see text or details.
			Indigo dotted and brown dot-dashed lines are the same as, respectively, turquoise solid and orange dashed lines, but removing all imaginary parts from all potentials.
		}
	\end{figure}

	It is also relevant to stress that the lack of a clearly visible peak in the computed excitation function
	is ultimately connected %
	to the choice on %
	the adopted interactions, %
	rather than an intrinsic limitation of the structure model (once deformations are introduced). In particular, the potentials employed in this work are mainly focused on describing %
	the sub-Coulomb energy range without introducing %
	spurious resonances, and not on reproducing %
	accurately the region relevant for the peak. %
	To understand to what extent better agreement in the resonant region is possible, %
	an alternative \nuclide[3]{He}--\nuclide{\alpha} potential was generated adjusting it on %
	the elastic scattering phase-shifts involving the resonant waves ($f_{5/2}$ and $f_{7/2}$), degrading at the same time the agreement with the total elastic cross-section. It was not possible to obtain agreement with all experimental phase-shifts using, for all waves, a single energy-independent potential with only spin-orbit coupling%
	\footnote{Adding an explicit dependence of the potential on angular momentum %
		is an effective way to add non-central components, which pose computational difficulties in the transfer calculation, see the discussion in \cref{secDWBAPriorPost}. %
		Limiting the potentials to standard spin-orbit components appeared as a good compromise between keeping the numerical approximations under control and providing the desired properties for the distorted waves.}.
	The results on the transfer calculation obtained using this %
	potential and the other one presented in \cref{sec3HeaOpticalPotential} (fitted on elastic cross-sections) are compared in \cref{figdTransferPotenziale3HeaAlternativo}, assigning the more modest %
	\nuclide[6]{Li} deformation suggested by reproducing the electric quadrupole moment ($c_2 \approx \num{-0.091}$, see the discussion in \cref{secCostruzioneGroundState6LiDeformato}).
	If the same deformation adopted in \cref{figdTransferConOSenzaL2} is instead employed, the resonant contribution is enhanced in all calculations, but the qualitative conclusions remain unaltered.
	From \cref{figdTransferPotenziale3HeaAlternativo} (brown dot-dashed line), it can be seen that a prominent structure is found using the \nuclide[3]{He}--\nuclide{\alpha} interaction fitted on phase-shifts and including only the real parts of all projectile-target potentials (also for \nuclide[6]{Li}--\nuclide{p}). This structure peaks at about the experimental resonance energy and has the expected quantum numbers, as found by performing a partial-wave expansion (analogous to the one in \cref{figdTransferPartialWaveExpansions}). %
	If the calculation is rescaled by an appropriate constant, a qualitatively reasonable energy trend is found throughout the energy range under study. However, the computed absolute cross-section overestimates the data at all energies. The issue is not solved by adding imaginary components to the projectile-target optical potentials (orange dashed line in \scref{figdTransferPotenziale3HeaAlternativo}), as these tend to primarily erode the resonant contribution to the cross-section.
	As commented previously, %
	a similar resonant structure is found also using the potentials %
	already employed in \cref{figdTransferConOSenzaL2,figdTransferPartialWaveExpansions} (turquoise solid line in \scref{figdTransferPotenziale3HeaAlternativo}), %
	but its magnitude is much more modest, even after removing all imaginary components %
	(indigo dotted line in \scref{figdTransferPotenziale3HeaAlternativo}). At the same time, the calculations using the \nuclide[3]{He}--\nuclide{\alpha} potential discussed in \cref{sec3HeaOpticalPotential} %
	compare better with experimental cross-sections regarding both the elastic scattering (on which the potential %
	was fitted) and the transfer channel.

	Finally, it can be important to consider that the breakup channel, %
	$\nuclide[6]{Li}+\nuclide{p} \to \nuclide{\alpha} + \nuclide{d} + \nuclide{p}$, %
	is already open at the resonance energy, thus %
	an explicit coupling to this channel could be useful %
	to describe the transfer cross-section, as was suggested in \cite{Soukeras2017}. More in general, the system may %
	achieve the change in projectile-target orbital angular momentum (required to populate the resonance) passing through an excited state, rather than by virtue of the ground-state deformed components studied here. See also the preliminary results in \cref{secCDCCCalcoloPratico} on this topic.

\subsubsection{Non-central optical potentials and prior-post equivalence}\label{secDWBAPriorPost} %
	
	When performing the couplings inducing transfer processes, %
	the \textsc{Fresco} code neglects all non-central terms (in particular spin-orbit, spin-spin, and tensor) in the core-core and projectile-target optical potentials (precisely, those appearing in the remnant term).
	This originates from the necessity of performing a coordinate change to practically compute the cross-sections, which is particularly difficult to implement for non-central terms. Such terms are however accounted for correctly when computing standard distorted waves, as this requires no coordinate change %
	(wave-function and potential refer to the same partition),
	allowing to construct more rich and realistic wave-functions. %
	One may rely on a partial cancellation between non-central terms in the projectile-target and core-core potential, %
	to argue that the calculation is still reasonably accurate. However, the cancellation cannot be exact, and in general does not take place to the same degree in both prior and post forms.
	This implies that, whenever non-central projectile-target potentials are introduced in a transfer calculation, the results will not be identical in prior and post (as would be otherwise expected, see \cref{sezDWBA}), and that one form yields more accurate cross-sections.
	
	It may be assumed %
	that the cancellation between %
	non-central components in the projectile-target and core-core potential %
	is greater when %
	these components bear a more similar form (each with respect to its own coordinates).
	On this basis, the calculation shown in \cref{figdTransferConOSenzaL2} was performed in post form, since only the $\nuclide[6]{Li}+\nuclide{p}$ potential has a spin-spin component. It was also tested that, if the $\nuclide[6]{Li}+\nuclide{p}$ potential is set to include only spin-orbit terms, the prior-post agreement is significantly improved.
	As expected, there is practically perfect agreement between the two forms when only central potentials are employed.

\section{DWBA \texorpdfstring{$\nuclide{p}+\nuclide{n}$}{p+n} transfer}\label{secCalcoliDWBApnTransfer}
	
	As done in \cref{secCalcoliDWBADeuteronTransfer}, %
	this \namecref{secCalcoliDWBApnTransfer} opens with a presentation and discussion of the physical ingredients required to perform the DWBA two-nucleon transfer calculation of the $\nuclide[6]{Li} + \nuclide{p} \to \nuclide[3]{He} + \nuclide{\alpha}$ reaction, while the resulting cross-sections are shown in \cref{secTwoParticleTransfer6Lip3He4HeCrossSection}. %
	\Cref{secTNTCalcoliPraticiDescrizioneWFsingleparticle} lists the adopted potentials and covers the construction of all single-nucleon bound-states of interest, which are later employed in \cref{secTNTCalcoliPraticiDescrizioneWFSimultaneo} to generate three-particle wave-functions for both \nuclide[3]{He} and \nuclide[6]{Li}.
	
	For consistency, the calculation was performed employing the same %
	ingredients as %
	in the deuteron transfer case, where possible. In particular, the \nuclide[6]{Li}--\nuclide{p} and \nuclide[3]{He}--\nuclide{\alpha} projectile-target optical potentials and the \nuclide{\alpha}--\nuclide{p} core-core interaction (employed in the simultaneous calculation) are the same adopted in \cref{secCalcoliDWBADeuteronTransfer}.
	Other instances are discussed in detail below.

\subsection{Single-nucleon states}\label{secTNTCalcoliPraticiDescrizioneWFsingleparticle}

	In this work, the two-nucleon-transfer calculations are performed using the ``two-body'' approximations discussed in \cref{secReactionTheoryTwoParticleTransfer}, and in particular the ``core-composite'' scheme in \cref{secTNTSimultaneoTeoriaSchemaCoreComposite} for the simultaneous term.
	This involves, for each reactant, the definition of bound states for the motion between only one nucleon and the core, to form an intermediate partition, and similarly of bound states for the motion between the second nucleon and the intermediate system. %
	The adopted ``single-nucleon'' wave-functions are described here. %
	The first parts %
	of this \namecref{secTNTCalcoliPraticiDescrizioneWFsingleparticle} focus on the %
	simultaneous calculation, %
	while the last paragraph %
	discusses  
	the ingredients and the considerations relevant for the sequential contribution. %

	Since the two transferred nucleons are not identical, there are two possible intermediate partitions for the system, $\nuclide[5]{Li} + \nuclide{d}$ and $\nuclide[5]{He} + \nuclide[2]{He}$.
	In the simultaneous calculation, it is however always sufficient to consider only one partition. %
	For instance, in the construction of the \nuclide[6]{Li} wave-function, the choice on the adopted intermediate partition amounts to choosing between the Hamiltonian in \cref{eqTwoNucTranHamiltExplicitPrior} and the one obtained swapping $\mu$ and $\nu$ in the expression: the system can be fully described through a set of solutions of either Hamiltonian. %
	The intermediate partition to use %
	is consequently selected as the %
	most appropriate in light of the further approximations involved in the calculation, in particular considering that only a very small number of intermediate states is included in the formalism. %
	In the present case, there is little difference between the \nuclide[5]{Li} and \nuclide[5]{He} structure, thus the $\nuclide[5]{Li} + \nuclide{d}$ partition was employed, given the advantage in treating internal states and interactions involving a deuteron rather than a di-proton. %
	In practice, %
	the overlap involved in the simultaneous contribution is here %
	constructed assuming %
	that the core, the transferred system and the composite nucleus %
	all have definite isospin modulus $T$. %
	In particular, \nuclide{\alpha} and \nuclide[6]{Li} are assigned $T=0$, %
	so that the transferred system has $T=0$ as well, and \nuclide[3]{He} has $T=1/2$.

\subsubsection{\texorpdfstring{\boldmath{\nuclide[3]{He}}}{3He} system}

	The three-particle wave-function for \nuclide[3]{He} ground state is constructed taking into account the configuration suggested by the independent-particle shell model.
	Thus,
	the internal motion of the \nuclide{d} intermediate system
	is constructed as the $1s$ (orbital angular momentum zero and no nodes) bound state of a phenomenological central potential obtained from the \textsc{Fr2in} code \cite{BrownReactionCodes}, whose parameters are listed in \cref{tabParametriNumericiPotenziali3He}.
	Since the two-body ``core-composite'' approximation discussed in \cref{secTNTSimultaneoTeoria} is being employed, a \nuclide{d}--\nuclide{p} single-particle $1s$ state is additionally defined %
	as %
	the spherical component %
	of the overlap function discussed in \cref{sec3HepPotential}, while the $1d$ component was here neglected for simplicity.
	
	Note that it is not necessary to consider a %
	state with isospin $T=1$ for the intermediate system \nuclide{d}.
	From the considerations above, the overlap $\Braket{\nuclide{\alpha}|\nuclide[6]{Li}}$ is a wave-function for the \nuclide{d} state including only a $T=0$ component. Since the simultaneous transfer is a one-step process and conserves the transferred system state, the \nuclide[3]{He} wave-function is thus truncated so that $\Braket{\nuclide{p}|\nuclide[3]{He}}$ includes only a $T=0$ state as well. This in turn implies that the overlap $\Braket{\nuclide{d}(T=1)|\nuclide[3]{He}}$ is zero for the truncated state. %

\subsubsection{\texorpdfstring{\boldmath{\nuclide[6]{Li}}}{6Li} system}
	
	In both possible intermediate partitions, the residual of \nuclide[6]{Li} is unbound. At present, for simplicity, the \nuclide[5]{Li} is described through %
	fictitious $\nuclide{\alpha}+\nuclide{p}$ bound states, with binding energy equal to a fraction of the $Q$-value for $\nuclide{\alpha}+\nuclide{p}+\nuclide{n}\to\nuclide[6]{Li}$, which is \SI{3.70}{\MeV}.
	The impact of the intermediate nucleus binding energy on the results is discussed in \cref{secThreeParticleWF6LiImpactIntermediateBindingEnergy}. %
	Keeping %
	into account the way the single-nucleon state is treated in the simultaneous calculation (the $\nuclide[5]{Li}+\nuclide{n}$ state actually represents the $\nuclide{\alpha}+\nuclide{n}$ motion), in the calculations shown in this work, %
	unless otherwise stated, the \nuclide[5]{Li} binding energy was %
	assigned half of the aforementioned $Q$-value. %
	
	The adopted \nuclide{\alpha}--\nuclide{p} potential was the same mentioned in \cref{secapPotential}, but with the volume term depth adjusted to reproduce the desired binding energy. For the \nuclide[5]{Li}--\nuclide{n}, again the same potential was employed, additionally rescaling all radii by $(6/5)^{1/3}$ to empirically account for the different size of the system. The parameters of the potential are listed in \cref{tabParametriNumericiPotenziali6Li}.
	
\begin{figure}[tb]%
		\centering
		\includegraphics[width=\textwidth,keepaspectratio=true]{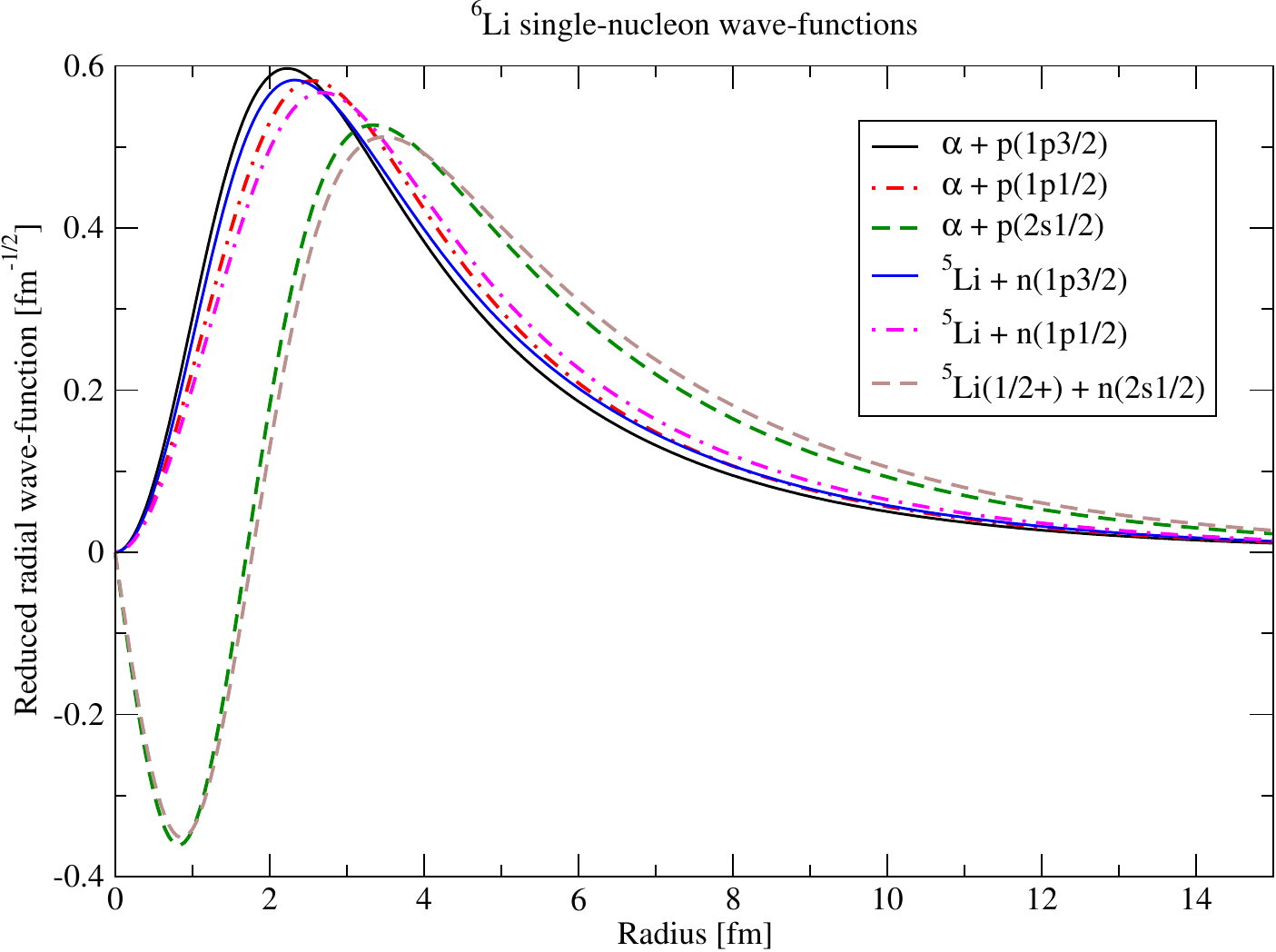}%
		\caption[\texorpdfstring{\nuclide[6]{Li}}{6Li} single-nucleon radial wave-functions]{\label{figapnSingleParticleWF}%
			Single-nucleon reduced radial wave-functions for the \texorpdfstring{\nuclide[6]{Li}}{6Li} system. Each line represents a $\nuclide{\alpha}+\nuclide{p}$ or $\nuclide[5]{Li}+\nuclide{n}$. %
			In legend, the first number between parenthesis is the number of nodes plus 1, ``s'' and ``p'' refer to a relative orbital angular momentum, $l_i$, of 0 or 1, and the final number is the modulus of the sum of $l_i$ and the nucleon spin.
		}
	\end{figure}

	Several distinct single-nucleon states were considered for both $\nuclide{\alpha}+\nuclide{p}$ and $\nuclide[5]{Li}+\nuclide{n}$ systems. This allows to build a more realistic wave-function for \nuclide[6]{Li}. In a shell-model picture, %
	the two transferred nucleons on top of the \nuclide{\alpha} core may be found in several shells: in this work, the $1p_{3/2}$, $1p_{1/2}$ and $2s_{1/2}$ shells were considered (with ``$1p$'' referring to the lowest shell with the desired angular momentum, and so on). Correspondingly, three states for \nuclide[5]{Li} were constructed, bearing the same binding energy, and a spin-parity of $3/2^-$, $1/2^-$ and $1/2^+$ respectively.
	The radial parts of the corresponding relative-motion wave-functions are shown in \cref{figapnSingleParticleWF}. %
	As can be seen, the difference between corresponding $\nuclide{\alpha}+\nuclide{p}$ and $\nuclide[5]{Li}+\nuclide{n}$ states is very small, as well as the difference between the radial parts of $1p_{3/2}$ and $1p_{1/2}$ states. The $2s_{1/2}$ wave-function instead has a distinct shape.

\subsubsection{Use of the single-nucleon wave-functions in the sequential calculation}
	
	Let $X(J^\pi)$ be a state of nucleus $X$ with spin-parity $J^\pi$. %
	In principle, the calculation of the sequential process requires the knowledge of all possible overlap functions involving both projectile and target, for instance all the possible
	$\Braket{\nuclide{\alpha}(0^+) \, \Phi^{\nuclide{p}}_{J} |\nuclide[5]{Li}(J^\pi)}$ and $\Braket{\nuclide[5]{Li}(J^\pi) \, \Phi^{\nuclide{n}}_{j} |\nuclide[6]{Li}(1^+)}$, %
	where %
	$\Phi$ is defined as in \cref{eqFractionalParentageExpansionGenerica}, having omitted the index $T_x=1/2$ (see also \scref{eqOverlapFunctionProiettataSuSingoloTrasferito}). %
	This also includes the information on the associated spectroscopic amplitudes, which for definiteness are here denoted as $\mathcal{A}^{\nuclide{p}}_{0J}$ and $\mathcal{A}^{\nuclide{n}}_{Jj}$ respectively.
	
	In an exact treatment, all ingredients required for the sequential calculation could be deduced from the full four-particle Hamiltonian, %
	see \cref{secTeoriaTransferContributoSequenziale}).
	For all intermediate partitions and states of interest, the overlap functions appearing in each transfer step would be computed explicitly by projecting the full three-particle wave-functions. Even within such exact framework, the set of intermediate states to include in the calculation is not uniquely determined and has to be supplied as part of the model (see again the discussion in \cref{secTeoriaTransferContributoSequenziale}). Furthermore, in practice all sequential calculations are performed under the two-body approximations discussed in \cref{secTeoriaTransferContributoSequenziale}. %
	
	In this work, as often done in literature (e.g.~in~\cite{Ayyad2017,Lay2022}, see also the implementation in the \textsc{Fr2in} code \cite{BrownReactionCodes}), %
	the intermediates states considered in the sequential process are defined in terms of the single-nucleon wave-functions already adopted in the simultaneous calculation. For instance, since the \nuclide[6]{Li} three-particle wave-function was constructed including a $3/2^-$, a $1/2^-$ and a $1/2^+$ state for the \nuclide{\alpha}--\nuclide{p} system, the same states are included for the intermediate \nuclide[5]{Li} nucleus in the sequential calculation, assigning for the inter-cluster motion in each intermediate (\nuclide{\alpha}--\nuclide{p}) and final (\nuclide[5]{Li}--\nuclide{n}) state (i.e.~the normalised overlap functions) the same wave-functions in \cref{figapnSingleParticleWF}.
	Similarly, %
	since the \nuclide[3]{He} three-particle wave-function for the simultaneous calculation was constructed including only a $T=0$ state for the intermediate \nuclide{d} system (see the discussion earlier in this \namecref{secTNTCalcoliPraticiDescrizioneWFsingleparticle}), the same choice is made in the sequential calculation. %
	The sequential transfer path %
	featuring an intermediate \nuclide[2]{He} system (which has necessarily $T=1$) is %
	excluded. %

	Regarding the spectroscopic amplitudes, %
	for the purpose of computing the sequential transfer cross section within the two-step DWBA formalism, it is not necessary to compute all $\mathcal{A}^{\nuclide{p}}_{0J}$ and $\mathcal{A}^{\nuclide{n}}_{Jj}$ separately. Rather,
	it is sufficient to just fix the product %
	$\mathcal{A}^{\nuclide{p}}_{0J} \mathcal{A}^{\nuclide{n}}_{Jj}$ %
	for each possible ``path'' in the sequential process, that is, each possible value%
	\footnote{%
		In general, there may also be several states with equal $J$, $\pi$ and $j$, to be distinguished by an additional quantum number, see for example \cref{eqAmpiezzaSpettroscpicaSingolaComponenteMotoValenza}.}
	of $J$, $\pi$ and $j$. %
	Consistently with the way the overlap functions themselves were chosen, these %
	were fixed in terms of the weights and amplitudes employed in the simultaneous calculation, as follows. %
	For each possible $J$, $\pi$ and $j$, let $\mathcal{A}^{\nuclide{d}}_{0(Jj)}$ be the spectroscopic amplitude for the $\Braket{\nuclide{\alpha}(0^+) \, \(\Phi^{\nuclide{p}}_{J} \, \Phi^{\nuclide{n}}_{j}\) |\nuclide[6]{Li}(1^+)}$ overlap, %
	where the state of the valence two-nucleon system is given by the product of the two one-nucleon wave-functions under consideration.
	$\mathcal{A}^{\nuclide{d}}_{0(Jj)}$ is expressed %
	as %
	the amplitude of the configuration of interest within the three-particle wave-function employed in the simultaneous (for instance the values displayed later in \cref{tabAmplitudes6LiThreeParticlejj}),
	multiplied by the square-root of the total spectroscopic factor %
	of the $\Braket{\nuclide{\alpha}(0^+)|\nuclide[6]{Li}(1^+)}$ overlap, which is taken to be same employed in the one-particle transfer calculation, see \cref{secCalcoloDWBAOverlapFunctions}.
	The relation between $\mathcal{A}^{\nuclide{p}}_{0J} \mathcal{A}^{\nuclide{n}}_{Jj}$ and $\mathcal{A}^{\nuclide{d}}_{0(Jj)}$ is then assumed to be the same holding within the independent-particle shell model%
	\footnote{The absolute value of the amplitudes computed using the independent-particle shell model %
		is not adopted directly in the calculation, as it is not deemed to be sufficiently accurate.}. %
	In particular, from the calculations performed in \cref{secOverlapSpectroscopicfactorswithinExtremeshellmodel} it can be seen that, for each possible transfer path, it is simply $\mathcal{A}^{\nuclide{p}}_{0J} \mathcal{A}^{\nuclide{n}}_{Jj} = \mathcal{A}^{\nuclide{d}}_{0(Jj)}$, and the same holds for the \nuclide[3]{He} overlaps of interest%
	\footnote{The result is instead different when considering the $\Braket{\nuclide{p}|\nuclide{t}}$ overlap (in particular, it is $\mathcal{A}^{\nuclide{n}}_{0J} \mathcal{A}^{\nuclide{n}}_{JJ} = \sqrt{2} \mathcal{A}^{\nuclide{2n}}_{0(JJ)}$), thus in $(\nuclide{p},\nuclide{t})$ transfer calculations the %
		amplitudes for the sequential term carry an extra factor.}.
	The spectroscopic amplitudes computed in this manner were compared with those found through variational Monte Carlo (VMC) calculations, reported in \cite{VMCresults}, for negative-parity \nuclide[5]{He} states%
	\footnote{The \nuclide[5]{He} and \nuclide[5]{Li} are expected to display similar properties. %
		No $\Braket{\nuclide[5]{He}(1/2^+) |\nuclide[6]{Li}(1^+)}$ overlap is reported in \cite{VMCresults}, thus the transfer path involving $\nuclide[5]{Li}(1/2^+)$ could not be compared.}: %
	excellent agreement is found for the sum of all spectroscopic factors, supporting the appropriateness of the choice on the overall weights. The values of the single amplitudes instead show some difference, but this may be %
	due to a difference in the adopted \nuclide[6]{Li} state. %

	Finally, the computation of the sequential contribution,  %
	within the present formalism, %
	requires the definition of %
	additional phenomenological potentials, whose parameters are given in \cref{tabParametriNumericiPotenzialiOttici,tabParametriNumericiPotenziali6Li}.
	The \nuclide[5]{Li}--\nuclide{p} core-core potential, required for the first sequential step, is taken from a generic parametrisation in \cite{Varner1991}.
	The \nuclide{d}--\nuclide[5]{Li} projectile-target optical potential (for the second step of the sequential transfer) is taken from a generic parametrisation in \cite{Daehnick1980}.
	The \nuclide{d}--\nuclide{\alpha} potential, representing the core-core interaction in the second sequential step, is taken to be the same discussed in \cref{secCostruzioneGroundState6LiDeformato}, %
	but with the depth of the volume term rescaled to match the potential in \cite{Gammel1960} at zero distance (whose numerical value is given in \cite[fig.~2]{Kubo1972}).

\subsection{Three-particle wave-functions}\label{secTNTCalcoliPraticiDescrizioneWFSimultaneo}

	Starting from the single-particle wave-functions described in \cref{secTNTCalcoliPraticiDescrizioneWFsingleparticle}, three-particle wave-functions to be employed in the simultaneous calculation are constructed as discussed in \cref{secTNTSimultaneoTeoria}, in particular using the scheme in \cref{eqHamiltonianaSimultaneoSchemaColonnaConCoordinateCoreCore} with the phenomenological potentials already listed earlier in this \namecref{secCalcoliDWBApnTransfer}. The wave-function is then expressed in ``t'' coordinates, namely in terms of the internal motion within the transferred system and relative motion between core and centre-of-mass of the transferred system, using the Moshinsky transformation \cite{Moshinsky1959}. %

\subsubsection{Single-particle and t-coordinates basis}\label{secTNTCalcoliPraticiDescrizioneWFSimultaneoCoordinatevt} %
	
	Note that there is not a one-to-one mapping between the angular components %
	in ``t'' %
	coordinates and in the single-particle basis.
	Let the vectors $\v l_1$ and $\v l_2$ represent the orbital angular momenta (or the corresponding operators) between the core of a nucleus (here, \nuclide{\alpha} or \nuclide{p}) and valence particle ``1'' or ``2'' (here, proton or neutron). Also let $\v{\mathcal L} = \v l_1 + \v l_2$.
	Additionally, let $\v s_1$ and $\v s_2$ be the spins of each transferred particle, and $\v j_i = \v s_i + \v l_i$. The total transferred angular momentum, $\v j$, is thus  $\v j_1 + \v j_2$. Any single-particle state is characterised by the modulus quantum numbers $l_i, s_i, j_i$.
	In ``t'' coordinates, similarly let $\v l$ be the relative orbital angular momentum between the transferred particles, $\v L$ the core-transferred orbital angular momentum, $\v S = \v s_1 + \v s_2$ (the ``intrinsic'' spin of the transferred system) and
	$\v J_x = \v l + \v S$ (the total spin of the transferred system). %
	The total orbital angular momentum is certainly the same in both cases, $\v l + \v L = \v{\mathcal L}$, and similarly $\v j = \v S + \v{\mathcal L}$, but the components can differ. For instance, for \nuclide[3]{He}, the single-particle states were constructed with $l_1 = l_2 = {\mathcal L} = 0$, but the corresponding three-particle state for \nuclide[3]{He} includes both a component with quantum numbers $l = L = 0$, another small one with $l = L = 2$, and in principle infinitely many others%
	\footnote{Odd values of $l$ are not allowed in this specific case. Given the simple shell-model structure adopted for \nuclide[3]{He} single-particles states (both transferred nucleons are found only in the $1s1/2$ shell), as commented in \cref{secOverlapSpectroscopicfactorswithinExtremeshellmodel}, if the transferred system has total isospin 0, then %
	$j$ must be 1. Given that ${\mathcal L} = 0$, this implies %
	$S=1$. %
		Anti-symmetrisation of the transferred system state then imposes $l$ to be even.}.
	In the calculations shown here, all wave-functions were truncated to components with $l \leq 2$. In a test calculation where all terms with $l\leq 4$ were included, the cross-sections did not change appreciably.
	
	A similar issue arises regarding the radial components in the two coordinate systems. More specifically, the Moshinsky transformation is performed expanding the complete three-particle state in an harmonic oscillator basis, truncating the expansion to a finite number of terms, and then transforming each single term. In the figures shown in this thesis, %
	the wave-functions space in t coordinates was truncated to a maximum distance between the two transferred nucleons of \SI{28}{\femto\metre}, and further discretised so that, for each angular component (definite values of $l$, $L$, $S$, etc.), the adopted harmonic oscillator truncated basis included about 28 elements%
	\footnote{Components whose contribution in norm falls below a given threshold are discarded, thus the number of basis elements is not exactly the same for all wave-functions. Also note that, in the \textsc{Fresco} code, the integration over the distance between the two transferred particles is performed using Gaussian quadrature \cite[sec.~5.3.2]{Thompson2004}.}.
	Additional preliminary calculations showed that a small %
	variation in the precise values of the computed cross-sections is found when including about 140 harmonic oscillator basis elements for each angular component%
	\footnote{Further increase of the number of basis elements, up to about 560 elements for each angular component, did not cause any appreciable difference in the results.}. %
	However, the conclusions discussed in this work are not affected by these numerical details.

\subsubsection{\texorpdfstring{\boldmath{\nuclide[3]{He}}}{3He} system}\label{secTNTCalcoliPraticiDescrizioneWFSimultaneo3He}

	\Cref{tabNorme3HeThreeParticleTcoordinates} shows the decomposition in ``t''-coordinates angular components of the Moshinsky-transformed \nuclide[3]{He} three-particle wave-function. %
	As mentioned in \cref{secTNTCalcoliPraticiDescrizioneWFSimultaneoCoordinatevt}, %
	the complete state %
	is in fact constructed as the combination of a wide number of components, thus the table only includes the overall norm of each %
	radial wave-function%
	\footnote{The amplitude assigned to each single component is the one obtained applying the Moshinsky transformation \cite{Moshinsky1959} as implemented in the \textsc{Fresco} code \cite{Thompson2004}.}.
	\begin{table}[tbp]
		\caption[Decomposition of the \texorpdfstring{\nuclide[3]{He}}{3He} three-particle wave-function]{\label{tabNorme3HeThreeParticleTcoordinates}%
			Decomposition %
			of the Moshinsky-transformed $\nuclide{p} + \nuclide{n} + \nuclide{p}$ three-particle wave-function (in ``t'' coordinates) employed to describe the ground state of \nuclide[3]{He}. Each row lists the total norm of all components %
			with definite quantum numbers $l$, $S$, $J_x$, $L$ (notation as in \cref{secTNTCalcoliPraticiDescrizioneWFSimultaneoCoordinatevt}). Regarding the total norm of the wave-function, see the discussion in \cref{secTNTCalcoliPraticiDescrizioneWFSimultaneoCoordinatevt,secTNTCalcoliPraticiDescrizioneWFSimultaneo3He}.}
		\centering
		\begin{tabular}{ccccS[table-format=2.4]}%
			$l$ & $S$ & $J_x$ & $L$	& {Norm [$10^{-2}$]} \\ \toprule %
			0	& 1	  & 1	  & 0	& 94.28	\\ %
			2	& 1	  & 1	  & 2	& 0.4327	\\ %
			2	& 1	  & 2	  & 2	& 0.7212	\\ %
			2	& 1	  & 3	  & 2	& 1.010 %
		\end{tabular}
	\end{table}
	It can be seen that the transformed %
	wave-function has a total norm of about $0.964$: the missing norm is connected to the truncation of the space mentioned in \cref{secTNTCalcoliPraticiDescrizioneWFSimultaneoCoordinatevt}. %
	The transformed wave-function is \emph{not} rescaled to fix the total norm to 1: doing so would amount to merely rescale the cross-section by a factor. Rather, it is expected that including the missing components %
	would cause only a small change, and not necessarily an increase, in the reaction cross-section. %
	Consequently, it appears to be a better approximation to simply neglect the norm associated with the discarded components.
	
\begin{figure}[tb]%
		\centering
		\includegraphics[width=\textwidth,keepaspectratio=true]{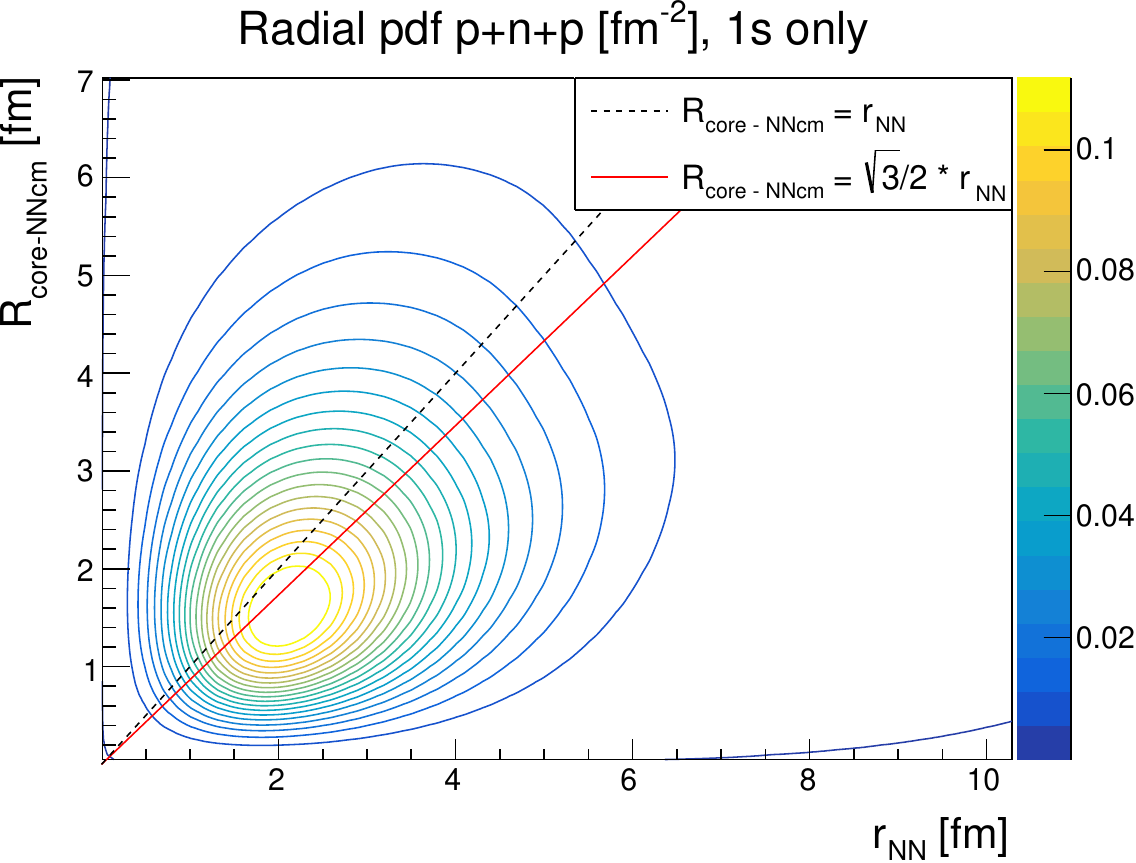}%
		\caption[\texorpdfstring{\nuclide{p}+\nuclide{n}+\nuclide{p}}{p+n+p} radial probability density function]{\label{figPDF3He}%
			Contour plot of the reduced radial probability function for the \nuclide{p}+\nuclide{n}+\nuclide{p} state constructed as described in text. The $x$-axis is the distance between the transferred nucleons, %
			the $y$-axis is the distance between core and centre-of-mass of the transferred system ($R_{ct}$ in text). The lines follow the equations given in the legend.
		}
	\end{figure}

	Finally, when calculating the transfer cross-section, the wave-function will be assigned an overall spectroscopic factor of \num{1.31}, the same value employed for the spherical component of \nuclide[3]{He} in the transfer of a structureless deuteron (see \cref{sec3HepPotential}). This factor is included only in the cross-section calculation, not when discussing the properties of the wave-function (e.g.~in \cref{tabNorme3HeThreeParticleTcoordinates} or in \cref{figPDF3He}).

	\Cref{figPDF3He} shows the radial probability density function (square-modulus of the wave-function integrated over all angular components) associated to the computed \nuclide[3]{He} wave-function.
	Let $\v r_{nn}$ be the distance between the transferred nucleons, and $\v R_{ct}$ the distance between core and centre-of-mass of the transferred system.
	In general, configurations with $r_{nn}$ significantly greater than $R_{ct}$ represent “cigar-like” structures, with the two nucleons found in anti-correlation. Conversely, the region with $r_{nn} < R_{ct}$ (or, even better, a structure with $r_{nn}$ approximately fixed and $R_{ct}$ running through a wide range) corresponds to correlated, or ``clustered'', components. Finally, configurations in which each particle is equally distant from the other two lie on the
	red solid line in \cref{figPDF3He}. Qualitatively, \nuclide[3]{He} dominant configurations are expected to resemble an equilateral triangle, apart from deformations due to the
	different electric charge and the nucleons spin projection. The result found here agrees with this expectation.
	
	The root-mean-square charge radius of the system can be evaluated through \cref{eqRaggioDiCaricaTreCluster}. Using for simplicity only the $l=L=0$ component of the wave-function (which incorporates most of the wave-function norm), the value of \SI{2.16}{\femto\metre} was found. This is slightly smaller than the di-cluster prediction given in \cref{sec3HepPotential}, but still \SI{10}{\percent} bigger than the experimental value.

\subsubsection{\texorpdfstring{\boldmath{\nuclide[6]{Li}}}{6Li} system}\label{secTNTCalcoliPraticiDescrizioneWFSimultaneo6Li}

	Since several single-particle states were defined for \nuclide[6]{Li}, the total wave-function will be a combination of all products of $\nuclide{\alpha}+\nuclide{p}$ and $\nuclide[5]{Li}+\nuclide{n}$ states coupling to the desired total angular momentum and parity. In particular, only single-particle states with equal parity can be paired. %
	For instance, there will be a component where the $\nuclide{\alpha}+\nuclide{p}$ system occupies the $1p_{3/2}$ state and $\nuclide[5]{Li}+\nuclide{n}$ occupies the $1p_{1/2}$ state, which for brevity will be denoted as ``$1p_{3/2} \times 1p_{1/2}$''; similarly, %
	a $1p_{3/2} \times 1p_{3/2}$ component is allowed, and so on, but no term such as $1p_{1/2} \times 2s_{1/2}$ is possible.
	
	The weights to be assigned to each product of single-particle wave-functions were deduced by comparing with the results of a three-body calculation in \cite{Bang1979}, %
	performed using the potential in \cite{DeTourreil1975} for the nucleon-nucleon interaction.
	Note that the \nuclide{\alpha}--nucleon interaction adopted in \cite{Bang1979} %
	is the same employed here to construct the single-particle states involving the \nuclide[6]{Li} (see again \cref{secTNTCalcoliPraticiDescrizioneWFsingleparticle}%
	\footnote{%
		As mentioned at the beginning of \cref{secCalcoliDWBApnTransfer}, the same potential is also used as core-core interaction in the simultaneous transfer.}).
	Hence, by additionally adopting the same weights, %
	it is expected that the calculated three-particle state %
	will be consistent %
	to some extent. %
	\cite[tab.~2]{Bang1979} (in particular under the column ``dTS'') %
	lists the norms of interest here, each associated to a component with (using the notation presented in the first part of this subsection) fixed $l_1$, $l_2$, $\mathcal L$ and $S$.
	These weights were then converted to the required weight for each product of specific single-particle states, in $jj$ coupling scheme, using the analogous of \cref{eqRelazioneMomentoAngolareInAccoppiamentoLSojj} (see below for more details). %
	
	Note that the three components involving a single-particle state occupying a $d$ shell ($l_1$ or $l_2$ equal to 2) appearing in \cite[tab.~2]{Bang1979} are discarded here, as it would not be possible to include them consistently in the sequential calculation without also adding all other possible $d$-shell components. Finally, since the present calculation is performed assuming that \nuclide[6]{Li} is found in a state with definite isospin modulus $T$ equal to 0, the $T=1$ component in \cite[tab.~2]{Bang1979} ($l_1 = l_2 = \mathcal L = S = 1$) %
	was discarded as well%
	\footnote{It is interesting to note that, of the remaining $p$-shell components, %
		all but the $\mathcal L=0$ one are either small or do not contribute to %
		the simultaneous transfer, thus a good account of the simultaneous calculation %
		could be obtained restricting to the $\mathcal L=0$ %
		$p$-shell component and the $s$-shell one.
		However, such approximation is not performed here.}. %
	The remaining components of the \nuclide[6]{Li} state, adopted here, %
	comprise a total norm of \num{0.936}. These components were not rescaled to fix the total norm, with the same rationale applying %
	regarding %
	the Moshinsky-transformed wave-function (see again the discussion in \cref{secTNTCalcoliPraticiDescrizioneWFSimultaneoCoordinatevt,secTNTCalcoliPraticiDescrizioneWFSimultaneo3He}).

\begin{figure}[tbp]%
		\centering
		\includegraphics[width=\textwidth,keepaspectratio=true]{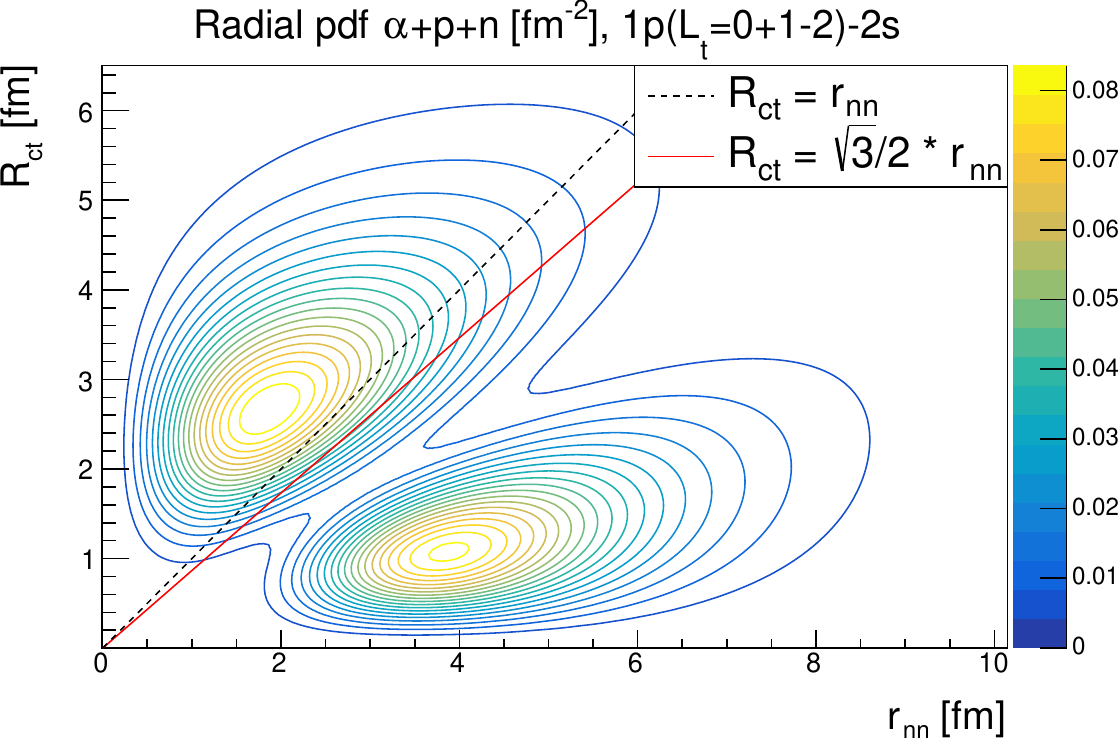}
		
		\includegraphics[width=\textwidth,keepaspectratio=true]{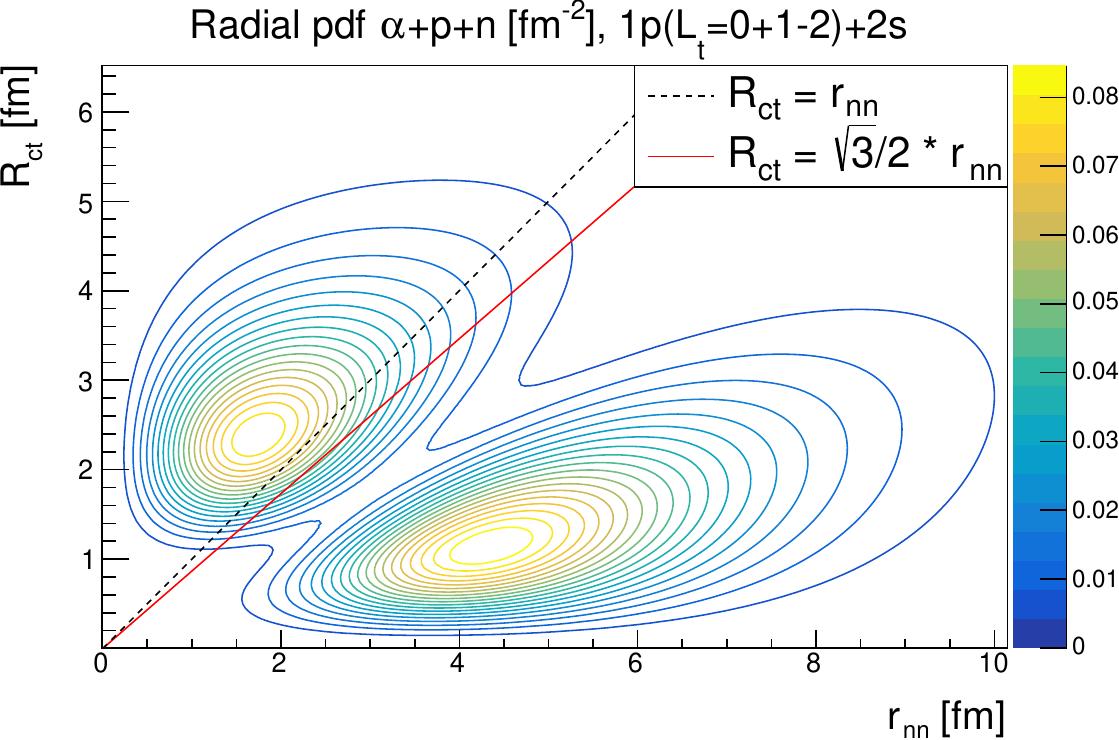}%
		\caption[\texorpdfstring{\nuclide{\alpha}+\nuclide{p}+\nuclide{n}}{a+p+n} radial probability density function]{\label{figPDF6Li}%
			Same as \cref{figPDF3He} for the \nuclide[6]{Li} three-particle system. Upper panel corresponds to the state constructed using the amplitudes given in \cref{tabAmplitudes6LiThreeParticlejj} (namely with the choice of relative signs yielding the best agreement with three-body calculations, see text for details). %
			Lower panel shows for comparison the opposite choice on the sign of the $2s_{1/2} \times 2s_{1/2}$ component. %
			In the figure titles, ``$2s$'' and ``$1p$'' refer to the two single-particle shells under consideration, ``L\textsubscript{t}'' is $\mathcal L$ (notation as in \cref{secTNTCalcoliPraticiDescrizioneWFSimultaneoCoordinatevt}), and the $\pm$ signs mark the relative phases adopted for each component of the wave-function.
		}
	\end{figure}

	In order to perform the conversion from the ``$\mathcal L S$'' basis adopted in \cite[tab.~2]{Bang1979} and the ``$j j$'' one required here, it is necessary %
	to know not only the absolute weight of each component, but also its sign. %
	However, the data included in \cite{Bang1979} %
	are not sufficient for extracting all phases in a straightforward manner%
	\footnote{These could be %
		found, for instance, by comparison of the radial components of the three-particle wave-function computed here (for each possible choice of signs) and in the original work.}.
	To find a set of reasonable signs, %
	the weight of each component was compared with those given by a three-body calculation in
	Hyperspherical Harmonics formalism \cite{Casal2021Private}, which appears to be in good agreement with the one in \cite{Bang1979}, meaning that, with an appropriate choice of signs, they predict very similar norms for each component%
	\footnote{Except for the relative importance of $p$- and $s$-shells. %
		The sign of the $2s_{1/2} \times 2s_{1/2}$ component %
		was thus fixed comparing the radial probability density function of the present calculation and of \cite{Casal2021Private}, which suggested to pick the sign generating a stronger clustered configuration.}.
	The signed amplitudes in $jj$ basis adopted in this work are reported in \cref{tabAmplitudes6LiThreeParticlejj}.
	\begin{table}[htbp]
		\caption[Decomposition of the \texorpdfstring{\nuclide[6]{Li}}{6Li} three-particle wave-function]{\label{tabPesi6LiThreeParticle}%
			Weights of the components of the $\nuclide{\alpha} + \nuclide{p} + \nuclide{n}$ three-particle wave-function employed to describe the ground state of \nuclide[6]{Li}, %
			deduced from \cite[tab.~2]{Bang1979} as detailed in text. Regarding the total norm of the wave-function, see the discussion in \cref{secTNTCalcoliPraticiDescrizioneWFSimultaneoCoordinatevt,secTNTCalcoliPraticiDescrizioneWFSimultaneo6Li}.}
	\begin{subtable}{\textwidth}
		\caption{\label{tabAmplitudes6LiThreeParticlejj}%
			Each row refers to a pair of single-particle states in ``$jj$'' angular momentum coupling scheme, as specified in the first two columns, marking the proton and neutron state respectively. The last column is the amplitude associated to the corresponding component of the \nuclide[6]{Li} wave-function.} %
		\centering
		\begin{tabular}{ccS[table-format=+1.4]}%
			\nuclide{p} shell & \nuclide{n} shell & {Amplitude} \\ \toprule
			$1p_{3/2}$	& $1p_{3/2}$	&  0.7482	\\
			$1p_{3/2}$	& $1p_{1/2}$	& -0.4044	\\
			$1p_{1/2}$	& $1p_{3/2}$	&  0.4044	\\
			$1p_{1/2}$	& $1p_{1/2}$	& -0.1228	\\
			$2s_{1/2}$	& $2s_{1/2}$	& -0.1843	
		\end{tabular}
	\end{subtable}
\\[\baselineskip]
	\begin{subtable}{\textwidth}
		\caption{\label{tabNorme6LiThreeParticleTcoordinates} Each row lists the total norm of all components of the Moshinsky-transformed %
			wave-function (in ``t'' coordinates) %
			with definite quantum numbers $l$, $S$, $J_x$, $L$ (notation as in \cref{secTNTCalcoliPraticiDescrizioneWFSimultaneoCoordinatevt}).}
		\centering
		\begin{tabular}{ccccS[table-format=2.4]}%
			$l$ & $S$ & $J_x$ & $L$	& {Norm [$10^{-2}$]} \\ \toprule %
			0	& 1	  & 1	  & 0	& 78.13	\\ %
			0	& 1	  & 1	  & 2	& 0.1774	\\ %
			1	& 0	  & 1	  & 1	& 5.027	\\ %
			2	& 1	  & 1	  & 0	& 0.1774	\\ %
			2	& 1	  & 1	  & 2	& 1.075	\\
			2	& 1	  & 2	  & 2	& 1.415	\\
			2	& 1	  & 3	  & 2	& 1.262
		\end{tabular}
	\end{subtable}
	\end{table}
	\Cref{tabNorme6LiThreeParticleTcoordinates} instead %
	lists the corresponding weights in t coordinates for the Moshinsky-transformed wave-function, which has a total norm of \num{0.873}, to be compared with the initial value of \num{0.936} (the same considerations made in \cref{secTNTCalcoliPraticiDescrizioneWFSimultaneo3He} %
	for \nuclide[3]{He} apply here)%
	\footnote{%
		The precise amount of norm loss, being of numerical nature, %
		depends %
		on the details of the state. %
		For instance, %
		the pdf in the top panel of \cref{figPDF6Li} has a smaller norm than the other one %
		by less than \SI{1}{\percent}. All other properties %
		tested here induced smaller differences. %
		The issue does not appear to play any measurable role in the comparisons shown later.%
		}.
	When calculating the transfer cross-section, the wave-function will be additionally assigned an overall spectroscopic factor of \num{0.85} (not included here), %
	the same value employed in the transfer of a structureless deuteron, see \cref{secadPotential}.
	
	\Cref{figPDF6Li} shows the the radial probability density (pdf) function associated to the \nuclide[6]{Li} wave-function computed in this manner. In order to show the role of the $s$-shell component on the result, the same figure includes the probability density function obtained inverting the sign of such component. The impact of all other components is not visually distinguishable in the probability density function.
	The \nuclide[6]{Li} radial pdf shows two peaks. As commented earlier, one peak may be associated to clustered configurations, where the two transferred nucleons are more strongly correlated in space, while the other can on the contrary be connected to “cigar-like” configurations. The tail behaviour of both peaks is different for each %
	\nuclide[6]{Li} state considered in \cref{figPDF6Li}. The choice of signs yielding the best agreement with three-body calculations, which is thought to represent a more realistic description of the system, is the one favouring the clustered peak (top panel in \scref{figPDF6Li}).
	The qualitative situation is similar to the one found when computing wave-functions related to two-neutron overlaps, for instance in \cite{Catara1984}. %

	Considering %
	only the $l=L=0$ component for simplicity, a root-mean-square charge radius of \SI{2.35}{\femto\metre} was found
	for both wave-functions shown in \cref{figPDF6Li}. This is \SI{9}{\percent} smaller than the experimental value in \cref{tabExperimentalGroundStateData} (compare also with the value found from the di-cluster model in \cref{secChargeRadius6LiDiclusterDeformato}). In order to evaluate the electric quadrupole moment, it is instead necessary to include %
	all angular components. Additionally, since the wave-function in use mimics the one in \cite{Bang1979}, which cannot reproduce the quadrupole moment (which is standard for three-body models of \nuclide[6]{Li}, see \cite{Tilley2002}), there is no reason to expect a better result here.

\subsection{Transfer cross-section}\label{secTwoParticleTransfer6Lip3He4HeCrossSection}
	
\begin{figure}[tbp]%
		\centering
		\includegraphics[width=\textwidth,keepaspectratio=true]{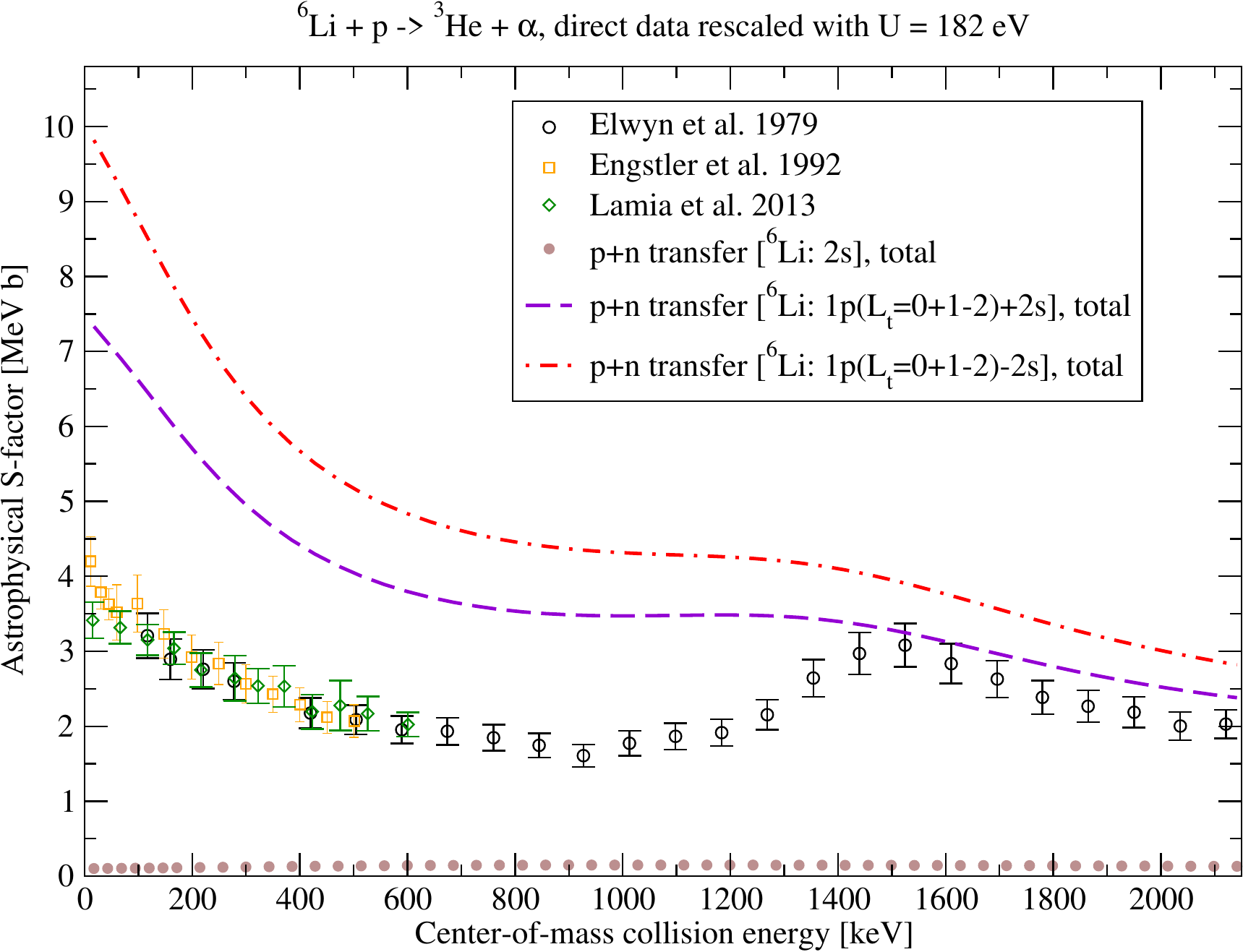}%
		\caption[Role of \texorpdfstring{\nuclide[6]{Li} $2s$}{6Li 2s} shell on \texorpdfstring{$\nuclide[6]{Li} + \nuclide{p} \to \nuclide[3]{He} + \nuclide{\alpha}$}{6Li+p->3He+4He} two-nucleon transfer]{\label{figpnTransferRuoloShell2s}%
			Points are the same in \cref{figdTransferConOSenzaL2}: direct data was rescaled as in \cref{figDatiCorrettiAdiabatic}. Red dot-dashed and violet dashed lines are the second-order DWBA $\nuclide{p}+\nuclide{n}$ transfer calculation performed as detailed in text, using the \nuclide[6]{Li} wave-function shown respectively in the top and bottom panel of \cref{figPDF6Li}. Brown dotted line is the same calculation but using as \nuclide[6]{Li} wave-function only the component with transferred nucleons in the $2s$ shell. %
			Symbols in the legend have the same meaning in \cref{figPDF6Li}. %
		}
	\end{figure}

	\Cref{figpnTransferRuoloShell2s} presents the computed astrophysical factors for the $\nuclide[6]{Li} + \nuclide{p} \to \nuclide[3]{He} + \nuclide{\alpha}$ reaction described as a two-nucleon transfer process, %
	which are commented in detail in the following paragraphs.
	Similarly to what was found in the deuteron-transfer case, the predicted absolute value for the cross-sections is approximately of the correct order of magnitude throughout the energy range under study, but the experimental cross-sections are not reliably reproduced.

\subsubsection{Role of \texorpdfstring{$2s$}{2s} shell on the cross-section}\label{secTNTCrossSectionRoleOf2s}
	
	Even though a reasonable assignment for all weights and phases in the components of the \nuclide[6]{Li} wave-function was found in \cref{secTNTCalcoliPraticiDescrizioneWFSimultaneo}, %
	it is interesting to consider alternative choices, as a mean to study the impact of each component on %
	the final result. %
	\Cref{figpnTransferRuoloShell2s} %
	compares %
	the calculations performed using either of the two \nuclide[6]{Li} wave-functions whose probability density function is reported in \cref{figPDF6Li}. 
	While the results are qualitatively similar in both cases, it can be seen that the \nuclide[6]{Li} featuring a longer “clustered” tail (upper panel in \scref{figPDF6Li}, red dot-dashed line in \scref{figpnTransferRuoloShell2s}) yields higher absolute cross-sections and enhances more the reaction at lower energies (see also \cref{figCalcoliRiscalatiSuiDati} later). Similarly, it is found that the opposite sign choice for the $2s$-shell component of \nuclide[6]{Li} wave-function (lower panel in \scref{figPDF6Li}, violet dashed line in \scref{figpnTransferRuoloShell2s}), which favours “cigar-like” configurations, also hinders the
	transfer reaction, especially at lower energies, with respect to the case where the $2s$ shell is excluded from the structure. %
	This result was found to hold in %
	all two-particle transfer calculations performed for the present project %
	(also those not shown in this thesis), %
	including different choices for the form (prior or post) employed, the adopted potentials, the components included in the reactants structure.
	Similar conclusions are also often drawn when studying two-neutron transfer processes,
	where correlations between the transferred nucleons affect the reaction cross-section and are connected to pairing phenomena, %
	see e.g.~\cite{Oertzen2001}.
	
	\Cref{figpnTransferRuoloShell2s} %
	also shows the cross-section associated to only the $2s$-shell component of \nuclide[6]{Li} (brown dotted line): note that such component has a total norm of \SI{3.4}{\percent}. %
	The total transition amplitudes can be expressed as the coherent sum of the amplitudes connected to the $1p$ and $2s$ components separately. %
	The astrophysical factors are in turn proportional to the amplitude square-modulus, through \cref{eqDefinizioneFattoreAstrofisico,eqScritturaSezioneDurtoDifferenzialeReazioneInTerminiDiAmpiezzaDiScattering}.
	Consequently, the cross-section for the complete wave-functions in \cref{figPDF6Li}, which are obtained combining the same components with different relative signs, can differ only because the phases of the amplitudes for each component show some coherence.
	This is particularly relevant considering that the $2s$-only astrophysical factor, around \SI{0.10}{\MeV \barn} at low energies, is rather small with respect the $1p$ contribution, %
	requiring a significant degree of coherence in order to yield the observed relative difference in the total cross-sections.
	It can also be seen that the phases are more coherent at lower energies, where greater differences between the two complete calculations are found, notwithstanding the fact that the
	$2s$-only absolute astrophysical factor does not change dramatically %
	throughout the energy range under study, reaching a maximum of \SI{0.15}{\MeV \barn} at about \SI{1}{\MeV} centre-of-mass energy. %

\subsubsection{Interplay between simultaneous and sequential contributions}

\begin{figure}[tbp]%
		\centering
		\includegraphics[width=\textwidth,keepaspectratio=true]{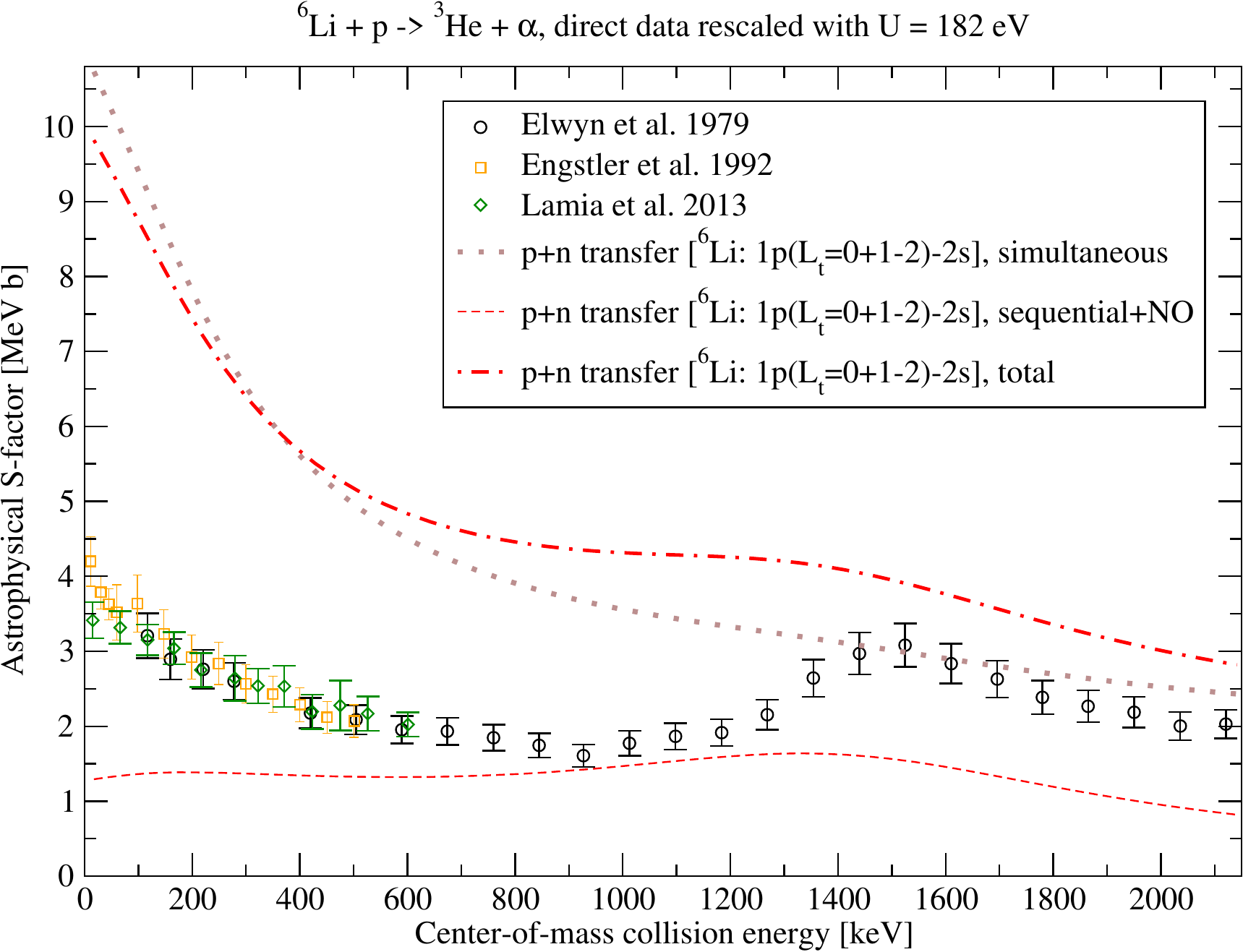}%
		\caption[Simultaneous and sequential contributions to \texorpdfstring{$\nuclide[6]{Li} + \nuclide{p} \to \nuclide[3]{He} + \nuclide{\alpha}$}{6Li+p->3He+4He}]{\label{figpnTransferSimSeqTot}%
			Points and red dot-dashed line are the same in \cref{figpnTransferRuoloShell2s}: direct data was rescaled as in \cref{figDatiCorrettiAdiabatic}. %
			Brown dotted and red dashed lines are the simultaneous and sequential (including non-orthogonality) contribution to the same calculation.
		}
	\end{figure}

\begin{figure}[tbp]%
		\centering
		\includegraphics[width=0.7\textwidth,keepaspectratio=true]{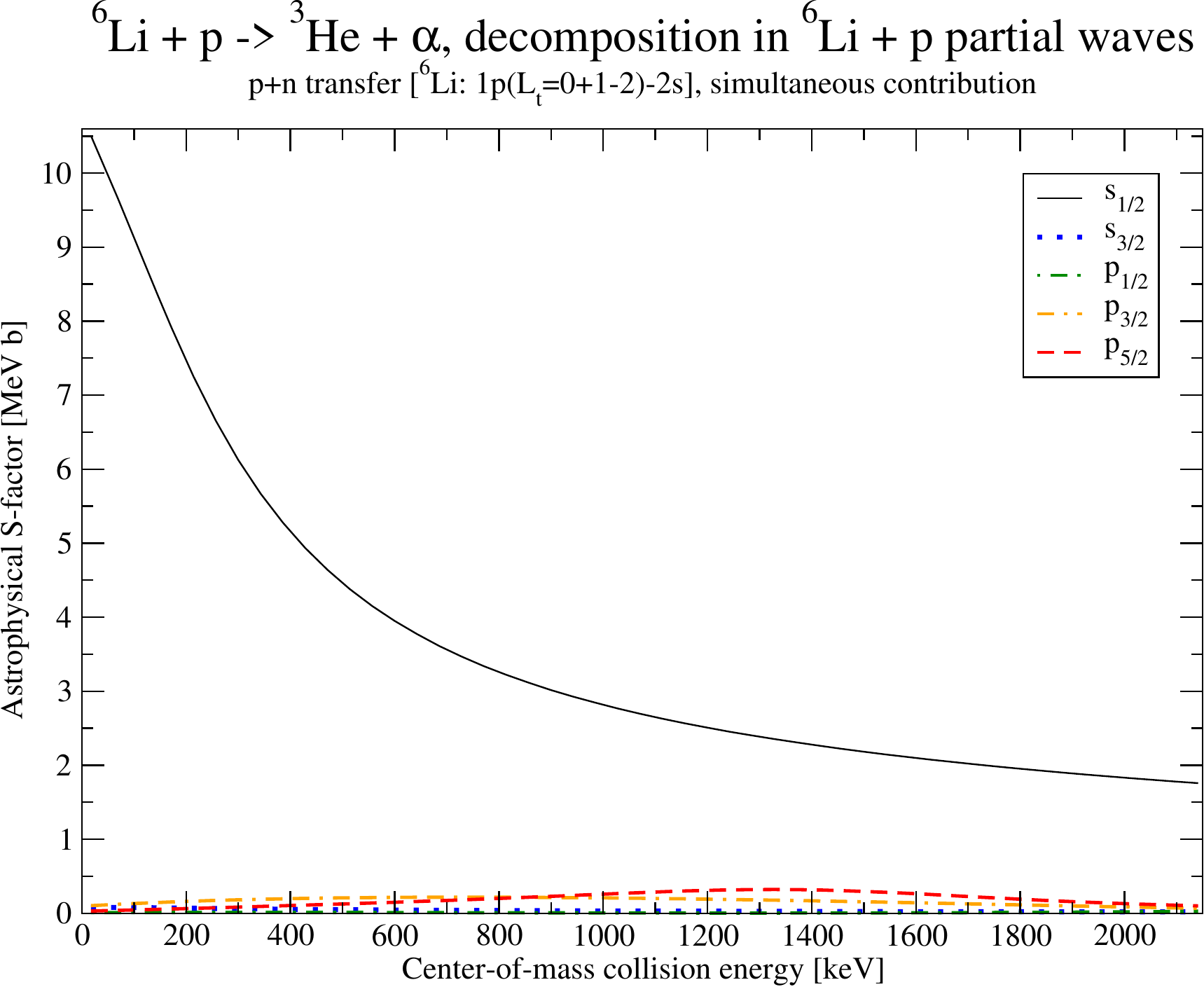} %
		
		\includegraphics[width=0.7\textwidth,keepaspectratio=true]{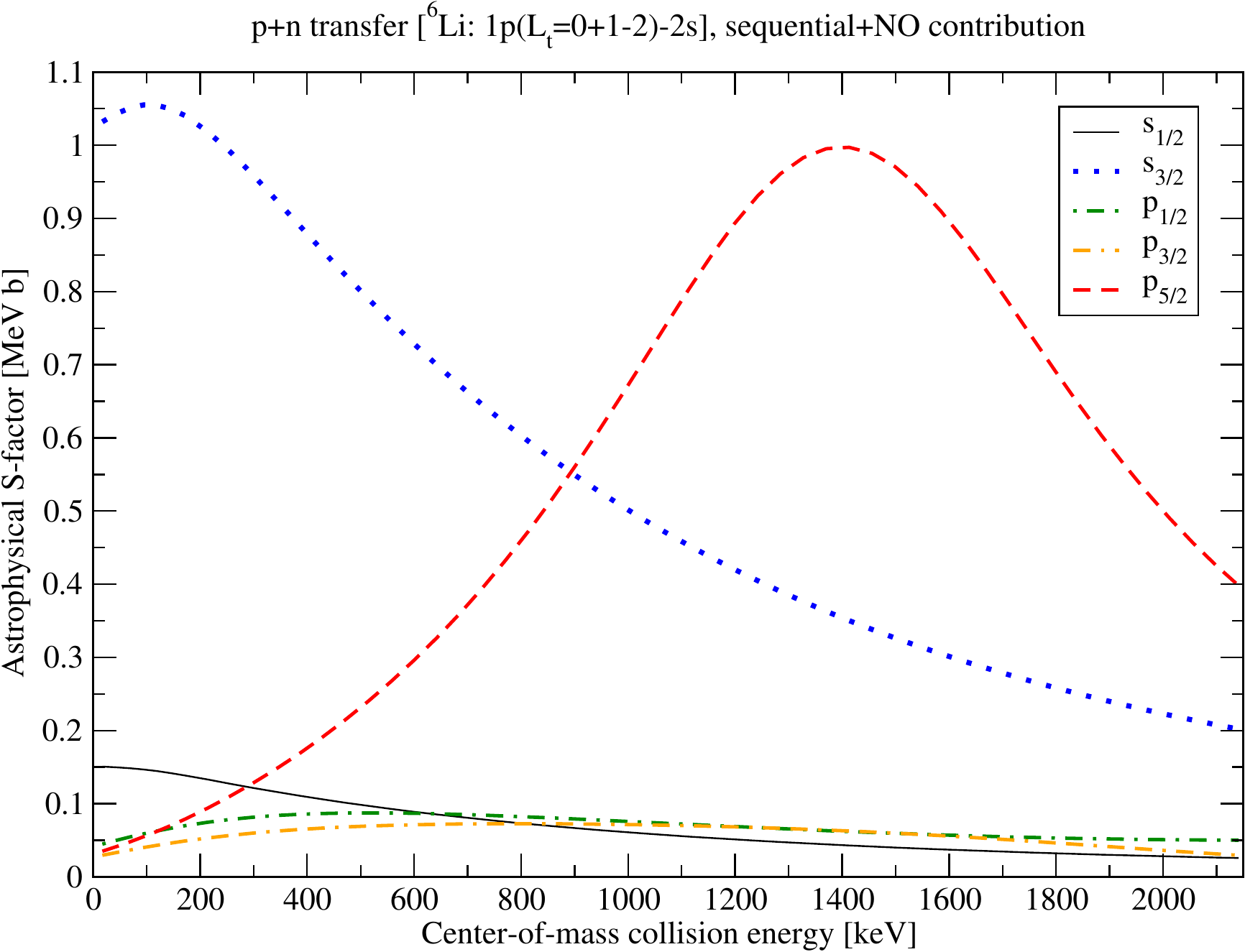}
		
		\includegraphics[width=0.7\textwidth,keepaspectratio=true]{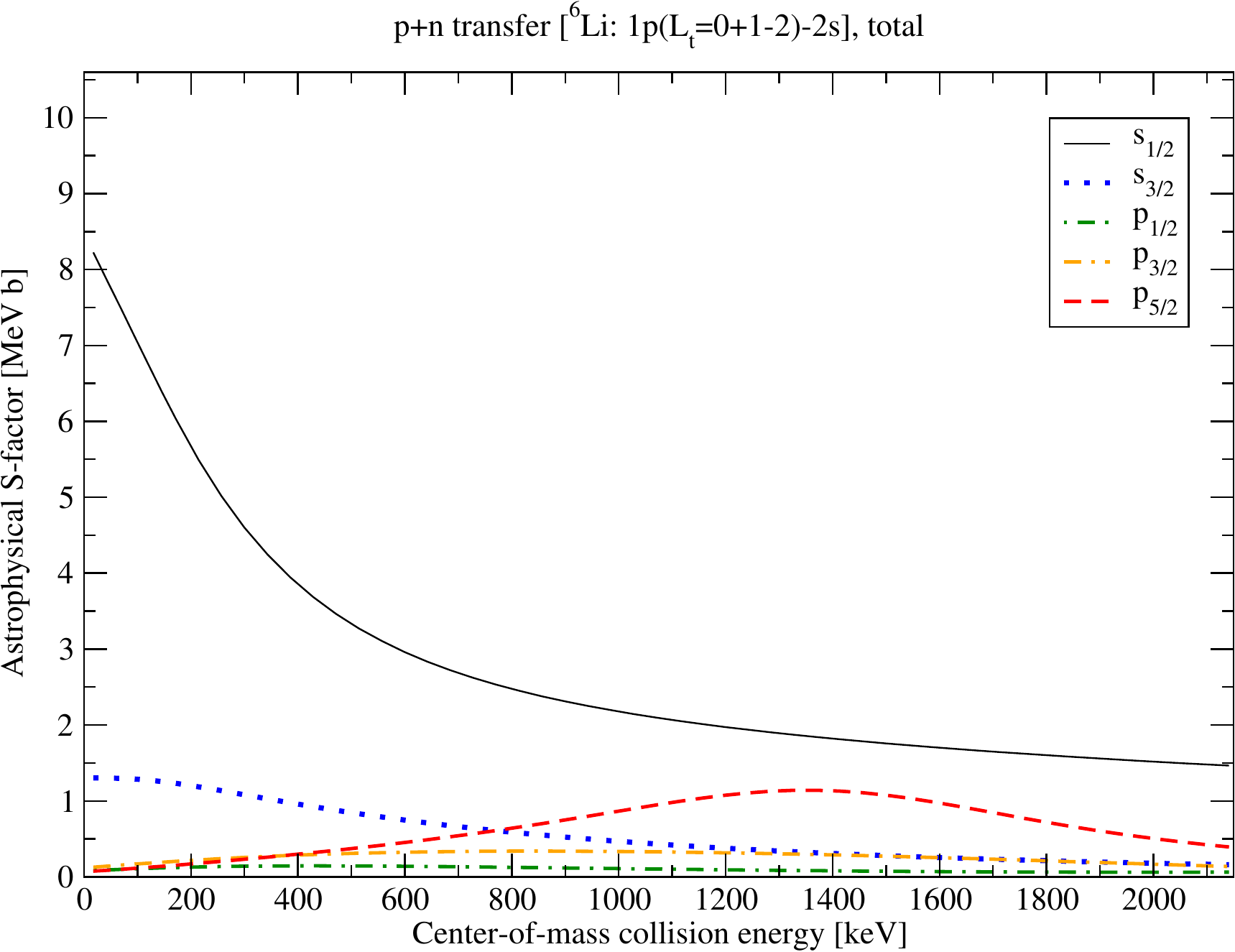}%
		\caption[\texorpdfstring{$\nuclide[6]{Li} + \nuclide{p} \to \nuclide[3]{He} + \nuclide{\alpha}$}{6Li+p->3He+4He} two-nucleon-transfer partial-wave expansion]{\label{figpnTransferPartialWaveExpansions}%
			Same as \cref{figdTransferPartialWaveExpansions} for the astrophysical factors in \cref{figpnTransferSimSeqTot}
			Upper, middle and bottom panel refer respectively to the simultaneous, sequential, and total calculations. %
		}
	\end{figure}

\begin{figure}[tbp]%
		\centering
		\includegraphics[width=0.7\textwidth,keepaspectratio=true]{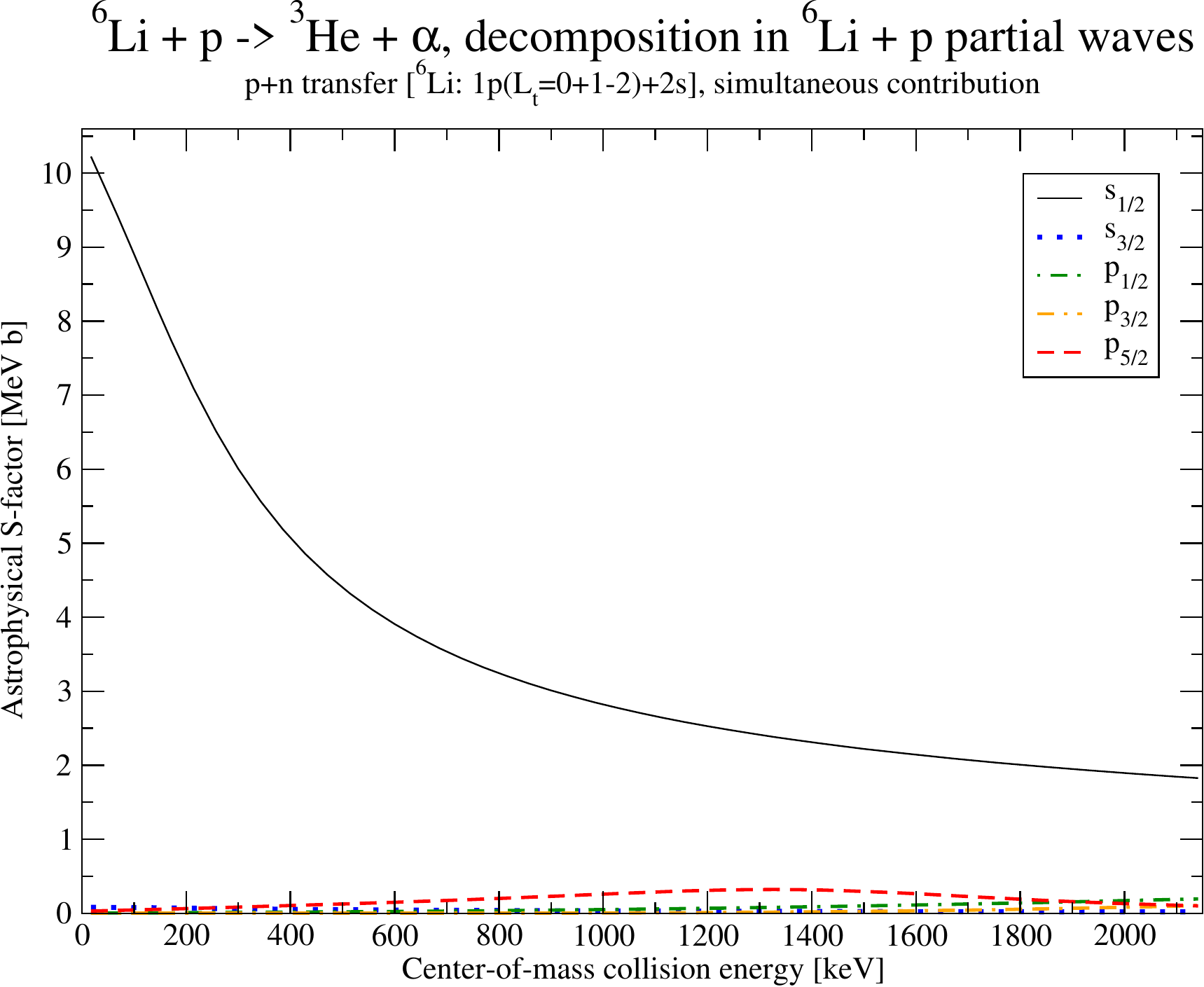} %
		
		\includegraphics[width=0.7\textwidth,keepaspectratio=true]{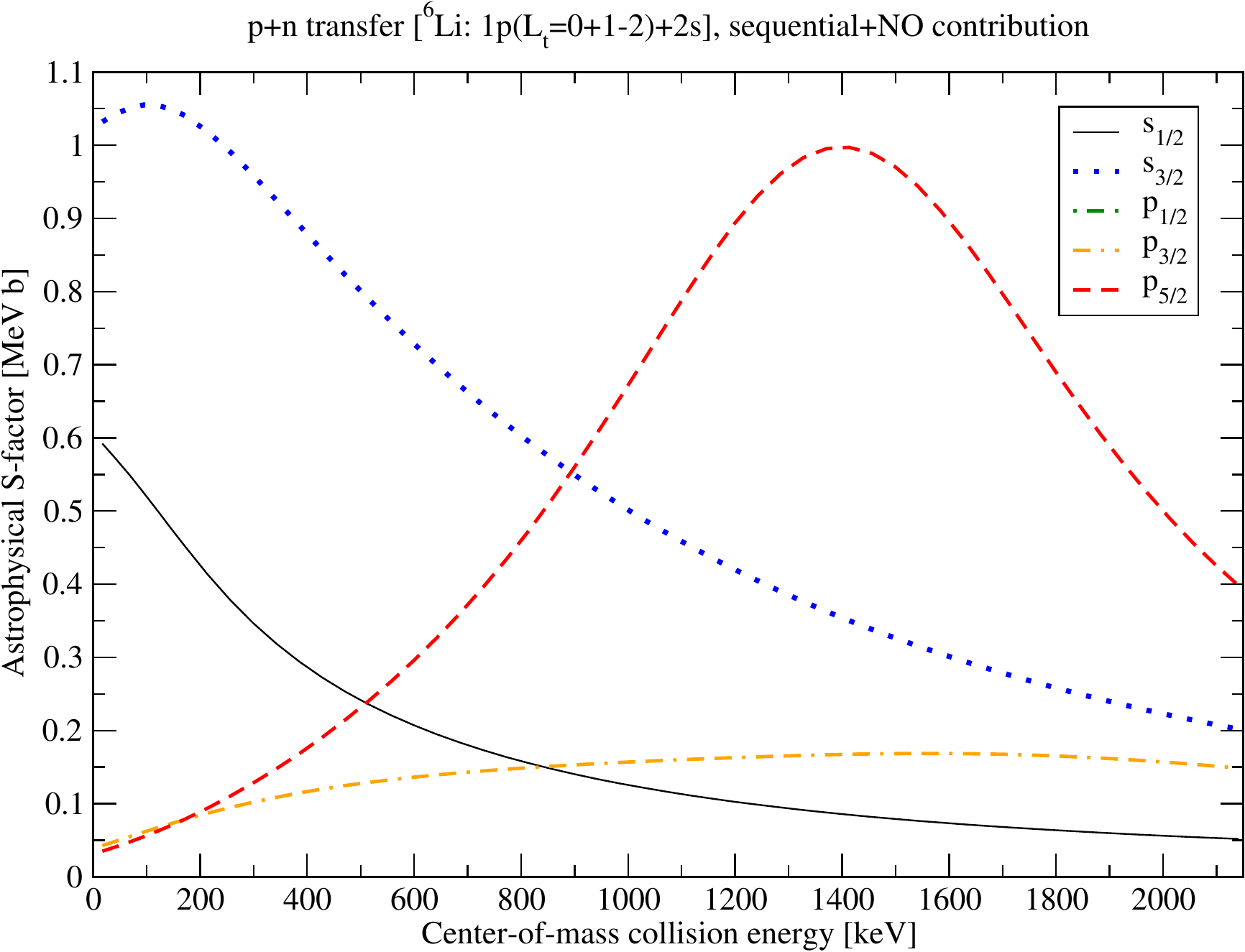}
		
		\includegraphics[width=0.7\textwidth,keepaspectratio=true]{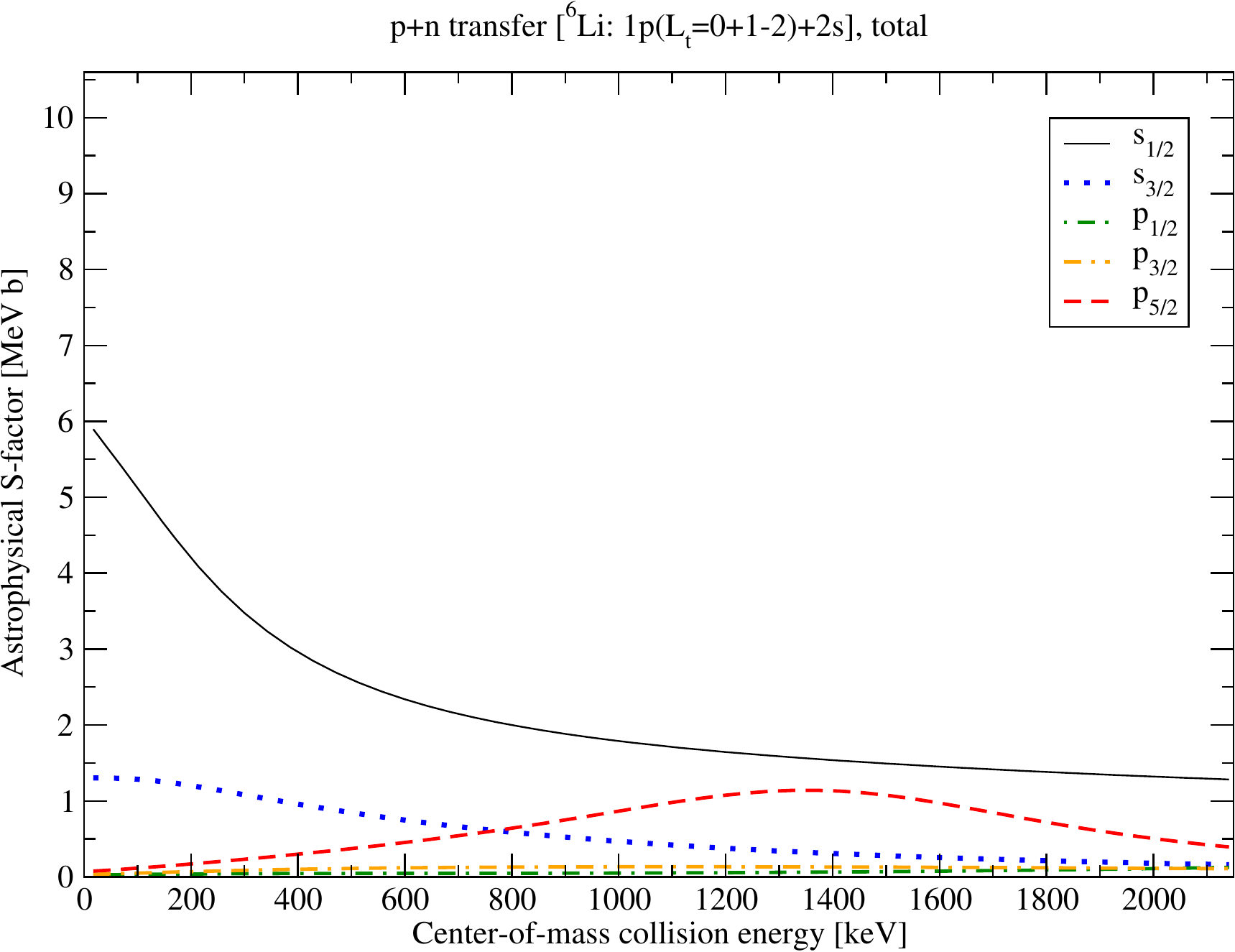}%
		\caption[Two-nucleon-transfer partial-wave expansion for alternative \texorpdfstring{\nuclide[6]{Li}}{6Li}]{\label{figpnTransferPartialWaveExpansionsCaso1pPiu2s}%
			Same as \cref{figpnTransferPartialWaveExpansions} but for the astrophysical factor in \cref{figpnTransferRuoloShell2s} associated to the less-strongly-clustered \nuclide[6]{Li} wave-function shown in the bottom panel of \cref{figPDF6Li}.
		}
	\end{figure}

	\Cref{figpnTransferSimSeqTot} %
	shows the simultaneous and sequential contributions to the calculation already shown as the red dot-dashed line in \cref{figpnTransferRuoloShell2s} (more-strongly clustered \nuclide[6]{Li} wave-function). %
	The total transition amplitude is %
	the coherent sum of the amplitudes of each component. %
	It can be seen that the amplitudes for simultaneous and sequential terms are in %
	phase opposition at low collision energy, resulting in a %
	destructive interference. %
	Such trend partially changes at higher energies, particularly in the energy range around the resonance at \SI{1.5}{\MeV}, where the two contributions are almost in phase quadrature (but still in slight opposition), %
	forming a modest but visible bump.
	The role of interference can be analysed in greater detail by checking
	the expansion into the initial-channel partial-waves, %
	shown in \cref{figpnTransferPartialWaveExpansions}.
	The only wave generating a non-negligible contribution to the simultaneous term (top panel in figure) is the $s1/2$ (\nuclide[6]{Li}--\nuclide{p} with relative orbital angular momentum 0 and total angular momentum $1/2$).
	By comparing the magnitude of this contribution between the simultaneous, sequential (middle panel) and complete (bottom panel) calculations, %
	it can be seen that the simultaneous and sequential $s1/2$ transition amplitudes are in very strong phase opposition.
	All other waves feel no significant interference as they do not contribute appreciably to the simultaneous term.
	The same analysis can be carried out for the calculation involving the less-strongly-clustered \nuclide[6]{Li} wave-function (violet dashed line in \cref{figpnTransferRuoloShell2s}): the partial-wave expansion of simultaneous and sequential contributions and their total is shown in \cref{figpnTransferPartialWaveExpansionsCaso1pPiu2s}.
	The situation is qualitatively the same found in \cref{figpnTransferPartialWaveExpansions}, but the simultaneous contribution is slightly smaller and, most importantly, the $s1/2$ wave of the sequential term is much stronger, causing a greater destructive interference at low energies.

\subsubsection{Allowed couplings in sequential and simultaneous terms}

	In passing, \cref{figpnTransferPartialWaveExpansions,figpnTransferPartialWaveExpansionsCaso1pPiu2s} %
	display the different character of simultaneous and sequential contributions with regard to the allowed couplings. %
	In the simultaneous, one-step process, the internal state of the transferred system is conserved, thus the only contributions to the transfer are related to components of the \nuclide[6]{Li} and \nuclide[3]{He} wave-functions sharing the same quantum numbers for the transferred internal degrees of freedom, in particular $l$, $S$ (notation as in \cref{secTNTCalcoliPraticiDescrizioneWFSimultaneoCoordinatevt}) and the total transferred isospin. At the same time, as in the deuteron-transfer case (see \cref{secCalcolidDWBARisonanza7Be}), the initial-channel waves $s3/2$ and $p5/2$ can be accessed only through a coupling between components with different core-transferred orbital angular momentum, $L$. In particular, since the \nuclide[3]{He} wave-function employed here includes only components with $S=1$ and either ($l=0$, $L=0$) or ($l=2$, $L=2$), a coupling to the aforementioned partial waves is possible only if the \nuclide[6]{Li} wave-function includes ($S=1$, $l=0$, $L=2$) or ($S=1$, $l=2$, $L=0$) components, and thus $\mathcal L = 2$ (this then couples with the projectile-target orbital angular momentum to form the total ``\nuclide[7]{Be}'' orbital angular momentum, which is conserved). Such components, albeit small, are indeed included in the wave-functions in \cref{figPDF6Li}, and a small $p5/2$ contribution is visible in the simultaneous contribution in \cref{figpnTransferPartialWaveExpansions,figpnTransferPartialWaveExpansionsCaso1pPiu2s}%
	\footnote{It was verified that $s3/2$ and $p5/2$ contributions disappear %
		if the $\mathcal L = 2$ component is removed.}. %
	
	The sequential process is not bound by the aforementioned constraint. Each transfer step proceeds independently, coupling all terms which can yield the desired initial, intermediate and final states. All partial waves can thus be observed regardless of the precise admixture of single-particle wave-functions, although the details of the corresponding cross-section still depend on the adopted state. %

\subsubsection{Impact of the intermediate \texorpdfstring{\boldmath{\nuclide[5]{Li}}}{5Li} state binding energy}\label{secThreeParticleWF6LiImpactIntermediateBindingEnergy} %
	
	A second calculation was performed changing the mass of the fictitious \nuclide[5]{Li} state (together with the corresponding $Q$-value for the intermediate partition), and rescaling the volume term of the \nuclide{\alpha}--\nuclide{p} and \nuclide[5]{Li}--\nuclide{n} binding potentials to adjust the %
	binding energies accordingly.
	
	Let $B_{\nuclide{\alpha} \nuclide{p}}$ be the binding energy assigned to the $\nuclide{\alpha}+\nuclide{p}$ system, and similarly $B_{\nuclide{Li} \nuclide{n}}$ the binding energy of the $\nuclide[5]{Li}+\nuclide{n}$ system.
	Their sum %
	must coincide with the $Q$-value for $\nuclide{\alpha}+\nuclide{p}+\nuclide{n} \to \nuclide[6]{Li}$ (\SI{3.70}{\MeV}), for consistency with the \nuclide[6]{Li} and \nuclide{\alpha} assigned masses. %
	Let $B_{r}$ be the ratio $B_{\nuclide{Li}\nuclide{n}} / B_{\nuclide{\alpha}\nuclide{p}}$.
	If \nuclide[5]{Li} were bound, the most natural choice for $B_{r}$ %
	would have been simply the physical value (as is done in this work when constructing the \nuclide[3]{He} state). In the present case, however, said ratio is essentially arbitrary
	\footnote{Also note that fixing the physical value for the \nuclide[6]{Li} root-mean-square radius does not fix the ratio between the binding energies, as the %
		binding potentials (which must be fictitious as the intermediate state they describe) depend on several parameters.},
	and may represent a phenomenological way to control the clustering strength.
	
	\cite[app.~A]{Potel2013} discusses, from a formal point of view, %
	the connection between the binding energy of the transferred nucleons, their degree of correlation, and %
	the relative importance of simultaneous and sequential terms.
	The case $B_{\nuclide{Li}\nuclide{n}} = B_{\nuclide{\alpha}\nuclide{p}}$ %
	is connected with the ``independent-particle'' limit for the two transferred nucleons, meaning that the removal of the first nucleon has no influence on the energy required to remove the second one.
	This choice %
	appears to be the most appropriate in the present construction, %
	where, when computing the simultaneous term, the wave-functions supplied in input describe the motion between the core and a single nucleon, independently of the second one (in ``v'' coordinates, see \cref{secTNTSimultaneoTeoriaSchemaCoreComposite}).
	In the limit $B_{\nuclide{\alpha}\nuclide{p}} \ll B_{\nuclide{Li}\nuclide{n}}$, %
	instead, %
	once a nucleon %
	is removed from \nuclide[6]{Li}, the other will drip almost spontaneously, implying a correlation between the transferred particles.
	In \cite[app.~A.2]{Potel2013}, it is discussed that in this limit the transition amplitude is expected to be dominated by the simultaneous contribution.
	If it is possible to supply a composite wave-function expressed in the appropriate Jacobi coordinates, %
	this configuration is also more consistent with the physical one for this work, since \nuclide[5]{Li} is actually unbound. In any case, %
	as long as the intermediate state is treated as a fictitious bound state, it is not possible to assign it exceedingly small binding energies, as that would produce $\nuclide{\alpha}+\nuclide{p}$ one-body wave-functions with abnormally large radii, generating inaccurate results.
	
\begin{figure}[tbp]%
		\centering
		\includegraphics[width=\textwidth,keepaspectratio=true]{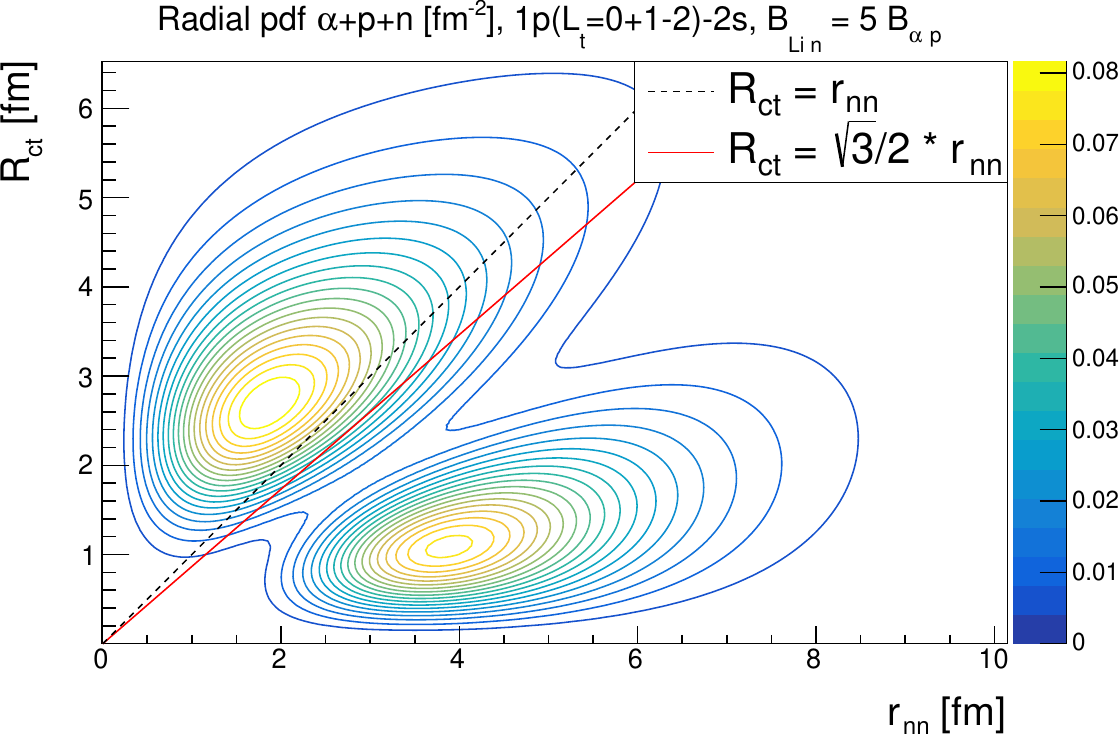}
		\caption[\texorpdfstring{\nuclide{\alpha}+\nuclide{p}+\nuclide{n}}{a+p+n} radial probability density function]{\label{figPDF6LiBr5}%
			Same as the upper panel of \cref{figPDF6Li}, but with ratio of single-nucleon binding energies $B_{\nuclide{Li}\nuclide{n}} / B_{\nuclide{\alpha}\nuclide{p}}$ equal to 5 (instead of 1).
		}
	\end{figure}

	The test discussed here was %
	carried out setting $B_r$ to either 3 or 5. %
	The peak structure of the three-particle \nuclide[6]{Li} wave-function remains qualitatively unaltered between the two calculations (in particular, the position of the peak maxima remains identical). \Cref{figPDF6LiBr5} shows the \nuclide[6]{Li} radial probability density function for $B_r = 5$ and the combination of components yielding the best agreement with three-body calculation (the one labelled ``$1p-2s$''): this can be compared with the upper panel of \cref{figPDF6Li}. The differences are difficult to notice visually, but by comparing the %
	expectation value of $R_{ct}^2$ and $r_{nn}^2$ for each wave-function,
	it can be seen that for greater $B_r$ both peaks are slightly wider. %
	At the same time, the maximum of both peaks is slightly smaller for greater $B_r$.

\begin{figure}[tbp]%
		\centering
		\includegraphics[width=\textwidth,keepaspectratio=true]{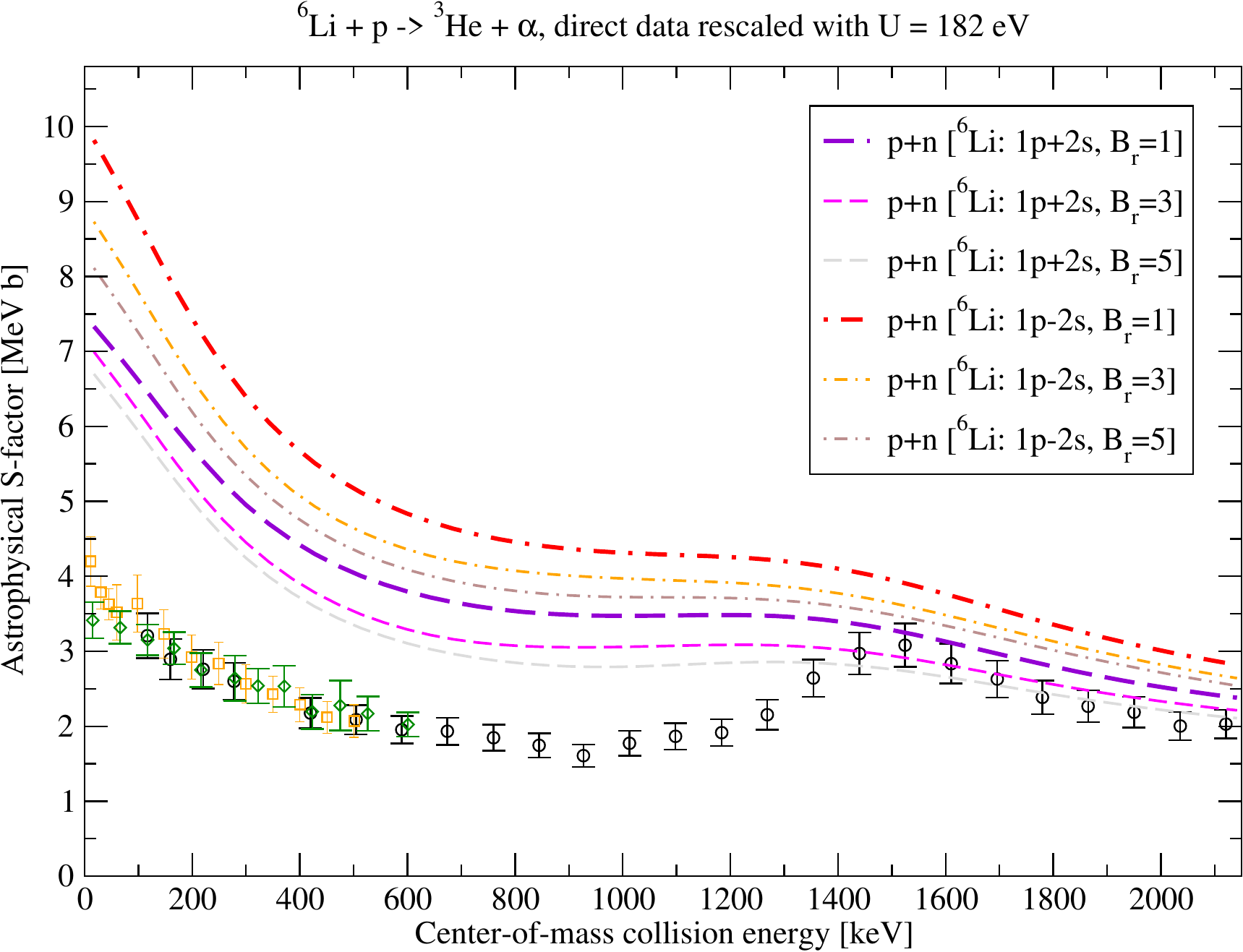}%
		\caption[Impact of the \texorpdfstring{\nuclide[5]{Li}}{5Li} binding energy on the \texorpdfstring{$\nuclide[6]{Li} + \nuclide{p} \to \nuclide[3]{He} + \nuclide{\alpha}$}{6Li+p->3He+4He} cross-section]{\label{figpnTransferImpattoBE5Li}%
			Points, violet thick dashed line and red thick dot-dashed line are the same in \cref{figpnTransferRuoloShell2s}. Direct data was rescaled as in \cref{figDatiCorrettiAdiabatic}. Thin lines represent the same calculation as the thick line with equal shape (dashed or dot-dashed) but adopting $B_r = 3$ (magenta and orange lines) or 5 (grey and brown lines), see text for details.
		}
	\end{figure}

	\Cref{figpnTransferImpattoBE5Li} shows the %
	total astrophysical $S$-factor for each of the aforementioned choices of $B_r$, %
	using both the \nuclide[6]{Li} wave-function constructions previously discussed in relation with %
	\cref{figpnTransferRuoloShell2s}, which differ by the sign assigned to the $2s_{1/2} \times 2s_{1/2}$ component in \cref{tabAmplitudes6LiThreeParticlejj}.
	It can be seen that in both cases the cross-section is smaller for greater $B_r$, somewhat in contrast with the expectation that a greater value of $B_r$ represents configuration where the transferred nucleons are more correlated. The relative variation in the cross-section is not negligible, but it is noted that the difference between setting $B_r = 1$ or 5 %
	is smaller than the difference induced by inverting the phase of the $2s_{1/2} \times 2s_{1/2}$ contribution. Additionally, the general result discussed when commenting \cref{figpnTransferRuoloShell2s} applies for all values of $B_r$.
	The qualitative impact %
	of the intermediate-state binding energy on the cross-section is different for the contribution of %
	each component of the wave-function. For example, truncating the \nuclide[6]{Li} state to the $1p_{3/2} \times 1p_{3/2}$ component (not shown in figure), the cross-section for $B_r = 5$ at a centre-of-mass collision energy of \SI{17}{\keV} is greater than the one for $B_r = 1$, by less than \SI{4}{\percent}. On the contrary, considering only the $2s_{1/2} \times 2s_{1/2}$ component, the $B_r = 5$ cross-section is smaller than the $B_r = 1$ one by about \SI{25}{\percent}. This complicates the process of predicting the overall effect for the total wave-function starting from the analysis of the single-nucleon wave-functions provided in input. %
	For instance, it can be seen from \cref{figpnTransferImpattoBE5Li} that the ``$1p-2s$'' configuration is more sensitive, with respect to the ``$1p+2s$'' one, to the value of $B_r$. %
	This is explained considering that increasing $B_r$ reduces the importance of the constructive interference between the $1p$- and $2s$-shell contributions taking place in the ``$1p-2s$'' case, and similarly reduces the impact of the destructive interference found in the ``$1p+2s$'' configuration. %
	Such results in turn arises because %
	the $2s_{1/2} \times 2s_{1/2}$ component is more sensitive to $B_r$, with respect to the other components.

\begin{figure}[tbp]%
		\centering
		\includegraphics[width=\textwidth,keepaspectratio=true]{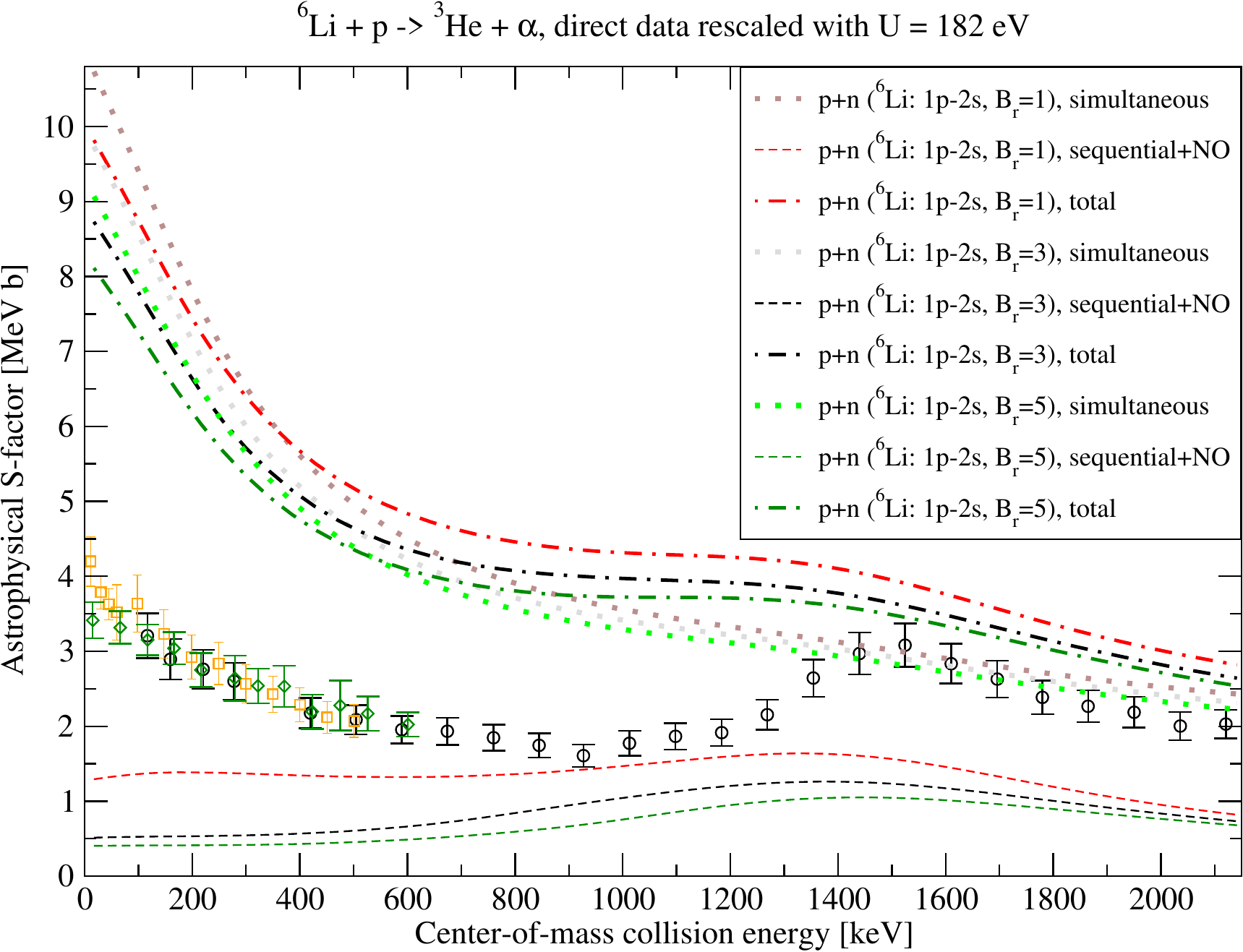}%
		\caption[Impact of the \texorpdfstring{\nuclide[5]{Li}}{5Li} binding energy: simultaneous and sequential contributions]{\label{figpnTransferSimSeqTotBr5}%
			Points and red and brown lines are the same in \cref{figpnTransferSimSeqTot}. Direct data was rescaled as in \cref{figDatiCorrettiAdiabatic}. The corresponding grey and black lines refere to the same calculation but setting $B_r = 3$ (see text), while light and dark green lines refer to $B_r = 5$. All dotted, dashed and dot-dashed lines are respectively the the simultaneous, sequential (including non-orthogonality) and total cross-section.
		}
	\end{figure}

	In general, the change in the binding energies is expected to influence the cross-section at both first and second order.
	First of all, the $B_r$ value %
	influences the intermediate partition $Q$-value, %
	which is especially relevant for the sequential term. %
	The reaction under study is such that the $Q$-value for $\nuclide[5]{Li}+\nuclide{d} \to \nuclide{\alpha} + \nuclide[3]{He}$ %
	is always positive (for $B_{\nuclide{Li}\nuclide{n}} > 0$) and $\gtrsim \SI{2}{\MeV}$, while the $Q$-value for the first step, $\nuclide[6]{Li}+\nuclide{p} \to \nuclide[5]{Li} + \nuclide{d}$, here labelled $Q_{1}$ for brevity, can take both signs. %
	In particular, $Q_{1}$ is greater for smaller values of $B_{r}$, and it is about zero for $B_{r} \approx \num{1.51}$. %
	The $Q$-value for the second step will then be just the difference between the $Q$-value for the complete process and $Q_1$.
	Thus, changing $B_r$ is expected to affect the amplitude of the two steps in an opposite manner.
	Furthermore, changing the binding energies also influences the root-mean-square radius and the asymptotic normalisation coefficient of the one-body wave-functions, which affect both simultaneous and sequential terms.
	\Cref{figpnTransferSimSeqTotBr5} shows the same calculations in \cref{figpnTransferSimSeqTot} and compares them with the results for $B_r = 3$ and 5. %
	It can be seen that the sequential contribution is smaller for higher $B_r$, particularly when comparing the cases $B_r = 1$ and 3, in which the $Q$-value for the intermediate step changes sign. However, also the simultaneous cross-section is reduced, and the importance of the interference between the two terms is qualitatively the same for all cases, in contrast with the aforementioned expectation that for high $B_r$ the sequential contribution would become negligible.
	The observed effects may be connected to the different asymptotic trend and scaling of the wave-function in each case. The issue is currently under study. %

\subsubsection{Prior-post equivalence} %
	
	\paragraph{Impact of the three-particle wave-function construction scheme for central optical potentials}
	The calculations in \cref{figpnTransferRuoloShell2s,figpnTransferSimSeqTot,figpnTransferPartialWaveExpansions,figpnTransferPartialWaveExpansionsCaso1pPiu2s} were performed adopting the post form for the simultaneous calculation and the post-post form for the sequential process, including the associated non-orthogonality term.
	It was verified that, if all optical potentials are taken to be central, %
	then the computed cross-sections are very similar %
	for all forms (see the discussion in \cref{secDWBAPriorPost}). %

\begin{figure}[tbp]%
		\centering
		\includegraphics[width=\textwidth,keepaspectratio=true]{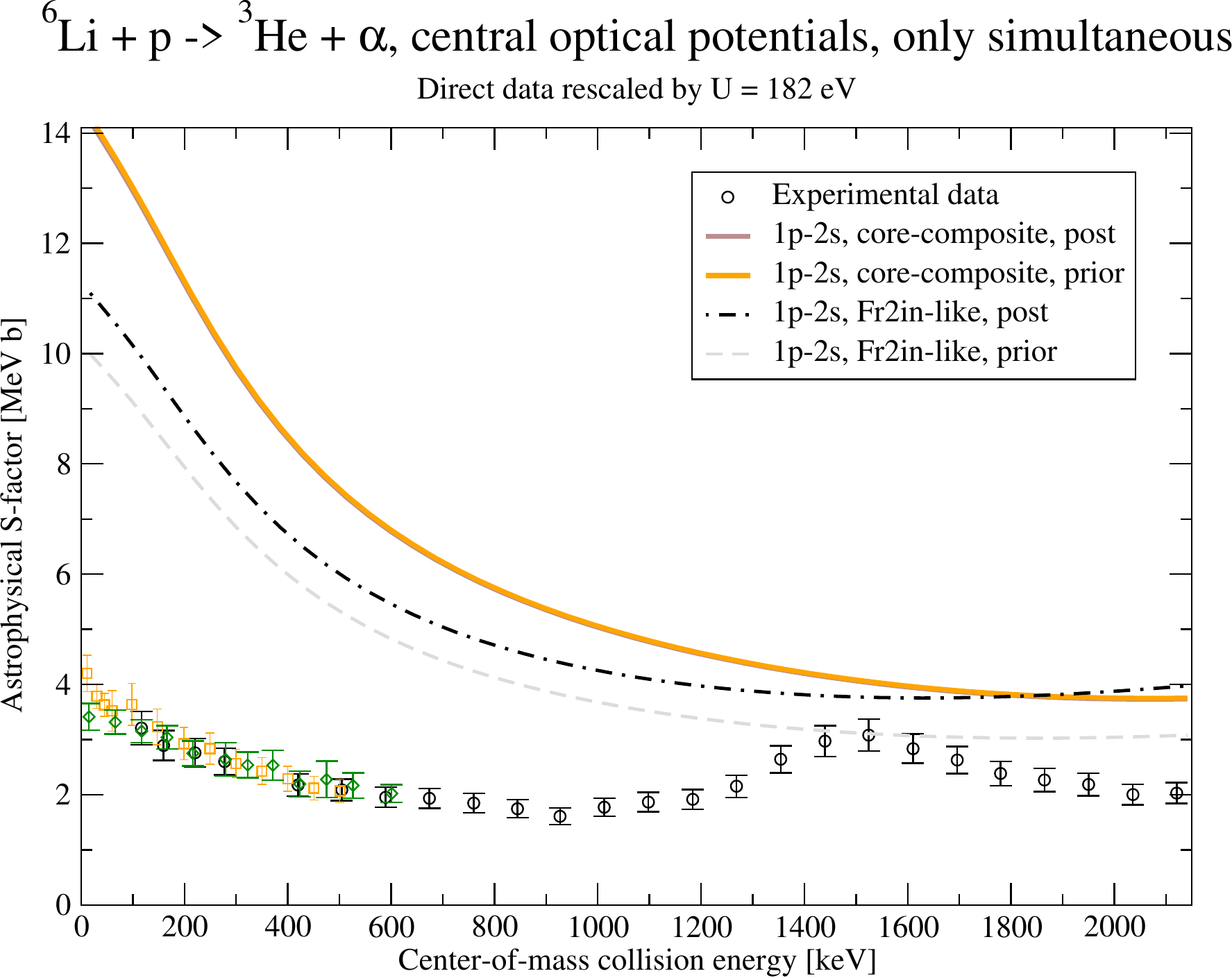}%
		\caption[Impact of the \texorpdfstring{\nuclide[5]{Li}}{5Li} binding energy: simultaneous and sequential contributions]{\label{figpnTransferSimPriorPostOtticiCentrali}%
			Points are the same in \cref{figpnTransferSimSeqTot}. Direct data was rescaled as in \cref{figDatiCorrettiAdiabatic}. All lines refer to simultaneous-only calculations performed removing all non-central components from projectile-target potentials. Brown solid line (%
			overlapping with the orange line) is the analogous of the brown dotted line in \cref{figpnTransferSimSeqTot}. Orange solid line  is the same calculation in prior form. Black dot-dashed and grey dashed lines are the analogous post- and prior-form calculations in the ``\textsc{Fr2in}-like'' scheme described in the text.%
		}
	\end{figure}

	As an example, the brown dot-dashed line in \cref{figpnTransferSimPriorPostOtticiCentrali} represents the same simultaneous-only post-form calculation shown as brown dotted line in \cref{figpnTransferSimSeqTot}, %
	but removing all non-central components from all projectile-target potentials
	\footnote{In \textsc{Fresco}, core-core potentials never include non-central terms, %
		see again \cref{secDWBAPriorPost}.}.
	The orange dot-dashed line in \scref{figpnTransferSimPriorPostOtticiCentrali} shows instead the same calculation in prior form: %
	the difference between the two cases is always smaller than \SI{0.5}{\percent}.
	It is underlined that the %
	simplification of the potentials considered in \scref{figpnTransferSimPriorPostOtticiCentrali}
	alters significantly the predictions on elastic scattering, thus the corresponding cross-sections %
	are not expected to be accurate, and only serve the purpose of investigating sources of prior-post discrepancy. %
	The aforementioned very good agreement %
	between prior and post forms is due to the use of the ``core-composite-like'' scheme %
	proposed in this work, and discussed in \cref{secTNTSimultaneoTeoriaSchemaCoreComposite}, %
	in which the kinetic terms of the Hamiltonian are treated more accurately than in the ``heavy-ion'' schemes (discussed in \cref{secHeavyIonScheme}) commonly employed in literature. This feature makes the approach adopted here %
	especially suitable for calculations involving light ions. %
	As shown in \cref{eqDifferenzaHamiltonianaPriorPostPerSchemaAdaptedCoreCompositeLike}, the adopted prior and post Hamiltonians are not strictly identical, %
	but are found to be sufficiently similar to yield consistent results, despite the relatively small mass of the core particles. %
	
	For comparison, the black and grey lines in \cref{figpnTransferSimPriorPostOtticiCentrali} represent the same post- and prior-form calculations, performed employing a construction of the three-particle wave-functions similar %
	to the one adopted in the \textsc{Fr2in} code \cite{BrownReactionCodes}.
	For the \nuclide[3]{He} nucleus, the two single-nucleon wave-functions in the ``v'' coordinate system (representing the motion between the core and each transferred nucleon) are taken to be identical, and are both assigned a kinetic term and a potential referring to the \nuclide{d}--\nuclide{p} motion (thus, this is a variation of the ``heavy-ion composite-composite'' scheme in \cref{secHeavyIonScheme}). For greater consistency, in this test calculation the adopted potential was the same \nuclide{d}--\nuclide{p} interaction employed in the other calculations (column ``\nuclide{p} -- \nuclide{d} ($L=0$)'' in \cref{tabParametriNumericiPotenziali3He}). 
	For the \nuclide[6]{Li} nucleus, the two single-nucleon wave-functions in the ``v'' coordinate system are computed as in the ``heavy-ion core-core scheme'' in \cref{secHeavyIonScheme}, assigning kinetic and potential terms pertaining to the \nuclide{\alpha}--\nuclide{p} and \nuclide{\alpha}--\nuclide{n} motion. The single-nucleon potential is the same for both wave-functions, except for the Coulomb interaction which was excluded for the \nuclide{\alpha}--\nuclide{n} motion, and the volume term which is rescaled separately in each case to obtain the desired binding energy.
	In this calculation, the same \nuclide{\alpha}--\nuclide{p} potential employed in the other calculations (reported in \cref{tabParametriNumericiPotenziali6Li}) was adopted.
	All other parameters of the calculation were the same employed to compute the brown and orange lines in \cref{figpnTransferSimPriorPostOtticiCentrali}. %
	Note that %
	the binding potentials entering this ``\textsc{Fr2in}-like'' calculation %
	are different than %
	those employed above in the ``core-composite-like'' scheme:
	this contributes to the difference in the computed simultaneous-only cross-sections.
	The prior-post consistency is instead entirely attributed to the choice of kinetic terms. In particular, the black and grey lines in \scref{figpnTransferSimPriorPostOtticiCentrali} differ by more than \SI{10}{\percent} at low energies, and more than \SI{20}{\percent} above the Coulomb barrier. %

	\paragraph{Prior-post discrepancy induced by non-central potentials}
	
\begin{figure}[tbp]%
		\centering
		\includegraphics[width=\textwidth,keepaspectratio=true]{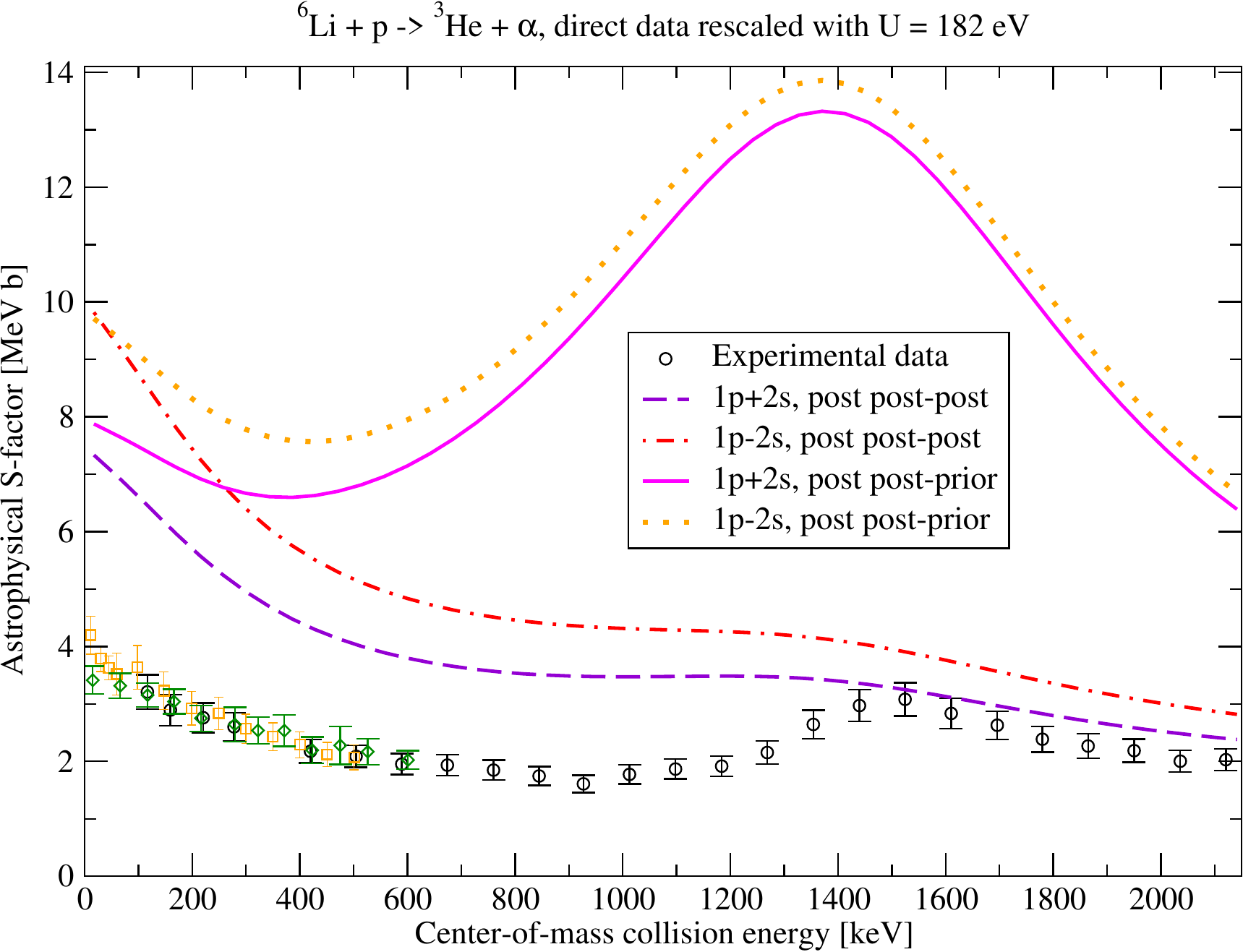}%
		\caption[Prior-post comparison for \texorpdfstring{$\nuclide[6]{Li} + \nuclide{p} \to \nuclide[3]{He} + \nuclide{\alpha}$}{6Li+p->3He+4He} two-nucleon transfer]{\label{figpnTransferPriorPost}%
			Points are the same in \cref{figdTransferConOSenzaL2}, with their %
			legend suppressed for readability: direct data was rescaled as in \cref{figDatiCorrettiAdiabatic}. Red dot-dashed and violet dashed lines are the same in \cref{figpnTransferRuoloShell2s}. Orange dotted and magenta solid lines are the same calculations but adopting the post-prior form for the sequential contribution. All calculations include the appropriate non-orthogonality terms. In the legend, $1p-2s$ and $1p+2s$ mark respectively the \nuclide[6]{Li} wave-function in the top and bottom panel of \cref{figPDF6Li}. ``post post-prior'' stands for ``simultaneous term in post form, sequential term in post-prior form'', and so on.
		}
	\end{figure}

	The projectile-target potentials employed for the calculations in \cref{figpnTransferRuoloShell2s,figpnTransferSimSeqTot,figpnTransferPartialWaveExpansions,figpnTransferPartialWaveExpansionsCaso1pPiu2s} include non-central terms. As a consequence, the results computed using different forms (prior or post) are never equivalent%
	\footnote{The overall discrepancy will still be smaller if the contribution due to the three-particle wave-functions construction is reduced.}.
	In analogy with was done in the deuteron-transfer calculation,
	the ``post'' choice in the simultaneous and first sequential step prevents the \nuclide[6]{Li}--\nuclide{p} optical potential to appear in the transition operator, which is desirable due to its strong non-central components (in particular the spin-spin one). %
	Regarding the form for the second sequential step,
	note that the possible non-central components of the \nuclide{\alpha}--\nuclide{d} interaction (which is the core-core term in the second step) are limited by the fact that the \nuclide{\alpha} particle has spin zero.
	It is then conceivable that, if it were possible to employ a set of completely realistic potentials for the calculation, %
	such non-central components %
	would resemble more %
	those of the \nuclide{\alpha}--\nuclide[3]{He} potential (also involving the \nuclide{\alpha} particle) than those of the \nuclide[5]{Li}--\nuclide{d} one.
	Additionally, the approximation %
	involved in the post form for this step is more coherent with the analogous one made in the simultaneous transfer computed in post form (in both cases, the non-central part of the same \nuclide[3]{He}--\nuclide{\alpha} potential is being neglected), thus one may suppose that the sum of both contributions will generate a more consistent result adopting the post form. %
	For comparison, \cref{figpnTransferPriorPost} shows the same calculations in \cref{figpnTransferRuoloShell2s} together with the results found %
	adopting the post-prior form for the sequential transfer. %
	It can be seen that the difference between the two forms is modest for $E \to 0$, but at greater energies the trend is dominated by a grossly overestimated resonance (peaked at approximately the correct energy).
	It is also stressed that the conclusions drawn earlier regarding the impact of clustering still apply regardless of the form in which the calculations are performed.

\subsubsection{Comparison between deuteron and two-nucleon transfer}\label{secRisultatipnTransferComparisonDeuteronAndpnTransfer}

\begin{figure}[tb]%
		\centering
		\includegraphics[width=\textwidth,keepaspectratio=true]{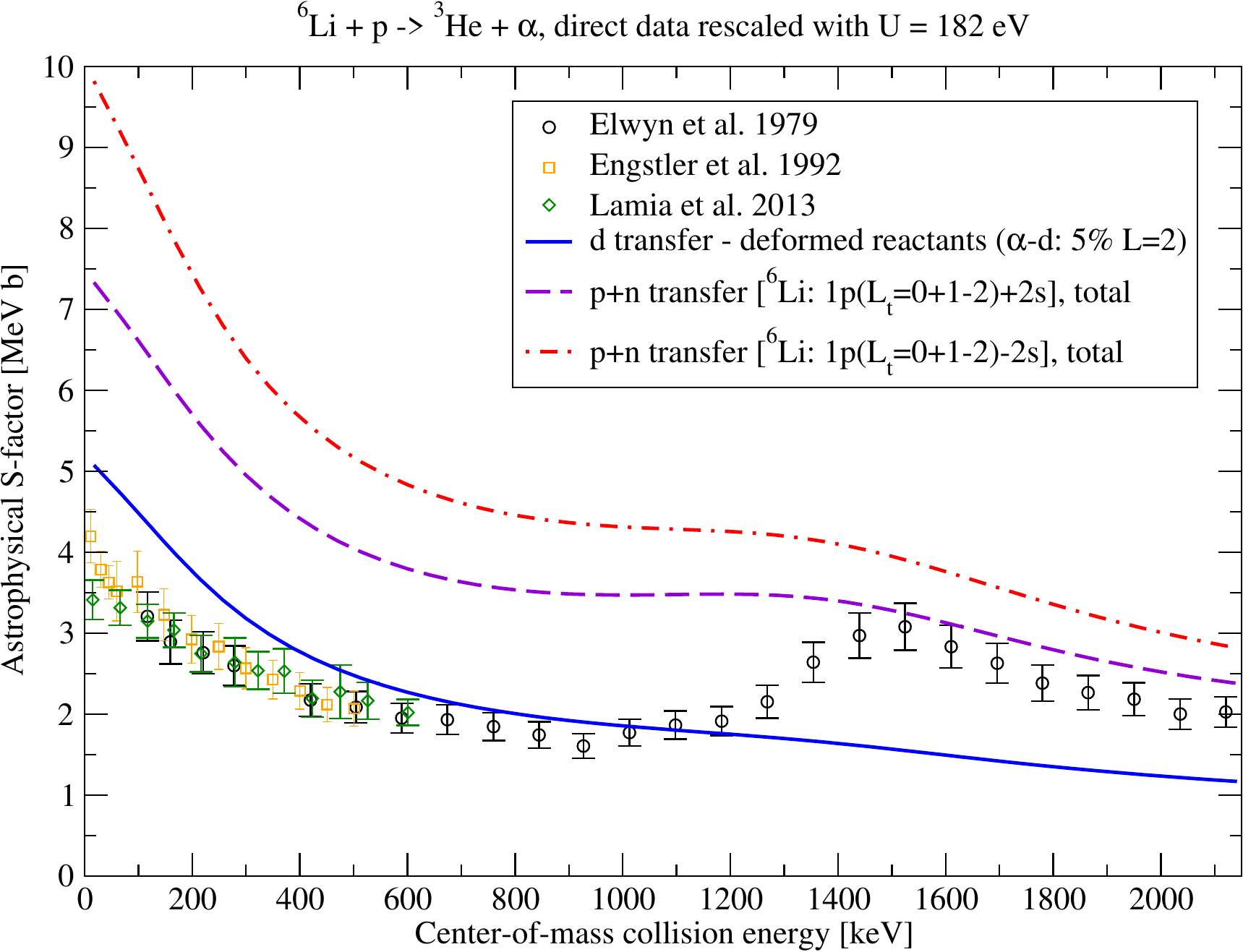}%
		\caption[Comparison of \texorpdfstring{$\nuclide[6]{Li} + \nuclide{p} \to \nuclide[3]{He} + \nuclide{\alpha}$}{6Li+p->3He+4He} one- and two-particle transfer]{\label{figConfrontoOneTwoParticleTransfer}%
			Points and blue dashed line are the same in \cref{figdTransferConOSenzaL2}: direct data was rescaled as in \cref{figDatiCorrettiAdiabatic}. Other lines are the same in \cref{figpnTransferRuoloShell2s}.
		}
	\end{figure}

	\Cref{figConfrontoOneTwoParticleTransfer} compares the two-nucleon-transfer calculations already shown in \cref{figpnTransferRuoloShell2s} with the deuteron-transfer astrophysical factor (with deformed nuclei) reported in \cref{figdTransferConOSenzaL2}.
	While the general trend of the three calculations is comparable, the two-nucleon-transfer ones display higher absolute cross-sections. An investigation was performed to understand whether this can be connected to the different structure adopted for the reactants in each case.
	The $\nuclide{\alpha}+\nuclide{p}+\nuclide{n}$ and $\nuclide{p}+\nuclide{n}+\nuclide{p}$ wave-functions were averaged over the coordinate of the transferred internal motion, to obtain effective \nuclide{\alpha}--\nuclide{d} and \nuclide{p}--\nuclide{d} wave-functions, which were then employed in the one-particle transfer calculation performed as in \cref{secCalcoliDWBADeuteronTransfer}. In particular, the same binding potentials found in \cref{secCalcoliDWBADeuteronTransfer} are employed, which implies that the reactants bound states are not eigenfunctions of such potentials any more. The corresponding cross-sections are in general significantly higher than those found in \cref{figdTransferConOSenzaL2}: the result is sensitive to the form (prior or post) employed for the calculation and the precise procedure adopted to generate the effective wave-function, but enhancements of about \SI{20}{\percent} or more at astrophysical energies were often found. While this is certainly significant, note that it is not sufficient by itself to reproduce the difference seen in \cref{figConfrontoOneTwoParticleTransfer}.
	
\begin{figure}[tbp]%
		\centering
		\includegraphics[keepaspectratio = true, width=\linewidth]{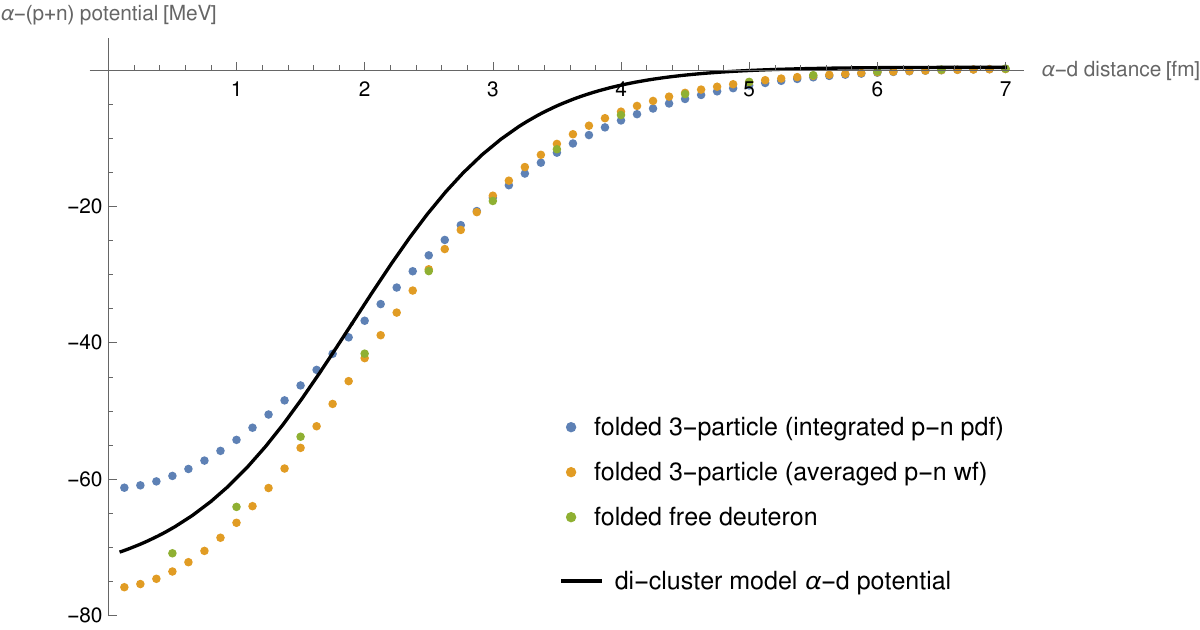}
		
		\includegraphics[keepaspectratio = true, width=\linewidth]{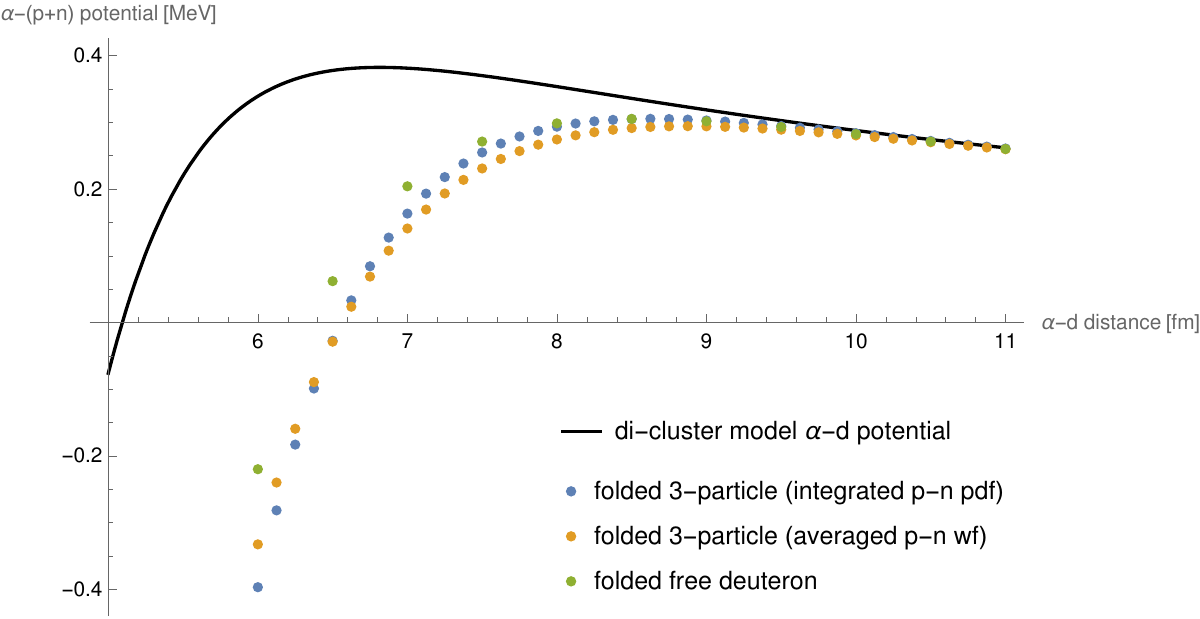}%
		\caption[\nuclide{\alpha}--\nuclide{d} (effective) potential]{\label{figPotenzialealphadEfficace}%
			Solid line is the \nuclide{\alpha}--\nuclide{d} potential employed for the di-cluster model \nuclide[6]{Li} wave-function, see \cref{secadPotential,secDeformazioneGroundState6LiQuadrupoleMoment}.
			Points are a folding of the ``$V_{\nuclide{\alpha}\nuclide{p}} + V_{\nuclide{Li}\nuclide{n}}$'' potential defined in text in \cref{secRisultatipnTransferComparisonDeuteronAndpnTransfer} (central part of the $\nuclide{\alpha}+\nuclide{p}+\nuclide{n}$ potential specified in \scref{secTNTCalcoliPraticiDescrizioneWFsingleparticle}),
			over different deuteron distributions. %
			See text for details. %
			Upper and lower panel show different ranges of \nuclide{\alpha}--\nuclide{d} distance.}
	\end{figure}

	Additionally, the \nuclide{\alpha}--\nuclide{d} potential employed in the one-particle transfer, which is the one appearing in the post-form transition amplitude, was compared to the interaction between \nuclide{\alpha} and transferred system in the two-nucleon transfer calculation, as discussed in the following.
	First, a simple wave-function for the isolated deuteron ground state, with binding energy corresponding to the experimental one for a free deuteron, was constructed as the $1s$ solution of a phenomenological unpublished Gaussian potential with a depth of \SI{72.15}{\MeV} and a diffuseness of \SI{1.484}{\femto\metre}. Let $\phi_{\nuclide{d}}(r)$ be such wave-function, and $u(r) = r \sqrt{4 \pi} \cdot \phi_{\nuclide{d}}(r)$ be the associated reduced radial wave-function.
	Then, for simplicity, the \nuclide{\alpha}--\nuclide{p} and \nuclide{Li}--\nuclide{n} binding potentials, employed in the two-particle transfer calculation (see \cref{secTNTCalcoliPraticiDescrizioneWFsingleparticle}), were truncated to their central parts, yielding the functions $V_{\nuclide{\alpha}\nuclide{p}}(r_{\nuclide{\alpha}\nuclide{p}})$ and $V_{\nuclide{Li}\nuclide{n}}(r_{\nuclide{Li}\nuclide{n}})$. The sum $V = V_{\nuclide{\alpha}\nuclide{p}} + V_{\nuclide{Li}\nuclide{n}}$ was converted in \nuclide[6]{Li} ``t'' coordinates, namely the distance between the transferred nucleons, $\v r$, and the distance between core \nuclide{\alpha} and transferred system, $\v R$, using the same writing found in \cref{eqCompleteProjectileTargetPotential}. The form factor
	\begin{multline}\label{eqFormFactorPerPotenzialeDiBindingEfficacead}
	V_f(R) = \Braket{\phi_{\nuclide{d}}|V_{\nuclide{\alpha}\nuclide{p}}+V_{\nuclide{Li}\nuclide{n}}|\phi_{\nuclide{d}}} = \\
	= \int_0^{+\infty}\d r\ |u(r)|^2 \, \frac{1}{2} \int_{-1}^1\d\cos\theta_{rR}\ V(r,R,\cos\theta_{rR})
	\end{multline}
	was finally evaluated. Since each $V$ is central and $\phi$ is spherical, the form factor has only a central component. The radial profiles of this form factor and of the \nuclide{d}--\nuclide{\alpha} binding potential mentioned in \cref{secadPotential,secDeformazioneGroundState6LiQuadrupoleMoment} were then compared. The result is shown in \cref{figPotenzialealphadEfficace}.
	The calculation was repeated adopting a different choice for the transferred system wave-function, as follows. %
	Let $\psi(r,R)$ be the radial part of the spherical component ($l=L=0$) of the \nuclide[6]{Li} three-particle wave-function in ``t'' coordinates, %
	and %
	the $\tilde\phi_{nn}(r)$ the normalised simple average defined as follows:
	\begin{equation}\label{eqDefinizioneFunzioneDOndaDeutoneMediataSuWFTreCorpi}
		\tilde\phi_{nn}(r) = \frac{ \int_0^{+\infty} \psi(r,R) \d R }{ \sqrt{ \int \m{ \int \psi(r,R) \d R }^2 \d r } }
	\end{equation}
	This was adopted in place of $u$ in %
	\cref{eqFormFactorPerPotenzialeDiBindingEfficacead}.
	Additionally, it is possible to consider %
	the normalised simple average of the three-particle probability density function, %
	\begin{equation}
	\tilde P_{nn}(r) = \frac{ \int_0^{+\infty} \m{\psi(r,R)}^2 \d R }{ \int \m{\psi(r,R)}^2 \d r \d R }
	\end{equation}
	and employ it in place of $|u(r)|^2$ in \cref{eqFormFactorPerPotenzialeDiBindingEfficacead}.
	It can be seen in \cref{figPotenzialealphadEfficace} that, around the maximum of the barrier, the computed effective potential is approximately the same for all form factors considered here, and that the ``free deuteron'' wave-function and the one computed from \cref{eqDefinizioneFunzioneDOndaDeutoneMediataSuWFTreCorpi} yield a similar form factor at all radii. %
	At the same time, the barrier of the \nuclide{\alpha}--\nuclide{d} di-cluster-model potential is higher and wider than the effective one found using the three-cluster model potentials and wave-functions. %
	It is thought that this difference may account for a significant fraction of the discrepancy between the one- and two-particle transfer calculation. %
	A similar analysis could be repeated for \nuclide[3]{He}. In the future, the form-factors computed as shown here will be employed to perform a single-particle transfer calculation (including the generation of the reactant bound states), to evaluate quantitatively the importance of the difference in the potentials. %

\subsubsection{Calculations rescaled to transfer experimental data}
	
\begin{figure}[tbp]%
		\centering
		\includegraphics[width=\textwidth,keepaspectratio=true]{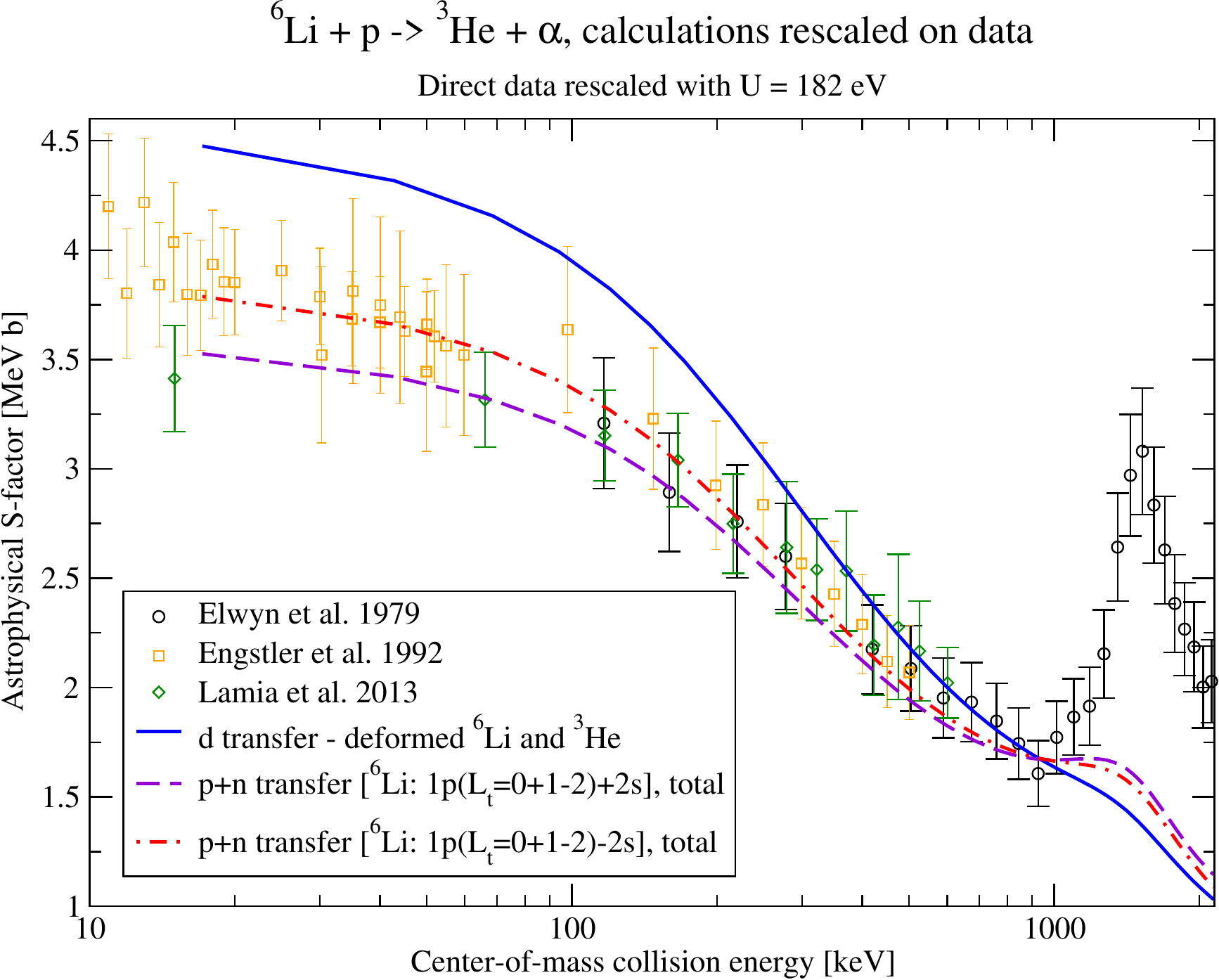}
		
		\includegraphics[width=\textwidth,keepaspectratio=true]{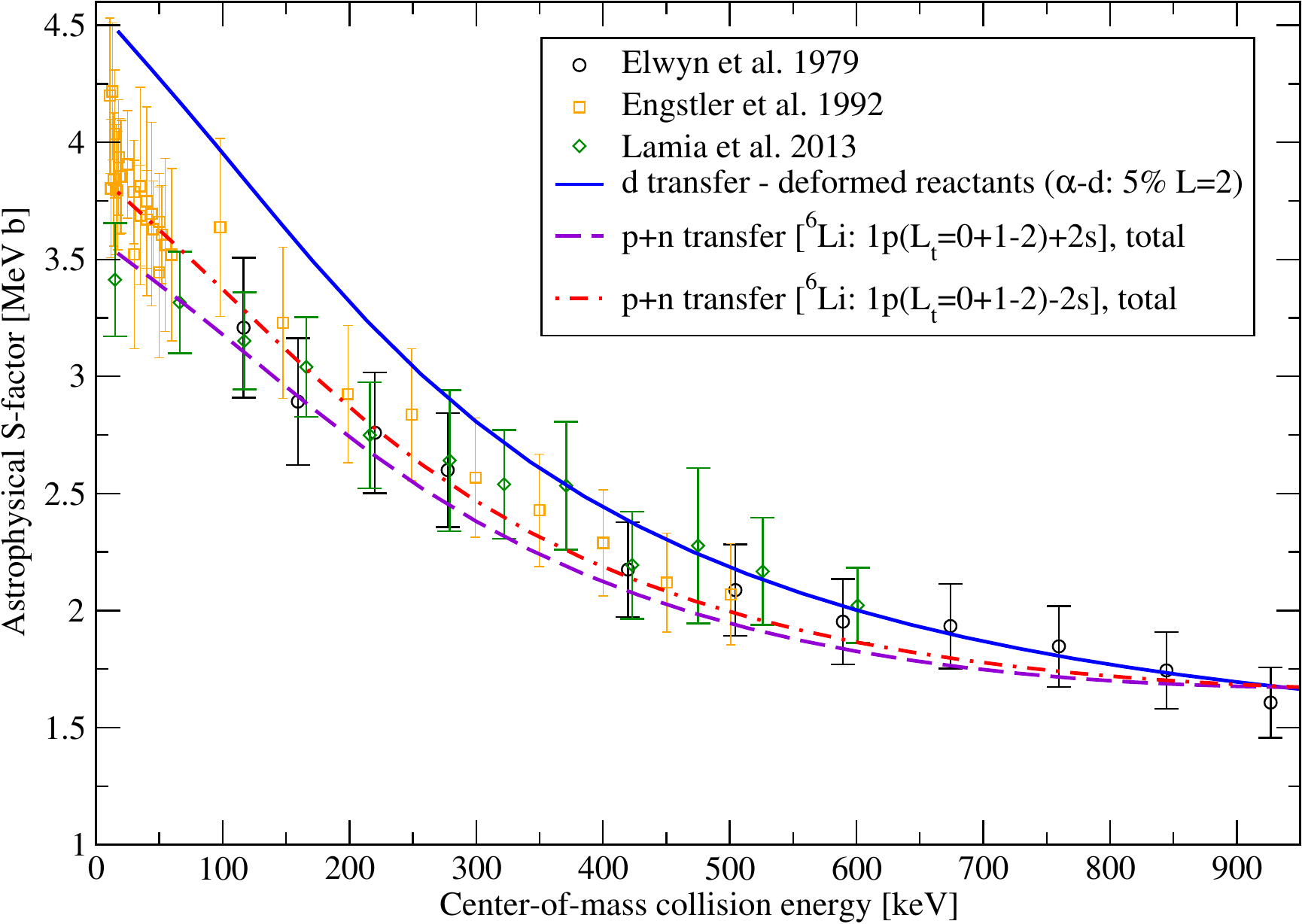}%
		\caption[\texorpdfstring{$\nuclide[6]{Li} + \nuclide{p} \to \nuclide[3]{He} + \nuclide{\alpha}$}{6Li+p->3He+4He} calculations rescaled on experimental data]{\label{figCalcoliRiscalatiSuiDati}%
			Same as \cref{figConfrontoOneTwoParticleTransfer},
			but the blue solid, violet dashed and red dot-dashed lines are rescaled respectively by factors of 0.88, %
			0.48 %
			and 0.39. %
			Upper and lower panel differ only by the $x$-axis scale. %
		}
	\end{figure}

	It was mentioned earlier that the choice between the more- or less-strongly clustered \nuclide[6]{Li} wave-function has a greater relative impact on the cross-sections at low energies. This may be seen visually in \cref{figpnTransferRuoloShell2s} but will now be analysed more carefully.
	In \cref{figCalcoliRiscalatiSuiDati}, the same calculations already shown in \cref{figConfrontoOneTwoParticleTransfer} are scaled in order to agree with experimental data at centre-of-mass collision energies slightly smaller than \SI{1}{\MeV}.
	This allows %
	to compare the predicted energy trends independently of the computed absolute values. The difference in the slopes of the astrophysical factors is more apparent, and it can additionally be seen that the impact of the choice of sign for the \nuclide[6]{Li} $2s$-shell contribution is of the right order of magnitude to be of interest for the experimental discrepancy discussed in \cref{secScreeningExperimentalData}.
	It is stressed that establishing which wave-function for \nuclide[6]{Li} is the ``correct'' one is not the issue at hand%
	\footnote{As mentioned in \cref{secTNTCalcoliPraticiDescrizioneWFSimultaneo6Li}, the state referenced in \cref{tabPesi6LiThreeParticle} and whose probability density function is shown in the upper panel of \cref{figPDF6Li} yields the best agreement with three-body calculations.}. %
	Rather, the point made here is that %
	an accurate treatment of the reactants static structure, in particular regarding the clustering strength, can %
	be important to obtain a good %
	understanding of the reaction dynamics between light ions even at very low energies and far from resonances.

	It is also interesting to point out that, after rescaling all calculations, the transfer of a structureless deuteron (blue solid line in \cref{figCalcoliRiscalatiSuiDati}) displays a significantly steeper slope than both \nuclide{p}+\nuclide{n} calculations. Qualitatively, %
	the \nuclide{d}-transfer process may be seen as the limit of a \nuclide{p}+\nuclide{n} transfer in which the transferred system is forced into a fully clustered configuration. %
	Nonetheless, it is to be reminded that, while the two shown \nuclide{p}+\nuclide{n}-transfer cross-sections correspond to the same calculation (apart from the commented change in the structure) and are certainly comparable, the \nuclide{d}-transfer process is not computed through the same formalism and does not employ the same physical ingredients, thus the attribution of a given difference in the results is less clear-cut. %
	
	While the rescaling procedure in \cref{figCalcoliRiscalatiSuiDati} allows an immediate and visual description, %
	the precise energy at which the calculations are fixed to match between each other and with the data is arbitrary, and the curves can appear slightly different if this is changed. %
	The trend suggested by each calculation at astrophysical energies may be compared in a more quantitative manner by computing the ratio $[\partial_E S(E)] / S(E)$ %
	and using it in \cref{eqEspansioneAlPrimoOrdineFattoreAstrofisicoWKBl0} to deduce the value of the effective nuclear radius associated to the calculated line (or similarly taking the logarithm of the astrophysical factor and using \cref{eqEspansioneAlPrimoOrdineEsponenzialeFattoreAstrofisicoWKBl0}).
	Such analysis is a possible future development for this work. %

\section{GGW deuteron transfer}\label{secCDCCCalcoloPratico} %

\begin{figure}[tbp]%
		\centering	\includegraphics[width=\textwidth,keepaspectratio=true]{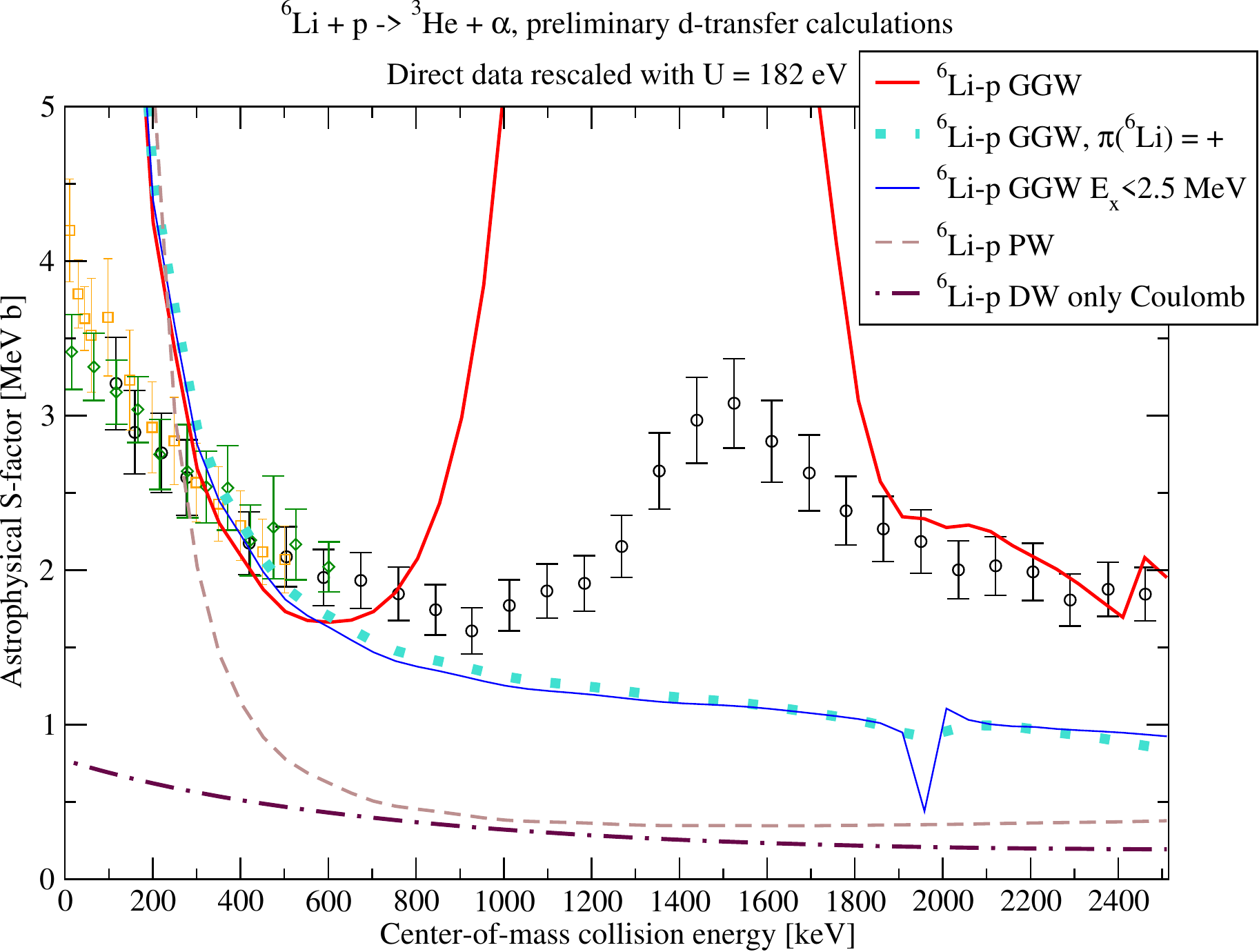}%
		\caption[Preliminary \texorpdfstring{$\nuclide[6]{Li} + \nuclide{p} \to \nuclide[3]{He} + \nuclide{\alpha}$}{6Li+p->3He+4He} GGW deuteron-transfer calculation]{\label{figCalcoloCDCCdTransfer}%
			Points are the same in \cref{figdTransferConOSenzaL2} (legend suppressed for readability): direct data was rescaled as in \cref{figDatiCorrettiAdiabatic}. Tick red solid line is the preliminary GGW deuteron transfer calculation discussed in \cref{secCDCCCalcoloPratico}. Turquoise dotted and blue thin solid lines are the same calculation including \nuclide[6]{Li} continuum levels with, respectively, only positive parity or excitation energy below \SI{2.5}{\MeV} above the ground state. Brown dashed line is a DWBA calculation with \nuclide[6]{Li}--\nuclide{p} optical potential equal to zero. Maroon dot-dashed line is the same calculation with \nuclide[6]{Li}--\nuclide{p} potential equal to the electrostatic repulsion.
		}
	\end{figure}

	One of the calculation schemes discussed in \cref{secOneParticleTransferCCApproaches} was applied to the $\nuclide[6]{Li} + \nuclide{p} \to \nuclide[3]{He} + \nuclide{\alpha}$ one-particle transfer reaction, as follows%
	\footnote{The study reported in this \namecref{secCDCCCalcoloPratico} was developed in collaboration with A.~Moro.}. %
	The generalised-distorted-wave formalism in \cref{sezGeneralizedDistortedWaves} is applied in prior form with the choice of generalised auxiliary potential yielding the ``GGW'' amplitude in \cref{eqAmpiezzaTransizioneGGW}, and the full scattering solution is approximated by the final-partition standard distorted wave as in \cref{sezDWBA},
	The \nuclide[6]{Li} structure includes a %
	continuum of inert di-cluster model \nuclide{\alpha}--\nuclide{d} states, with either parity, total angular momentum modulus quantum number up to 3, and excitation energies up to about \SI{3.5}{\MeV} above the ground state, %
	discretised using the binning method \cite{Austern1987}.
	For \nuclide[3]{He}, only the ground state is currently included. %
	All bound states considered here are eigenstates of the core-valence orbital angular momentum ($L$ in \cref{secadPotential}), in particular the ground states include no deformed components.
	The adopted %
	potentials are the same mentioned in \cref{secCalcoliDWBADeuteronTransfer},
	with two exceptions. %
	First, the \nuclide{\alpha}--\nuclide{d} binding potential was modified by including a spin-orbit component, adjusted to reproduce the energy of the $3^+$ first excited state of \nuclide[6]{Li}: see the ``\nuclide{\alpha} -- \nuclide{d} (GGW)'' column of \cref{tabParametriNumericiPotenziali6Li}. Second, there is no %
	\nuclide[6]{Li}--\nuclide{p} optical potential, since a generalised auxiliary potential is employed in its place. %
	Remind that, as discussed in \cref{secDWBAPriorPost}, non-central components of the core-core \nuclide{\alpha}--\nuclide{p} are still being neglected; however, no projectile-target optical potential appears in the transition amplitude operator, thus no other term is here approximated.
	\Cref{figCalcoloCDCCdTransfer} shows the preliminary astrophysical factor computed in this way (tick red solid line).
	
	Two very large peaks are visible in the calculation. The first one, at lower energies, is discussed later.
	The second peak has its maximum at an energy similar (but slightly smaller)
	to the experimental resonance connected to the \nuclide[7]{Be} $5/2^-$ level at an excitation energy of about \SI{7.2}{\MeV}. Decomposition in initial-channel partial-waves %
	reveals that the peak is indeed primarily fed by $p$-waves (\nuclide[6]{Li}--\nuclide{p} orbital angular momentum 1), and thus states with negative total parity, as expected, see the discussion in \cref{secCalcolidDWBARisonanza7Be}, but the total angular momentum is incorrectly reproduced, since the contribution of the $p5/2$ wave is negligible.
	On this regard, remind that %
	the present formalism %
	does not include
	the \nuclide[6]{Li}--\nuclide{p} distorted wave obtained from optical potential discussed in \cref{sec6LipOpticalPotential}, which in the DWBA calculations %
	was sufficient to exclude all significant spurious contributions from the transfer channel. %
	Here, the \nuclide[6]{Li} generalised distorted wave is generated through \cref{eqAusiliariaOndeGeneralizzatePrior} by the \nuclide{\alpha}--\nuclide{d} binding potential, which currently possesses a rather simple and phenomenological form, and the \nuclide{\alpha}--\nuclide{p} core-core potential, which however must be central in the current numerical implementation, %
	The other component of the complete \nuclide[6]{Li}--\nuclide{p} interaction, the \nuclide{d}--\nuclide{p} binding potential, in this calculation is the transition amplitude operator (see again \cref{eqAmpiezzaTransizioneGGW}) and thus can be always applied to %
	the \nuclide[3]{He} state, which currently is just a spherical wave-function for the ground state, %
	so that any non-central component of the potential would be effectively irrelevant.
	Additionally,
	the only projectile-target optical potential employed in this calculation, the \nuclide[3]{He}--\nuclide{\alpha} one (discussed in \cref{sec3HeaOpticalPotential}), was not adjusted on the elastic scattering phase-shifts. %
	A better description of the experimental resonance might thus be obtained by including more %
	appropriate interactions, %
	or by computing both the initial- and final-partition %
	wave-functions through a CDCC approach. %

	The influence of continuum states of \nuclide[6]{Li} on the complete result was studied by repeating the calculation after removing selected states from the structure. Two examples are displayed in \cref{figCalcoloCDCCdTransfer} (blue thin solid and turquoise dotted lines). Note that all states are coupled in a non-trivial manner, and furthermore their contributions %
	on the transition amplitude interfere, thus it is not possible to clearly assign fractions of the angle-integrated cross-section to any specific state.
	Nonetheless, it is seen that the peak at about \SI{1.3}{\MeV} completely disappears if all \nuclide[6]{Li} negative-parity states are removed, or if the continuum is truncated to a maximum excitation energy of about \SI{2.5}{\MeV} above the ground state. %
	Note that \SI{2.5}{\MeV} is approximately the collision energy at which the calculations shown in \cref{figCalcoloCDCCdTransfer} were stopped, thus this is an indication that virtual excitations to closed channels can play a relevant role in the reaction.
	
	However, regardless of the way the \nuclide[6]{Li} continuum is modified, it was not possible to remove or significantly alter the strongly rising trend %
	appearing in the computed cross-section at low energies.
	In fact, such trend persists even if all excited states are removed%
	\footnote{Note that removing all excited states within the GGW %
		formalism is equivalent to reduce the calculation to a DWBA with $\op U_{Ab} = \Braket{\psi_\alpha|\op V_{ab}|\psi_\alpha}$ (compare \cref{eqEquazioneOndaDistortaStandard,eqProblemaCDCCAccoppiatoSpazioCompleto}) in zero-remnant approximation (namely, $V_{ab} - U_{Ab} \approx 0$).}.%
	To get an insight into the possible origin of this trend, the following test calculations were performed.
	The brown dashed line in \cref{figCalcoloCDCCdTransfer} is a DWBA calculation, including only a spherical ground state for each composite nucleus, in which the (central) remnant is fully taken into account, but the \nuclide[6]{Li}--\nuclide{p} optical potential is set to zero: the astrophysical factor shows %
	the same low-energy trend as the GGW calculations.
	Finally, the maroon dot-dashed line in \cref{figCalcoloCDCCdTransfer} is another DWBA calculation where the \nuclide[6]{Li}--\nuclide{p} potential is set equal to just the Coulomb potential between the two nuclei%
	\footnote{For convenience in performing the numerical computation, the potential actually has the form in \cref{eqCoulombWoodsSaxonDefinition} with $R_C = \SI{0.1}{\femto\metre}$. This has no influence on the discussion.}.
	With this choice, the long-range Coulomb part of the transition potential, which is $V_{\nuclide{\alpha}\nuclide{p}} + V_{\nuclide{d}\nuclide{p}} - U_{\nuclide{Li}\nuclide{p}}$ (see \cref{eqAmpiezzaDiTransizioneDWBAPrimoOrdine}), %
	is cancelled, and it is seen in the figure that the divergence toward low energies disappears.
	It is thought that, in the GGW calculations, the long-range Coulomb potential between core and transferred systems
	similarly causes the observed feature in the low-energy cross-section, which is thus deemed as an artefact of the calculation. %
	It is also relevant to point out that, in general, calculations involving coupled-channels schemes at low energies are complicated by numerical instabilities arising in the inclusion of %
	virtual couplings to states at excitation energies which are too high with respect to the collision energy.
	The issue, which currently prevents to draw conclusions %
	on the energy range of astrophysical interest, is still under study.
	Also note that a full analysis of the convergence of the result for all numerical parameters of the calculation was not completed yet.
	In addition, %
	the numerical integrations are currently performed using the Numerov method: it may turn out to be beneficial to employ %
	instead the ``calculable'' $R$-Matrix method, see e.g.~\cite{Descouvemont2010}. %
	Nonetheless, it is interesting to note that, far from the two peaks, the agreement between experimental data and calculation is remarkable, given also the simple structure currently adopted for \nuclide[3]{He}.

\chapter[Impact of \texorpdfstring{$\nuclide[6]{Li}$}{6Li} ground-state deformation on barrier penetrability]{Impact of \texorpdfstring{$\nuclide[6]{Li}$}{6Li} ground-state deformation on barrier penetrability}\markboth{\Cref{secLiDeformation}. Impact of $\nuclide[6]{Li}$ ground-state deformation}{\Cref{secLiDeformation}. Impact of $\nuclide[6]{Li}$ ground-state deformation}\label{secLiDeformation}%

	The impact of clustering on barrier penetrability has been explored in literature within semi-classical models \cite{Spitaleri2016}.
	In the present \namecref{secLiDeformation}, this kind of approach was investigated, expanded, and applied to some cases of interest.

	\Cref{secLiDeformationClassicalModel} treats the classical cluster model %
	already discussed in \cite{Spitaleri2016}. By assigning a defined spatial position to the clusters forming each nucleus, it is possible to deduce an effective projectile-target potential which depends on the reactants orientation. The radial barrier penetrability associated to each possible configuration is then estimated in WKB approximation. The average result over all configurations, possibly weighted over some distribution suggested by more microscopic models, can then be compared with other calculations to explore the role of the assigned cluster structure. %
	The practical implementation of the model was here improved, and an alternative way to analyse the results is proposed. %
	The features were applied to the study of \nuclide[6]{Li}--\nuclide[6]{Li} barrier penetrability, to make a comparison with results found in literature. %
	
	In \cref{secLiDeformationQuantumModel}, the same underlying ideas are then reformulated within the framework of a quantum-mechanical cluster model. The effective projectile-target potential for a pair of composite particles can be expressed as the form factor of the ``microscopic'' interaction between the elementary components of the system evaluated on the reactants internal state. In this way, the details of the reactants structure can be connected to the barrier penetrability for a reactant pair subject to the aforementioned effective potential. %
	In particular, the formalism is employed to study the impact of ground-state quadrupole deformations in the \nuclide[6]{Li} wave-function, %
	considering for convenience %
	its scattering with a proton, which can be regarded as structureless.
	As will be seen in \cref{sec6LiDeformationsConstructionoftheprojectiletargetpotential}, the deformed components in the wave-function give rise to tensor components in the form factor: %
	the anisotropic projectile-target interaction is then treated precisely as in the classical model. %

\section{Classical cluster model}\label{secLiDeformationClassicalModel}

	Two nuclei $P$ and $T$, undergoing a collision, %
	are assigned a classical frozen di-cluster state, meaning that %
	$P$ is described as a system of two particles $P_1$ and $P_2$ fixed (during the whole collision) at a given relative displacement $\v d_P$ (and similarly for $T$). The coordinate system may be chosen e.g.~to fix the orientation of $\v d_P$. The projectile-target relative orbital angular momentum, and thus the impact parameter, is set to zero (which is expected to be the dominant component). %

	The projectile-target potential, $V_{TP}$, is defined as the sum of the potentials between each cluster, $\sum_{i,j} V_{P_i T_j}$, and thus depends on the orientation and extension of each reactant, so it may be written as $V_{TP}(\v R, d_P, \v d_T)$, where $\v R$ is the projectile-target displacement. For any fixed value of the inter-cluster displacements, %
	and for any projectile-target orientation,
	the $s$-wave barrier penetrability is computed in WKB (using \cref{eqDefinizionePenetrabilitaWKB} with the factor in front of the exponential approximated to a constant), assuming a spherically symmetric potential $V(R)$ equal to $V_{TP}(\v R, d_P, \v d_T)$.
	The transmission coefficient for each reactants orientation may then be weighted on some distribution phenomenologically accounting for the particles structure or dynamical effects. In \cite{Spitaleri2016}, all orientations were averaged with equal weight to check the implications of the model in the absence of any alignment.

\subsection{Present implementation of the classical model}
	
	In the calculation shown in \cite{Spitaleri2016}%
	\footnote{The source code of the algorithm employed in \cite{Spitaleri2016} was provided by L.~Fortunato, who also contributed to the work discussed in this \namecref{secLiDeformation}.},
	the lower bound of the integral in \cref{eqDefinizionePenetrabilitaWKB} was chosen as the inner classical turning point for the system. The same procedure was adopted here.
	It should be considered that the WKB approximation is not expected to be accurate around turning points (see \cref{eqCondizioneValiditaWKB}), however, %
	the integrand in \cref{eqDefinizionePenetrabilitaWKB} is small around the same points,
	thus the errors induced by such approach are not expected to be large%
	\footnote{Nonetheless, in the future the calculations may be repeated using the prescription for $R_n$ mentioned in \cref{secCoulombbarrierpenetrabilityastrophysicalFactor}.}.
	In order to practically find the classical turning points, %
	a numerical root-finding algorithm was employed. The algorithm employed in \cite{Spitaleri2016} in some cases failed and returned an incorrect result. %
	In this work, a more robust numerical algorithm was implemented, %
	improving the practical results of the calculations.
	
	Furthermore, the potentials adopted for the calculations were slightly modified. In \cite{Spitaleri2016}, point-like Coulomb potentials between each pair of clusters are employed, which however generate spikes in the total potential, and thus degrade %
	the numerical computation whenever they appear in the integration domain. Here, the electrostatic repulsion potentials were parametrised as generated by an uniformly charged sphere, overcoming the issue%
	\footnote{It was verified that this change brought negligible differences at sufficiently high energies, where the divergences of the Coulomb potential did not appear in the integration domain.}.

\subsection{Results for \texorpdfstring{$\nuclide[6]{Li}+\nuclide[6]{Li}$}{6Li+6Li} barrier penetration}

	To allow for a comparison with results in \cite{Spitaleri2016}, the same system, $\nuclide[6]{Li}+\nuclide[6]{Li}$, was considered, adopting the same parameters for the potentials (with the already mentioned modification on the Coulomb repulsion).

\begin{figure}[tbp]%
		\centering
		\includegraphics[keepaspectratio = true, width=\linewidth]{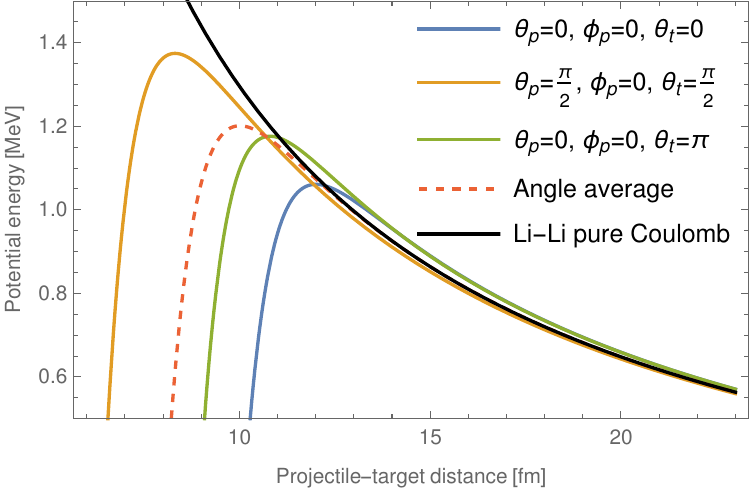}%
		\caption[\texorpdfstring{\nuclide[6]{Li}--\nuclide[6]{Li}}{6Li-6Li} potential within classical cluster model]{\label{figPotDeformatoClassicoVariAngoli}%
			Potential energy for the \nuclide[6]{Li}-\nuclide[6]{Li} classical model as a function of the projectile-target distance $r$. $\theta_p$ and $\phi_p$ are the polar and azimuthal angle of the projectile orientation, while $\theta_t$ is the target polar angle. The black solid line is $V = 9 \alpha_e \hbar c / r$.
			Every other solid line is the projectile-target potential, defined as explained in text, at a different reactants orientation. The dashed line is the average of the same potential over all orientations.
		}
	\end{figure}

	In the practical calculation performed in this work, the coordinate system was chosen by fixing the projectile-target distance on the $z$ axis, and fixing the $x$ axis so that the azimuthal angle of the target orientation is always zero. Varying the projectile orientation and the target polar angle suffices to properly account for all possible configurations. Note that this is not the coordinate system adopted in \cite{Spitaleri2016}. \Cref{figPotDeformatoClassicoVariAngoli} shows a sample of the radial profiles of the projectile-target potential obtained in selected configurations.
	
\begin{figure}[tbp]%
		\centering
		\includegraphics[keepaspectratio = true, width=\linewidth]{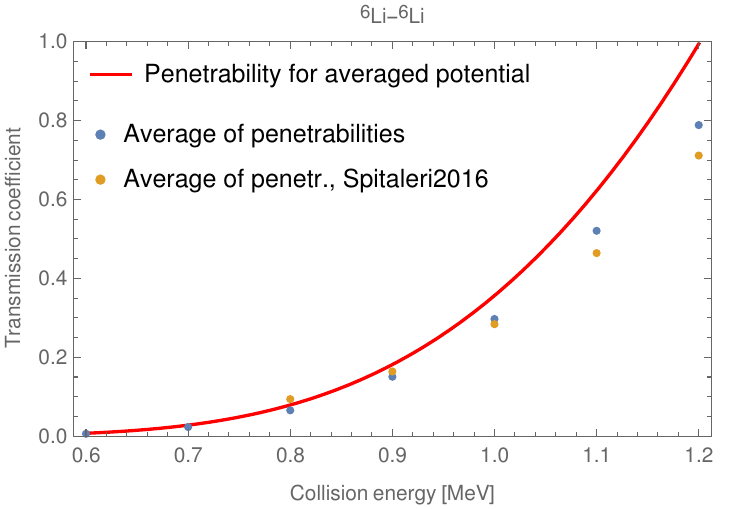}
		
		\includegraphics[keepaspectratio = true, width=\linewidth]{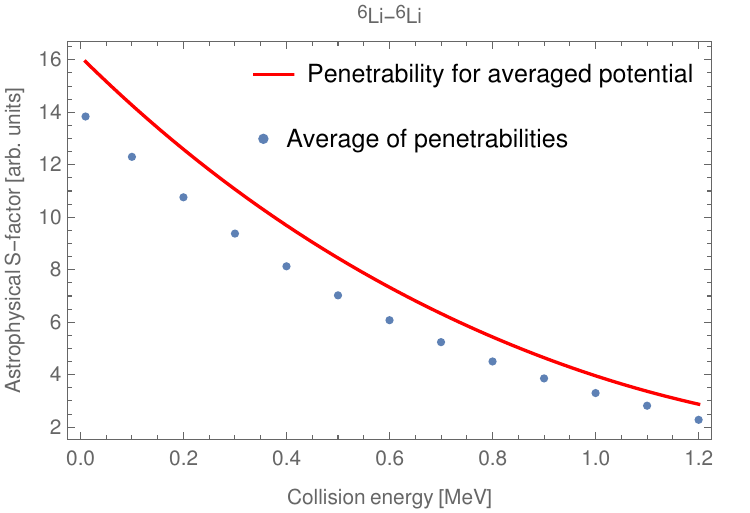}%
		\caption[\texorpdfstring{\nuclide[6]{Li}--\nuclide[6]{Li}}{6Li-6Li} average barrier penetrability within classical cluster model]{\label{figTransmissionPotSpitaleri2016PLB}%
			Top panel: average WKB transmission coefficient for \nuclide[6]{Li}-\nuclide[6]{Li} scattering using the same parameters for the cluster-cluster potentials employed in \cite{Spitaleri2016} (see text for details). The values computed from \cref{eqDefinizionePenetrabilitaWKB} are rescaled to set the coefficient in front of the exponential to one. Orange points are the ``averaged dicluster--dicluster system'' values in \cite[fig.~1]{Spitaleri2016}. Blue points are the corresponding values for the present calculation (average over all angles of the transmission coefficient at a given collision energy). The red line is the transmission coefficient for the averaged potential (see text).\\
			Bottom panel: same transmission coefficient converted in astrophysical $S$-factor using \cref{eqDecomposizioneSezioneDurtoPenetrabilitaPotenzialeCentrale,eqDefinizioneFattoreAstrofisico}.
		}
	\end{figure}

	The average of the penetrabilities computed on the potential for each orientation is shown in \cref{figTransmissionPotSpitaleri2016PLB}, and compared with the analogous calculation in  \cite[fig.~1]{Spitaleri2016}.
	The improvements on the computation implemented in this work are not negligible (especially in the limit of small collision energies), %
	but do not alter the qualitative conclusion in the original paper.
	The same figure includes the results of the same calculation performed at lower energies (down to \SI{10}{\keV}). In order to show a plot including the whole investigated energy range, %
	the transmission coefficient was converted into an astrophysical factor. %
	
	Again in \cref{figTransmissionPotSpitaleri2016PLB}, it is also shown (red line) the penetrability under a potential %
	obtained averaging the complete potential $V_{TP}$ over all directions. In this way, the non-spherical components induced by the adopted structure model are integrated away, and the averaged potential may be thought as representing a more ``uniform'' (less strongly clustered) system. A different comparison was performed in \cite[fig.~1]{Spitaleri2016}, where the ``sphere-on-sphere'' calculation refers to the penetrability computed using a different central, energy dependent, optical-model potential reproducing the \nuclide[6]{Li}--\nuclide[6]{Li} elastic scattering \cite{Potthast1997}. %
	In the approach shown in this work, all calculations are performed starting from an unique set of potential: this has the advantage of providing a consistent framework and reducing the required external inputs, %
	but implies that both averaging schemes ultimately derive from the cluster model. %
	The ``sphere-on-sphere'' penetrability in \cite{Spitaleri2016} is always lower than the averaged penetrability in the cluster model,
	which in turn is always lower than the penetrability for the averaged potential computed here%
	\footnote{While this may be coincidental, it is noted that an analogous result was found in \cref{figPotenzialealphadEfficace}, where the potential employed for the pair of \nuclide{\alpha} and \nuclide{d} structureless clusters was found to feature an higher barrier with respect to the effective potential obtained from the form factor of the interactions involving the two nucleons composing the deuteron.}. %
	
	Due to the approximations involved in the calculations and the way each penetrability was constructed, the physical meaning of the result is to some extent subject to interpretation.
	In spite of this, the comparison in \cref{figTransmissionPotSpitaleri2016PLB} between the average penetrability and the penetrability for the average potential is insightful, as it shows that a relevant feature found in atomic-like screening effects does not hold in the case of purely nuclear effects. %
	Consider a statistical ensemble of systems subject to the same potential, which is however shifted uniformly (i.e.\ equally at all distances) by a value (the screening potential in \cref{sezScreeningPotentialDefinizione})
	different for each element of the ensemble.
	It can be shown (see \cite[sec.~1.2.A]{PerrottaTesiSSC} for a discussion) that, for not too small collision energies,
	the WKB penetrability averaged over all configurations is greater than the WKB penetrability for the average shift. In the present calculation the opposite situation is found. The reason is that the use of an alteration to the potential energy which is constant at all distances is (often) a good approximation for non-nuclear screening phenomena, but is certainly not applicable here, where the projectile-target potential depends significantly on the direction only around and below the barrier, while the long-range Coulomb tail is essentially unmodified.

\section{Quantum-mechanical cluster model}\label{secLiDeformationQuantumModel}

	The goal of this \namecref{secLiDeformationQuantumModel} is to reframe %
	the model discussed in \cref{secLiDeformationClassicalModel} within a quantum-mechanical cluster model.
	This involves %
	two main tasks. First, the construction of a sufficiently realistic cluster-model wave-function for the nuclei under study, %
	which has to include some features of interest %
	which can affect the barrier penetrability estimation.
	Here, in particular, the deformation of \nuclide[6]{Li} was studied, thus its wave-function will be constructed taking into account the cluster-model predictions for the nucleus magnetic dipole and electric quadrupole moments, which are sensitive to such property. In the present work, this was done in \cref{secCostruzioneGroundState6LiDeformato} following an approach close to the one reported in \cite{Nishioka1984}. %
	The other references mentioned at the beginning of \cref{secLegameProprietaCompositoEFunzioneDonda} are also relevant. %

	The second step, in the same spirit of the classical model, %
	is then to combine the information on the adopted reactants structure and the assumed elementary interactions between their components, to deduce an effective projectile-target interaction in terms of which the barrier penetrability of the system can be evaluated. %
	Here, this is done following the formalism adopted in \cite{Nishioka1984,Merchant1985}.
	As will be seen in \cref{sec6LiDeformationsConstructionoftheprojectiletargetpotential}, even if the elementary interaction between the clusters is taken to be central, the non-spherical components in the inter-cluster motion generate a tensor projectile-target interaction.
	
	As was mentioned at the beginning of the \namecref{secLiDeformation}, the barrier penetrability is finally computed using the same approach already shown in \cref{secLiDeformationClassicalModel}. The results of the calculation are presented %
	in \cref{secResults6LipFusion}.
	A very brief summary of the contents of this \namecref{secLiDeformationQuantumModel} can be found in \cite{PerrottaSIF2021}.

\subsection{Construction of the \texorpdfstring{$\nuclide[6]{Li}$}{6Li} di-cluster model deformed state}\label{secCostruzioneGroundState6LiDeformato}

	In the following paragraphs, %
	some structure properties of the \nuclide[6]{Li} ground state (listed in \cref{tabExperimentalGroundStateData})
	are studied modelling the nucleus
	as a bound state of two inert clusters, \nuclide{\alpha} and \nuclide{d}. Let $\Ket{\nuclide{Li}_{J, M}}$ be a state of \nuclide[6]{Li} with total spin modulus and projection quantum numbers $J$ and $M$.
	By assumption, each cluster is in the ground state of the respective isolated system, in particular the deuteron has spin $S=1$ and positive parity, while the $\nuclide{\alpha}$ particle has spin-parity $0^+$.
	All low-lying \nuclide[6]{Li} states have positive parity, thus for these states %
	the inter-cluster relative motion must take place in states with even orbital angular momentum modulus quantum number, $L$.
	For all positive-parity %
	states with total spin %
	smaller than 3, %
	the only possible wave-function components have $L = 0$ or $2$.
	These angular momenta are the same ones allowed by the Wildermuth connection (see \cref{eqWildermuthconnection} and commenting text), which also suggests the number of radial nodes $n$ to assign to each wave-function, $n=1$ for $L=0$ and $n=0$ for $L=2$. %
	In more compact notation, these states are referred to as ``$2s$'' and ``$1d$'' (where the number is $n+1$ and the letter marks the value of $L$).
	It is expected a priori (see text commenting \cref{eqWildermuthconnection}), that the $2s$ component will be the dominant one. %

	Proceeding as for \cref{eqDefinizioneFunzioneRadialeRidotta,eqDecomposizioneClebschGordanEsempioSpecificoLOrbitale}, the ground state can thus be written as follows:
	\begin{multline}\label{eqFunzione6LiGroundStateDeformataGenerica}
		\Braket{\v r|\nuclide{Li}_{1, M}} =\\= \sum_{L,m} c_L \frac{u_{L}(r)}{r} \Braket{(L,m),(1,M-m)|1,M} \Ket{\Psi_{1, M-m}} Y_{L,m}(\theta,\phi)
	\end{multline}
	where $\Ket{\Psi_{S,\sigma}}$ is the state for the internal motion of the isolated clusters (\nuclide{\alpha} and \nuclide{d}), with their spins coupled to an eigenstate of the total intrinsic spin modulus and projection, with quantum numbers $S$ and $\sigma$, see \cref{eqDecomposizioneClebschGordan} and commenting text. If $\Ket{\nuclide{Li}_{1, M}}$ is a bound state, all $u_L$ and $c_L$ can be taken to be real.
	Even after this choice, there is an ambiguity in the signs of these quantities%
	\footnote{For any $L$, if the sign of both $c_L$ and $u_L$ is inverted, $\ket{\nuclide{Li}_{1, M}}$ remains unaltered. Additionally, the global sign of $\ket{\nuclide{Li}_{1, M}}$ is irrelevant, thus the signs of all $c_L$ can be inverted without altering any result.},
	which can be fixed imposing that each $u_L(r)$ is non-negative for $r \to \infty$, and that $c_0$ is positive.
	Furthermore, if $\Ket{\nuclide{Li}_{1, M}}$, $\Ket{\Psi_{S,\sigma}}$ and $\Ket{u_L}$ are normalised to 1, then $c_0$ can be set to $\sqrt{1 - \m{c_2}^2}$. %
	
	In the language of \cref{secOverlapFunctionSpectroscopicFactors}, the state in \cref{eqFunzione6LiGroundStateDeformataGenerica} is the normalised overlap function of \nuclide[6]{Li} state on the ground states of the two clusters (thus fixing all intrinsic quantum numbers, $J^\pi_i$, $T_i$, $\tau_i$, $\nu$).
	if the irrelevant state of the isolated \nuclide{\alpha} core is discarded, $\Ket{\nuclide{Li}_{1, M}}$ is the projection of $\Ket{\Phi^x_{T_x=0,j=1}}$ in \cref{eqEspansioneModelloAClusterGenerica} on states with fixed $J_x=1$ (and $\nu$ corresponding to the deuteron ground state), where the summation index $l$ is here called $L$. The weights $A_{\nu,l,1,0,1}$ are the $c_L$ in \cref{eqFunzione6LiGroundStateDeformataGenerica}, and $\phi_{\nu,1,l,m}(\v r)$ corresponds to $\frac{u_{L}(r)}{r} Y_{L,m}(\theta,\phi)$.

\begin{figure}[tbp]%
		\centering
		\includegraphics[keepaspectratio = true, width=\linewidth]{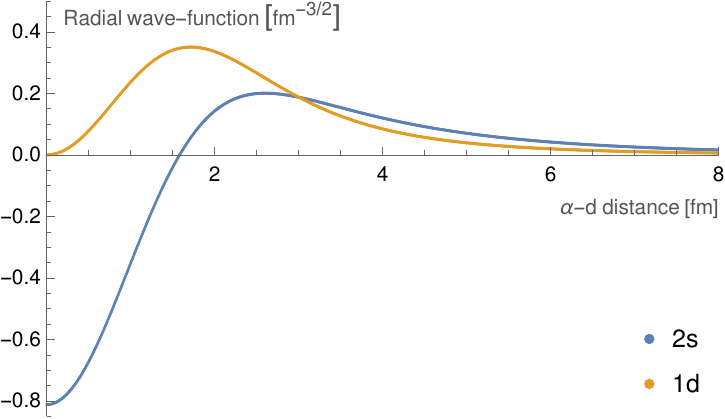}%
		\caption[\texorpdfstring{\nuclide{\alpha}--\nuclide{d}}{4He-d} radial wave-functions]{\label{figFunzioniDondaRadialiAlphad}%
			Non-reduced radial wave-functions, normalised to 1, for the inter-cluster motion within \nuclide[6]{Li} employed in \cref{secCalcoliDWBADeuteronTransfer,secLiDeformationQuantumModel}. ``2s'' and ``1d'' refer to $u_0/r$ and $u_2/r$, respectively (the components with inter-cluster orbital angular momentum $L=0$ and 2).
		}
	\end{figure}

	As in \cite[sec.~2, 5]{Nishioka1984}, each reduced radial wave-function $u_L$ was chosen as an eigenfunction of an Hamiltonian with the form in \cref{eqEffectiveRadialHamiltonianDefinition}, with the appropriate value for the orbital angular momentum. The potential ($V$ in \scref{eqEffectiveRadialHamiltonianDefinition}) is chosen to be the \nuclide{\alpha}--\nuclide{d} interaction given in \cite[eq.~(13)]{Kubo1972}: this was %
	obtained as an analytical approximation of the central part of a potential derived in \cite{Gammel1960}, which %
	in turn is a folding of a phenomenological \nuclide{\alpha}--nucleon potential on a deuteron wave-function including $L=0$ and 2 components%
	\footnote{As an aside, by keeping the non-central parts of the folding potential in \cite{Gammel1960}, including tensor components, it would be possible to find the complete cluster-model solution $\ket{\nuclide{Li}_{1, M}}$ as an eigenfunction of such potential. This approach is followed in \cite{Merchant1985}, and it may be implemented here in the future to obtain a more realistic wave-function and binding potential for $\nuclide[6]{Li}$.}.
	The Woods-Saxon potential depth ($V_v$ in \cref{eqGeneralPotentialParametrization}) was rescaled so that both the $2s$ and $1d$
	components match the $\nuclide{\alpha} + \nuclide{d} \to \nuclide[6]{Li}$ experimental binding energy ($\approx \SI{1474}{\keV}$): this implies that the scaling is different for each component. %
	No information about the Coulomb repulsion term is given in \cite{Kubo1972}, %
	but if the shape in \cref{eqCoulombWoodsSaxonDefinition} is used with $R_C = R_v$, then the value of $V_v$ reproducing the aforementioned binding energy for $L=0$
	is %
	close to the one given in \cite[eq.~(13b)]{Kubo1972}, \SI{77.2}{\MeV}. The parameters of the potential are reported in column ``\nuclide{\alpha} -- \nuclide{d}'' of \cref{tabParametriNumericiPotenziali6Li}.
	A plot of the radial wave-functions is shown in \cref{figFunzioniDondaRadialiAlphad}.
	The $L=0$ and $2$ components %
	have a single-particle asymptotic normalisation coefficient (defined in \cref{secANCdefinition}) of \SI[per-mode=power]{2.54}{\femto\metre\tothe{-1/2}} and \SI[per-mode=power]{0.356}{\femto\metre\tothe{-1/2}} %
	respectively.
	The weight assigned to each component (the values of $c_0$ and $c_2$) then fixes the ``asymptotic $d/s$ ratio'', namely the ratio of the $L=2$ and $L=0$ asymptotic normalisation coefficients. In principle, an independent determination of such ratio could be employed to constraint $c_2$, but in \cite{Tilley2002} (published in 2002) it is mentioned that the value is not well determined in literature.

	The following paragraphs discuss the determination of the weight $c_2$ by comparison with experimental ground-state structure data, for the given assignment of $u_L$. %

\subsubsection{Charge electric quadrupole moment}\label{secDeformazioneGroundState6LiQuadrupoleMoment}

	The electric quadrupole moment is, among those listed in \cref{tabExperimentalGroundStateData}, the most sensitive quantity (in terms of relative variation) to the deformed component of the state.
	For brevity, let $C_{L,m} = \Braket{(L,m),(1,1-m)|1,1}$, and
	$B = 4 \sqrt{\frac{\pi}{5}} \left( Z_{\nuclide{d}} A_{\nuclide{\alpha}}^2 + Z_{\nuclide{\alpha}} A_{\nuclide{d}}^2 \right) / \left(A_{\nuclide{d}} + A_{\nuclide{\alpha}} \right)^2$, %
	with $A_i$ and $Z_i$ being the mass and charge numbers of particle $i$.
	Taking into account that the quadrupole moment of the \nuclide{\alpha} particle ground-state is zero, the expectation value of such observable on the \nuclide[6]{Li} state in \cref{eqFunzione6LiGroundStateDeformataGenerica} is, using \cref{eqValAspChargeQuadrupoleMomentDueCluster} for each component involving a different cluster internal state $\Ket{\Psi_{1, \sigma}}$,
	\begin{multline}
		\Braket{\nuclide{Li}_{11}|Q_{\text{ch}}|\nuclide{Li}_{11}}
		= (1 - |c_2|^2) Q_{\nuclide{d},1} +\\+
		|c_2|^2 \left[ \sum_{m=0}^2 |C_{2,m}|^2 Q_{\nuclide{d},1-m} + B \sum_{m=0}^2 |C_{2,m}|^2 \Braket{Y_{2m}|Y_{20}|Y_{2m}} \Braket{u_2|r^2|u_2} \right] +\\+
		2 B \Re\left( c_2 \sqrt{1 - |c_2|^2} \, C_{2,0} \Braket{Y_{20}|Y_{20}|Y_{00}} \Braket{u_2|r^2|u_0} \right)
	\end{multline}
	where $Q_{\nuclide{d},\sigma}$ is the electric quadrupole moment expectation value for a deuteron in the ground state with spin projection $\sigma$, which can be deduced from the quoted experimental value in \cref{tabExperimentalGroundStateData} (referring to $\sigma=1$) using \cref{eqTeoremaWignerEckartMomentoQuadrupoloElettrico}. Note how, since $\Ket{\nuclide{Li}_{1, 1}}$ include states of its clusters with different spin projections, it was necessary to take into account the contribution of each of these to the expectation value.
	Substituting the values for all Clebsch-Gordan coefficients and all integrals on spherical harmonics, using \cref{eqBraketTreArmonicheSferiche}, it is found
	\begin{multline}\label{eqEspressioneMomentoQuadrupolo6LiDueCluster}
		\Braket{\nuclide{Li}_{11}|Q_{\text{ch}}|\nuclide{Li}_{11}}
		= \(1 - \frac{9}{10} c_2^2\) Q_{\nuclide{d},1} +\\+ \frac{Z_{\nuclide{d}} A_{\nuclide{\alpha}}^2 + Z_{\nuclide{\alpha}} A_{\nuclide{d}}^2}{(A_{\nuclide{d}} + A_{\nuclide{\alpha}})^2} \frac{1}{5} \left[ \sqrt{8} \, c_2 \sqrt{1 - c_2^2} \Braket{u_2|r^2|u_0} - c_2^2 \Braket{u_2|r^2|u_2} \right]
	\end{multline}
	which is the same as %
	\cite[eq.\ (5.4)]{Nishioka1984} (accounting for the differences in the adopted notation).
	Using the radial wave-functions in \cref{figFunzioniDondaRadialiAlphad} (constructed as commented earlier), one finds %
	$\Braket{u_2|r^2|u_2} = \SI{85.3}{\milli\barn}$ %
	and $\Braket{u_2|r^2|u_0} = \SI{104}{\milli\barn}$. %
	Substituting these values, and the data in \cref{tabExperimentalGroundStateData}%
	\footnote{Note that the currently recommended value for the quadrupole moment of \nuclide[6]{Li} is not the same employed in \cite{Nishioka1984} (published in 1984).}, %
	two possible solutions for $c_2$ are found, \num{-0.0909} and \num{0.969}. %

\subsubsection{Magnetic dipole moment}
	
	As will be seen below, the magnetic dipole moment is less sensitive than the quadrupole moment to the amplitude of the deformation, but it is independent from the form of the radial wave-functions, and very high-precision measurements are available (see \cref{tabExperimentalGroundStateData}), thus a comparison with experimental data can be fruitful.
	Let $\op{\v\mu}$ be the magnetic dipole operator.
	The experimental determinations in \cref{tabExperimentalGroundStateData} refer to the expectation value of the $z$-projection of $\op{\v\mu}$, %
	labelled $\op\mu^{(z)}$, evaluated on the eigenstate with maximum $z$-projection of the nucleus total spin (analogously to the convention adopted for the quadrupole moment, see \cref{secElectricQuadrupoleMoment}).
	From \cite[eq.~(A.9)]{Mason2008}, the magnetic dipole moment of a system of two clusters is given by %
	\begin{equation}\label{eqMomentoDipoloMagneticoSistemaDueCluster}
		\op \mu^{(z)} = \op\mu^{(z)}_1 + \op\mu^{(z)}_2 + \mu_N \frac{Z_1 A_2^2 + Z_2 A_1^2}{(A_1+A_2) A_1 A_2} \op L_z %
	\end{equation}
	where $\op\mu^{(z)}_i$ is the corresponding operator acting only on cluster $i$, $\mu_N$ the nuclear magneton (the explicit value can be found e.g.~in \cite{NISTcodata}), and $\op L_z$ the $z$-projection of the orbital angular momentum operator for the relative motion between the clusters.
	
	Since %
	$\op{\v\mu}$ is defined as the composition of orbital angular momentum and intrinsic spin operators (see \cite[eq.~(A.8)]{Mason2008}), it is itself a vector operator %
	(meaning that it transforms as a vector under spatial rotations), and in particular $\op\mu^{(z)}$ transforms as the spherical harmonic $Y_{10}$, %
	hence, using the Wigner-Eckart theorem,
	\begin{equation}\label{eqTeoremaWignerEckartMomentoDipoloMagnetico}
	\Braket{J, M |\op\mu^{(z)}| J, M}
	= \frac{\Braket{(J,M), (1,0)|J,M}}{\Braket{(J,J), (1,0)|J,J}} \Braket{J, J |\op\mu^{(z)}| J, J}
	\end{equation}
	Combining \cref{eqFunzione6LiGroundStateDeformataGenerica,eqMomentoDipoloMagneticoSistemaDueCluster,eqTeoremaWignerEckartMomentoDipoloMagnetico,eqRelazioniOrtonormalitaCoefficientiClebschGordan}, it is found that
	\begin{equation}\label{eqEspressioneMomentoDiDipoloMagnetico6LiADueCluster}
		\Braket{\nuclide{Li}_{1, 1} |\op\mu^{(z)}| \nuclide{Li}_{1, 1}} %
		= \( 1 - \frac{3}{2} c_2^2 \) \mu_{\nuclide{d}} + \mu_N \frac{Z_1 A_2^2 + Z_2 A_1^2}{A A_1 A_2} \frac{3}{2} c_2^2
	\end{equation}
	where $\mu_{\nuclide{d}}$ is the $z$-projection of the magnetic dipole moment evaluated on a deuteron with maximum spin projection. %
	Similarly, for brevity, let $\mu_{\nuclide[6]{Li}} = \Braket{\nuclide{Li}_{1, 1} |\op\mu^{(z)}| \nuclide{Li}_{1, 1}}$.
	Using the experimental values in \cref{tabExperimentalGroundStateData} for $\mu_{\nuclide{d}}$ and $\mu_{\nuclide[6]{Li}}$, from \cref{eqEspressioneMomentoDiDipoloMagnetico6LiADueCluster} one finds $|c_2| = 0.257$.

\subsubsection{Charge root-mean-square radius}\label{secChargeRadius6LiDiclusterDeformato}

	The radius of the system is much more sensitive to the form of the radial wave-functions than to the magnitude of the deformed component, thus it is not very important in determining $c_2$. Notwithstanding this, it is studied here for completeness. %
	Using the square-modulus of $\Braket{\v r|\nuclide{Li}_{1, M}}$ in \cref{eqFunzione6LiGroundStateDeformataGenerica} as probability density in \cref{eqRaggioDiCaricaDueCluster}, it is simply
	\begin{multline}
		r_{\nuclide{Li}}^2 = \frac{Z_{\nuclide{d}}}{Z} r_{\nuclide{d}}^2 + \frac{Z_{\nuclide{\alpha}}}{Z} r_{\nuclide{\alpha}}^2 +\\+ \frac{Z_{\nuclide{d}} A_{\nuclide{\alpha}}^2 + Z_{\nuclide{\alpha}} A_{\nuclide{d}}^2}{Z A^2} \left[ (1-|c_2|^2) \Braket{u_0|r^2|u_0} + |c_2|^2 \Braket{u_2|r^2|u_2} \right]
	\end{multline}
	where $A = A_{\nuclide{d}} + A_{\nuclide{\alpha}}$ etc.
	Considering again the functions $u_L$ in \cref{figFunzioniDondaRadialiAlphad}, it is $\Braket{u_0|r^2|u_0} = \SI{165}{\milli\barn}$, %
	which is almost twice $\Braket{u_2|r^2|u_2}$, meaning that the deformed component is more spatially compact than the spherical one (this is due to the different number of nodes of each function). As a consequence, higher values of $|c_2|$ yield smaller radii for \nuclide[6]{Li}. In order to precisely reproduce experimental values in \cref{tabExperimentalGroundStateData}, $|c_2| = \num{0.459}$ would be required. However, setting instead $|c_2| = \num{0.257}$ (the value reproducing the magnetic dipole moment) yields $r_{\nuclide{Li}} = \SI{2.64}{\femto\metre}$, %
	which is still acceptable in comparison with the experimental value of \SI{2.59(4)}{\femto\metre} \cite{Angeli2013}. Finally, choosing $c_2=0$ one finds $r_{\nuclide{Li}} = \SI{2.66}{\femto\metre}$, %
	which is not very different than before, but at the same time it would be sufficient to reduce $\Braket{u_0|r^2|u_0}$ by \SI{10}{\percent} to adjust the radius to \SI{2.59}{\femto\metre}.
	In summary, the predictions on the root-mean-square charge radius are useful to confirm that the adopted radial wave-functions are qualitatively reasonable. However, given the accuracy on experimental radii, and on the form of the radial wave-functions adopted in this calculation, this observable cannot provide a stringent constraint on the deformation of the state. %

\subsubsection{Discussion on the weight of the deformed component in \nuclide[6]{Li}}

	For each of the two electromagnetic moments considered above, there are two distinct values $c_2$ reproducing %
	experimental data.
	In the case of the quadrupole moment, the most realistic solution can be guessed from the general expectation that the composite system wave-function is dominated by the component with smaller $L$ (i.e.~$c_2^2 \ll 1/2$).
	Furthermore, even though the predictions connected to each observable are in disagreement, they can be employed together to constraint the amplitude of the deformed component, as follows. %

	First, the two predictions for $c_2$ given by the quadrupole moment can be applied to compute the dipole moment. Using the experimental value in \cref{tabExperimentalGroundStateData} for $\mu_{\nuclide{d}}$, the reconstructed value for $\mu_{\nuclide[6]{Li}}$ %
	is, respectively, \SI{0.853}{\mu_N} %
	for $c_2 = \num{-0.0909}$ and \SI{0.354}{\mu_N} %
	for $c_2 = \num{0.969}$. Comparison with the experimental value of $\mu_{\nuclide[6]{Li}}$, \SI{0.822043}{\nuclearmagneton} \cite{IAEAElectroMagneticMoments},
	clearly favours $c_2 = \num{-0.0909}$. %
	Such value corresponds to an asymptotic $d/s$ ratio (defined in \cref{secCostruzioneGroundState6LiDeformato}) %
	of \num{-0.0128}. %
	Analogously, among the two predictions for $c_2$ obtained adjusting on the dipole moment, the choice $c_2 = \num{-0.257}$ %
	(i.e.\ an asymptotic $d/s$ ratio of \num{-0.0372}) %
	is the one better reproducing the quadrupole moment.
	Consequently, it can be expected %
	that a realistic model of \nuclide[6]{Li} would predict an overlap on the ground states of \nuclide{\alpha} and \nuclide{d} with two components %
	bearing spectroscopic amplitudes with opposite sign, and such that the deformed term comprises less than \SI{7}{\percent} of the overall spectroscopic factor. %
	The variational Monte Carlo (VMC) wave-function %
	in \cite[sec.~V.C]{Forest1996} has %
	(in the notation of the present work) $c_2 = \num{-0.158}$, %
	which is in between the two values suggested above. %
	However, note that %
	both radial components of the VMC wave-function have one node, thus the predictions on the observables as a function of $c_2$ are expected to be rather different than in the present calculation%
	\footnote{Additionally, %
		in \cite[sec.~V.C]{Forest1996} it is stated that the computed quadrupole moment strongly depends on the wave-function asymptotic behaviour, which is not determined with high accuracy from VMC and Green's function Monte Carlo methods \cite[sec.~V, sec.~V.C]{Forest1996}.}.
	As can be seen graphically plotting \cref{eqEspressioneMomentoQuadrupolo6LiDueCluster,eqEspressioneMomentoDiDipoloMagnetico6LiADueCluster}, %
	the magnetic dipole moment is much less sensitive to the \nuclide[6]{Li} deformation than the electric quadrupole moment. %
	For instance, as quoted above, the value, precisely reproducing the quadrupole moment, $c_2 = \num{-0.0909}$ predicts a dipole moment $\mu_{\nuclide[6]{Li}}$ which differs by \SI{3.8}{\percent} with respect to the experimental value. %
	Conversely, substituting $c_2 = \num{-0.257}$ (the value precisely reproducing the dipole moment) in \cref{eqEspressioneMomentoQuadrupolo6LiDueCluster}, using the same radial wave-functions discussed above, yields an electric quadrupole moment for \nuclide[6]{Li} which is almost ten times bigger than the experimental value.
	Consequently, within the model adopted here, where the form of the radial wave-functions is fixed, the value for $c_2$ %
	yielding better overall agreement with experimental data is %
	the one %
	reproducing precisely the electric quadrupole moment and approximately the magnetic dipole moment%
	\footnote{It would also be possible to constraint the form of the radial wave-functions (or the tensor potential generating them) %
	combining the information on both measured moments, and possibly the radius, but this goes beyond the model in use here.}.
	However, also note that %
	\cref{eqEspressioneMomentoDiDipoloMagnetico6LiADueCluster}, differently than \cref{eqEspressioneMomentoQuadrupolo6LiDueCluster}, does not depend on the radial wave-functions form. As a result, any di-cluster-model construction of $\Ket{\nuclide{Li}_{1M}}$ leading to fully compatible predictions for both electromagnetic moments (and thus presumably more realistic) would certainly yield the value of $c_2$ suggested by the magnetic dipole moment. %
	Within the present approximate model, the ``best'' choice may in summary depend on what features of the system are more important for the phenomena of interest. Regarding specifically the study in %
	\cref{secResults6LipFusion}, the qualitative results and the conclusions do not change by adjusting $c_2$ on either observable. The pictures shown in the following %
	refer to the choice $c_2 = \num{-0.257}$ (reproducing the magnetic dipole moment). %

\subsection{Construction of the projectile-target potential}\label{sec6LiDeformationsConstructionoftheprojectiletargetpotential} %

	Let $m_1$ and $m_2$ be the masses of the two elementary clusters composing the \nuclide[6]{Li} target, and $\mu_{12} = m_1 m_2 / (m_1 + m_2)$.
	The second reactant %
	(“projectile” in the following, a proton in the calculation discussed in \cref{secResults6LipFusion}) is treated as an elementary particle.
	Let $V_{1p}(\v r_{1p})$ and $V_{2p}(\v r_{2p})$ be phenomenological potentials for the interaction between the projectile $p$ and each cluster composing the target. For simplicity, and given that the present work is only concerned with barrier penetrability, assume they are central potentials. The complete projectile-target potential, depending both on the distance from cluster 1 to cluster 2, $\v r$, and the distance between projectile and target centre-of-mass, $\v R$, is then
	\begin{equation}\label{eqCompleteProjectileTargetPotential}
		V_{tp}(\v r, \v R) = V_{1p}\left(\left|\v R - \frac{m_2}{m_1+m_2} \v r \right|\right) + V_{2p}\left(\left|\v R + \frac{m_1}{m_1+m_2} \v r \right|\right) %
	\end{equation}
	Using König's decomposition, and going to the system centre-of-mass coordinate system, the total system Hamiltonian can be written as
	\begin{equation}\label{eqCompleteCLusterClusterProjectileHamiltonian}
		\mathcal H = H_{\nuclide{Li}} + K_{tp} + V_{tp} %
	\end{equation}
	where $K_{tp}$ is the kinetic energy for the projectile-target relative motion, and $H_{\nuclide{Li}}$ is the Hamiltonian for the internal motion of \nuclide[6]{Li} in the inert-cluster model in use, of which the deformed state $\Ket{\nuclide{Li}_{1M}}$ discussed in \cref{secCostruzioneGroundState6LiDeformato} is the ground state, with eigenvalue $E_{\nuclide{Li}}$. %
	
	The complete potential $V_{tp}$ conserves the relative orbital angular momentum between the projectile and each cluster (since each $V_{ip}$ is central), and thus the total orbital angular momentum. However, the potential allows to change the target internal state by altering the orbital angular momentum of both the inter-cluster motion within the target and the projectile-target motion. %
	In order to focus on the impact of \nuclide[6]{Li} ground-state deformation, neglect the contribution of dynamical excitations, assuming that during the interaction %
	the target remains fixed in the ground state.
	Then, the eigenvalue problem for the inter-cluster motion can be decoupled from the projectile-target motion problem. %
	To this end, it is useful to define
	\begin{multline}\label{eqCompleteHamiltonianTargetStatesMatrixElements}
		\tilde{\mathcal H}_f(M, M', \v R) = \Braket{ \nuclide{Li}_{1 M'} | \mathcal H | \nuclide{Li}_{1 M} } = \\
		= (K_{tp} + E_{\nuclide{Li}}) \delta_{M,M'} + \Braket{ \nuclide{Li}_{1M'} | V_{tp} | \nuclide{Li}_{1M} }
	\end{multline}
	where the matrix element of $V_{tp}$ is the \emph{form factor}. As a function of $M, M'$, this is a tensor.
	To see this explicitly, let
	\begin{equation}\label{eqDefinitiontildeVtp}
		\Braket{ \nuclide{Li}_{1M'} | V_{tp} | \nuclide{Li}_{1M} } = \Braket{1,M'| \tilde V_{tp} |1,M}
	\end{equation}
	where $\Ket{J,M}$ is a state describing only the target spin degree of freedom ($\tilde V_{tp}$ incorporates the integration on $\v r$).
	As will be shown in the following (see \cref{eqWignerEckartTheoremFormFactorExpansion}), $\tilde V_{tp}$ is constituted by the sum of a central and a tensor component, and can be expressed as: %
	\begin{equation}\label{eqScomposizioneGenericaPotenzialeCentraleTensoriale}
		\tilde V_{tp} = U_C(R) + U_T(R) \frac{1}{\hbar^2} \left[ \left( \op{\v J}_t \cdot \frac{\v R}{R} \right)^2 - \frac{1}{3} \op{J}_t^2 \right]
	\end{equation}
	where $\op{\v J}_t$ %
	is the target intrinsic spin vector \emph{operator} %
	and $\v R / R$ is the direction of the projectile-target distance, thus, $\op{\v J}_t \cdot \frac{\v R}{R}$ is the operator of the projection of the target spin in $\v R$ direction. $U_C(R)$ and $U_T(R)$ are central functions that will be computed in the following.
	
	$\tilde{\mathcal H}_f$ was purposely constructed to not operate on the inter-cluster motion degrees of freedom, thus it conserves the clusters relative orbital angular momentum. %
	However, the projectile-target orbital angular momentum can still be altered by flipping the target spin projection, as allowed by the tensor term. %

\subsubsection{Explicit form of the tensor component}
	
	A rank-$K$ irreducible spherical tensor, $\boldsymbol{T}_K(\boldsymbol{A}_{k_1}, \boldsymbol{B}_{k_2})$, can be obtained from the composition of two other such tensors $\boldsymbol{A}_{k_1}$ and $\boldsymbol{B}_{k_2}$ (of rank $k_1$ and $k_2$ respectively) as in \cite{BrinkSatchler1968}[eq.~(4.6)]: %
	\begin{equation}\label{eqDefinizioneComposizioneTensoriSferici}
		T_{KQ}(\boldsymbol{A}_{k_1}, \boldsymbol{B}_{k_2}) = \sum_{q} \Braket{(k_1, q), (k_2, Q-q)| K, Q} A_{k_1,q} B_{k_2,Q-q}
	\end{equation}
	The scalar product of two tensors of equal and integer rank, $\boldsymbol{A}_{k}$ and $\boldsymbol{B}_{k}$, is defined to be proportional to the rank-0 tensor $\boldsymbol{T}_0(\boldsymbol{A}_{k}, \boldsymbol{B}_{k})$, as in \cite{BrinkSatchler1968}[eq.~(4.7)]: %
	\begin{equation}
		\boldsymbol{A}_k \cdot \boldsymbol{B}_k = \sum_Q (-1)^{K-Q} A_{K,Q} B_{K,-Q}
	\end{equation}
	Finally, the rank-1 spherical tensor $\boldsymbol{a}_1$ associated to a Cartesian vector $\vec a = (a^{(x)}, a^{(y)}, a^{(z)})$ has components
	\cite[eq.~(4.10)]{BrinkSatchler1968} \cite[eq.~(XIII.124)]{Messiah2014quantum}%
	\footnote{The analogous expression in \cite{Satchler1983Direct} (see text before eq.~(4.73)) has a typo.}
	\begin{equation}\label{eqDefinizioneComponentiTensoreSfericoDiUnVettore}
		a_{1,0} = a^{(z)} \quad , \quad a_{1,\pm1} = \mp \[ a^{(x)} \pm i a^{(y)} \] / \sqrt{2}
	\end{equation}
	With these definitions, \cite[eq.~(4.83)]{Satchler1983Direct} can be deduced. %
	For brevity, %
	let $\boldsymbol{T}_2\left(\v a\right) = \boldsymbol{T}_2(\boldsymbol{a}_{1}, \boldsymbol{a}_{1})$, where the tensor $\boldsymbol{a}_1$ is constructed in terms of $\vec a$ as in \cref{eqDefinizioneComponentiTensoreSfericoDiUnVettore}.
	Then,
	\begin{equation}
		\left( \op{\v J}_t \cdot \frac{\v R}{R} \right)^2 - \frac{1}{3} \op{\v J}_t^2 = \boldsymbol{T}_2\left( \op{\v J}_t \right) \cdot \boldsymbol{T}_2\left(\v R/R\right)
	\end{equation}
	Additionally, as in \cite[eq.~(4.77)]{Satchler1983Direct},
	\begin{equation}
		T_{2Q}\left( \v R/R \right) = \sqrt{\frac{8 \pi}{15}} Y_{2Q}(\v R)
	\end{equation}
	Thus:
	\begin{multline}
	\frac{1}{\hbar^2} \Braket{ J,M' |\left( \op{\v J}_t \cdot \frac{\v R}{R} \right)^2 - \frac{1}{3} \op{J}_t^2 | J,M} = \\
	= \frac{(-1)^{M-M'}}{\hbar^2} \sqrt{\frac{8 \pi}{15}} Y_{2,M-M'}(\v R) \Braket{ 1, M' | T_{2,M'-M}\left( \op{\v J}_t \right) | 1, M }
	\end{multline}
	Using the Wigner-Eckart theorem, and the definition in \cref{eqDefinizioneComposizioneTensoriSferici}, the matrix element can be evaluated as
	\begin{multline}
		\Braket{ 1, M' | T_{2,M'-M}\left( \op{\v J}_t \right) | 1, M } = \\
		= \frac{ \Braket{(1,M),(2,M'-M)|1,M'} \Braket{(1, 1), (1, 1)| 2, 2} }{ \Braket{(1,-1),(2,2)|1,1} } \Braket{ 1, 1 | \op{J}_{t,1 \, 1}^2 | 1, -1 }
	\end{multline}
	where $\op{J}_{t,1 \, 1}$ (the component $+1$ of the tensor associated to $\op{\v J}_t$ as per \cref{eqDefinizioneComponentiTensoreSfericoDiUnVettore}) is %
	proportional to the creation operator for the target spin: it is $\op{J}_{t,1 \, 1} \Ket{J, M} = -\frac{\hbar}{\sqrt{2}} \sqrt{J(J+1) - M (M + 1)} \Ket{J, M + 1}$. Thus:
	\begin{multline}\label{eqTensorForceExpansion}
	\frac{1}{\hbar^2} \Braket{ J,M' |\left( \op{\v J}_t \cdot \frac{\v R}{R} \right)^2 - \frac{1}{3} \op{J}_t^2 | J,M} = \\
	= (-1)^{M-M'} \frac{2}{3} \sqrt{2 \pi} Y_{2,M-M'}(\v R) \Braket{(1,M),(2,M'-M)|1,M'}
	\end{multline}

\subsubsection{Form factor expression}\label{secExplicitRelationElementaryPotentialWithFOrmFactor}

	$V_{tp}(\v r, \v R)$ can be decomposed in multipoles of both coordinates using \cref{eqComplexConjugateSphericalHarmonics,eqMultipoleExpansionRotatedCoordinates}. %
	Let each component be $V_L(r,R) = \Braket{Y_{L0}(\Omega_{Rr})|V_{tp}(\v r, \v R)}$.
	By definition, each multipole is a spherical tensor %
	in the associated coordinate. Thus, by Wigner-Eckart theorem (see also \cite[eq.~(4.87), (4.88)]{Satchler1983Direct}), %
	\begin{multline}\label{eqWignerEckartTheoremFormFactorExpansion}
	\Braket{\nuclide{Li}_{1, M'} | V_{tp} | \nuclide{Li}_{1, M} }
	= \sum_{Lm} \sqrt{\frac{4 \pi}{2L+1}} \coniugato{Y}_{L,m}(\v R) \Braket{\nuclide{Li}_{1, M'}|V_L(r,R) Y_{Lm}(\v r)|\nuclide{Li}_{1, M}} = \\
		\shoveleft{= (-1)^{M-M'} \sum_{L} \sqrt{\frac{4 \pi}{2L+1}} Y_{L,M-M'}(\v R) \cdot}\\\cdot \frac{ \Braket{(1,M),(L,M'-M)|1,M'} }{ \Braket{(1,1),(L,0)|1,1} } \Braket{\nuclide{Li}_{1, 1}|V_L(r,R) Y_{L,0}(\v r)|\nuclide{Li}_{1, 1}}
	\end{multline}
	Note that non-diagonal elements ($M \neq M'$) are non-zero, implying that $V_{tp}$ is capable of flipping \nuclide[6]{Li} spin. %

	The complete potential $V_{tp}(\v r, \v R)$ will, in general, include multipoles of all $L$. However, in the form factor in \cref{eqWignerEckartTheoremFormFactorExpansion},
	the angular integration will be a sum of terms such as $\Braket{Y_{l'\lambda} | Y_{L0} Y_{l\lambda}}$ where $l,l' = 0,2$ by construction of $\Ket{\nuclide{Li}_{1, 1}}$, thus only terms with $L =$ 0, 2, or 4 give contribution.
	All $L = 4$ terms are bound to vanish, since a spin-1 system is being studied, and $\Braket{(1,M),(4,M'-M)|(1,M')} = 0$. The $L=0$ contribution (appearing only for $M = M'$) forms a central potential $U_C(R)$. Using \cref{eqFunzione6LiGroundStateDeformataGenerica}, it is %
	\begin{multline}\label{eqEspressioneEsplicitaTermineCentraleFormFactor6LiDeformato}
		U_C(R) = \frac{1}{\sqrt{4 \pi}} \Braket{\nuclide{Li}_{1M}|V_0(r,R)|\nuclide{Li}_{1M}} = \\ %
		= \frac{1}{\sqrt{4 \pi}} \int_0^{+\infty} \left[ (1 - |c_2|^2) |u_0(r)|^2 + |c_2|^2 |u_2(r)|^2 \right] V_0(r,R) \d r
	\end{multline}
	
	The $L=2$ contribution, as can be seen by comparison of \cref{eqTensorForceExpansion,eqWignerEckartTheoremFormFactorExpansion}, is the tensor component in \cref{eqScomposizioneGenericaPotenzialeCentraleTensoriale}. The scalar coefficient $U_T$ is, using \cref{eqFunzione6LiGroundStateDeformataGenerica,eqBraketTreArmonicheSferiche},
	\begin{multline}
	U_T(R)
	= 3 \Braket{\nuclide{Li}_{11}|V_2(r,R) Y_{2,0}(\v r)|\nuclide{Li}_{11}} = \\
		= \frac{3}{\sqrt{10 \pi}} \[ c_2 \sqrt{1 - c_2^2} \Braket{u_0|V_2|u_2} - \frac{c_2^2}{\sqrt{8}} \Braket{u_2|V_2|u_2} \]
	\end{multline}

\begin{figure}[tb]%
		\centering
		\includegraphics[keepaspectratio = true, width=\linewidth]{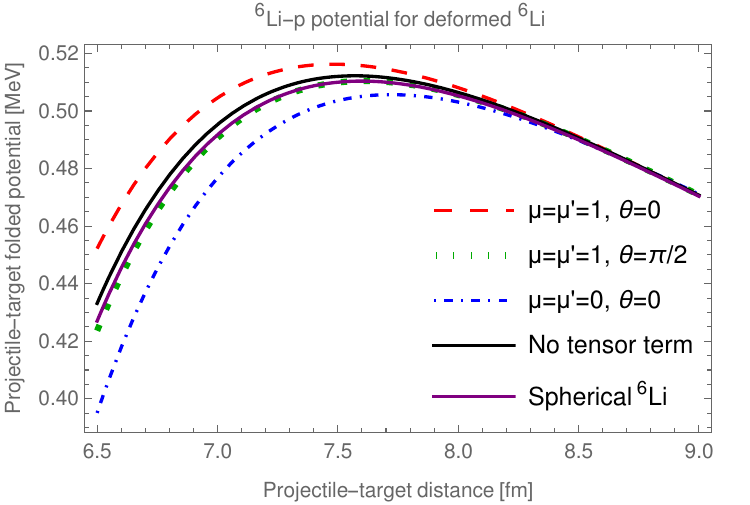}%
		\caption[\texorpdfstring{\nuclide[6]{Li}--\nuclide{p}}{6Li-p} form factor for a deformed \texorpdfstring{\nuclide[6]{Li}}{6Li} wave-function]{\label{figFormFactorQuantistico6LipVariAngoli}%
			Form factor in \cref{eqScomposizioneGenericaPotenzialeCentraleTensoriale} employed in \cref{secResults6LipFusion} for the \nuclide[6]{Li}-\nuclide{p} system, as a function of the projectile-target distance $r$, for different \nuclide[6]{Li} spin projections $\mu, \mu'$, and angles $\theta$ between projectile-target momentum and \nuclide[6]{Li} quantization axis. The black solid line is the average over all orientations (i.e.~just $U_C$ in \scref{eqScomposizioneGenericaPotenzialeCentraleTensoriale}). The purple solid line is the potential generated by a spherical \nuclide[6]{Li} ($c_2=0$ in \cref{eqEspressioneEsplicitaTermineCentraleFormFactor6LiDeformato}).
		}
	\end{figure}

	A sample of the radial profiles of the form factor computed in this manner are shown in \cref{figFormFactorQuantistico6LipVariAngoli} for the \nuclide[6]{Li}-\nuclide{p} system.
	The \nuclide{\alpha}--\nuclide{p} potential adopted for the calculation is the central part of the potential discussed in \cref{secapPotential}.
	The \nuclide{d}--\nuclide{p} interaction %
	is a phenomenological potential adjusted on \nuclide[3]{He} ground-state basic structure properties, %
	obtained %
	from the \textsc{Fr2in} code \cite{BrownReactionCodes} and whose parameters are listed in \cref{tabParametriNumericiPotenziali3He}. Such potential %
	has a simpler and smoother form with respect to the other \nuclide{d}--\nuclide{p} potential introduced in \cref{sec3HepPotential}, and thus appears to be more suitable %
	for the present application. %
	\Cref{figFormFactorQuantistico6LipVariAngoli} %
	also shows the central part of the form factor separately (black solid line), and the form factor obtained including only the spherical component of \nuclide[6]{Li} cluster-model wave-function (purple solid line). It can be seen that the central potential generated by the deformed \nuclide[6]{Li} is slightly \emph{higher} than the one found in the spherical case.
	This happens because the $L=2$ radial wave-function in \cref{figFunzioniDondaRadialiAlphad} is on average more compact than the spherical one, due to the difference in the number of nodes of the two functions. %
	It is stressed that the difference in the central part of the potential is small, possibly enough to be comparable with the inaccuracy originated by the use of phenomenological forms for the radial wave-functions, as described in \cref{secCostruzioneGroundState6LiDeformato}. In this respect, a more consistent construction, based on a \nuclide{\alpha}--\nuclide{d} potential with spin-orbit and tensor components, could thus turn out to be relevant. %

\subsection{The \texorpdfstring{$\nuclide[6]{Li}+\nuclide{p}$}{6Li+p} barrier penetrability}\label{secResults6LipFusion} %

	The same idea discussed in \cref{secLiDeformationClassicalModel} was here employed to study the impact of \nuclide[6]{Li} ground state deformation in the $\nuclide[6]{Li}+\nuclide{p}$ barrier penetrability. For each spin projection $M$ of \nuclide[6]{Li} and each orientation of the projectile-target momentum with respect to the quantization axis for \nuclide[6]{Li} spin, the $s$-wave WKB barrier penetrability is computed for a central interaction equal to the corresponding diagonal component of the projected Hamiltonian in \cref{eqCompleteHamiltonianTargetStatesMatrixElements}, $\tilde{\mathcal H}_f(M, M, \v R)$. %
	All effects of non-conservation of orbital angular momentum are thus neglected%
	\footnote{Note that the non-diagonal components of \cref{eqCompleteHamiltonianTargetStatesMatrixElements} are not purely real and do not include kinetic terms, so they cannot be treated within the rather simple model in use at present.}. %
	It is stressed that, just as in \cref{secLiDeformationClassicalModel}, treating the complete non-central problem phenomenologically as a set of many central problems (as a function of the direction) greatly simplifies the computations, but may destroy some features of the system which are relevant for the process of interest. %
	This is thus considered to be the most relevant approximation performed in the calculation discussed here. %
	
\begin{figure}[tb]%
		\centering
		\includegraphics[keepaspectratio = true, width=\linewidth]{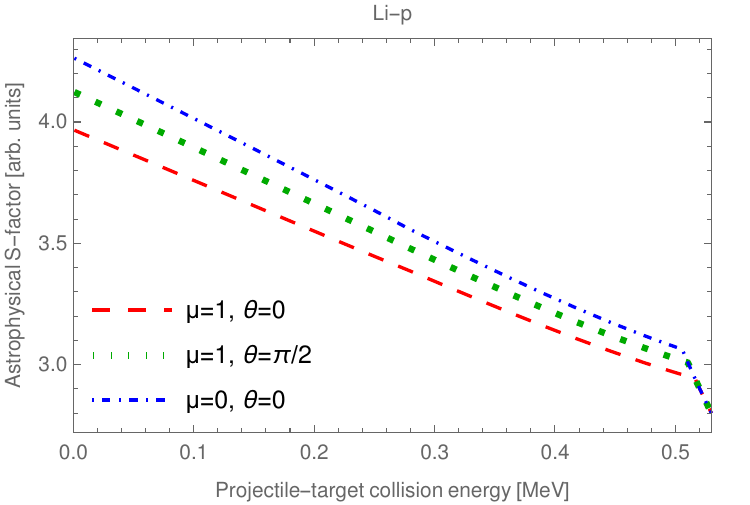}%
		\caption[\texorpdfstring{\nuclide[6]{Li}--\nuclide{p}}{6Li-p} barrier penetrability for a deformed \texorpdfstring{\nuclide[6]{Li}}{6Li} wave-function]{\label{figSFactorDeformazioneQuantistica6LipVariAngoli}%
			Each line is the astrophysical factor for WKB radial barrier penetration, computed as described in text, for the radial potential shown as a line with equal colour and pattern in \cref{figFormFactorQuantistico6LipVariAngoli}. %
		}
	\end{figure}

\begin{figure}[tb]%
		\centering
		\includegraphics[keepaspectratio = true, width=\linewidth]{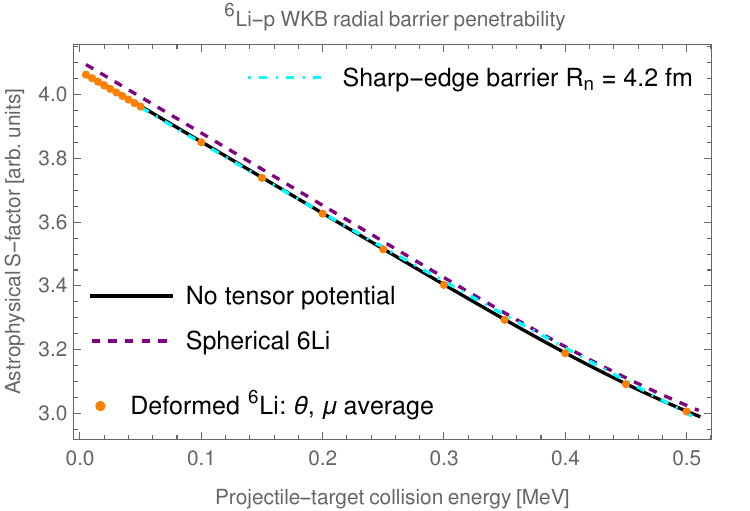}%
		\caption[\texorpdfstring{\nuclide[6]{Li}--\nuclide{p}}{6Li-p} average barrier penetrability for a spherical or deformed \texorpdfstring{\nuclide[6]{Li}}{6Li} wave-function]{\label{figSFactorDeformazione6LiMediato}%
			Orange points are the average of calculations performed as in \cref{figSFactorDeformazioneQuantistica6LipVariAngoli} over all angles and spin projections. Black solid line is the astrophysical factor for WKB radial barrier penetration using only the central part of the potential in \cref{eqScomposizioneGenericaPotenzialeCentraleTensoriale}. Purple dashed line is the calculation for the potential obtained fixing $c_2=0$ in \cref{eqFunzione6LiGroundStateDeformataGenerica}. For comparison, cyan dot-dashed line is \cref{eqSFactorWKBL0} with parameters fixed to match the other calculations.
		}
	\end{figure}

	\Cref{figSFactorDeformazioneQuantistica6LipVariAngoli} represents a sample of the associated astrophysical factors obtained in this manner. %
	\Cref{figSFactorDeformazione6LiMediato} displays, as orange points, the average of the computed astrophysical factor %
	over all orientations and spin projections, %
	while the black solid line is %
	the astrophysical factor associated to the average of the potential in \cref{eqScomposizioneGenericaPotenzialeCentraleTensoriale}, which coincides with just the central part of the same potential (shown as the black solid line in \cref{figFormFactorQuantistico6LipVariAngoli}).
	These calculations are the analogous of points and line in \cref{figTransmissionPotSpitaleri2016PLB} regarding the classical cluster model. Again, it is found that the calculation involving the averaged potential yields higher transmission coefficients, but in this case the difference between the two averaging schemes is negligible (less than 1\textperthousand),
	and would be even smaller if the %
	\nuclide[6]{Li} deformation (in particular the parameter $c_2$) were adjusted on the \nuclide[6]{Li} quadrupole moment (see the discussion in \cref{secCostruzioneGroundState6LiDeformato}).
	The difference between classical and quantum model may in part be due to the different reaction considered here (an effect due to \nuclide[6]{Li} structure is presumably more important in \nuclide[6]{Li}--\nuclide[6]{Li} scattering than in \nuclide[6]{Li}--\nuclide{p}), but it is thought that the most prominent difference lies in the way the non-central potential was computed. In particular, in the classical case the complete cluster-model potential contains all multipoles, while in the quantum form factor in \cref{eqCompleteHamiltonianTargetStatesMatrixElements} only multipoles compatible with the composite nucleus structure are allowed (here, in particular, only central and quadrupolar terms).
	
	It can also be seen from the figure that the calculated penetrability is very well reproduced by a sharp-edge barrier model fit (similar to those shown in \cref{chaPhenomenology}), computed in WKB approximation for consistency.
	This implies that the central part of the adopted nuclear potential is not causing relevant ``non-standard'' effects in the low-energy barrier penetrability. It is however stressed %
	that the precise form of the potential %
	is still important in determining the parameters of the phenomenological model. For instance, %
	the effective nuclear radius found from the fit (\SI{4.2}{\femto\metre}) is smaller than both the peak radius (\SI{7.5}{\femto\metre}) and the zero-crossing radius (\SI{5.5}{\femto\metre}) of the average form factor, meaning that the adopted potential causes a less sharp energy trend of the astrophysical factor with respect to a pure-Coulomb potential truncated at the aforementioned zero-crossing radius.
	
	Finally, \cref{figSFactorDeformazione6LiMediato} includes a calculation (purple dashed line) performed again using the projected Hamiltonian in \cref{eqCompleteHamiltonianTargetStatesMatrixElements}, adopting an inert di-cluster model for the \nuclide[6]{Li} state, but in this case including only the spherical component (i.e.~setting $c_2=0$ in \cref{eqFunzione6LiGroundStateDeformataGenerica}): the corresponding form factor is the purple solid line in \cref{figFormFactorQuantistico6LipVariAngoli}. The comparison with the other calculations thus gives an estimation of the impact of \nuclide[6]{Li} ground state deformation. Here, %
	given the small %
	difference in the potentials shown in \cref{figFormFactorQuantistico6LipVariAngoli} (see comment to the figure in \cref{secExplicitRelationElementaryPotentialWithFOrmFactor}), the spherical case yields slightly higher transmission coefficients.
	The qualitative result would be the same adopting a smaller value for $c_2$ (namely a less deformed \nuclide[6]{Li}), %
	but the aforementioned difference would become smaller.
\chapter*{Conclusions}\addcontentsline{toc}{chapter}{Conclusions}\markboth{Conclusions}{}

	This thesis investigated several %
	aspects of nuclear reactions induced by light charged particles at energies below or around the Coulomb barrier.
	Special emphasis %
	was put on the use of cluster models for the description of the reactants structure,
	in order to probe its impact %
	on reaction dynamics.
	It was possible to examine separately the effect %
	of several properties characterising %
	the involved nuclei, including ground-state deformations, spatial correlations between the elementary constituents, %
	and the couplings induced by reactant interactions, %
	evaluating the importance of including each feature %
	in the calculations.
	
	Of particular interest was the application of this study to the regime relevant for astrophysical processes, also in view of the experimental anomalies constituting the electron screening problem for atomic systems, %
	namely the %
	excess of enhancement on cross-sections measured directly in fixed-target experiments %
	at sufficiently low energies %
	which is often observed in %
	several reactions of astrophysical interest (see %
	\cref{tabPotScreeningAltreReazioni} for some references, and \cref{secScreeningExperimentalData} for the discussion on the $\nuclide[6]{Li}+\nuclide{p}\to\nuclide[3]{He}+\nuclide{\alpha}$ case). %
	In this regard,
	in \cref{chaPhenomenology} it was seen how directly measured data can be corrected by the expected screening enhancement (using \cref{eqFattoreEnhancementPotScreeningCasoGenerale} and the theoretical estimations of the screening potential in \cref{secScreeningByAtomicElectrons}), which is important to later perform a sensible comparison with calculations modelling the bare-nucleus reaction process %
	at low energies.
	It was also shown in \cref{secScreeningExperimentalData} that experimental observations on atomic screening can give rise to different, and possibly misleading, interpretations depending on the way they are analysed, implying that %
	sufficient %
	care is required in their study, %
	in particular
	for the purpose of addressing %
	the open issue on the consistency between observations from different experiments and theoretical expectations. %

	In \cref{secCalcoliDiTransfer}, the $\nuclide[6]{Li} + \nuclide{p} \to \nuclide[3]{He} + \nuclide{\alpha}$ reaction was explicitly studied as a direct transfer process, adopting several formalisms.
	The one-step deuteron-transfer distorted-wave Born approximation (DWBA) calculation presented in \cref{secCalcoliDWBADeuteronTransfer}
	takes into account only the core--deuteron relative motion in both \nuclide[6]{Li} and \nuclide[3]{He} ground states, and the interactions between \nuclide{\alpha}, \nuclide{p} and \nuclide{d}.
	Such model, which is well-established in literature, was found here capable of capturing %
	the qualitative features of the process %
	at sub-Coulomb energies and reproduce the correct order of magnitude for absolute cross-sections. This is also a result of the work that was carried out to adequately fix the physical ingredients of the calculation, outlined in \cref{secDWBAOpticalPotentials,secCalcoloDWBAOverlapFunctions}, in terms of available %
	structure and elastic scattering data. %
	The calculated trend of cross-sections toward astrophysical energies is however somewhat higher than what is found in experimental data. The same feature is routinely found in microscopic calculations \cite{Arai2002,Vasilevsky2009,Solovyev2018}. In particular, the blue solid line %
	in \cref{figdTransferConOSenzaL2} (see also \cref{figCalcoliRiscalatiSuiDati}) computed in this work resembles very closely the resonating group method (RGM) calculation shown as the solid line in \cite[fig.~8]{Arai2002}, %
	in both absolute value and trend. %
	Better (but still not full) agreement was later found in the algebraic RGM calculation in \cite{Solovyev2018}, where it is hypothesised (in line with the comments in \cite{Arai2002})
	that the improvement may be due to the use of a parametrisation for the nuclear interaction including tensor %
	components. %
	
	The one-particle DWBA framework %
	was also employed to study the impact of \nuclide[6]{Li} and \nuclide[3]{He} ground-state quadrupole deformation on the transfer cross-section.
	The wave-function deformed components introduce new partial waves for the transfer process which are otherwise forbidden (see \cref{figdTransferPartialWaveExpansions}),
	but no significant effect was observed on the transfer %
	cross-section for $s$-wave collisions, %
	which is the dominant one at the energies of astrophysical interest. %
	This notwithstanding the fact that the blue solid line in \cref{figdTransferConOSenzaL2} emphasises \nuclide[6]{Li} deformation by including a rather strong quadrupolar component in the wave-function %
	(see the discussion in \cref{secadPotential,secCostruzioneGroundState6LiDeformato}). %
	On the other hand, %
	static deformations are %
	strictly required to account for one-step ground-state-to-ground-state (namely, without coupling to intermediate states or virtual excitations) population of the experimental resonance at about \SI{1.5}{\MeV} centre-of-mass collision energy, associated to a $5/2^-$ \nuclide[7]{Be} level.
	The importance of such resonant mechanism, as well as other features of the computed cross-section, is sensitive to the choice of the adopted optical potentials, as was studied in \cref{figdTransferPotenziale3HeaAlternativo}.
	Nonetheless, %
	it is reasonable to expect that the resonance can be %
	additionally fed %
	through excitations to unbound states,
	given also that the breakup channel is already open at the resonance energy. The preliminary Greider-Goldberger-Watson (GGW) calculation shown in \cref{figCalcoloCDCCdTransfer} seems to support this interpretation.
	In order to obtain quantitative agreement in absolute value and trend with experimental data, both above and below the Coulomb barrier, it may thus be relevant to adopt a description going beyond the %
	one-step DWBA.

	Some research %
	was performed on %
	the importance of the approximation, dictated by computational limitations, %
	of neglecting non-central components of all optical potentials appearing in the remnant term of the transition amplitude (``$V_{ab} - U$'' in the notation of e.g.~\cref{eqAmpiezzaDiTransizioneDWBAPrimoOrdine}),
	which was applied to all calculations in \cref{secCalcoliDiTransfer} (see the discussion in \cref{secDWBAPriorPost}).
	The discrepancy between results in prior and post forms, which was checked for both one- and two-particle transfer calculations performed in this work %
	(see \cref{figpnTransferPriorPost} for an example),
	clearly shows that, in order to obtain reliable results, %
	a careful study and judicious choice on the form to be employed is necessary, %
	and suggests that the error committed on computed cross-sections might %
	be significant%
	\footnote{Note on the other hand that prior-post agreement is neither necessary nor sufficient for an accurate calculation adopting a well-chosen form.}.
	At the same time, non-central components of projectile-target potentials are important to generate realistic shapes for the distorted-waves. %
	For instance, it was observed that the inclusion of spin-spin components in the \nuclide[6]{Li}--\nuclide{p} interaction (see \cref{sec6LipOpticalPotential}) was decisive to remove spurious resonances from both the elastic scattering and the %
	transfer cross-section.
	More in general, the inclusion of non-central components was on the whole seen to improve the description of the transfer process, and %
	it is thought to be preferable to manage the associated numerical limitations, rather than impoverishing the physical description reducing all optical potentials to central forms.
	In \cref{secCalcoliDWBApnTransfer}, the $\nuclide[6]{Li} + \nuclide{p} \to \nuclide[3]{He} + \nuclide{\alpha}$ reaction was %
	analysed as a two-nucleon-transfer process %
	in second-order DWBA, thus explicitly taking into account the transferred system internal degrees of freedom.
	The work focused on the emergence of clustering in the \nuclide[6]{Li} structure, observed as the formation of spatially correlated configurations for the pair of transferred nucleons (see \cref{figPDF6Li} and commenting text). %
	It was shown in \cref{figpnTransferRuoloShell2s} that more-strongly-clustered wave-functions give rise to higher transfer cross-sections, a result which was found to be very robust with respect to the choice of physical ingredients for the calculations, %
	and which is in fact often observed in literature when studying the transfer of two neutrons \cite{Oertzen2001}. %
	Regarding in particular the present calculation, the result formally arose because the simultaneous and sequential %
	transition amplitudes are in phase opposition, and the less-strongly-clustered \nuclide[6]{Li} wave-function enhances the importance of the $s$-wave sequential contribution, emphasising the destructive interference (compare \cref{figpnTransferPartialWaveExpansions,figpnTransferPartialWaveExpansionsCaso1pPiu2s}). The same mechanism caused the cross-section in the two analysed cases to differ not just through an uniform rescaling, but also in its energy trend. %
	This study also demonstrated %
	that an accurate and detailed description of the reactants internal-motion wave-function can be paramount for a correct modelling of the reaction process. %
	To evaluate quantitatively %
	this statement, %
	it can be considered that the aforementioned
	change in the clustering strength %
	was obtained by simply %
	inverting the relative phase of a selected component %
	in the \nuclide[6]{Li} wave-function (see again \cref{figPDF6Li}), contributing to the norm of the state by less than \SI{4}{\percent}, and thus altering its interference with other components. %
	Even ignoring the effects on the cross-section absolute value, %
	the change in the energy trend %
	induced by this procedure
	is seen %
	to be of the same order of magnitude required to model the discrepancy between direct and indirect data at astrophysical energies (see \cref{figCalcoliRiscalatiSuiDati}).
	The comparison between results obtained in the one- and two-particle transfer formalisms was somewhat puzzling. %
	Even though both the \nuclide{\alpha}--\nuclide{d} interaction, employed in the one-particle transfer (see \cref{secCostruzioneGroundState6LiDeformato}), and the \nuclide{\alpha}--nucleon one, adopted to construct the three-body wave-function %
	(see \cref{secTNTCalcoliPraticiDescrizioneWFsingleparticle,secTNTCalcoliPraticiDescrizioneWFSimultaneo}), yield a reasonable description of \nuclide[6]{Li} basic structure properties, %
	the two-nucleon transfer calculation yields greater cross-sections (see \cref{figConfrontoOneTwoParticleTransfer}),
	and the deviation seems in fact to be connected %
	to the different adopted binding potentials and internal-motion wave-functions (see the discussion in \cref{secRisultatipnTransferComparisonDeuteronAndpnTransfer} and in particular \cref{figPotenzialealphadEfficace}).
	Considering also %
	the comparison with experimental data, this could be a suggestion that the reactants description %
	in the two-particle-transfer picture could benefit from an improvement. In particular, the intermediate \nuclide[5]{Li} system was always treated as a fictitious bound nucleus, to simplify the practical calculation, %
	and %
	the interaction between the transferred nucleons is in the present formulation always neglected, or at most reabsorbed phenomenologically.

	In \cref{secCDCCCalcoloPratico}, %
	a generalised-distorted-wave
	deuteron-transfer calculation was attempted, using the formalism of the Greider-Goldberger-Watson (GGW) transition amplitude.
	The use of generalised-distorted-waves approaches is interesting, in perspective, because it offers the possibility of describing virtual couplings to excited states, to all orders (rather than as a multi-step process), within both the initial and final partitions in the same calculation. Such excitations can describe quantum-mechanically processes of dynamical deformation or reorientation of reactants.
	The obtained preliminary results suggest %
	that the present calculation %
	is not accurate at low collision energies because of %
	the long-range Coulomb tail of the transition operator, which is not cancelled when the transferred system is a charged particle. %
	An improvement over the original GGW formulation appears to be %
	required in order to apply it %
	to generic transfer reactions %
	in the regime here of interest, and %
	is currently under study.
	Above the Coulomb barrier, virtual excitations to closed channels, especially to \nuclide[6]{Li} negative parity states, seem to be cardinal %
	in populating the resonance visible in the transfer excitation function. %
	In \cref{secLiDeformation}, the role of clustering and ground-state deformations in \nuclide[6]{Li} was studied %
	by evaluating the Coulomb barrier penetrability in reactions involving this nuclide. The penetrability is in turn a prominent %
	characteristic in the description of reactions taking place at energies below the barrier.
	In particular, in \cref{secLiDeformationQuantumModel} a tensor \nuclide[6]{Li}--\nuclide{p} interaction was constructed in terms of the properties of a \nuclide[6]{Li} di-cluster model state including quadrupole deformations, %
	and the corresponding barrier penetrability was compared with the one found %
	assuming a spherical shape for reactants.
	The \nuclide[6]{Li} static deformation is important to describe its observed structure properties, %
	nonetheless, similarly to what was seen when studying the transfer process, %
	it does not appear to induce significant effects in the barrier penetration process (see \cref{figSFactorDeformazione6LiMediato}), %
	at least in the relatively simple calculation performed in this work.

	\paragraph{} The investigation that was carried out so far opens up interesting perspectives under several points of view.
	First of all, each of the theoretical frameworks taken into account in this work was only applied to the study of the \nuclide[6]{Li} nucleus, and in particular the \nuclide[6]{Li}+\nuclide{p} interaction and its transfer channel. It would be interesting to extend these studies to other systems and compare the conclusions. %
	For example, an investigation on the \nuclide[9]{Be} nucleus appears intriguing from both the nuclear structure %
	point of view, given its Borromean cluster configuration \cite{Casal2014}, and in relation with its astrophysical applications in both primordial and stellar nucleosynthesis (see references in \cite{Brune1998}), and the relevance for the electron screening problem (see e.g.~\cite{Fang2018}). In the study of such system, one may expect excitations to continuum states to play a particularly prominent role, given the small binding energy %
	of the nucleus: their description is particularly challenging because \nuclide[9]{Be} has a marked three-body cluster structure which includes two charged particles \cite{Casal2014}, making a simple binning discretisation procedure for the continuum (as the one employed here in \cref{secCDCCCalcoloPratico}) unsuitable.
	
	Regarding instead possible improvements of the studies already performed here,
	it was argued %
	that a detailed description of reactants as three-particle systems (core and two transferred nucleons) %
	is important to understand the strength and role of clustered configurations, and to generate an accurate modelling of the two-nucleon transfer process. %
	In particular, it seems relevant to upgrade the construction of the three-particle wave-functions %
	from the phenomenological approach adopted in \cref{secCalcoliDWBApnTransfer} %
	to a more microscopical, full three-body calculation. One of the difficulties %
	involved in such improvement is to provide a construction not only of the wave-functions of interest, but also the associated vertex functions (namely, the product of the binding potential and the wave-function) required for the transfer calculation.
	Another challenging but possibly important aspect is %
	the treatment %
	of the unbound intermediate states of the sequential process in a more consistent manner, which would involve the inclusion of a continuum spectrum at least in the intermediate partitions.
	Additionally, it is significant for the analysis carried out here %
	to evaluate %
	in a more quantitative manner the strength and features of the clustered configurations of a given structure. %
	Within two-cluster models, the main directly comparable quantity would be the spectroscopic factor for a chosen configuration, while other parameters, as the shape of inter-cluster potentials, are affected only in a rather indirect manner.
	Within three-particle models, instead, a more explicit analysis is possible %
	in terms of the computed internal-motion %
	wave-functions, for instance estimating the norm and tail behaviour associated to each peak in \cref{figPDF6Li}, or adopting an approach as the one discussed in \cite[sec.~5]{Bang1979}.
	
	As discussed earlier, the impact of dynamical excitations, especially below the Coulomb barrier, %
	is still under study.
	Apart from the aforementioned improvements in the GGW %
	formalism, %
	in this respect it is doubtless that a good description of the reactants continuum structure can be very useful in improving the predictions given by the calculation.
	The system spectrum in each partition, in turn, is essentially determined by the adopted inter-cluster potentials, which are often fixed phenomenologically to reproduce selected features of potentially relevant resonances (as done in \cref{secCDCCCalcoloPratico} for the \nuclide{\alpha}--\nuclide{d} interaction).
	Since the most complete and detailed available structure information regards the ground-state properties of nuclei (e.g.~their deformations), a promising route for improvement could consist in using these properties
	not just in the description of the ground-state wave-functions themselves, %
	but mainly as a constraint and benchmark for the form of the system interactions.
	It is thus appealing %
	to expand the formalism discussed in \cref{sec6LiDeformationsConstructionoftheprojectiletargetpotential} to include spin-coupling terms to the elementary interactions (%
	see e.g.~\cite{Merchant1985}), in order to examine its application to the generation of %
	more realistic and consistent binding potentials. For instance, starting from only the \nuclide{\alpha}--nucleon interaction and the deuteron ground-state wave-function, an \nuclide{\alpha}--\nuclide{d} potential with spin-orbit and tensor components could be constructed, which can be adjusted to reproduce radius and electromagnetic moments of \nuclide[6]{Li} ground state, and then constrained by the available information on %
	the \nuclide[6]{Li} low-lying continuum structure.

	Finally, to improve the barrier penetrability calculations reported in \cref{secLiDeformation}, it may be fruitful to implement a formalism similar to the one treated in \cite{Hagino2012}, %
	going beyond %
	the standard single-channel radial WKB approximation currently in use. %
	In perspective, it would be desirable to achieve a full quantum-mechanical (but practically computable) formulation of the problem, where both the anisotropies and the non-diagonal components of the projectile-target %
	form factor can be consistently taken into account.

\appendix%
\numberwithin{equation}{chapter}%
\chapter{Angular momentum and spherical harmonics}\label{secAppendiceSphericalHarmonics}%
	Let $Y_{lm}(\theta,\phi)$ be the spherical harmonic of degree $l$ and order $m$ (where $l$ is a non-negative integer and $m$ an integer $\in [-l,l]$) %
	computed at angles $(\theta,\phi)$ in some spherical coordinate system with the $z$ axis as polar axis.
	For brevity, let $\Omega$ be a shorthand for $(\theta,\phi)$. %
	Spherical harmonics are eigenfunctions of the angular momentum square-modulus operator, $\op L^2$, with eigenvalue $\hbar^2 l (l+1)$, and of the angular momentum $z$-projection operator, $\op L_z$, with eigenvalue $\hbar m$ \cite[sec.~VI, eq.~(D-8)]{Cohen1977}.
	
\paragraph{Complex conjugate of spherical harmonics}%
	Any spherical harmonic $Y_{lm}$ can be computed from $Y_{l,-m}$ using
	\begin{equation}\label{eqComplexConjugateSphericalHarmonics}
	\coniugato{Y}_{l,m}(\Omega) = (-1)^m Y_{l,-m}(\Omega)
	\end{equation}
	
\paragraph{Spherical harmonics expansion}%
	Spherical harmonics (being eigenvectors of angular momentum with distinct eigenvalues) are orthonormal, meaning that \cite[sec.~VI, eq.~(D-23)]{Cohen1977}
	\begin{equation}\label{eqOrtonormalitaArmonicheSferiche}
	\Braket{Y_{\lambda\mu}|Y_{lm}} = \int_{4 \pi} \coniugato{Y}_{\lambda \mu}(\Omega) Y_{l m}(\Omega) \d\Omega = \delta_{l\lambda} \delta_{m\mu}
	\end{equation}
	they also form a complete set, meaning that any %
	function $F$ defined on the $(\theta,\phi)$ domain can be written as a superposition of spherical harmonics
	\cite[sec.~VI, eq.~(D-24)]{Cohen1977}, $F(\theta,\phi) = \sum_{lm} c_{lm} Y_{lm}(\theta,\phi)$,
	where each coefficient can be found projecting on the desired component, $c_{lm} = \Braket{Y_{lm}|F}$. Clearly, if the function also depends on other parameters, the coefficients $c$ of the expansion will in general be functions of each of those.
	The most common occurrence is %
	that of a function $F(r, \theta, \phi)$ depending both on modulus and direction of a vector $\v r$: for this case, it is convenient to introduce a specific notation,
	\begin{equation}\label{eqDefinizioneFunzioneRadialeRidotta}
	F(\v r) = \sum_{ml} c_{lm} \frac{u_{lm}(r)}{r} Y_{lm}(\Omega)
	\end{equation}
	where $u_{lm}$ is %
	a \emph{reduced} radial function and $c_{lm}$ is a numerical coefficient. %
	They are defined so that, if $F$ has norm 1, then also each $\int_0^{+\infty} \m{u_{lm}(r)}^2 \d r = 1$, and $\sum_{lm} \m{c_{lm}}^2 = 1$.

\paragraph{Laplacian}%
	The Laplacian operator $\nabla^2$, which is $\sum_{k \in \{x,y,z\}} \partial_k^2$ in Cartesian coordinates, in spherical coordinates can be decomposed as follows \cite[sec.~VII, eq.~(A-16)]{Cohen1977}:
	\begin{equation}\label{eqEspressioneLaplacianoComeMomentoAngolare}
	\nabla^2 = \frac{1}{r^2} \partial_r \left(r^2 \partial_r\right) - \frac{\op L^2}{\hbar^2 r^2}
	\end{equation}
	Note that, given a function $F(\v r)$ written as in \cref{eqDefinizioneFunzioneRadialeRidotta}, it is
	\begin{equation}\label{eqApplicazioneLaplacianoScritturaRadialeRidotta}
		\nabla^2 F(\v r) = \sum_{ml} c_{lm} \frac{1}{r} \[ \partial_r^2 - \frac{l (l+1)}{r^2} \] u_{lm}(r) Y_{lm}(\Omega)
	\end{equation}

\paragraph{Coupling of two angular momenta}%
	Let $\Ket{j_1, m_1}$ be an eigenvector of some angular momentum modulus and projection operators $\op j_1^2$ and $\op j_{1,z}$, with quantum numbers $j_1$ and $m_1$ (either integer or half-integer).
	Let $\Braket{(j_1, m_1), (j_2,m_2) | J, M}$ be the Clebsch-Gordan coefficient for the coupling of two distinct angular momenta, with modulus and projection quantum numbers $j_i$ and $m_i$, into a total angular momentum denoted by quantum numbers $J$ and $M$, meaning that \cite[sec.~X, eq.~(C-66)]{Cohen1977}
	\begin{equation}\label{eqDecomposizioneClebschGordan}
		\Ket{j_1,j_2, (J,M)} = \sum_{m_1,m_2} \Braket{(j_1, m_1), (j_2,m_2) | J, M} \, \Ket{j_1,m_1} \Ket{j_2,m_2}
	\end{equation}
	Clebsch-Gordan coefficients are real, and yield zero whenever the related coupling would be impossible (for instance, if $m_1 + m_2 \neq M$).
	They are also defined so that the set of $\Ket{j_1,j_2, (J,M)}$ for each value of $J$ and $M$ is orthonormal, provided that the set of $\Ket{(j_1,m_1),(j_2,m_2)}$ for each value of $m_1$ and $m_2$ is orthonormal, and vice-versa, %
	which implies \cite[sec.~$\text{B}_\text{X}$, eq.~(7), (9)]{Cohen1977} %
	\begin{equation}\label{eqRelazioniOrtonormalitaCoefficientiClebschGordan}\begin{aligned}
	& \sum_{m_1,m_2} \Braket{(j_1, m_1), (j_2,m_2) | J', M'} \Braket{(j_1, m_1), (j_2,m_2) | J, M} = \delta_{J,J'} \delta_{M,M'} \\ %
	& \sum_{M,J} \Braket{(j_1, m'_1), (j_2,m'_2) | J, M} \Braket{(j_1, m_1), (j_2,m_2) | J, M} = \delta_{m_1,m_1'} \delta_{m_2,m_2'}
	\end{aligned}\end{equation}

	The sign of Clebsch-Gordan coefficients depends on the order in which angular momenta are coupled. In particular, it is
	\begin{equation}\label{eqRelazioneSegnoClebschGordanOrdineAccoppiamento}
		\Braket{(j_1,m_1), (j_2, m_2) | J, m_J} = (-1)^{J - j_1 - j_2} \Braket{(j_2,m_2), (j_1,m_1) | J, m_J}
	\end{equation}
	
	A common occurrence where such a coupling is useful is that of a physical system whose Hamiltonian $\mathcal H$ commutes with the moduli of two angular momenta, $\op j_i^2$ with $i = 1,2$, and the modulus and projection of their composition, $\op J^2$ and $\op J_z$, but not with each $\op j_{i,m}$ separately, so that there exist a basis of eigenstates of $\mathcal H$ %
	with definite $j_1, j_2, J$ and $M$.
	Since $\mathcal H$ usually also acts on some other space (in addition to the angular momentum one), here generically denoted by coordinate $x$, the elements of the aforementioned eigenbasis %
	can be denoted as $\Ket{F_{j_1,j_2,J,M,n}}$, where %
	$n$ is a generic quantum number enumerating %
	the different states %
	at fixed $j_1,j_2,J,M$, which together form a basis of $x$ subspace.
	In turn, the basis element $\Ket{F_{j_1,j_2,J,M,n}}$ may be expressed as $\Ket{f_{j_1,j_2,J,M,n}} \Ket{j_1,j_2, (J,M)}$, where $\Ket{f}$ is a state in the $x$ subspace only.
	
	Often, $\mathcal H$ is the identity with respect to the total angular momentum projection, %
	in which case $\Ket{f_{j_1,j_2,J,M,n}}$ is actually the same for any $M$.
	As explicit example, if $j_1$ is the orbital angular momentum of some spatial coordinate, $l$, and $\mathcal H$ acts (also) on the associated radial coordinate $r$, %
	a given basis element, using the same notation in \cref{eqDefinizioneFunzioneRadialeRidotta}, would take the form
	\begin{multline}\label{eqDecomposizioneClebschGordanEsempioSpecificoLOrbitale}
		\Braket{\v r | F_{l,j_2,J,M,n}} =\\= \frac{u_{l,j_2,J,n}(r)}{r} \sum_{m,m_2} \Braket{(l, m), (j_2,m_2) | J, M} \, \Ket{j_2,m_2} Y_{lm}(\Omega)
	\end{multline}

\paragraph{Change of coupling scheme in the sum of four angular momenta}%
	
	In analogy with the preceding subsection, let $\Ket{l_1, m_{l_1}} \Ket{l_2, m_{l_2}} \Ket{s_1, m_{s_1}} \Ket{s_2,m_{s_2}}$ be an eigenstate of modulus and projection of four distinct angular momenta, with quantum numbers $l_1$ and $m_{l_1}$ for modulus and projection of the first angular momentum, $\v l_1$, and so on.
	Consider an eigenspace with fixed modulus quantum numbers for all angular momenta.
	A basis of such eigenspace is given by the set of states in the form $\Ket{l_1, m_{l_1}} \Ket{l_2, m_{l_2}} \Ket{s_1, m_{s_1}} \Ket{s_2,m_{s_2}}$ %
	for all possible values of each $m$. %
	
	Consider now the sum of angular momenta $\v s_1$ and $\v s_2$ into a total angular momentum $\v S$, and the coupling of the set of states $\Ket{s_1, m_{s_1}} \Ket{s_2, m_{s_2}}$ into the set $\Ket{s_1,s_2, (S, m_{S})}$ precisely as in \cref{eqDecomposizioneClebschGordan}.
	Similarly let $\v{\mathcal L} = \v l_1 + \v l_2$ and $\v j_i = \v l_i + \v s_i$ and consider the respective couplings.
	Finally, let $\v j = \v j_1 + \v j_2$, and similarly consider the coupling of states $\Ket{l_1,s_1, (j_1, m_{j_1})} \Ket{l_2,s_2, (j_2, m_{j_2})}$ into eigenstates of only the modulus of $\v j_1$ and $\v j_2$ and of both modulus and projection of $\v j$, denoted here in compact form%
		\footnote{A more complete notation would specify also the other modulus quantum numbers, e.g.~$\Ket{(l_1, s_1) j_1, (l_2, s_2) j_2; j, m}$.}
	as $\Ket{j_1, j_2, (j,m)}$.
	The set of such eigenstates forms another basis of the eigenspace under study. This is referred to as the ``$j_1 j_2$ angular momentum coupling scheme''.
	It also must be $\v j = \v{\mathcal L} + \v S$, and the corresponding coupling (``$\mathcal L S$ scheme'') can be considered as well.

	Each state in the $\mathcal L S j$ basis can be expanded in the $j_1 j_2 j$ basis (and vice-versa) as %
	\begin{equation}\label{eqRelazioneMomentoAngolareInAccoppiamentoLSojj}
		\Ket{\mathcal L,S,(j,m)} = \sum_{j_1,j_2} \Braket{j_1, j_2, (j, m)|\mathcal L, S, (j, m)} \Ket{j_1, j_2, (j,m)}
	\end{equation}
	The coefficients of the expansion %
	can be found by expanding the corresponding states in the basis with fixed projections $m_{l_i}$ and $m_{s_i}$. %
	The result can be expressed through the 9-$j$ symbols, see e.g.~\cite[eq.~C.12, C.40]{Messiah2014quantum},
	but it is here given %
	more explicitly in terms of Clebsch-Gordan coefficients.
	For brevity, %
	let $\Braket{m_1, m_2 |j, m}$ be a shorthand for $\Braket{(j_1, m), (j_2,m_2) | j, m}$. %
	It is: %
	\begin{multline}
		\Braket{j_1, j_2, (j, m)|\mathcal L, S, (j, m)} = \\
		\shoveleft{= \sum_{m_S, m_{j_1}, m_{s_1}} \Braket{j, m|m_{j_1}, m_{j_2}} \Braket{j_1, m_{j_1}|m_{l_1}, m_{s_1}} \Braket{j_2, m_{j_2}|m_{l_2}, m_{s_2}} \cdot}\\\cdot \Braket{m_{l1}, m_{l_2}|L, m_L} \Braket{m_{s_1}, m_{s_2} |S, m_S} \Braket{m_L, m_S |j, m}
	\end{multline}
	Such relation is convenient for practical calculations, but note that the coefficients $\Braket{j_1, j_2, (j, m)|\mathcal L, S, (j, m)}$ actually do not depend on $m$ (in agreement with the properties of 9-$j$ symbols), as can be seen by applying on both sides of \cref{eqRelazioneMomentoAngolareInAccoppiamentoLSojj} a ladder operator acting on the total angular momentum~$j$.
	
	Note that the sign of $\Braket{j_1, j_2, (j)|\mathcal L, S, (j)}$ depends on the order in which the angular momenta are coupled, due to \cref{eqRelazioneSegnoClebschGordanOrdineAccoppiamento}. Here, the adopted convention is that ``1'' and ``2'' refer to states regarding the proton and neutron respectively, and that $\v j_i$ is obtained summing $\v l_i$ and $\v s_i$ in this order, and similarly $\mathcal{L}$ and $S$ are combined in this order.

\paragraph{Product of two spherical harmonics}%
	The product of two spherical harmonics, computed at the same direction $\Omega$, can be conveniently expressed expanding it in the spherical harmonics basis \cite[eq.~(3.7.72), (3.7.73)]{SakuraiModern1994}:
	\begin{equation}
	Y_{l_1 m_1}(\Omega) Y_{l_2 m_2}(\Omega) = \sum_{LM} \Braket{Y_{L M} | Y_{l_2 m_2} | Y_{l_1 m_1}} Y_{LM}(\Omega)
	\end{equation}
	where
	\begin{multline}\label{eqBraketTreArmonicheSferiche}
	\Braket{Y_{L M} | Y_{l_2 m_2} | Y_{l_1 m_1}} = \int_{4 \pi} \coniugato{Y}_{L M}(\Omega) Y_{l_2 m_2}(\Omega) Y_{l_1 m_1}(\Omega) \d\Omega =\\= \sqrt{\frac{(2 l_1 + 1) (2 l_2 + 1)}{4 \pi (2 L + 1)}} \Braket{(l_1, 0),(l_2,0)|L, 0} \Braket{(l_1, m_1), (l_2,m_2) | L, M}
	\end{multline}
\paragraph{Spherical harmonics addition theorem}%
	Let $\Omega_{Rz}$ be the direction of some vector $\v R$ in a spherical coordinate system having the direction of vector $\v z$ as polar axis. Let $\v r$ be another vector.
	The spherical harmonic $Y_{l0}(\Omega_{Rr})$, computed using $\v r$ as quantisation axis,
	can be expressed using \cite[sec.~$\text{A}_\text{VI}$, eq.~(57), (70)]{Cohen1977} and \cref{eqComplexConjugateSphericalHarmonics} as:
	\begin{equation}\label{eqAdditionTheoremSphericalHarmonics}
		Y_{l0}(\Omega_{Rr})
		= \sqrt{\frac{4 \pi}{2l+1}} \sum_{\lambda} Y^*_{l\lambda}(\Omega_{rz}) Y_{l\lambda}(\Omega_{Rz})
	\end{equation}

\paragraph{Multipoles expansion in rotated coordinates}%
	Consider a function $F(\theta_{Rr})$ depending only on the angle between two vectors $\v R$ and $\v r$. As seen earlier, and using the properties of spherical harmonics, it is possible to express $F(\theta_{Rr})$ as $\sum_l F_l Y_{l0}(\theta_{Rr})$. Such expansion can be converted to a rotated coordinate system, with $\v z$ as polar axis, using \cref{eqAdditionTheoremSphericalHarmonics}:
	\begin{equation}\label{eqMultipoleExpansionRotatedCoordinates}
	F(\theta_{Rr}) = \sum_l F_l \sqrt{\frac{4 \pi}{2l+1}} \sum_{\lambda} Y^*_{l\lambda}(\Omega_{rz}) Y_{l\lambda}(\Omega_{Rz})
	\end{equation}
	where
	\begin{equation}
	F_l = \int_{4\pi} \coniugato{Y}_{l0}(\theta_{Rr}) F(\theta_{Rr}) \d\Omega_{Rr}
	= 2 \pi \int_{-1}^1 \coniugato{Y}_{l0}(\theta_{Rr}) F(\theta_{Rr}) \d\cos\theta_{Rr}
	\end{equation}

\chapter{Computational aspects} %
	
\section{Experimental data sources and treatment}\label{appExperimentaldatasourcesandtreatment}

	This appendix collects some notes regarding the source and treatment of some experimental data which is employed throughout this work.

	\paragraph{Nuclear masses and excited states} %
	Nuclear ground-state masses (and associated $Q$-values) are taken from \cite{NISTcodata}, where available, or otherwise deduced using atomic masses in \cite{AME2016} %
	and electron binding energies in \cite{NISTasd}.
	
	Energy, spin and parity of excited states were found in \cite{NNDCnudat}.

	\paragraph{Cross-sections and scattering phase-shifts} Numerical values of experimental cross-sections employed in this work are taken from \cite{Exfor}, while the papers on which the data was first published are referenced in the text.
	Phase-shifts were digitised directly from figures in the original papers. %
	
	Information found %
	in the original papers was employed to check the data. %
	In principle, systematic errors are best treated separately from statistical ones, but in the present work, for simplicity, the uncertainty assigned to each measurement is the sum in quadrature of all relevant indetermination sources quoted in the original papers, and (if applicable) digitization errors given in \cite{Exfor}.
	Results from data fits and error bars shown in pictures reflect this assignment. %
	If the estimation of some error source is uncertain, and only an interval is given for the indetermination, the worst case is assumed here.

	\paragraph{Radii and multiple moments} Root-mean-square charge radii are taken from \cite{Angeli2013}. Electric quadrupole moments were found in \cite{IAEAElectroMagneticMoments}. Magnetic dipole moments were obtained from \cite{NISTcodata} if available, or from \cite{IAEAElectroMagneticMoments} otherwise. For convenience, some ground-state properties useful in the practical calculations performed in the present work are listed in \cref{tabExperimentalGroundStateData}.
	\begin{table}[tbp]
		\caption[Experimental ground-state structure data]{Experimental ground-state structure data of several nuclides of interest in the present work. %
			The table lists: spin and parity, root-mean-square charge radius $<r^2>$, %
			charge quadrupole moment $Q$, %
		magnetic dipole moment $\mu$ (in nuclear magneton units). %
		See text for data sources.}%
		\centering
		\sisetup{uncertainty-mode=compact,retain-explicit-plus}
		\begin{tabular}{ccS[table-format=1.4(3)]S[table-format=+1.3(1)]S[table-format=+1.10(2)]}\label{tabExperimentalGroundStateData} %
			Nuclide & Spin-parity & {$\sqrt{<r^2>}$ [\si{\femto\metre}]} & {$Q$ [\si{\milli\barn}]} & {$\mu$ [$\mu_N$]} \\ \toprule
			\nuclide[6]{Li} & 1+   & 2.5890(390) & -0.806(6) & +0.822043(3)		\\ %
			\nuclide{\alpha}& 0+   & 1.6755(28)  & 0	& 0				\\
			\nuclide{d}		& 1+   & 2.1421(88)  & +2.86(2) & +0.8574382338(22)	\\
			\nuclide{p} & 1/2+	& 0.8783\pm0.0086 & 0 & +2.7928473446(8) \\
			\nuclide{n} & 1/2+	& {$\sqrt{-\num{0.1149\pm0.0027}}$} & 0	& -1.91304273(45) \\
			\nuclide[3]{He} & 1/2+ & 1.9661(30)  & 0	& -2.127625307(25)	\\
			\bottomrule
		\end{tabular}
	\end{table}

	\paragraph{Other constants} Unless otherwise noted, values for ``notable'' physical constants %
	(e.g.~the fine-structure constant)
	are taken from \cite{NISTcodata}. %

\section{Adopted potentials}\label{secPotentialsDefinition}

	All two body potentials employed in this work are energy-independent and have the following form: %
	\begin{multline}\label{eqGeneralPotentialParametrization}
		V_C(r, R_C) - V_v \, f(r,R_v,a_v) - V_g \, \exp\(-\[r/a_g\]^2\) -\\
		- i V_w \, f(r,R_w,a_w)
			+ i 4 V_x a_x \, \frac{\d}{\d r} f(r,R_x,a_x) %
		+\\+ V_o \left[ j(j+1) - l(l+1) - s_1(s_1+1) \right] \frac{\SI{2}{\femto\metre\squared}}{r} \, \frac{\d}{\d r} f(r,R_o,a_o) + \\
		+ V_s \left[ s(s+1) - s_1(s_1+1) - s_2(s_2+1) \right] f(r,R_s,a_s)
	\end{multline}
	where:
	\begin{equation}\label{eqCoulombWoodsSaxonDefinition}\begin{aligned}
	V_C(r, R_C) &= \hbar c \alpha_e Z_1 Z_2 \left\lbrace\begin{aligned}
		&\frac{ 3 - r^2 / R_C^2 }{2 R_C} & r \leq R_C \\
		&\frac{1}{r} & r \geq R_C
	\end{aligned}\right.
	\\
	f(r,R,a) &= \frac{1}{1 + \exp\left(\frac{r-R}{a}\right)}
	\end{aligned}\end{equation}
	$r$ represents the distance between the interacting particles, while $l, s, j$ are the modulus quantum numbers for, respectively, %
	their relative orbital angular momentum, total intrinsic spin, total angular momentum.
	$s_1$ and $s_2$ similarly refer to the intrinsic spins of, respectively, the lightest and heaviest particle in the pair (e.g.~in the \nuclide{p}--\nuclide{d} potential, $s_1$ is the proton spin).
	In all potentials taken from literature, %
	the spin-orbit term couples only the lightest particle spin (often because the other particle has spin zero). Only in the fitted \nuclide[6]{Li}--\nuclide{p} potential, the spin-orbit term involves the total intrinsic spin, and thus has the form $V_o \left[ j(j+1) - l(l+1) - s(s+1) \right]$, contrary to what is stated in \cref{eqGeneralPotentialParametrization}.
	$Z_1$ and $Z_2$ are the reactants charge numbers, $\alpha_e$ is the fine-structure constant, $\hbar$ the reduced Planck's constant and $c$ the speed of light in vacuum. %
	Each specific potential is defined by setting the values of all parameters ($V_i$, $R_i$, $a_i$), which are reported in table \cref{tabParametriNumericiPotenziali}.
	A discussion on the construction of each potential and the relevant bibliography %
	is instead given in the main text, see in particular
	\cref{secCalcoliDWBADeuteronTransfer,secTNTCalcoliPraticiDescrizioneWFsingleparticle,secCDCCCalcoloPratico,secCostruzioneGroundState6LiDeformato,secExplicitRelationElementaryPotentialWithFOrmFactor}. %
	\begin{table}[hbp]
		\caption[Parameters of the adopted potentials]{\label{tabParametriNumericiPotenziali}%
			List of all potentials employed in this work.
			Each column of each table refers to a different potential, identified in the column header by the pair of particles it refers to. %
			Each line refers to parameter in \cref{eqGeneralPotentialParametrization} (with same notation).
			For brevity, parameters whose value is zero are sometimes omitted. %
		}
	\begin{subtable}{\textwidth}
		\caption{\label{tabParametriNumericiPotenzialiOttici}
			Optical potentials employed only for transfer calculations. Only for the \nuclide[6]{Li}--\nuclide{p} potential, the spin-orbit term involves the total intrinsic spin (see text). ``\nuclide{\alpha} -- \nuclide[3]{He} (alt.)'' refers to the alternative potential fitted on phase-shifts discussed in \cref{secTransferDeuterioTransferCrossSection}.}
		\centering
		\sisetup{table-format=2.4}
		\begin{tabular}{lSSSSS} %
& {\nuclide[6]{Li} -- \nuclide{p}} & {\nuclide{\alpha} -- \nuclide[3]{He}} & {\nuclide{\alpha} -- \nuclide[3]{He} (alt.)} & {\nuclide[5]{Li} -- \nuclide{d}} & {\nuclide[5]{Li} -- \nuclide{p}} \\ \toprule
\vphantom{$\left(\nuclide[6]{Li}\right)$}$R_C$ [\si{\femto\metre}]	&  2.326  &  0.1000 &  0.1000 &  2.223  &  2.233 \\ %
			\midrule
\vphantom{$\left(\nuclide[6]{Li}\right)$}$V_v$ [\si{\MeV}]			& 48.20   & 66.08   & 62.00   & 90.04   & 50.97   \\ %
\vphantom{$\left(\nuclide[6]{Li}\right)$}$R_v$ [\si{\femto\metre}]	&  1.908  &  2.649  &  2.500 &  2.001  &  1.912  \\ %
\vphantom{$\left(\nuclide[6]{Li}\right)$}$a_v$ [\si{\femto\metre}]	&  0.6700 &  0.7175 &  0.7175 &  0.7090  &  0.6900 \\
			\midrule
\vphantom{$\left(\nuclide[6]{Li}\right)$}$V_w$ [\si{\MeV}]			&  0	  &  0		&  1.200  & 0		  &  0.6908 \\
\vphantom{$\left(\nuclide[6]{Li}\right)$}$R_w$ [\si{\femto\metre}]	& {--}	  & {--}	&  2.500  & {--}	  &  1.854  \\
\vphantom{$\left(\nuclide[6]{Li}\right)$}$a_w$ [\si{\femto\metre}]	& {--}	  & {--}	&  0.7175 & {--}	  &  0.6900 \\
			\midrule
\vphantom{$\left(\nuclide[6]{Li}\right)$}$V_x$ [\si{\MeV}]			&  0.2246 &  1.089  & 0       & 12.20	  &  4.723  \\ %
\vphantom{$\left(\nuclide[6]{Li}\right)$}$R_x$ [\si{\femto\metre}]	&  3.634  &  2.102  & {--}	  &  2.266  &  1.854  \\ %
\vphantom{$\left(\nuclide[6]{Li}\right)$}$a_x$ [\si{\femto\metre}]	&  2.715  &  0.7386 & {--}	  &  0.6497 &  0.6900 \\
			\midrule
\vphantom{$\left(\nuclide[6]{Li}\right)$}$V_o$ [\si{\MeV}]			&  1.000  &  4.162  &  1.500 &  7.330  &  0 \\
\vphantom{$\left(\nuclide[6]{Li}\right)$}$R_o$ [\si{\femto\metre}]	&  1.817  &  2.644  &  2.500 &  1.830  &  {--} \\
\vphantom{$\left(\nuclide[6]{Li}\right)$}$a_o$ [\si{\femto\metre}]	&  0.700  &  0.2078 &  0.2078 &  0.6600 &  {--} \\
			\midrule
\vphantom{$\left(\nuclide[6]{Li}\right)$}$V_s$ [\si{\MeV}]			& 18.00   & 0 & 0 & 0 & 0 \\
\vphantom{$\left(\nuclide[6]{Li}\right)$}$R_s$ [\si{\femto\metre}]	&  1.853  & {--} & {--} & {--} & {--} \\
\vphantom{$\left(\nuclide[6]{Li}\right)$}$a_s$ [\si{\femto\metre}]	&  0.200  & {--} & {--} & {--} & {--}
		\end{tabular}
	\end{subtable}
	\end{table}
	\begin{table}[p]\ContinuedFloat%
		\caption[]{%
			List of all potentials employed in this work (continued).
		Whenever a potential was employed to construct a bound wave-function, $V_v$ was rescaled to obtain the desired binding energy for the system. Whenever a potential was employed as core-core interaction, all non-central terms were discarded.}
	\begin{subtable}{\textwidth}
		\caption{\label{tabParametriNumericiPotenziali3He}
			Potentials related to the construction of %
			the \nuclide[3]{He} state. ``{\nuclide{p} -- \nuclide{d} ($L=0$)}'' or ``$L=2$'' refers to the two potentials employed in \cref{secCalcoloDWBAOverlapFunctions} to construct the \nuclide[3]{He} state. ``\nuclide{p} -- \nuclide{d} (\textsc{Fr2in})'' refers to the potential employed in \cref{secExplicitRelationElementaryPotentialWithFOrmFactor}.}
		\centering
		\begin{tabular}{lS[table-format=3.4]S[table-format=+3.4]S[table-format=+4.4]S[table-format=2.4]} %
& {\nuclide{p} -- \nuclide{n}} & {\nuclide{p} -- \nuclide{d} ($L=0$)}  & {\nuclide{p} -- \nuclide{d} ($L=2$)} & {\nuclide{p} -- \nuclide{d} (\textsc{Fr2in})}	\\ \toprule
\vphantom{$\left(\nuclide[6]{Li}\right)$}$R_C$ [\si{\femto\metre}]	& {--}    &  2.000 &  2.000  &  1.280	\\ %
\midrule
\vphantom{$\left(\nuclide[6]{Li}\right)$}$V_v$ [\si{\MeV}]			&165.4    &179.9   &8155   & 87.30	\\
\vphantom{$\left(\nuclide[6]{Li}\right)$}$R_v$ [\si{\femto\metre}]	&  0.4000 &  0.5400  &-2.190   & 0.9500	\\
\vphantom{$\left(\nuclide[6]{Li}\right)$}$a_v$ [\si{\femto\metre}]	&  0.6000 &  0.6800 &  0.9100 & 0.6500	\\
\midrule
\vphantom{$\left(\nuclide[6]{Li}\right)$}$V_g$ [\unit{\MeV}]		&  0   &-203.3    &-8400 & 0 \\
\vphantom{$\left(\nuclide[6]{Li}\right)$}$a_g$ [\si{\femto\metre}]	& {--} & 0.6400 & 0.3500 & {--} \\
\midrule
\vphantom{$\left(\nuclide[6]{Li}\right)$}$V_o$ [\si{\MeV}]			&  0      & 0 &  1.47  & 0	\\
\vphantom{$\left(\nuclide[6]{Li}\right)$}$R_o$ [\si{\femto\metre}]	& {--}    & {--} &  2.07  &  {--}	\\
\vphantom{$\left(\nuclide[6]{Li}\right)$}$a_o$ [\si{\femto\metre}]	& {--}    & {--} &  0.06 &  {--}
		\end{tabular}
	\end{subtable}
\\[\baselineskip]
		\begin{subtable}{\textwidth}
		\caption{\label{tabParametriNumericiPotenziali6Li}
			Potentials related to the construction of the \nuclide[6]{Li} state. ``\nuclide{\alpha} -- \nuclide{d} (GGW)'' refers to the alternative potential with phenomenological spin-orbit employed in \cref{secCDCCCalcoloPratico}.}
		\centering
		\begin{tabular}{lSSSS} %
& {\nuclide{\alpha} -- \nuclide{d}} & {\nuclide{\alpha} -- \nuclide{d} (GGW)} & {\nuclide{\alpha} -- \nuclide{p}} & {\nuclide[5]{Li} -- \nuclide{n}}	\\ \toprule
\vphantom{$\left(\nuclide[6]{Li}\right)$}$R_C$ [\si{\femto\metre}]	&  1.900  & 1.900  &  2.900  & {--}	\\ %
			\midrule
\vphantom{$\left(\nuclide[6]{Li}\right)$}$V_v$ [\si{\MeV}]			& 80.09   & 80.09  & 43.00    & 43.00	\\
\vphantom{$\left(\nuclide[6]{Li}\right)$}$R_v$ [\si{\femto\metre}]	&  1.900  & 1.900  &  2.000  &  2.125	\\
\vphantom{$\left(\nuclide[6]{Li}\right)$}$a_v$ [\si{\femto\metre}]	&  0.6500 & 0.6500 &  0.7000 &  0.7000	\\
			\midrule
\vphantom{$\left(\nuclide[6]{Li}\right)$}$V_o$ [\si{\MeV}]			&  0      & 2.450  & 10.00    & 10.00	\\
\vphantom{$\left(\nuclide[6]{Li}\right)$}$R_o$ [\si{\femto\metre}]	& {--}    & 1.900  &  1.500  &  1.594	\\
\vphantom{$\left(\nuclide[6]{Li}\right)$}$a_o$ [\si{\femto\metre}]	& {--}    & 0.6500 &  0.3500 &  0.3500
		\end{tabular}
	\end{subtable}
	\end{table}

\printbibliography[heading=bibintoc]%
\end{document}